\documentclass[11pt]{article}

\usepackage{latexsym}
\usepackage{amsmath}
\usepackage{amssymb}

\newtheorem{theorem}{Theorem}

\newtheorem{lemma}{Lemma}
\newtheorem{corollary}{Corollary}
\newtheorem{definition}{Definition}
\newtheorem{conjecture}{Conjecture}

\newtheorem{example}{Example}

\title{Characterization of the Gittins index \\ 
for sequential multistage jobs}

\author{Samuli Aalto \\ 
Department of Communications and Networking \\
Aalto University, Finland
}

\begin{document}
	
\date{}
	
\maketitle

\begin{abstract}
The optimal scheduling problem in single-server queueing systems is a classic 
problem in queueing theory. The Gittins index policy is known to be the 
optimal preemptive nonanticipating policy (both for the open version of the 
problem with Poisson arrivals and the closed version without arrivals) 
minimizing the expected holding costs \cite{Git89,Git11}. While the Gittins 
index is thoroughly characterized for ordinary jobs whose state is described 
by the attained service, it is not at all the case with jobs that have more 
complex structure. 
Recently, a class of such jobs, the multistage jobs, were introduced, and 
it was shown that the computation of Gittins index of a multistage job reduces 
into separable computations for the individual stages \cite{Scu18arXiv}. 
The characterization is, however, indirect in the sense that it relies 
on the recursion for an auxiliary function (so called SJP function) and not 
for the Gittins index itself. In this paper, we answer the natural question: 
Is it possible to compute the Gittins index for a multistage job more directly 
by recursively combining the Gittins indexes of its individual stages? 
According to our results, it seems to be possible, at least, for sequential 
multistage jobs that have a fixed (deterministic) sequence of stages. We prove 
this for sequential two-stage jobs that have monotonous hazard rates in both 
stages, but our numerical experiments give an indication that the result could 
possibly be generalized to any sequential multistage jobs. Our approach, in this 
paper, is based on the Whittle index originally developed in the context of 
restless bandits \cite{Whi88JAP}.
\end{abstract}

\newpage

\section{Introduction}
\label{sec:intro}

Consider the following optimal scheduling problem related to the {M/G/1} 
queueing model. There is a single-server queue with $K$ job classes. 
For each class, new jobs arrive according to independent Poisson processes 
and service times are independent and identically distributed with 
a class-wise service time distribution with a finite mean. Assume that 
the system is stable for any work-conserving scheduling discipline 
(i.e., the traffic load is strictly less than 1), and let $N^\pi_k$ 
denote the steady-state number of jobs in class $k$ when scheduling 
policy $\pi$ is applied. Let $\Pi$ refer to the family of all 
{\em nonanticipating}\footnote{
A nonanticipating scheduling policy does not have any knowledge of 
the remaining service times of jobs.}
scheduling policies that allow preemption. The aim is to find the optimal 
policy $\pi^* \in \Pi$ that minimizes the expected {\em average} holding 
costs per time unit, 
\begin{equation}
\sum_{k = 1}^K h_k E[N^{\pi^*}_k] 
= \min_{\pi \in \Pi} 
\sum_{k = 1}^K h_k E[N^\pi_k], 
\label{eq:optimal-scheduling-problem-open}
\end{equation}
where $h_k$ is the weight of class $k$. If $h_k = 1$ for all $k$, then the 
problem is, by Little's result, equivalent with the minimization of the mean 
delay (a.k.a. sojourn time or response time).

What we described above is the {\em open} version of the problem. In the 
{\em closed} version, there is a single-server queue with $K$ jobs in the 
beginning and no new arrivals are allowed. The aim is to find the optimal 
policy $\pi^* \in \Pi$ that minimizes the expected {\em total} holding costs, 
\begin{equation}
\sum_{k = 1}^K h_k E[C^{\pi^*}_k] 
= \min_{\pi \in \Pi} 
\sum_{k = 1}^K h_k E[C^{\pi}_k], 
\label{eq:optimal-scheduling-problem-closed}
\end{equation}
where $C^\pi_k(t)$ is the service completion time (a.k.a.\ flow time) 
of job $k$.

Among the nonanticipating policies, 
the optimal scheduling policy (both for the open version and the closed 
version of the problem) is known to be the {\em Gittins index policy} 
\cite{Git89,Git11,Aal09QS,Aal11PEIS}. J.C.~Gittins derived this result as 
a by-product of his ground-breaking results on the 
{\em multi-armed bandit}\footnote{
Multi-armed bandit problem refers to a system with a finite number of bandits. 
At every time slot, the gambler chooses to activate one of the bandits, which 
will then yield a reward and undergo a Markovian state transition, while all 
the other bandits are passive, i.e., their states remain frozen. The aim is 
to find the optimal activating policy that maximizes the expected profit.}
problems \cite{Git89,Git11}. These problems remained unsolved for long, until 
Gittins proved that the structure of the optimal policy is surprisingly simple 
based on an index associated to each bandit separately depending just the state 
of the bandit \cite{Whi02OpnsRes}.

In our scheduling problem, bandits are represented by jobs. The state of a job 
is described by its {\em attained service}\footnote{
Attained service (a.k.a.\ age) is the amount of service that the job has 
already received.}
$a$, which increases at rate $1$ whenever served, until the service is completed. 
As presented in 
\cite{Git89,Git11,Aal09QS,Aal11PEIS}, the corresponding {\em Gittins index} 
$G(a)$ (both for the open version and the closed version of the scheduling 
problem) can be characterized as follows: 
\begin{equation}
G(a) = 
\sup_{\Delta \ge 0} 
\frac{P\{S - a \le \Delta \mid S > a\}}{E[\min \{S - a, \Delta\} \mid S > a]}, 
\label{eq:Aalto-Gittins-index-single-stage}
\end{equation}
where $S$ denotes the (original) service time of the job.

An alternative method to compute the Gittins index has recently 
been presented by Scully et al.\ in \cite{Scu17PER,Scu18arXiv} 
related to their {\em SJP (single-job profit) framework}: 
\begin{equation}
G(a) = 
\frac{1}{\inf \{r \ge 0 : V^{\mathrm{SJP}}(r;a) > 0\}}, 
\label{eq:Scully-Gittins-index-single-stage}
\end{equation}
where $V^{\mathrm{SJP}}(r;a)$ denotes the so-called SJP function defined by 
\begin{equation}
V^{\mathrm{SJP}}(r;a) = 
\sup_{\Delta \ge 0} 
\Big( 
r P\{S - a \le \Delta \mid S > a\} - E[\min \{S - a, \Delta\} \mid S > a] 
\Big).
\label{eq:Scully-V-function-single-stage}
\end{equation}

Note that Equations~(\ref{eq:Aalto-Gittins-index-single-stage}) 
and (\ref{eq:Scully-Gittins-index-single-stage}) refer to the minimization 
of the mean delay, i.e., $h_k = 1$ for all $k$. If this is not the case, 
the Gittins index is simply multiplied by the corresponding weight $h_k$, 
where $k$ denotes the class of the job.

\subsection{Multistage jobs}
\label{subsec:multistage-jobs}

While the Gittins index is nicely characterized for ordinary jobs whose state 
is described by the attained service, it is not at all the case with jobs that 
have more complex structure. A class of such jobs, the {\em multistage jobs}, 
were introduced and analyzed by Scully et al.\ in \cite{Scu18arXiv}. According 
to their definition, a multistage job progresses through a sequence of stages, 
each of which has its own stochastic service requirement, which is independent 
of everything else, and the job completes when its last stage completes. The 
sequence of stages may even be stochastic, and the server cannot influence 
the sequence of stages. It is natural to assume that a nonanticipating 
scheduler is aware of the current stage of such a job and the attained service 
in the current stage.

As noted by Scully et al.\ in \cite{Scu18arXiv}, it is possible to define 
the Gittins index of a multistage job, and the resulting Gittins policy 
minimizes the mean delay (or the expected average holding costs) among the 
nonanticipating scheduling policies, but computing the Gittins index 
requires solving a multidimensional optimization problem, which seems to 
make the computation of the Gittins indexes intractable. However, they 
managed to tame this complexity by utilizing the SJP framework mentioned 
above. More specifically, they stated and proved a {\em composition law}, 
which reduces the computation of Gittins index of a multistage job into 
separable SJP computations for the individual stages.

As an example, consider a {\em sequential}\footnote{
In this paper, multistage jobs with a fixed (i.e., deterministic) sequence 
of stages are briefly called sequential multistage jobs.}
two-stage job (called AB) that consists of stages A and B (in this order). 
According to \cite{Scu18arXiv}, the Gittins index $G_{\mathrm{AB}}(j,a)$ 
of job AB in state $(j,a)$, where $j \in \{\mathrm{A},\mathrm{B}\}$ and 
$a \ge 0$ denotes the attained service in the current stage, satisfies 
\begin{equation}
G_{\mathrm{AB}}(j,a) = 
\frac{1}{\inf \{r \ge 0 :V_{\mathrm{AB}}^{\mathrm{SJP}}(r;j,a) > 0\}}, 
\label{eq:Scully-Gittins-index-two-stage}
\end{equation}
where $V_{\mathrm{AB}}^{\mathrm{SJP}}(r;j,a)$ denotes the SJP function of 
job AB in state $(j,a)$. If the current stage is B, then clearly 
\begin{equation}
V_{\mathrm{AB}}^{\mathrm{SJP}}(r;B,a) = 
V_{\mathrm{B}}^{\mathrm{SJP}}(r;a), 
\label{eq:Scully-V-function-stage-B}
\end{equation}
where $V_{\mathrm{B}}^{\mathrm{SJP}}(r;a)$ refers to the SJP function of the 
single-stage job that consists only of stage B. But if the current stage 
is A, then the new composition law can be applied according to which 
\begin{equation}
V_{\mathrm{AB}}^{\mathrm{SJP}}(r;A,a) = 
V_{\mathrm{A}}^{\mathrm{SJP}}(V_{\mathrm{B}}^{\mathrm{SJP}}(r;0);a), 
\label{eq:Scully-V-function-stage-A}
\end{equation}
where $V_{\mathrm{A}}^{\mathrm{SJP}}(r;a)$ refers to the SJP function of the 
single-stage job that consists only of stage A. Thus, to compute the Gittins 
index, it suffices to determine the SJP functions of the two stages separately.

\subsection{Our contribution and the structure of the paper}
\label{subsec:contribution}

While 
Equations~(\ref{eq:Scully-Gittins-index-two-stage})-(\ref{eq:Scully-V-function-stage-A}) 
characterize the Gittins index for a multistage job indirectly via the 
SJP function, it is natural to ask whether the characterization is possible 
to do {\em more directly}: 
\begin{itemize}
\item 
Is it possible to compute the Gittins index for a multistage job by 
recursively combining the Gittins indexes of its individual stages?
\end{itemize}
Scully et al.\ \cite{Scu18arXiv} mention that there is no known way to 
do this, and they also seem to believe that it is not possible at all. 
In this paper, we, however, reveal a way to do it, at least in some cases. 
For the sequential two-stage jobs (such as job AB above) that have 
{\em monotonous hazard rates} in both stages, we prove that 
\begin{equation}
\begin{split}
& 
G_{\mathrm{AB}}(\mathrm{B},a) = 
G_\mathrm{B}(a), \\
& 
G_{\mathrm{AB}}(\mathrm{A},a) = 
\sup_{\Delta \ge 0} 
\left( 
\frac{P\{S_\mathrm{B} \le \Delta\}}
{\frac{1}{G_\mathrm{A}(a)} + E[\min\{S_\mathrm{B},\Delta\}]} 
\right), 
\end{split}
\label{eq:Aalto-Gittins-index-two-stage}
\end{equation}
where $S_\mathrm{B}$ refers to the (original) service requirement in stage~B 
and $G_j(a)$ to the Gittins index of the single-stage job that consists 
only of stage $j \in \{\mathrm{A},\mathrm{B}\}$ and has attained service 
an amount of $a$.\footnote{
Equations~(\ref{eq:Scully-Gittins-index-two-stage}) and 
(\ref{eq:Aalto-Gittins-index-two-stage}) refer to the minimization 
of the mean delay, i.e., $h_k = 1$ for all $k$. If this is not the case, 
the Gittins index is simply multiplied by the corresponding weight $h_k$, 
where $k$ denotes the class of the job, see 
Theorem~\ref{thm:Gittins-index-average} in Section~\ref{sec:average-cost}.}

So, Equation~(\ref{eq:Aalto-Gittins-index-two-stage}) gives a {\em recursive} 
way to compute the Gittins index for a sequential two-stage job when we know the 
Gittins indexes $G_j(a)$ separately for each individual stage~$j$. Moreover, our 
numerical experiments give an indication that this result could possibly be 
generalized to any sequential two-stage jobs, and even to the sequential 
multistage jobs with more than two stages.

Our approach to derive the Gittins index for multistage jobs is based on 
the {\em Whittle index} originally developed by P.~Whittle \cite{Whi88JAP} 
in the context of {\em restless bandits}.\footnote{
A restless bandit is a generalization of a multi-armed bandit such that 
the bandits continue to change state even when they are not being chosen, 
i.e., their states are no longer frozen, which makes the related optimization 
problem even more complicated.}
The point here is that, for multi-armed bandit problems, the resulting 
Whittle index coincides with the Gittins index \cite{Whi88JAP,Git11}.

We start with the closed version of a {\em discrete-time} single-server 
scheduling problem, where there are $K$ {\em sequential two-stage} jobs in 
the beginning and no new arrivals are allowed. The aim is to complete the 
service of these $K$ jobs with minimal expected {\em discounted} holding 
costs. The problem is formulated in Section~\ref{sec:original-problem}. It can 
be considered as a constrained optimization problem, since at most one job can 
be chosen in service in each time slot.

Next, in Section~\ref{sec:relaxed-problem}, we apply the Whittle index 
approach to make the problem {\em separable}. The idea is that the 
constrained optimization problem is relaxed by allowing for the server to 
serve any number of jobs in one time slot and only requiring that at most 
one job is served per time slot on average. This makes the problem much more 
tractable by decomposing it to separate subproblems per each job.

In Sections~\ref{sec:DHR-DHR}-\ref{sec:IHR-DHR}, we solve the subproblem 
related to a single sequential two-stage job and derive the related 
discrete-time Whittle indexes in the four cases where the single-stage hazard 
rates are {\em monotonous}: 
\begin{itemize}
\item
{\em DHR-DHR}: \\
Decreasing hazard rate in both stages.
\vskip 6pt
\item
{\em IHR-IHR}: \\
Increasing hazard rate in both stages.
\vskip 6pt
\item
{\em DHR-IHR}: \\
Decreasing hazard rate in the first stage and 
increasing hazard rate in the second stage.
\vskip 6pt
\item
{\em IHR-DHR}: \\
Increasing hazard rate in the first stage and 
decreasing hazard rate in the second stage.
\end{itemize}

In Section~\ref{sec:special-cases}, we give the Whittle indexes for the 
special cases of the sequential two-stage jobs where the service time 
distribution in one of stages is {\em geometric} with a constant hazard 
rate and monotonous in the other stage. All the results are direct 
corollaries of the results given in Sections~\ref{sec:DHR-DHR}-\ref{sec:IHR-DHR}.

In Section~\ref{sec:average-cost}, we finally move from discounted to 
{\em undiscounted} costs (either average or total costs depending on whether 
we talk about the open or closed version of the problem, respectively). 
In addition, we move from the discrete time setup to single server 
scheduling problems in {\em continuous time}. Based on above mentioned 
Whittle index results, we derive the recursive equation 
(\ref{eq:Aalto-Gittins-index-two-stage}) that allows us to compute the Gittins 
index for a sequential two-stage job with monotonous hazard rates in both 
stages when we know the Gittins indexes separately for each individual stage. 
Section~\ref{sec:conclusions} summarizes the paper and discusses briefly for 
possible further work.

The proofs of all the main theorems are given in 
Appendixes~\ref{app:DHR-DHR-A-proof}-\ref{app:average-continuous-proof}. 
In Appendix~\ref{app:whittle-index-examples-discounted-costs}, we give 
numerical examples on the Whittle index for discounted costs in various 
cases of a sequential two-stage job with monotonous hazard rates in both 
stages, and in Appendix~\ref{app:whittle-index-examples-average-multistage}, 
examples on the Whittle index for sequential multistage jobs when 
the hazard rate in a stage is nonmonotonous and/or there are more than 
two stages.

\subsection{Related work}
\label{subsec:related-work}

The optimal scheduling problem in single-server queueing systems is a classic 
problem in queueing theory. The optimal anticipating scheduling policy, 
Shortest-Remaining-Processing-Time (SRPT), minimizes the number of jobs even 
sample-path-wise for general service times and arrival processes \cite{Sch68,Smi78}. 
As already mentioned, the optimal nonanticipating policy with respect to 
the mean delay (or expected holding costs) is known to be the Gittins index 
policy \cite{Git89,Git11,Aal09QS,Aal11PEIS}. For exponential service times, 
the Gittins index policy reduces to the well-known $c \mu$-rule, which is 
also the optimal nonpreemptive policy for general service time distributions 
\cite{Cox61}. If there is just one class of jobs and the target is to minimize 
the mean delay, then the Gittins index policy can be characterized as follows 
\cite{Aal09QS,Aal11PEIS}: For the service time distributions that belong to 
the New Better than Used in Expectation (NBUE) class of distributions (and 
only for those ones), the Gittins index policy coincides with any nonpreemptive 
scheduling policy (including the well-known First-Come-First-Served (FCFS) discipline), whereas for the service time distributions that belong to 
the Decreasing Hazard Rate (DHR) class of distributions (and only for those 
ones), it is equivalent with the Foreground–Background (FB) policy, 
in which the job with the least attained service is scheduled. These 
optimality results of FCFS and FB do not even require the Poisson arrivals 
assumption \cite{Rig90JAP,Yas87QS,Rig89PEIS}. However, for 
multistage jobs, the Gittins index is studied only in \cite{Scu18arXiv}, 
as far as we know.

As also already mentioned, the Whittle index approach was originally developed 
in the context of restless bandits \cite{Whi88JAP}. It has successfully been 
applied, e.g., in optimal scheduling problems in wireless systems, where the 
channel state of all users typically varies randomly independent of the 
scheduling decisions, see, e.g., 
\cite{Aye10PEVA,Jac11PEVA,Tab14PEVA,Cec16PEVA,Aal16QS,Aal17PEVA,Ana18POMACS,Aal19bPEVA}. 
Another application area concerns dispatching (a.k.a.\ task assignment) 
problems in parallel queuing systems, where the state of the parallel queues 
is modified not only by the assigned jobs but also the completed ones, see, e.g., 
\cite{Nin02MP,Arg09PEIS,Lar16TNET,Aal19aPEVA}. The application of the Whittle 
index approach is even more challenging when the state space of a bandit is 
multidimensional (as in our scheduling problem in this paper\footnote{
The state space of a multistage job is two-dimensional with one element indicating 
the stage and the other one the attained service in the current stage.})
since there is no natural {\em order} of indexes available but it has to be 
discovered while solving the problem. In 
\cite{Aal16QS,Aal17PEVA,Aal19aPEVA,Aal19bPEVA}, we have, however, managed to 
derive the Whittle index for some problems where the state space of the bandit 
is multidimensional.

\section{Optimal scheduling problem in discrete time with discounted costs}
\label{sec:original-problem}

We consider the following closed version of the single server optimal scheduling 
problem in discrete time. At time $0$ (i.e., in the beginning of the 
first time slot), there are $K$ sequential two-stage jobs. Thus, each job consists 
of two consecutive {\em stages} such that the service of stage $j = 1$ should be 
completed before the service of stage $j = 2$ can be initiated. Let $S_{k,j}$ 
denote the random service time of stage $j$ of job $k$ taking values in 
$\{1,2,\ldots\}$. Let $\mu_{k,j}(n)$, $n \in \{0,1,\ldots\}$, denote the 
corresponding {\em discrete hazard rate} \cite{Sha95COR}, i.e., the conditional 
probability that the service time of stage $j$ of job $k$ is equal to $n + 1$, 
given that it is strictly greater than $n$, 
\begin{equation}
\mu_{k,j}(n) = P\{ S_{k,j} = n + 1 \mid S_{k,j} \ge n + 1 \}.
\label{eq:hazard-rate}
\end{equation}
In addition, when $\lim_{n \to \infty} \mu_{k,j}(n)$ exists, 
we denote this limit by 
\begin{equation}
\mu_{k,j}(\infty) = \lim_{n \to \infty} \mu_{k,j}(n).
\label{eq:hazard-rate-limit}
\end{equation}

In this paper, we are, in particular, interested in the sequential two-stage 
jobs with {\em monotonous} hazard rates in both stages. We say that the 
service time distribution in stage $j$ of job $k$ belongs to the class of 
\begin{itemize}
\item[(i)]
{\em Decreasing Hazard Rate (DHR) distributions} 
if the hazard rate is decreasing, i.e., 
\begin{equation}
\mu_{k,j}(n) \ge \mu_{k,j}(n+1) 
\; \hbox{for all $n \in \{0,1,\ldots\}$}; 
\label{eq:DHR}
\end{equation}
\item[(ii)]
{\em Increasing Hazard Rate (IHR) distributions} 
if the hazard rate is increasing, i.e., 
\begin{equation}
\mu_{k,j}(n) \le \mu_{k,j}(n+1) 
\; \hbox{for all $n \in \{0,1,\ldots\}$}; 
\label{eq:IHR}
\end{equation}
\item[(iii)]
{\em Constant Hazard Rate (CHR) distributions}\footnote{
CHR distributions are also known as {\em geometric (GEO) distributions}.}
if the hazard rate is constant, i.e., 
\begin{equation}
\mu_{k,j}(n) = \mu_{k,j}(n+1) 
\; \hbox{for all $n \in \{0,1,\ldots\}$}.
\label{eq:CHR}
\end{equation}
\end{itemize}
Note that the limit $\mu_{k,j}(\infty)$ is well-defined according to 
(\ref{eq:hazard-rate-limit}) whenever the hazard rate $\mu_{k,j}(n)$ 
in stage $j$ is monotonous.

Jobs are served according to a nonanticipating scheduling discipline $\pi$ 
that allows preemptions. Let $\Pi$ denote the family of such disciplines. 
For any time instant $t \in \{0,1,\ldots\}$, the scheduler chooses at most 
one of the jobs for service (during that time slot). Denote $A^\pi_k(t) = 1$ 
if job~$k$ is chosen at time instant~$t$, where $\pi$ refers to the scheduling 
policy used; otherwise $A^\pi_k(t) = 0$. Thus, for any policy $\pi$ and time 
instant~$t$, we have the constraint 
\begin{equation}
\sum_{k = 1}^K A^\pi_k(t) \le 1.
\label{eq:capacity-constraint}
\end{equation}
From the scheduler point of view, the state of job $k$ is described by the 
pair $(j,n)$, where $j$ refers to the current stage and $n$ to the amount of 
attained service in the current stage $j$. Thus, at time instant $0$, 
the state of each job is equal to $(1,0)$. If the service of job $k$ has 
already been completed, its state is marked by symbol $*$. Let $X^\pi_k(t)$ 
denote the state of job $k$ at time instant $t$ taking values in 
\[
{\mathcal S} = \big\{ (j,n), j \in \{1,2\}, n \in \{0,1,\ldots\} \big\} 
\cup \{*\}.
\]
For any job $k$, holding costs are accumulated at rate $h_k > 0$ until 
the whole job is completed. The costs are discounted with factor 
$\beta \in (0,1)$. The objective function in our scheduling problem is, 
thus, given by 
\begin{equation}
E\left[ \sum_{t = 0}^\infty \sum_{k = 1}^K \beta^t h_k 1_{\{X^\pi_k(t) \ne *\}} \right].
\label{eq:original-expected-discounted-holding-costs}
\end{equation}
The aim is to find the optimal scheduling policy $\pi$ that minimizes the 
expected discounted holding costs 
(\ref{eq:original-expected-discounted-holding-costs}) subject to the strict 
capacity constraint (\ref{eq:capacity-constraint}) for all $t$ and assuming 
that the scheduling decisions at each time slot~$t$ are based on the states 
$X^\pi_k(t)$ of jobs.

\section{Whittle index approach to the scheduling problem}
\label{sec:relaxed-problem}

The optimal scheduling problem described in the previous section belongs 
to the class of multi-armed bandit problems, for which the optimal 
policy is described by the Gittins index \cite{Git89,Git11}. One option to 
determine the Gittins index is to follow Whittle's approach \cite{Whi88JAP} 
developed for restless bandit problems, which is a strictly wider class 
of problems, since it is known that, for multi-armed bandit problems, the 
resulting Whittle index coincides with the Gittins index~\cite{Whi88JAP,Git11}.

According to Whittle's approach \cite{Whi88JAP}, the original problem is 
modified by replacing the strict capacity constraint (\ref{eq:capacity-constraint}) 
by an averaged one and handling the relaxed problem by Lagrangian methods. 
In this paper, we use the same approach, which results in the following 
separate subproblems for each job~$k$: Find the optimal policy $\pi$ that 
minimizes the objective function 
\begin{equation}
f_{k,\beta}^{\pi} + \nu g_{k,\beta}^{\pi}, 
\label{eq:separable-discounted-costs}
\end{equation}
where $\nu$ can be interpreted as the unit price of work, $f^\pi_{k,\beta}$ 
as the expected discounted holding costs of job~$k$, and $g^\pi_{k,\beta}$ as 
the expected discounted amount of work needed for job~$k$, 
\[
f^\pi_{k,\beta} = E\left[ \sum_{t = 0}^\infty \beta^t h_k 1_{\{X^\pi_k(t) \ne *\}} \right], \quad 
g^\pi_{k,\beta} = E\left[ \sum_{t = 0}^\infty \beta^t A^\pi_k(t) \right].
\]

The separable subproblems of the Lagrangian version of the relaxed scheduling 
problem are now considered in the context of Markov decision processes. The 
possible actions $a \in {\mathcal A} = \{0,1\}$ are ``to schedule'' ($a = 1$) 
and ``not to schedule'' ($a = 0$).

Let $q_k(y|x,a) \ge 0$ denote the transition probability from state 
$x \in {\mathcal S}$ to state $y \in {\mathcal S}$ after action 
$a \in {\mathcal A}$. It follows from the previous discussion that 
the non-zero transition probabilities are as follows: 
\begin{equation}
\begin{array}{ll}
q_k(x|x,0) = \; 1, 
& x \in {\mathcal S}, \\
q_k((2,0)|(1,n),1) = \; \mu_{k,1}(n), 
& n \in \{0,1,\ldots\}, \\
q_k((1,n+1)|(1,n),1) = \; 1 - \mu_{k,1}(n), 
& n \in \{0,1,\ldots\}, \\
q_k(*|(2,n),1) = \; \mu_{k,2}(n), 
& n \in \{0,1,\ldots\}, \\
q_k((2,n+1)|(2,n),1) = \; 1 - \mu_{k,2}(n), 
& n \in \{0,1,\ldots\}, \\
q_k(*|*,1).
& 
\end{array}
\label{eq:transition-probabilities}
\end{equation}
Note that $*$ is an absorbing state for any policy.

Finally, let $c_k(x,a)$ denote the immediate cost in state~$x \in {\mathcal S}$ 
after action~$a \in {\mathcal A}$. In our model, 
\begin{equation}
\begin{array}{ll}
c_k(x,a) = h_k + a \nu, & x \in {\mathcal S} \setminus \{*\}, \\
c_k(*,a) = a \nu.       & 
\end{array}
\label{eq:immediate-costs}
\end{equation}
Thus, the state space~${\mathcal S}$ is discrete, the action 
space~${\mathcal A}$ is finite, and the immediate costs are bounded. 
It follows that the optimal policy belongs to the class of stationary 
policies \cite[Thm.~6.3]{Ros70}. For each stationary policy $\pi$, 
the scheduling decisions are deterministic depending just on the 
current state of job~$k$, 
\[
A_k^{\pi}(t) = 
\left\{ 
\begin{array}{ll}
1, & \quad \hbox{if $X_k^{\pi}(t) \in {\mathcal B}^\pi$}, \\
0, & \quad \hbox{otherwise},
\end{array}
\right.
\]
where ${\mathcal B}^\pi \subset {\mathcal S}$ is the {\em activity set} of 
policy $\pi$.

Let $V_{k,\beta}(x;\nu)$ denote the {\em value function} for state 
$x \in {\mathcal S}$ related to the minimization of the expected discounted 
costs (\ref{eq:separable-discounted-costs}) with Lagrangian parameter~$\nu$. 
The corresponding {\em optimality equations} \cite[Thm.~6.1]{Ros70} read as 
follows: 
\begin{equation}
\begin{split}
& 
V_{k,\beta}(1,n;\nu) = 
h_k + \min \big\{ \beta V_{k,\beta}(1,n;\nu), \\
& \quad 
\nu + \beta \mu_{k,1}(n) V_{k,\beta}(2,0;\nu) + 
\beta (1 - \mu_{k,1}(n)) V_{k,\beta}(1,n+1;\nu) \big\}, \\
& 
V_{k,\beta}(2,n;\nu) = 
h_k + \min \big\{ \beta V_{k,\beta}(2,n;\nu), \\
& \quad 
\nu + \beta \mu_{k,2}(n) V_{k,\beta}(*;\nu) + 
\beta (1 - \mu_{k,2}(n)) V_{k,\beta}(2,n+1;\nu) \big\}, \\
& 
V_{k,\beta}(*;\nu) = 
\min\{ 0, \nu \} + \beta V_{k,\beta}(*;\nu).
\end{split}
\label{eq:opt-eqs-discounted-general}
\end{equation}
In addition, let $V_{k,\beta}^{\pi}(x;\nu)$ denote the corresponding 
value function for policy~$\pi$. The policy $\pi_k^*$ for which 
\[
V_{k,\beta}^{\pi_k^*}(x;\nu) = V_{k,\beta}(x;\nu)
\]
for all $x \in {\mathcal S}$ is said to be {\em $(\nu,\beta)$-optimal} 
for job~$k$.

From optimality equations (\ref{eq:opt-eqs-discounted-general}), we see 
that if $\nu \ge 0$, then the minimum expected discounted cost 
$V_{k,\beta}(*;\nu)$ for the absorbing state $*$ clearly equals $0$ and 
is achieved by the policies $\pi$ that choose action $0$ in state $*$ 
(or even action $1$ if $\nu = 0$). On the other hand, if $\nu < 0$, 
the minimum expected discounted cost $V_{k,\beta}(*;\nu)$ for state 
$*$ equals $\nu/(1 - \beta)$ and is achieved by those policies $\pi$ that 
choose action $1$ in state $*$.

Let us conclude this section by defining the {\em indexability} property, 
which is not automatically guaranteed for genuine restless bandit problems 
\cite{Whi88JAP}. However, for the subclass of multi-armed bandit problems, 
which our problem belongs to, the relaxed optimization problem is indexable 
for sure.

\begin{definition}
\label{def:indexability}
The relaxed optimization problem (\ref{eq:separable-discounted-costs}) 
related to job~$k$ is {\em indexable} if, for any belief state 
$x \in {\mathcal S} \setminus \{*\}$, there exists 
$W_{\beta,k}(x) \in [-\infty,\infty]$ such that 
\begin{itemize}
\item[(i)]
decision $a = 1$ (to schedule job~$k$) is optimal in belief state~$x$ 
if and only if $\nu \le W_{\beta,k}(x)$; 
\item[(ii)]
decision $a = 0$ (not to schedule job~$k$) is optimal in belief state~$x$ 
if and only if $\nu \ge W_{\beta,k}(x)$.
\end{itemize}
If the problem is indexable, the corresponding index $W_{\beta,k}(x)$ is 
called the {\em Whittle index}.
\end{definition}

Note that, according to this definition, the two actions are equally good 
(and, thus, optimal) in state $x$ if and only if $\nu = W_{\beta,k}(x)$.

In Sections~\ref{sec:DHR-DHR}-\ref{sec:IHR-DHR} below, we derive the 
Whittle index values $W_{\beta,k}(x)$ of a sequential two-stage job 
for all states $x \in {\mathcal S}$ in the following four cases that are 
related to the monotonicity properties of the service time distribution 
in the two stages: 
\begin{itemize}
\item
{\em DHR-DHR}: \\
Decreasing hazard rate in both stages.
\vskip 6pt
\item
{\em IHR-IHR}: \\
Increasing hazard rate in both stages.
\vskip 6pt
\item
{\em DHR-IHR}: \\
Decreasing hazard rate in the first stage and 
increasing hazard rate in the second stage.
\vskip 6pt
\item
{\em IHR-DHR}: \\
Increasing hazard rate in the first stage and 
decreasing hazard rate in the second stage.
\end{itemize}
Moreover, in Section~\ref{sec:special-cases}, we give the Whittle indexes 
for the special cases of the sequential two-stage jobs where the service 
time distribution in one of stages is geometric and monotonous in the other 
stage. All the results are direct corollaries of the results given in 
Sections~\ref{sec:DHR-DHR}-\ref{sec:IHR-DHR}. 

In these sections, we are all the time considering a single job, say job~$k$, 
and, thus, leave out the related subscript $k$ to lighten the notation. 
In addition, we use the following shorthand notation for the (conditional) 
probabilities for the service time in stage $j \in \{1,2\}$: 
\begin{equation}
\begin{split}
& 
p_j(i) = P\{S_j = i + 1\}, \quad 
p_j(i|n) = P\{S_j = n + i + 1 \mid S_j \ge n + 1\}, \\
& 
\bar p_j(i) = P\{S_j \ge i + 1\}, \quad 
\bar p_j(i|n) = P\{S_j \ge n + i + 1 \mid S_j \ge n + 1\}.
\end{split}
\label{eq:pjin-shorthand-notation}
\end{equation}
Note also that 
\[
\begin{split}
& 
\bar p_j(i) \mu_j(i) = p_j(i) = \bar p_j(i) - \bar p_j(i+1), \\
& 
\bar p_j(i|n) \mu_j(n+i) = p_j(i|n) = \bar p_j(i|n) - \bar p_j(i+1|n).
\end{split}
\]

\section{Whittle index for the DHR-DHR case}
\label{sec:DHR-DHR}

In this section, we assume that both stages of job~$k$ belong to class DHR, 
which is the DHR-DHR case defined in Section~\ref{sec:relaxed-problem}. 
Under this assumption, we derive the Whittle index values $W_{\beta,k}(x)$ 
for any state $x$ by solving the relaxed optimization problem 
(\ref{eq:separable-discounted-costs}) for any $\nu$. Before the main result 
given in Theorem~\ref{thm:Whittle-index-discounted-DHR-DHR}, we present some 
auxiliary lemmas that are needed in the proof of the main result.

\begin{lemma}
\label{lem:Whittle-index-discounted-DHR-DHR-lemma-1}
Assume the DHR-DHR case. 
Let us define the following functions: 
\begin{equation}
w_2(n_2) = h \mu_2(n_2) \, \frac{\beta}{1 - \beta}, 
\quad n_2 \in \{0,1,\ldots\}, 
\label{eq:w2-discounted-DHR-DHR}
\end{equation}
and 
\begin{equation}
\begin{split}
& 
\psi(n_1,n_2) = 
h \mu_1(n_1) \, 
\frac{\beta \sum_{i=0}^{n_2} \beta^i p_2(i)}{1 + \beta \mu_1(n_1) \sum_{i=0}^{n_2} \beta^i \bar p_2(i)} \, 
\frac{\beta}{1 - \beta}, \\
& \quad 
n_1, n_2 \in \{0,1,\ldots\}.
\end{split}
\label{eq:psi-discounted-DHR-DHR}
\end{equation}
These functions have the following properties: 
\begin{itemize}
\item[(i)]
Function $w_2(n_2)$ is decreasing with respect to $n_2$ converging to 
\begin{equation}
w_2(\infty) = \lim_{n_2 \to \infty} w_2(n_2) = 
h \mu_2(\infty) \, 
\frac{\beta}{1 - \beta}; 
\label{eq:w2-infty-discounted-DHR-DHR}
\end{equation}
\item[(ii)]
Function $\psi(n_1,n_2)$ is decreasing with respect to $n_1$ converging to 
\begin{equation}
\begin{split}
& 
\psi(\infty,n_2) = \lim_{n_1 \to \infty} \psi(n_1,n_2) = \\
& \quad 
h \mu_1(\infty) \, 
\frac{\beta \sum_{i=0}^{n_2} \beta^i p_2(i)}{1 + \beta \mu_1(\infty) \sum_{i=0}^{n_2} \beta^i \bar p_2(i)} \, 
\frac{\beta}{1 - \beta}; 
\end{split}
\label{eq:psi-infty-n2-discounted-DHR-DHR}
\end{equation}
\item[(iii)]
$\psi(0,0) < w_2(0)$; 
\item[(iv)]
$\psi(n_1,n_2) \le w_2(n_2+1)$ if and only if $\psi(n_1,n_2+1) \le w_2(n_2+1)$; 
\item[(v)]
For any $n_2 \in \{0,1,\ldots\}$ and $n \in \{0,1,\ldots,n_2\}$, 
\begin{equation}
w_2(n_2) \le 
h \, 
\frac{\sum_{i = 0}^{n_2-n} \beta^i p_2(i|n)}
{\sum_{i = 0}^{n_2-n} \beta^i \bar p_2(i|n)} \, 
\frac{\beta}{1 - \beta}.
\label{eq:w2-upper-bound-DHR-DHR}
\end{equation}
\end{itemize}
\end{lemma}

\paragraph{Proof} 
{\em (i)} 
This follows immediately from the monotonicity of $\mu_2(n_2)$.

\vskip 3pt
{\em (ii)} 
This follows immediately from the monotonicity of $\mu_1(n_1)$.

\vskip 3pt
{\em (iii)} 
This follows from the definitions of $\psi(0,0)$ and $w_2(0)$: 
\[
\psi(0,0) = 
h \mu_1(0) \, 
\frac{\beta \mu_2(0)}{1 + \beta \mu_1(0)} \, 
\frac{\beta}{1 - \beta} < 
h \mu_2(0) \, \frac{\beta}{1 - \beta} = 
w_2(0).
\]

\vskip 3pt
{\em (iv)} 
This follows from the following equivalencies: 
\[
\begin{split}
& 
\psi(\infty,n_2+1) \le 
w_2(n_2+1) \quad \Longleftrightarrow \\
& 
\beta \mu_1(n_1) \sum_{i=0}^{n_2+1} \beta^i \bar p_2(i) \mu_2(i) \\
& \quad \le \; 
\mu_2(n_2+1) \left( 1 + \beta \mu_1(n_1) \sum_{i=0}^{n_2+1} \beta^i \bar p_2(i) \right) 
\quad \Longleftrightarrow \\
& 
\beta \mu_1(n_1) \sum_{i=0}^{n_2} \beta^i \bar p_2(i) \mu_2(i) \\
& \quad \le \; 
\mu_2(n_2+1) \left( 1 + \beta \mu_1(n_1) \sum_{i=0}^{n_2} \beta^i \bar p_2(i) \right) 
\quad \Longleftrightarrow \\
& 
\psi(n_1,n_2) \le 
w_2(n_2+1).
\end{split}
\]

\vskip 3pt
{\em (v)} 
Since $p_2(i|n) = \bar p_2(i|n) \mu_2(n+i)$ and $\mu_2(n_2)$ is a decreasing 
function of $n_2$, we have, for any $n \in \{0,1,\ldots,n_2\}$, 
\[
\sum_{i = 0}^{n_2-n} \beta^i p_2(i|n) = 
\sum_{i = 0}^{n_2-n} \beta^i \bar p_2(i|n) \mu_2(n+i) \ge 
\sum_{i = 0}^{n_2-n} \beta^i \bar p_2(i|n) \mu_2(n_2), 
\]
from which (\ref{eq:w2-upper-bound-DHR-DHR}) clearly follows. 
This completes the proof of Lemma~\ref{lem:Whittle-index-discounted-DHR-DHR-lemma-1}.
\hfill $\Box$
\vskip 6pt

\begin{lemma}
\label{lem:Whittle-index-discounted-DHR-DHR-lemma-2}
Assume the DHR-DHR case. 
Let us define the following function: 
\begin{equation}
w_1(n_1) = \psi(n_1,\phi(n_1)), 
\quad n_1 \in \{0,1,\ldots\}, 
\label{eq:w1-discounted-DHR-DHR}
\end{equation}
where function $\psi(n_1,n_2)$ is defined in (\ref{eq:psi-discounted-DHR-DHR}) 
and function $\phi(n_1)$ as follows: 
\begin{equation}
\begin{split}
&
\phi(n_1) = 
\min\{ n_2 \in \{0,1,\ldots\} \cup \{\infty\} : \psi(n_1,n_2) > w_2(n_2+1) \}, \\
& \quad 
n_1 \in \{0,1,\ldots\}, 
\end{split}
\label{eq:phi-discounted-DHR-DHR}
\end{equation}
where function $w_2(n_2)$ is defined in (\ref{eq:w2-discounted-DHR-DHR}) 
and we interpret that $\phi(n_1) = \infty$ if $\psi(n_1,n_2) \le w_2(n_2+1)$ 
for all $n_2$, in which case we naturally define 
\begin{equation}
\begin{split}
& 
\psi(n_1,\infty) = 
\lim_{n_2 \to \infty} \psi(n_1,n_2) = \\
& \quad 
h \mu_1(n_1) \, 
\frac{\beta \sum_{i=0}^{\infty} \beta^i p_2(i)}
{1 + \beta \mu_1(n_1) \sum_{i=0}^{\infty} \beta^i \bar p_2(i)} \, 
\frac{\beta}{1 - \beta}.
\end{split}
\label{eq:psi-n1-infty-discounted-DHR-DHR}
\end{equation}
These functions have the following properties: 
\begin{itemize}
\item[(i)]
Function $\phi(n_1)$ is increasing with respect to $n_1$; 
\item[(ii)]
Function $w_1(n_1)$ is decreasing with respect to $n_1$; 
\item[(iii)]
If $\phi(n_1) < \infty$, then 
\begin{equation}
w_2(\phi(n_1)) \ge w_1(n_1) > w_2(\phi(n_1)+1); 
\label{eq:Whittle-index-order-DHR-DHR}
\end{equation}
otherwise $\phi(n_1) = \infty$ and 
\begin{equation}
w_2(\infty) \ge w_1(n_1). 
\label{eq:Whittle-index-order-infty-DHR-DHR}
\end{equation}
\end{itemize}
\end{lemma}

\paragraph{Proof} 
{\em (i)} 
Assume first that $\phi(n_1) < \infty$. 
Thus, $\psi(n_1,n) \le w_2(n+1)$ for all $n \in \{0,1,\ldots,\phi(n_1)-1\}$. 
Now it follows from 
Lemma~\ref{lem:Whittle-index-discounted-DHR-DHR-lemma-1}(ii) that 
$\psi(n_1+1,n) \le w_2(n+1)$ for all $n \in \{0,1,\ldots,\phi(n_1)-1\}$, which 
implies that $\phi(n_1+1) \ge \phi(n_1)$.

Assume now that $\phi(n_1) = \infty$. 
Thus, $\psi(n_1,n) \le w_2(n+1)$ for all $n \in \{0,1,\ldots\}$. 
Now it follows from 
Lemma~\ref{lem:Whittle-index-discounted-DHR-DHR-lemma-1}(ii) that 
$\psi(n_1+1,n) \le w_2(n+1)$ for all $n \in \{0,1,\ldots\}$, which 
implies that $\phi(n_1+1) = \infty = \phi(n_1)$.

\vskip 3pt
{\em (ii)} 
Let $n_2 = \phi(n_1)$ and $n'_2 = \phi(n_1+1)$. Now $n_2 \le n'_2$ by (i). 
First, if $n_2 = n'_2 \le \infty$, then 
$w_1(n_1+1) = \psi(n_1+1,n_2) \le \psi(n_1,n_2) = w_1(n_1)$ 
by Lemma~\ref{lem:Whittle-index-discounted-DHR-DHR-lemma-1}(ii). 
Secondly, if $n_2 < n'_2 < \infty$, then it follows from the definition of 
$w_1(n_1)$ that $\psi(n_1,n_2) > w_2(n_2+1)$ and from the definition of 
$w_1(n_1+1)$ that $\psi(n_1+1,n'_2-1) \le w_2(n'_2)$, which is equivalent 
with $\psi(n_1+1,n'_2) \le w_2(n'_2)$ by 
Lemma~\ref{lem:Whittle-index-discounted-DHR-DHR-lemma-1}(iv). 
Thus, we have 
\[
w_1(n_1+1) = \psi(n_1+1,n'_2) \le w_2(n'_2) \le 
w_2(n_2+1) < \psi(n_1,n_2) = w_1(n_1).
\]
Thirdly, if $n_2 < n'_2 = \infty$, then it follows from the definition of 
$w_1(n_1)$ that $\psi(n_1,n_2) > w_2(n_2+1)$ and from the definition of 
$w_1(n_1+1)$ that $\psi(n_1+1,n) \le w_2(n+1)$ for all $n$, which implies 
that $\psi(n_1+1,\infty) \le w_2(\infty)$. Thus, we have 
\[
w_1(n_1+1) = \psi(n_1+1,\infty) \le w_2(\infty) \le 
w_2(n_2+1) < \psi(n_1,n_2) = w_1(n_1).
\]

\vskip 3pt
{\em (iii)} 
Assume first that $\phi(n_1) < \infty$ and let $n_2 = \phi(n_1)$. 
Thus, $\psi(n_1,n_2-1) \le w_2(n_2)$, which is equivalent with 
$w_1(n_1) = \psi(n_1,n_2) \le w_2(n_2)$ by 
Lemma~\ref{lem:Whittle-index-discounted-DHR-DHR-lemma-1}(iv). 
On the other hand, $w_1(n_1) = \psi(n_1,n_2) > w_2(n_2+1)$. 
These results together justify (\ref{eq:Whittle-index-order-DHR-DHR}).

Assume now that $\phi(n_1) = \infty$. Thus, $\psi(n_1,n) \le w_2(n+1)$ 
for all $n$, which implies that $w_1(n_1) = \psi(n_1,\infty) \le 
w_2(\infty)$.
\hfill $\Box$
\vskip 6pt

Based on the functions $w_2(n_2)$ and $w_1(n_1)$ defined in the previous 
lemmas, we split the DHR-DHR case into the following three subcases 
(A, B, and C), since the proof of the main result presented in 
Theorem~\ref{thm:Whittle-index-discounted-DHR-DHR} 
below is slightly different in these three subcases: 
\begin{itemize}
\item
{\em DHR-DHR-A}: \\
For any $n_2 \in \{0,1,\ldots\}$, there is $n_1 \in \{0,1,\ldots\}$ such that 
\begin{equation}
w_1(n_1) \le w_2(n_2), 
\label{eq:DHR-DHR-A1}
\end{equation}
and, for any $n_1 \in \{0,1,\ldots\}$, there is $n_2 \in \{0,1,\ldots\}$ such that 
\begin{equation}
w_2(n_2) < w_1(n_1). 
\label{eq:DHR-DHR-A2}
\end{equation}
\vskip 6pt
\item
{\em DHR-DHR-B}: \\
There is $\bar n_2 \in \{0,1,\ldots\}$ such that, 
for any $n_1 \in \{0,1,\ldots\}$, 
\begin{equation}
w_2(\bar n_2+1) < w_1(n_1).
\label{eq:DHR-DHR-B}
\end{equation}
\vskip 6pt
\item
{\em DHR-DHR-C}: \\
There is $\bar n_1 \in \{-1\} \cup \{0,1,\ldots\}$ such that, 
for any $n_2 \in \{0,1,\ldots\}$, 
\begin{equation}
w_1(\bar n_1+1) \le w_2(n_2).
\label{eq:DHR-DHR-C}
\end{equation}
\end{itemize}

Numerical examples of these three subcases DHR-DHR-A, DHR-DHR-B, and DHR-DHR-C 
are given in Appendix~\ref{app:whittle-index-examples-discounted-costs}.

The following three lemmas give supplementary results concerning functions 
$\phi(n_1)$ and $w_1(n_1)$ for the three subcases, respectively.

\begin{lemma}
\label{lem:Whittle-index-discounted-DHR-DHR-A-lemma-3}
Assume the DHR-DHR-A subcase defined in (\ref{eq:DHR-DHR-A1}) and 
(\ref{eq:DHR-DHR-A2}). 
\begin{itemize}
\item[(i)]
Function $\phi(n_1)$ converges to 
\begin{equation}
\phi(\infty) = \lim_{n_1 \to \infty} \phi(n_1) = \infty; 
\label{eq:phi-infty-discounted-DHR-DHR-A}
\end{equation}
\item[(ii)]
Function $w_1(n_1)$ converges to 
\begin{equation}
\begin{split}
& 
w_1(\infty) = \lim_{n_1 \to \infty} w_1(n_1) = 
\psi(\infty,\infty) \; = \\
& \quad 
h \mu_1(\infty) \, 
\frac{\beta \sum_{i=0}^{\infty} \beta^i p_2(i)}
{1 + \beta \mu_1(\infty) \sum_{i=0}^{\infty} \beta^i \bar p_2(i)} \, 
\frac{\beta}{1 - \beta}, 
\end{split}
\label{eq:w1-infty-discounted-DHR-DHR-A}
\end{equation}
which satisfies 
\begin{equation}
w_1(\infty) = w_2(\infty), 
\label{eq:w1-w2-result-DHR-DHR-A}
\end{equation}
where $w_2(\infty)$ is defined in (\ref{eq:w2-infty-discounted-DHR-DHR}).
\end{itemize}
\end{lemma}

\paragraph{Proof} 
{\em (i)} 
This follows immediately from the definition of the DHR-DHR-A subcase.

\vskip 3pt
{\em (ii)} 
The first equation (\ref{eq:w1-infty-discounted-DHR-DHR-A}) is a direct 
consequence of part (i) since $w_1(n_1) = \psi(n_1,\phi(n_1))$, 
and the second equality (\ref{eq:w1-w2-result-DHR-DHR-A}) follows immediately 
from the definition of the DHR-DHR-A subcase.
\hfill $\Box$
\vskip 6pt

\begin{lemma}
\label{lem:Whittle-index-discounted-DHR-DHR-B-lemma-4}
Assume the DHR-DHR-B subcase defined in (\ref{eq:DHR-DHR-B}), and let 
$n_2^* \in \{0,1,\ldots\}$ denote the smallest $\bar n_2$ satisfying 
(\ref{eq:DHR-DHR-B}).
\begin{itemize}
\item[(i)]
Function $\phi(n_1)$ converges to 
\begin{equation}
\phi(\infty) = \lim_{n_1 \to \infty} \phi(n_1) = n_2^*; 
\label{eq:phi-infty-discounted-DHR-DHR-B}
\end{equation}
\item[(ii)]
Function $w_1(n_1)$ converges to 
\begin{equation}
\begin{split}
& 
w_1(\infty) = \lim_{n_1 \to \infty} w_1(n_1) = 
\psi(\infty,n_2^*) \; = \\
& \quad 
h \mu_1(\infty) \, 
\frac{\beta \sum_{i=0}^{n_2^*} \beta^i p_2(i)}
{1 + \beta \mu_1(\infty) \sum_{i=0}^{n_2^*} \beta^i \bar p_2(i)} \, 
\frac{\beta}{1 - \beta}, 
\end{split}
\label{eq:w1-infty-discounted-DHR-DHR-B}
\end{equation}
which satisfies 
\begin{equation}
w_2(n_2^*) \ge w_1(\infty) \ge w_2(\infty), 
\label{eq:w1-w2-result-DHR-DHR-B}
\end{equation}
where $w_2(\infty)$ is defined in (\ref{eq:w2-infty-discounted-DHR-DHR}).
\end{itemize}
\end{lemma}

\paragraph{Proof} 
{\em (i)} 
First, by the definition of the DHR-DHR-B subcase and 
Lemma~\ref{lem:Whittle-index-discounted-DHR-DHR-lemma-1}(i), 
there is $\bar n_1$ such that, for any $n_1 \ge \bar n_1$ and $i \ge 1$, 
\[
w_1(n_1) > w_2(n_2^*+i), 
\]
which implies, by (\ref{eq:Whittle-index-order-DHR-DHR}), that, 
for any $n_1 \ge \bar n_1$, 
\begin{equation}
\phi(n_1) \le n_2^*.
\label{eq:phi-result3a-DHR-DHR-B}
\end{equation}
On the other hand, since $n_2^*$ is the smallest $\bar n_2$ satisfying 
condition (\ref{eq:DHR-DHR-B}) and $w_2(n_2)$ is a decreasing function of 
$n_2$ by Lemma~\ref{lem:Whittle-index-discounted-DHR-DHR-lemma-1}(i), 
there is $n'_1$ such that, for any $n_1 \ge n'_1$ and $0 \le i \le n_2^*$, 
\[
w_1(n_1) \le w_2(n_2^*-i), 
\]
which implies, by (\ref{eq:Whittle-index-order-DHR-DHR}), that, 
for any $n_1 \ge n'_1$, 
\begin{equation}
\phi(n_1) \ge n_2^*.
\label{eq:phi-result3b-DHR-DHR-B}
\end{equation}
By combining (\ref{eq:phi-result3a-DHR-DHR-B}) and 
(\ref{eq:phi-result3b-DHR-DHR-B}), we finally conclude that, 
for any $n_1 \ge \max\{\bar n_1,n'_1\}$, 
\begin{equation}
\phi(n_1) = n_2^*, 
\label{eq:phi-result3-DHR-DHR-B}
\end{equation}
which justifies the claim.

\vskip 3pt
{\em (ii)} 
The first result (\ref{eq:w1-infty-discounted-DHR-DHR-B}) is a direct 
consequence of part (i) since $w_1(n_1) = \psi(n_1,\phi(n_1))$, 
and the second result (\ref{eq:w1-w2-result-DHR-DHR-B}) follows 
from the definition of $n_2^*$ combined with 
Lemma~\ref{lem:Whittle-index-discounted-DHR-DHR-lemma-1}(i) and 
Lemma~\ref{lem:Whittle-index-discounted-DHR-DHR-lemma-2}(iii).
\hfill $\Box$
\vskip 6pt

\begin{lemma}
\label{lem:Whittle-index-discounted-DHR-DHR-C-lemma-5}
Assume the DHR-DHR-C subcase defined in (\ref{eq:DHR-DHR-C}), and let 
$n_1^* \in \{-1\} \cup \{0,1,\ldots\}$ denote the smallest $\bar n_1$ 
satisfying (\ref{eq:DHR-DHR-C}).
\begin{itemize}
\item[(i)]
For any $m \ge 1$, 
\begin{equation}
\phi(n_1^*+m) = \infty.
\label{eq:phi-result4-DHR-DHR-C}
\end{equation}
In addition, function $\phi(n_1)$ converges to 
\begin{equation}
\phi(\infty) = \lim_{n_1 \to \infty} \phi(n_1) = \infty; 
\label{eq:phi-infty-discounted-DHR-DHR-C}
\end{equation}
\item[(ii)]
For any $m \ge 1$, 
\begin{equation}
\begin{split}
& 
w_1(n_1^*+m) = \psi(n_1^*+m,\infty) \; = \\
& \quad 
h \mu_1(n_1^*+m) \, 
\frac{\beta \sum_{i=0}^{\infty} \beta^i p_2(i)}
{1 + \beta \mu_1(n_1^*+m) \sum_{i=0}^{\infty} \beta^i \bar p_2(i)} \, 
\frac{\beta}{1 - \beta}
\end{split}
\label{eq:w1-result4-DHR-DHR-C}
\end{equation}
In addition, function $w_1(n_1)$ converges to 
\begin{equation}
\begin{split}
& 
w_1(\infty) = \lim_{n_1 \to \infty} w_1(n_1) = 
\psi(\infty,\infty) \; = \\
& \quad 
h \mu_1(\infty) \, 
\frac{\beta \sum_{i=0}^{\infty} \beta^i p_2(i)}
{1 + \beta \mu_1(\infty) \sum_{i=0}^{\infty} \beta^i \bar p_2(i)} \, 
\frac{\beta}{1 - \beta}, 
\end{split}
\label{eq:w1-infty-discounted-DHR-DHR-C}
\end{equation}
which satisfies 
\begin{equation}
w_1(n_1^*) \ge w_2(\infty) \ge w_1(\infty), 
\label{eq:w1-w2-result-DHR-DHR-C}
\end{equation}
where $w_2(\infty)$ is defined in (\ref{eq:w2-infty-discounted-DHR-DHR}).
\end{itemize}
\end{lemma}

\paragraph{Proof} 
{\em (i)} 
This follows from the definition of $n_1^*$ combined with 
Lemma~\ref{lem:Whittle-index-discounted-DHR-DHR-lemma-1}(ii).

\vskip 3pt
{\em (ii)} 
The first two results, (\ref{eq:w1-result4-DHR-DHR-C}) and 
(\ref{eq:w1-infty-discounted-DHR-DHR-C}), are direct consequences of 
part (i) since $w_1(n_1) = \psi(n_1,\phi(n_1))$. 
The last result (\ref{eq:w1-w2-result-DHR-DHR-C}) follows 
from the definition of $n_1^*$ combined with 
Lemma~\ref{lem:Whittle-index-discounted-DHR-DHR-lemma-2}(ii).
\hfill $\Box$
\vskip 6pt

\begin{theorem}
\label{thm:Whittle-index-discounted-DHR-DHR}
For the DHR-DHR case, the Whittle index $W_\beta(x)$ related to the relaxed 
optimization problem (\ref{eq:separable-discounted-costs}) is given by the 
following formulas: 
\begin{equation}
W_\beta(2,n_2) = w_2(n_2), 
\quad n_2 \in \{0,1,\ldots\}, 
\label{eq:Whittle-index-2n-discounted-DHR-DHR}
\end{equation}
where $w_2(n_2)$ is defined in (\ref{eq:w2-discounted-DHR-DHR}); 
\begin{equation}
W_\beta(1,n_1) = w_1(n_1), 
\quad n_1 \in \{0,1,\ldots\}, 
\label{eq:Whittle-index-1n-discounted-DHR-DHR}
\end{equation}
where $w_1(n_1)$ is defined in (\ref{eq:w1-discounted-DHR-DHR}).
\end{theorem}

\paragraph{Proof} 
The proof is slightly different for the three subcases (DHR-DHR-A, DHR-DHR-B, 
and DHR-DHR-C) and presented in Appendices~\ref{app:DHR-DHR-A-proof}, 
\ref{app:DHR-DHR-B-proof}, and \ref{app:DHR-DHR-C-proof}, respectively.
\hfill $\Box$
\vskip 6pt

Numerical examples of the Whittle index in the DHR-DHR case are given in 
Appendix~\ref{app:whittle-index-examples-discounted-costs}. 
In addition, the special cases DHR-GEO, GEO-DHR, and GEO-GEO 
are discussed in Section~\ref{sec:special-cases}.

\section{Whittle index for the IHR-IHR case}
\label{sec:IHR-IHR}

In this section, we assume that both stages of job~$k$ belong to class IHR, 
which is the IHR-IHR case defined in Section~\ref{sec:relaxed-problem}. 
Under this assumption, we derive the Whittle index values $W_{\beta,k}(x)$ 
for any state $x$ by solving the relaxed optimization problem 
(\ref{eq:separable-discounted-costs}) for any $\nu$. Before the main result 
given in Theorem~\ref{thm:Whittle-index-discounted-IHR-IHR}, we present some 
auxiliary lemmas that are needed in the proof of the main result.

\begin{lemma}
\label{lem:Whittle-index-discounted-IHR-IHR-lemma-1}
Assume the IHR-IHR case. 
Let us define the following function: 
\begin{equation}
w_2(n_2) = 
h \, 
\frac{\frac{1}{1 - \beta} - \sum_{i=0}^{\infty} \beta^i \bar p_2(i|n_2)}
{\sum_{i=0}^{\infty} \beta^i \bar p_2(i|n_2)}, 
\quad n_2 \in \{0,1,\ldots\}.
\label{eq:w2-discounted-IHR-IHR}
\end{equation}
Function $w_2(n_2)$ is increasing with respect to $n_2$ converging to 
\begin{equation}
w_2(\infty) = \lim_{n_2 \to \infty} w_2(n_2) = 
h \mu_2(\infty) \frac{\beta}{1 - \beta}. 
\label{eq:w2-infty-discounted-IHR-IHR}
\end{equation}
In addition, for any $n_2 \in \{0,1,\ldots\}$, 
\begin{equation}
w_2(n_2) \ge 
h \mu_2(n_2) \frac{\beta}{1 - \beta}.
\label{eq:w2-n2-result-discounted-IHR-IHR}
\end{equation}
\end{lemma}

\paragraph{Proof} 
Since $\bar p_2(i|n_2)$ is a decreasing function of $n_2$ in the IHR-IHR 
case, we see from (\ref{eq:w2-discounted-IHR-IHR}) that $w_2(n_2)$ is 
increasing with respect to $n_2$. In addition, 
\[
\begin{split}
& 
\lim_{n_2 \to \infty} w_2(n_2) = 
h \, 
\frac{\frac{1}{1 - \beta} - \sum_{i=0}^{\infty} \beta^i (1 - \mu_2(\infty))^i}
{\sum_{i=0}^{\infty} \beta^i (1 - \mu_2(\infty))^i} \\
& \quad 
= \; h \, 
\frac{\frac{1}{1 - \beta} - \frac{1}{1 - \beta(1 - \mu_2(\infty))}}
{\frac{1}{1 - \beta(1 - \mu_2(\infty))}} 
= h \mu_2(\infty) \frac{\beta}{1 - \beta}.
\end{split}
\]
Moreover, since 
\[
p_2(i|n) = \bar p_2(i|n) \mu_2(n+i) = \bar p_2(i|n) - \bar p_2(i+1|n), 
\]
we have 
\[
\begin{split}
& 
1 - (1 - \beta) \sum_{i=0}^{\infty} \beta^i \bar p_2(i|n_2) 
= \beta \sum_{i=0}^{\infty} \beta^i \bar p_2(i|n_2) 
- \sum_{i=1}^{\infty} \beta^i \bar p_2(i|n_2) \\
& \quad 
= \; \beta \sum_{i=0}^{\infty} \beta^i p_2(i|n_2) 
= \beta \sum_{i=0}^{\infty} \beta^i \bar p_2(i|n_2) \mu_2(n_2+i).
\end{split}
\]
Thus, $w_2(n_2)$ given in (\ref{eq:w2-discounted-IHR-IHR}) can be 
written in the following form: 
\begin{equation}
\begin{split}
& 
w_2(n_2) = 
h \, 
\frac{\sum_{i=0}^{\infty} \beta^i p_2(i|n_2)}
{\sum_{i=0}^{\infty} \beta^i \bar p_2(i|n_2)} 
\frac{\beta}{1 - \beta} \; = \\
& \quad 
h \, 
\frac{\sum_{i=0}^{\infty} \beta^i \bar p_2(i|n_2) \mu_2(n_2+i)}
{\sum_{i=0}^{\infty} \beta^i \bar p_2(i|n_2)} 
\frac{\beta}{1 - \beta}.
\end{split}
\label{eq:w2-alt-discounted-IHR-IHR}
\end{equation}
Finally, since $\mu_2(n_2+i) \ge \mu_2(n_2)$ for all $i$, we get the 
required inequality (\ref{eq:w2-n2-result-discounted-IHR-IHR}), 
which completes the proof.
\hfill $\Box$
\vskip 6pt

\begin{lemma}
\label{lem:Whittle-index-discounted-IHR-IHR-lemma-2}
Assume the IHR-IHR case. 
Let us define the following function: 
\begin{equation}
\begin{split}
& 
w_1(n_1) = 
h \, 
\frac{\frac{1}{1 - \beta} - \sum_{i=0}^{\infty} \beta^i \bar p_1(i|n_1) 
\Big( 1 + \beta \mu_1(n_1+i) \sum_{j=0}^{\infty} \beta^{j} \bar p_2(j) \Big)}
{\sum_{i=0}^{\infty} \beta^i \bar p_1(i|n_1) 
\Big( 1 + \beta \mu_1(n_1+i) \sum_{j=0}^{\infty} \beta^{j} \bar p_2(j) \Big)}, \\
& \quad 
n_1 \in \{0,1,\ldots\}.
\end{split}
\label{eq:w1-discounted-IHR-IHR}
\end{equation}
Function $w_1(n_1)$ is increasing with respect to $n_1$ converging to 
\begin{equation}
w_1(\infty) = \lim_{n_1 \to \infty} w_1(n_1) = 
h \mu_1(\infty) \, 
\frac{\beta \sum_{i=0}^{\infty} \beta^i p_2(i)}
{1 + \beta \mu_1(\infty) \sum_{i=0}^{\infty} \beta^i \bar p_2(i)} \, 
\frac{\beta}{1 - \beta}.
\label{eq:w1-infty-discounted-IHR-IHR}
\end{equation}
In addition, for any $n_1 \in \{0,1,\ldots\}$, 
\begin{equation}
w_1(n_1) \ge 
h \mu_1(n_1) \, 
\frac{\beta \sum_{i=0}^{\infty} \beta^i p_2(i)}
{1 + \beta \mu_1(n_1) \sum_{i=0}^{\infty} \beta^i \bar p_2(i)} \, 
\frac{\beta}{1 - \beta}
\label{eq:w1-n1-result-discounted-IHR-IHR}
\end{equation}
and, furthermore, we have 
\begin{equation}
w_1(\infty) < w_2(0), 
\label{eq:w1-infty-result-discounted-IHR-IHR}
\end{equation}
where $w_2(0)$ is defined in (\ref{eq:w2-discounted-IHR-IHR}).
\end{lemma}

\paragraph{Proof} 
Since 
\[
\bar p_1(i|n) \mu_1(n+i) = p_1(i|n) = \bar p_1(i|n) - \bar p_1(i+1|n), 
\]
we have 
\[
\begin{split}
& 
\sum_{i=0}^{\infty} \beta^i \bar p_1(i|n_1) 
\Big( 1 + \beta \mu_1(n_1+i) \sum_{j=0}^{\infty} \beta^{j} \bar p_2(j) \Big) \\
& \quad 
= \; 1 + \beta \sum_{j=0}^{\infty} \beta^{j} \bar p_2(j) + 
\sum_{i=1}^{\infty} \beta^i \bar p_1(i|n_1) 
\Big( 1 - (1 - \beta) \sum_{j=0}^{\infty} \beta^{j} \bar p_2(j) \Big).
\end{split}
\]
From the right hand side, we see that this expression is decreasing with respect 
to $n_1$, since $\bar p_1(i|n_1)$ is a decreasing function of $n_1$ in the IHR-IHR 
case and 
\[
1 - (1 - \beta) \sum_{j=0}^{\infty} \beta^{j} \bar p_2(j) 
\ge 1 - (1 - \beta) \sum_{j=0}^{\infty} \beta^{j} \ge 0.
\]
It also implies that $w_1(n_1)$ is increasing with respect to $n_1$, since $w_1(n_1)$ 
given in (\ref{eq:w1-discounted-IHR-IHR}) can be written in the following form: 
\begin{equation}
w_1(n_1) = 
h \, 
\frac{\frac{1}{1 - \beta} - 
\Big[ 1 + \beta A + 
\sum_{i=1}^{\infty} \beta^i \bar p_1(i|n_1) (1 - (1 - \beta) A) \Big]}
{1 + \beta A + 
\sum_{i=1}^{\infty} \beta^i \bar p_1(i|n_1) 
(1 - (1 - \beta) A)}, 
\label{eq:w1-alt-discounted-IHR-IHR}
\end{equation}
where we have used local shorthand notation 
\[
A = \sum_{j=0}^{\infty} \beta^{j} \bar p_2(j).
\]

Secondly, by (\ref{eq:w1-discounted-IHR-IHR}), 
\[
\begin{split}
& 
\lim_{n_1 \to \infty} w_1(n_1) = 
h \, 
\frac{\frac{1}{1 - \beta} - 
\sum_{i=0}^{\infty} \beta^i (1 - \mu_1(\infty))^i (1 + \beta \mu_1(\infty) A)}
{\sum_{i=0}^{\infty} \beta^i (1 - \mu_1(\infty))^i (1 + \beta \mu_1(\infty) A)} \\
& \quad 
= \; h \, 
\frac{\frac{1}{1 - \beta} - 
\frac{1 + \beta \mu_1(\infty) A}
{1 - \beta(1 - \mu_1(\infty))}}
{\frac{1 + \beta \mu_1(\infty) A}
{1 - \beta(1 - \mu_1(\infty))}} 
= h \, \mu_1(\infty) \, 
\frac{\beta \sum_{i=0}^{\infty} \beta^i p_2(i)}
{1 + \beta \mu_1(\infty) \sum_{i=0}^{\infty} \beta^i \bar p_2(i)} \, 
\frac{\beta}{1 - \beta}.
\end{split}
\]

Thirdly, by (\ref{eq:w1-alt-discounted-IHR-IHR}) and the fact that 
in the IHR-IHR case 
\[
\bar p_1(i|n_1) \le (1 - \mu_1(n_1))^i, 
\]
we have 
\[
\begin{split}
& 
w_1(n_1) \ge 
h \, 
\frac{\frac{1}{1 - \beta} - 
\Big[ 1 + \beta A + 
\sum_{i=1}^{\infty} \beta^i (1 - \mu_1(n_1))^i (1 - (1 - \beta) A) \Big]}
{1 + \beta A + 
\sum_{i=1}^{\infty} \beta^i (1 - \mu_1(n_1))^i (1 - (1 - \beta) A)} \\
& \quad 
= \; 
h \, 
\frac{\frac{1}{1 - \beta} - 
\sum_{i=0}^{\infty} \beta^i (1 - \mu_1(n_1))^i (1 + \beta \mu_1(n_1) A)}
{\sum_{i=0}^{\infty} \beta^i (1 - \mu_1(n_1))^i (1 + \beta \mu_1(n_1) A)} \\
& \quad 
= \; h \, 
\frac{\frac{1}{1 - \beta} - 
\frac{1 + \beta \mu_1(n_1) A}
{1 - \beta(1 - \mu_1(n_1))}}
{\frac{1 + \beta \mu_1(n_1) A}
{1 - \beta(1 - \mu_1(n_1))}} 
= h \, \mu_1(n_1) \, 
\frac{\beta \sum_{i=0}^{\infty} \beta^i p_2(i)}
{1 + \beta \mu_1(n_1) \sum_{i=0}^{\infty} \beta^i \bar p_2(i)} \, 
\frac{\beta}{1 - \beta}.
\end{split}
\]

Finally, by (\ref{eq:w1-infty-discounted-IHR-IHR}) and 
(\ref{eq:w2-alt-discounted-IHR-IHR}), 
\[
\begin{split}
& 
w_1(\infty) = 
h \mu_1(\infty) \, 
\frac{\beta \sum_{i=0}^{\infty} \beta^i p_2(i)}
{1 + \beta \mu_1(\infty) \sum_{i=0}^{\infty} \beta^i \bar p_2(i)} \, 
\frac{\beta}{1 - \beta} \\
& \quad 
< \; h \, 
\frac{\sum_{i=0}^{\infty} \beta^i p_2(i)}
{\sum_{i=0}^{\infty} \beta^i \bar p_2(i)} 
\frac{\beta}{1 - \beta} 
= w_2(0), 
\end{split}
\]
which completes the proof.
\hfill $\Box$
\vskip 6pt

\begin{theorem}
\label{thm:Whittle-index-discounted-IHR-IHR}
For the IHR-IHR case, the Whittle index $W_\beta(x)$ related to the relaxed 
optimization problem (\ref{eq:separable-discounted-costs}) is given by the 
following formulas: 

\begin{equation}
W_\beta(2,n_2) = w_2(n_2), 
\quad n_2 \in \{0,1,\ldots\}, 
\label{eq:Whittle-index-2n-discounted-IHR-IHR}
\end{equation}
where $w_2(n_2)$ is defined in (\ref{eq:w2-discounted-IHR-IHR});\footnote{
See also Equation~(\ref{eq:w2-alt-discounted-IHR-IHR}).}
\begin{equation}
W_\beta(1,n_1) = w_1(n_1), 
\quad n_1 \in \{0,1,\ldots\}, 
\label{eq:Whittle-index-1n-discounted-IHR-IHR}
\end{equation}
where $w_1(n_1)$ is defined in (\ref{eq:w1-discounted-IHR-IHR}).
\end{theorem}

\paragraph{Proof} 
The proof is presented in Appendix~\ref{app:IHR-IHR-proof}.
\hfill $\Box$
\vskip 6pt

Numerical examples of the Whittle index in the IHR-IHR case are given in 
Appendix~\ref{app:whittle-index-examples-discounted-costs}. 
In addition, the special cases IHR-GEO, GEO-IHR, and GEO-GEO 
are discussed in Section~\ref{sec:special-cases}.

\section{Whittle index for the DHR-IHR case}
\label{sec:DHR-IHR}

In this section, we assume that the first stage of job~$k$ belongs to class 
DHR, and the second stage to class IHR, which is the DHR-IHR case defined in 
Section~\ref{sec:relaxed-problem}. 
Under this assumption, we derive the Whittle index values $W_{\beta,k}(x)$ 
for any state $x$ by solving the relaxed optimization problem 
(\ref{eq:separable-discounted-costs}) for any $\nu$. Before the main result 
given in Theorem~\ref{thm:Whittle-index-discounted-DHR-IHR}, we present some 
auxiliary lemmas that are needed in the proof of the main result.

\begin{lemma}
\label{lem:Whittle-index-discounted-DHR-IHR-lemma-1}
Assume the DHR-IHR case. 
Let us define the following function: 
\begin{equation}
w_2(n_2) = 
h \, 
\frac{\frac{1}{1 - \beta} - \sum_{i=0}^{\infty} \beta^i \bar p_2(i|n_2)}
{\sum_{i=0}^{\infty} \beta^i \bar p_2(i|n_2)}, 
\quad n_2 \in \{0,1,\ldots\}.
\label{eq:w2-discounted-DHR-IHR}
\end{equation}
Function $w_2(n_2)$ is increasing with respect to $n_2$ converging to 
\begin{equation}
w_2(\infty) = \lim_{n_2 \to \infty} w_2(n_2) = 
h \mu_2(\infty) \frac{\beta}{1 - \beta}. 
\label{eq:w2-infty-discounted-DHR-IHR}
\end{equation}
In addition, for any $n_2 \in \{0,1,\ldots\}$, 
\begin{equation}
w_2(n_2) \ge 
h \mu_2(n_2) \frac{\beta}{1 - \beta}.
\label{eq:w2-n2-result-discounted-DHR-IHR}
\end{equation}
\end{lemma}

\paragraph{Proof} 
This lemma can be proved just similarly as 
Lemma~\ref{lem:Whittle-index-discounted-IHR-IHR-lemma-1} 
in Section~\ref{sec:IHR-IHR}. Therefore we may omit the proof here. 
We just note that $w_2(n_2)$ given in (\ref{eq:w2-discounted-DHR-IHR}) 
can be written in the following form: 
\begin{equation}
\begin{split}
& 
w_2(n_2) = 
h \, 
\frac{\sum_{i=0}^{\infty} \beta^i p_2(i|n_2)}
{\sum_{i=0}^{\infty} \beta^i \bar p_2(i|n_2)} 
\frac{\beta}{1 - \beta} \; = \\
& \quad 
h \, 
\frac{\sum_{i=0}^{\infty} \beta^i \bar p_2(i|n_2) \mu_2(n_2+i)}
{\sum_{i=0}^{\infty} \beta^i \bar p_2(i|n_2)} 
\frac{\beta}{1 - \beta}, 
\end{split}
\label{eq:w2-alt-discounted-DHR-IHR}
\end{equation}
see Equation (\ref{eq:w2-alt-discounted-IHR-IHR}) in the proof of 
Lemma~\ref{lem:Whittle-index-discounted-IHR-IHR-lemma-1}.
\hfill $\Box$
\vskip 6pt

\begin{lemma}
\label{lem:Whittle-index-discounted-DHR-IHR-lemma-2}
Assume the DHR-IHR case. 
Let us define the following function: 
\begin{equation}
w_1(n_1) = 
h \mu_1(n_1) \, 
\frac{\beta \sum_{i=0}^{\infty} \beta^i p_2(i)}
{1 + \beta \mu_1(n_1) \sum_{i=0}^{\infty} \beta^i \bar p_2(i)} \, 
\frac{\beta}{1 - \beta}, 
\quad n_1 \in \{0,1,\ldots\}.
\label{eq:w1-discounted-DHR-IHR}
\end{equation}
Function $w_1(n_1)$ is decreasing with respect to $n_1$ converging to 
\begin{equation}
\begin{split}
& 
w_1(\infty) = \lim_{n_1 \to \infty} w_1(n_1) \; = \\
& \quad 
h \mu_1(\infty) \, 
\frac{\beta \sum_{i=0}^{\infty} \beta^i p_2(i)}
{1 + \beta \mu_1(\infty) \sum_{i=0}^{\infty} \beta^i \bar p_2(i)} \, 
\frac{\beta}{1 - \beta}, 
\quad n_1 \in \{0,1,\ldots\}.
\end{split}
\label{eq:w1-infty-discounted-DHR-IHR}
\end{equation}
In addition, 
\begin{equation}
w_2(0) > w_1(0). 
\label{eq:Whittle-index-order-0-DHR-IHR}
\end{equation}
\end{lemma}

\paragraph{Proof} 
The monotonicity of $w_1(n_1)$ follows immediately from the monotonicity of 
$\mu_1(n_1)$. Limit (\ref{eq:w1-infty-discounted-DHR-IHR}) is obvious, 
and claim (\ref{eq:Whittle-index-order-0-DHR-IHR}) follows immediately 
from (\ref{eq:w2-alt-discounted-DHR-IHR}) and (\ref{eq:w1-discounted-DHR-IHR}) 
when applied with $n_2 = 0$ and $n_1 = 0$, respectively.
\hfill $\Box$
\vskip 6pt

\begin{theorem}
\label{thm:Whittle-index-discounted-DHR-IHR}
For the DHR-IHR case, the Whittle index $W_\beta(x)$ related to the relaxed 
optimization problem (\ref{eq:separable-discounted-costs}) is given by the 
following formulas: 
\begin{equation}
W_\beta(2,n_2) = w_2(n_2), 
\quad n_2 \in \{0,1,\ldots\}, 
\label{eq:Whittle-index-2n-discounted-DHR-IHR}
\end{equation}
where $w_2(n_2)$ is defined in (\ref{eq:w2-discounted-DHR-IHR});\footnote{
See also Equation~(\ref{eq:w2-alt-discounted-DHR-IHR}).}
\begin{equation}
W_\beta(1,n_1) = w_1(n_1), 
\quad n_1 \in \{0,1,\ldots\}, 
\label{eq:Whittle-index-1n-discounted-DHR-IHR}
\end{equation}
where $w_1(n_1)$ is defined in (\ref{eq:w1-discounted-DHR-IHR}).
\end{theorem}

\paragraph{Proof} 
The proof is presented in Appendix~\ref{app:DHR-IHR-proof}.
\hfill $\Box$
\vskip 6pt

Numerical examples of the Whittle index in the DHR-IHR case are given in 
Appendix~\ref{app:whittle-index-examples-discounted-costs}. 
In addition, the special cases DHR-GEO, GEO-IHR, and GEO-GEO 
are discussed in Section~\ref{sec:special-cases}.

\section{Whittle index for the IHR-DHR case}
\label{sec:IHR-DHR}

In this section, we assume that the first stage of job~$k$ belongs to class 
IHR, and the second stage to class DHR, which is the IHR-DHR case defined in 
Section~\ref{sec:relaxed-problem}. 
Under this assumption, we derive the Whittle index values $W_{\beta,k}(x)$ 
for any state $x$ by solving the relaxed optimization problem 
(\ref{eq:separable-discounted-costs}) for any $\nu$. Before the main result 
given in Theorem~\ref{thm:Whittle-index-discounted-IHR-DHR}, we present some 
auxiliary lemmas that are needed in the proof of the main result.

\begin{lemma}
\label{lem:Whittle-index-discounted-IHR-DHR-lemma-1}
Assume the IHR-DHR case. 
Let us define the following functions: 
\begin{equation}
w_2(n_2) = h \mu_2(n_2) \, \frac{\beta}{1 - \beta}, 
\quad n_2 \in \{0,1,\ldots\}, 
\label{eq:w2-discounted-IHR-DHR}
\end{equation}
and 
\begin{equation}
\begin{split}
& 
\psi(n_1,n_2) \; = \\
& \quad 
h \, 
\frac{\frac{1}{1 - \beta} - \sum_{i=0}^{\infty} \beta^i \bar p_1(i|n_1) 
\left( 1 + \mu_1(n_1+i) \Big( 1 - \beta \sum_{j=0}^{n_2} \beta^{j} p_2(j) \Big) \, \frac{\beta}{1 - \beta} \right)}
{\sum_{i=0}^{\infty} \beta^i \bar p_1(i|n_1) \Big( 1 + \mu_1(n_1+i) \beta \sum_{j=0}^{n_2} \beta^{j} \bar p_2(j) \Big)}, \\
& \quad 
n_1, n_2 \in \{0,1,\ldots\}.
\end{split}
\label{eq:psi-discounted-IHR-DHR}
\end{equation}
These functions have the following properties: 
\begin{itemize}
\item[(i)]
Function $w_2(n_2)$ is decreasing with respect to $n_2$ converging to 
\begin{equation}
w_2(\infty) = \lim_{n_2 \to \infty} w_2(n_2) = 
h \mu_2(\infty) \, 
\frac{\beta}{1 - \beta}; 
\label{eq:w2-infty-discounted-IHR-DHR}
\end{equation}
\item[(ii)]
Function $\psi(n_1,n_2)$ is increasing with respect to $n_1$ converging to 
\begin{equation}
\begin{split}
& 
\psi(\infty,n_2) = \lim_{n_1 \to \infty} \psi(n_1,n_2) = \\
& \quad 
h \mu_1(\infty) \, 
\frac{\beta \sum_{i=0}^{n_2} \beta^i p_2(i)}{1 + \beta \mu_1(\infty) \sum_{i=0}^{n_2} \beta^i \bar p_2(i)} \, 
\frac{\beta}{1 - \beta}; 
\end{split}
\label{eq:psi-infty-n2-discounted-IHR-DHR}
\end{equation}
\item[(iii)]
$\psi(\infty,0) < w_2(0)$; 
\item[(iv)]
$\psi(n_1,n_2) \le w_2(n_2+1)$ if and only if $\psi(n_1,n_2+1) \le w_2(n_2+1)$; 
\item[(v)]
$\psi(\infty,n_2) \le w_2(n_2+1)$ if and only if $\psi(\infty,n_2+1) \le w_2(n_2+1)$; 
\item[(vi)]
For any $n_2 \in \{0,1,\ldots\}$ and $n \in \{0,1,\ldots,n_2\}$, 
\begin{equation}
w_2(n_2) \le 
h \, 
\frac{\sum_{i = 0}^{n_2-n} \beta^i p_2(i|n)}
{\sum_{i = 0}^{n_2-n} \beta^i \bar p_2(i|n)} \, 
\frac{\beta}{1 - \beta}; 
\label{eq:w2-upper-bound-IHR-DHR}
\end{equation}
\item[(vii)]
For any $n_1, n_2 \in \{0,1,\ldots\}$, 
\begin{equation}
\psi(n_1,n_2) \ge 
h \mu_1(n_1) \, 
\frac{\beta \sum_{i=0}^{n_2} \beta^i p_2(i)}
{1 + \beta \mu_1(n_1) \sum_{i=0}^{n_2} \beta^i \bar p_2(i)} \, 
\frac{\beta}{1 - \beta}.
\label{eq:psi-n1-n2-result-discounted-IHR-DHR}
\end{equation}
\end{itemize}
\end{lemma}

\paragraph{Proof} 
{\em (i)} 
This follows immediately from the monotonicity of $\mu_2(n_2)$.

\vskip 3pt
{\em (ii)} 
Since 
\[
\bar p_1(i|n) \mu_1(n+i) = p_1(i|n) = \bar p_1(i|n) - \bar p_1(i+1|n), 
\]
we have 
\[
\begin{split}
& 
\sum_{i=0}^{\infty} \beta^i \bar p_1(i|n_1) 
\left( 1 + \mu_1(n_1+i) \Big( 1 - \beta \sum_{j=0}^{n_2} \beta^{j} p_2(j) \Big) \, \frac{\beta}{1 - \beta} \right) \\
& \quad 
= \; 1 + \Big( 1 - \beta \sum_{j=0}^{n_2} \beta^{j} p_2(j) \Big) \, 
\frac{\beta}{1 - \beta} + 
\sum_{i=1}^{\infty} \beta^i \bar p_1(i|n_1) 
\beta \sum_{j=0}^{n_2} \beta^{j} p_2(j).
\end{split}
\]
From the right hand side, we see that this expression is decreasing with respect 
to $n_1$, since $\bar p_1(i|n_1)$ is a decreasing function of $n_1$ in the IHR-DHR 
case. Correspondingly, we have 
\[
\begin{split}
& 
\sum_{i=0}^{\infty} \beta^i \bar p_1(i|n_1) 
\Big( 1 + \beta \mu_1(n_1+i) \sum_{j=0}^{n_2} \beta^{j} \bar p_2(j) \Big) \\
& \quad 
= \; 1 + \beta \sum_{j=0}^{n_2} \beta^{j} \bar p_2(j) + 
\sum_{i=1}^{\infty} \beta^i \bar p_1(i|n_1) 
\Big( 1 - (1 - \beta) \sum_{j=0}^{n_2} \beta^{j} \bar p_2(j) \Big).
\end{split}
\]
From the right hand side, we see that also this expression is decreasing with respect 
to $n_1$, since $\bar p_1(i|n_1)$ is a decreasing function of $n_1$ in the IHR-DHR 
case and 
\[
1 - (1 - \beta) \sum_{j=0}^{n_2} \beta^{j} \bar p_2(j) 
\ge 1 - (1 - \beta) \sum_{j=0}^{\infty} \beta^{j} \ge 0.
\]
These results together imply that $\psi_1(n_1,n_2)$ is increasing with respect 
to $n_1$, since $\psi_1(n_1,n_2)$ given in (\ref{eq:psi-discounted-IHR-DHR}) 
can be written in the following form: 
\begin{equation}
\psi_1(n_1,n_2) = 
h \, 
\frac{\frac{1}{1 - \beta} - 
\Big[ 1 + \beta A + 
\sum_{i=1}^{\infty} \beta^i \bar p_1(i|n_1) (1 - (1 - \beta) A) \Big]}
{1 + \beta B + 
\sum_{i=1}^{\infty} \beta^i \bar p_1(i|n_1) (1 - (1 - \beta) B)}, 
\label{eq:psi-alt-discounted-IHR-DHR}
\end{equation}
where we have used local shorthand notations 
\[
A = \Big( 1 - \beta \sum_{j=0}^{n_2} \beta^{j} p_2(j) \Big) \, 
\frac{1}{1 - \beta}, 
\quad 
B = \sum_{j=0}^{n_2} \beta^{j} \bar p_2(j).
\]

In addition, by (\ref{eq:psi-discounted-IHR-DHR}), 
\[
\begin{split}
& 
\lim_{n_1 \to \infty} \psi_1(n_1,n_2) = 
h \, 
\frac{\frac{1}{1 - \beta} - 
\sum_{i=0}^{\infty} \beta^i (1 - \mu_1(\infty))^i (1 + \beta \mu_1(\infty) A)}
{\sum_{i=0}^{\infty} \beta^i (1 - \mu_1(\infty))^i (1 + \beta \mu_1(\infty) B)} \\
& \quad 
= \; h \, 
\frac{\frac{1}{1 - \beta} - 
\frac{1 + \beta \mu_1(\infty) A}
{1 - \beta(1 - \mu_1(\infty))}}
{\frac{1 + \beta \mu_1(\infty) B}
{1 - \beta(1 - \mu_1(\infty))}} 
= h \, \mu_1(\infty) \, 
\frac{\beta \sum_{i=0}^{n_2} \beta^i p_2(i)}
{1 + \beta \mu_1(\infty) \sum_{i=0}^{n_2} \beta^i \bar p_2(i)} \, 
\frac{\beta}{1 - \beta}.
\end{split}
\]

\vskip 3pt
{\em (iii)} 
This follows from the definitions of $\psi(\infty,0)$ and $w_2(0)$: 
\[
\psi(\infty,0) = 
h \mu_1(\infty) \, 
\frac{\beta \mu_2(0)}{1 + \beta \mu_1(\infty)} \, 
\frac{\beta}{1 - \beta} < 
h \mu_2(0) \, \frac{\beta}{1 - \beta} = 
w_2(0).
\]

\vskip 3pt
{\em (iv)} 
This follows from the following equivalencies: 
\[
\begin{split}
& 
\psi(n_1,n_2+1) \le 
w_2(n_2+1) \quad \Longleftrightarrow \\
& 
1 - \sum_{i=0}^{\infty} \beta^i \bar p_1(i|n_1) 
\left( 1 - \beta + \beta \mu_1(n_1+i) \Big( 1 - \beta \sum_{j=0}^{n_2+1} \beta^{j} \bar p_2(j) \mu_2(j) \Big) \right) \\
& \quad \le \; 
\beta \mu_2(n_2+1) 
\sum_{i=0}^{\infty} \beta^i \bar p_1(i|n_1) \Big( 1 + \beta \mu_1(n_1+i) \sum_{j=0}^{n_2+1} \beta^{j} \bar p_2(j) \Big) \quad \Longleftrightarrow \\
& 
1 - \sum_{i=0}^{\infty} \beta^i \bar p_1(i|n_1) 
\left( 1 - \beta + \beta \mu_1(n_1+i) \Big( 1 - \beta \sum_{j=0}^{n_2} \beta^{j} p_2(j) \Big) \right) \\
& \quad \le \; 
\beta \mu_2(n_2+1) 
\sum_{i=0}^{\infty} \beta^i \bar p_1(i|n_1) \Big( 1 + \beta \mu_1(n_1+i) \sum_{j=0}^{n_2} \beta^{j} \bar p_2(j) \Big) \quad \Longleftrightarrow \\
& 
\psi(n_1,n_2) \le 
w_2(n_2+1).
\end{split}
\]

\vskip 3pt
{\em (v)} 
This follows from the following equivalencies: 
\[
\begin{split}
& 
\psi(\infty,n_2+1) \le 
w_2(n_2+1) \quad \Longleftrightarrow \\
& 
\beta \mu_1(\infty) \sum_{i=0}^{n_2+1} \beta^i \bar p_2(i) \mu_2(i) \\
& \quad \le \; 
\mu_2(n_2+1) \left( 1 + \beta \mu_1(\infty) \sum_{i=0}^{n_2+1} \beta^i \bar p_2(i) \right) 
\quad \Longleftrightarrow \\
& 
\beta \mu_1(\infty) \sum_{i=0}^{n_2} \beta^i \bar p_2(i) \mu_2(i) \\
& \quad \le \; 
\mu_2(n_2+1) \left( 1 + \beta \mu_1(\infty) \sum_{i=0}^{n_2} \beta^i \bar p_2(i) \right) 
\quad \Longleftrightarrow \\
& 
\psi(\infty,n_2) \le 
w_2(n_2+1).
\end{split}
\]

\vskip 3pt
{\em (vi)} 
Since $p_2(i|n) = \bar p_2(i|n) \mu_2(n+i)$ and $\mu_2(n_2)$ is a decreasing 
function of $n_2$, we have, for any $n \in \{0,1,\ldots,n_2\}$, 
\[
\sum_{i = 0}^{n_2-n} \beta^i p_2(i|n) = 
\sum_{i = 0}^{n_2-n} \beta^i \bar p_2(i|n) \mu_2(n+i) \ge 
\sum_{i = 0}^{n_2-n} \beta^i \bar p_2(i|n) \mu_2(n_2), 
\]
from which (\ref{eq:w2-upper-bound-IHR-DHR}) clearly follows.

\vskip 3pt
{\em (vii)} 
By (\ref{eq:psi-alt-discounted-IHR-DHR}) and the fact that 
in the IHR-DHR case 
\[
\bar p_1(i|n_1) \le (1 - \mu_1(n_1))^i, 
\]
we have 
\[
\begin{split}
& 
\psi(n_1,n_2) \ge 
h \, 
\frac{\frac{1}{1 - \beta} - 
\Big[ 1 + \beta A + 
\sum_{i=1}^{\infty} \beta^i (1 - \mu_1(n_1))^i (1 - (1 - \beta) A) \Big]}
{1 + \beta B + 
\sum_{i=1}^{\infty} \beta^i (1 - \mu_1(n_1))^i (1 - (1 - \beta) B)} \\
& \quad 
= \; 
h \, 
\frac{\frac{1}{1 - \beta} - 
\sum_{i=0}^{\infty} \beta^i (1 - \mu_1(n_1))^i (1 + \beta \mu_1(n_1) A)}
{\sum_{i=0}^{\infty} \beta^i (1 - \mu_1(n_1))^i (1 + \beta \mu_1(n_1) B)} \\
& \quad 
= \; h \, 
\frac{\frac{1}{1 - \beta} - 
\frac{1 + \beta \mu_1(n_1) A}
{1 - \beta(1 - \mu_1(n_1))}}
{\frac{1 + \beta \mu_1(n_1) B}
{1 - \beta(1 - \mu_1(n_1))}} = 
h \, 
\mu_1(n_1) \, 
\frac{\beta \sum_{i=0}^{n_2} \beta^i p_2(i)}
{1 + \beta \mu_1(n_1) \sum_{i=0}^{n_2} \beta^i \bar p_2(i)} \, 
\frac{\beta}{1 - \beta}, 
\end{split}
\]
which completes the proof of Lemma~\ref{lem:Whittle-index-discounted-IHR-DHR-lemma-1}.
\hfill $\Box$
\vskip 6pt

\begin{lemma}
\label{lem:Whittle-index-discounted-IHR-DHR-lemma-2}
Assume the IHR-DHR case. 
Let us define the following function: 
\begin{equation}
w_1(n_1) = \psi(n_1,\phi(n_1)), 
\quad n_1 \in \{0,1,\ldots\}, 
\label{eq:w1-discounted-IHR-DHR}
\end{equation}
where function $\psi(n_1,n_2)$ is defined in (\ref{eq:psi-discounted-IHR-DHR}) 
and function $\phi(n_1)$ as follows: 
\begin{equation}
\begin{split}
& 
\phi(n_1) = 
\min\{ n_2 \in \{0,1,\ldots\} \cup \{\infty\} : \psi(n_1,n_2) > w_2(n_2+1) \}, \\
& \quad 
n_1 \in \{0,1,\ldots\}, 
\end{split}
\label{eq:phi-discounted-IHR-DHR}
\end{equation}
where function $w_2(n_2)$ is defined in (\ref{eq:w2-discounted-IHR-DHR}) 
and we interpret that $\phi(n_1) = \infty$ if $\psi(n_1,n_2) \le w_2(n_2+1)$ 
for all $n_2$, in which case we naturally define 
\begin{equation}
\begin{split}
& 
\psi(n_1,\infty) = 
\lim_{n_2 \to \infty} \psi(n_1,n_2) = \\
& \quad 
h \, 
\frac{\frac{1}{1 - \beta} - \sum_{i=0}^{\infty} \beta^i \bar p_1(i|n_1) 
\left( 1 + \mu_1(n_1+i) \Big( 1 - \beta \sum_{j=0}^{\infty} \beta^{j} p_2(j) \Big) \, \frac{\beta}{1 - \beta} \right)}
{\sum_{i=0}^{\infty} \beta^i \bar p_1(i|n_1) \Big( 1 + \mu_1(n_1+i) \beta \sum_{j=0}^{\infty} \beta^{j} \bar p_2(j) \Big)}.
\end{split}
\label{eq:psi-n1-infty-discounted-IHR-DHR}
\end{equation}

These functions have the following properties: 
\begin{itemize}
\item[(i)]
Function $\phi(n_1)$ is decreasing with respect to $n_1$ converging to 
\begin{equation}
\begin{split}
& 
\phi(\infty) = \lim_{n_1 \to \infty} \phi(n_1) = \\
& \quad 
\min\{ n_2 \in \{0,1,\ldots\} \cup \{\infty\} : \psi(\infty,n_2) > w_2(n_2+1) \}, 
\end{split}
\label{eq:phi-infty-discounted-IHR-DHR}
\end{equation}
where $\psi(\infty,n_2)$ is defined in (\ref{eq:psi-infty-n2-discounted-IHR-DHR}) 
and we interpret that $\phi(\infty) = \infty$ if $\psi(\infty,n_2) \le w_2(n_2+1)$ 
for all $n_2$;
\item[(ii)]
Function $w_1(n_1)$ is increasing with respect to $n_1$ converging to 
\begin{equation}
\begin{split}
& 
w_1(\infty) = \lim_{n_1 \to \infty} w_1(n_1) = \psi(\infty,\phi(\infty)) \; = \\
& \quad 
h \mu_1(\infty) \, 
\frac{\beta \sum_{i=0}^{\phi(\infty)} \beta^i p_2(i)}
{1 + \beta \mu_1(\infty) \sum_{i=0}^{\phi(\infty)} \beta^i \bar p_2(i)} \, 
\frac{\beta}{1 - \beta}; 
\end{split}
\label{eq:w1-infty-discounted-IHR-DHR}
\end{equation} 
\item[(iii)]
If $\phi(n_1) < \infty$, then 
\begin{equation}
w_2(\phi(n_1)) \ge w_1(n_1) > w_2(\phi(n_1)+1); 
\label{eq:Whittle-index-order-IHR-DHR}
\end{equation}
otherwise $\phi(n_1) = \infty$ and 
\begin{equation}
w_2(\infty) \ge w_1(n_1).
\label{eq:Whittle-index-order-infty-IHR-DHR}
\end{equation}
\end{itemize}
\end{lemma}

\paragraph{Proof} 
{\em (i)} 
Assume first that $\phi(n_1) < \infty$. 
Thus, $\psi(n_1,n) \le w_2(n+1)$ for all $n \in \{0,1,\ldots,\phi(n_1)-1\}$. 
Now it follows from 
Lemma~\ref{lem:Whittle-index-discounted-IHR-DHR-lemma-1}(ii) that 
$\psi(n_1-1,n) \le w_2(n+1)$ for all $n \in \{0,1,\ldots,\phi(n_1)-1\}$, which 
implies that $\phi(n_1-1) \ge \phi(n_1)$.

Assume now that $\phi(n_1) = \infty$. 
Thus, $\psi(n_1,n) \le w_2(n+1)$ for all $n \in \{0,1,\ldots\}$. 
Now it follows from 
Lemma~\ref{lem:Whittle-index-discounted-IHR-DHR-lemma-1}(ii) that 
$\psi(n_1-1,n) \le w_2(n+1)$ for all $n \in \{0,1,\ldots\}$, which 
implies that $\phi(n_1-1) = \infty = \phi(n_1)$.

\vskip 3pt
{\em (ii)} 
Let $n_2 = \phi(n_1)$ and $n'_2 = \phi(n_1-1)$. Now $n_2 \le n'_2$ by (i). 
First, if $n_2 = n'_2 \le \infty$, then 
$w_1(n_1-1) = \psi(n_1-1,n_2) \le \psi(n_1,n_2) = w_1(n_1)$ 
by Lemma~\ref{lem:Whittle-index-discounted-IHR-DHR-lemma-1}(ii). 
Secondly, if $n_2 < n'_2 < \infty$, then it follows from the definition of 
$w_1(n_1)$ that $\psi(n_1,n_2) > w_2(n_2+1)$ and from the definition of 
$w_1(n_1-1)$ that $\psi(n_1-1,n'_2-1) \le w_2(n'_2)$, which is equivalent 
with $\psi(n_1-1,n'_2) \le w_2(n'_2)$ by 
Lemma~\ref{lem:Whittle-index-discounted-IHR-DHR-lemma-1}(iv). 
Thus, we have 
\[
w_1(n_1-1) = \psi(n_1-1,n'_2) \le w_2(n'_2) \le 
w_2(n_2+1) < \psi(n_1,n_2) = w_1(n_1).
\]
Thirdly, if $n_2 < n'_2 = \infty$, then it follows from the definition of 
$w_1(n_1)$ that $\psi(n_1,n_2) > w_2(n_2+1)$ and from the definition of 
$w_1(n_1-1)$ that $\psi(n_1-1,n) \le w_2(n+1)$ for all $n$, which implies 
that $\psi(n_1-1,\infty) \le w_2(\infty)$. Thus, we have 
\[
w_1(n_1-1) = \psi(n_1-1,\infty) \le w_2(\infty) \le 
w_2(n_2+1) < \psi(n_1,n_2) = w_1(n_1), 
\]
which completes the proof of the monotonicity of $w_1(n_1)$.

Moreover, by (\ref{eq:psi-discounted-IHR-DHR}) and 
(\ref{eq:psi-n1-infty-discounted-IHR-DHR}), 
\[
\begin{split}
& 
w_1(\infty) = \lim_{n_1 \to \infty} w_1(n_1) \; = \\
& \quad 
\lim_{n_1 \to \infty} \psi_1(n_1,\phi(n_1)) = 
h \, 
\frac{\frac{1}{1 - \beta} - 
\sum_{i=0}^{\infty} \beta^i (1 - \mu_1(\infty))^i (1 + \beta \mu_1(\infty) A)}
{\sum_{i=0}^{\infty} \beta^i (1 - \mu_1(\infty))^i (1 + \beta \mu_1(\infty) B)} 
\; = \\
& \quad 
h \, 
\frac{\frac{1}{1 - \beta} - 
\frac{1 + \beta \mu_1(\infty) A}
{1 - \beta(1 - \mu_1(\infty))}}
{\frac{1 + \beta \mu_1(\infty) B}
{1 - \beta(1 - \mu_1(\infty))}} 
= h \, \mu_1(\infty) \, 
\frac{\beta \sum_{i=0}^{\phi(\infty)} \beta^i p_2(i)}
{1 + \beta \mu_1(\infty) \sum_{i=0}^{\phi(\infty)} \beta^i \bar p_2(i)} \, 
\frac{\beta}{1 - \beta}, 
\end{split}
\]
where we have used local shorthand notations 
\[
A = \Big( 1 - \beta \sum_{j=0}^{\phi(\infty)} \beta^{j} p_2(j) \Big) \, 
\frac{1}{1 - \beta}, 
\quad 
B = \sum_{j=0}^{\phi(\infty)} \beta^{j} \bar p_2(j).
\]

\vskip 3pt
{\em (iii)} 
Assume first that $\phi(n_1) < \infty$ and let $n_2 = \phi(n_1)$. 
Thus, $\psi(n_1,n_2-1) \le w_2(n_2)$, which is equivalent with 
$w_1(n_1) = \psi(n_1,n_2) \le w_2(n_2)$ by 
Lemma~\ref{lem:Whittle-index-discounted-IHR-DHR-lemma-1}(iv). 
On the other hand, $w_1(n_1) = \psi(n_1,n_2) > w_2(n_2+1)$. 
These results together justify (\ref{eq:Whittle-index-order-IHR-DHR}).

Assume now that $\phi(n_1) = \infty$. Thus, $\psi(n_1,n) \le w_2(n+1)$ 
for all $n$, which implies that $w_1(n_1) = \psi(n_1,\infty) \le 
w_2(\infty)$.
\hfill $\Box$
\vskip 6pt

Based on the functions $w_2(n_2)$ and $w_1(n_1)$ defined in the previous 
lemmas, we split the IHR-DHR case into the following two subcases (D and E), 
since the proof of the main result presented in 
Theorem~\ref{thm:Whittle-index-discounted-IHR-DHR} 
below is slightly different in these two subcases: 
\begin{itemize}
\item
{\em IHR-DHR-D}: \\
There is $\bar n_2 \in \{0,1,\ldots\}$ such that, 
for any $n_1 \in \{0,1,\ldots\}$, 
\begin{equation}
w_2(\bar n_2+1) < w_1(n_1).
\label{eq:IHR-DHR-D}
\end{equation}
\vskip 6pt
\item
{\em IHR-DHR-E}: \\
There is $\bar n_1 \in \{1,2,\ldots\}$ such that, 
for any $n_2 \in \{0,1,\ldots\}$, 
\begin{equation}
w_1(\bar n_1-1) \le w_2(n_2).
\label{eq:IHR-DHR-E}
\end{equation}
\end{itemize}

Numerical examples of these two subcases IHR-DHR-D and IHR-DHR-E 
are given in Appendix~\ref{app:whittle-index-examples-discounted-costs}.

The following two lemmas give supplementary results concerning functions 
$\phi(n_1)$ and $w_1(n_1)$ for the two subcases, respectively.

\begin{lemma}
\label{lem:Whittle-index-discounted-IHR-DHR-D-lemma-3}
Assume the IHR-DHR-D subcase defined in (\ref{eq:IHR-DHR-D}), and let 
$n_2^* \in \{0,1,\ldots\}$ denote the smallest $\bar n_2$ satisfying 
(\ref{eq:IHR-DHR-D}). 
Now $n_2^* = \phi(0)$, and condition (\ref{eq:IHR-DHR-D}) is equivalent with 
\begin{equation}
\phi(0) < \infty.
\label{eq:IHR-DHR-D-alt}
\end{equation}
In addition, 
\begin{equation}
w_1(\infty) \ge w_1(0) > w_2(n_2^*+1) \ge w_2(\infty), 
\label{eq:w1-w2-result-IHR-DHR-D}
\end{equation}
where $w_1(\infty)$ is defined in (\ref{eq:w1-infty-discounted-IHR-DHR}) and 
$w_2(\infty)$ in (\ref{eq:w2-infty-discounted-IHR-DHR}).
\end{lemma}

\paragraph{Proof} 
The results follow directly from the definition of $n_2^*$ combined with 
Lemma~\ref{lem:Whittle-index-discounted-IHR-DHR-lemma-1}(i) and 
Lemma~\ref{lem:Whittle-index-discounted-IHR-DHR-lemma-2}(ii)\&(iii).
\hfill $\Box$
\vskip 6pt

\begin{lemma}
\label{lem:Whittle-index-discounted-IHR-DHR-E-lemma-4}
Assume the IHR-DHR-E subcase defined in (\ref{eq:IHR-DHR-E}), and let 
$n_1^* \in \{1,2,\ldots\} \cup \{\infty\}$ denote the greatest $\bar n_1$ 
satisfying (\ref{eq:IHR-DHR-E}), where we interpret that $n_1^* = \infty$ 
if (\ref{eq:IHR-DHR-E}) is satisfied for any $\bar n_1 \in \{1,2,\ldots\}$. 
Now condition (\ref{eq:IHR-DHR-E}) is equivalent with 
\begin{equation}
\phi(0) = \infty.
\label{eq:IHR-DHR-E-alt}
\end{equation}
In addition, if $n_1^* < \infty$, then 
\begin{equation}
w_2(\infty) \ge w_1(n_1^*-1) \ge w_1(0), 
\label{eq:w1-w2-result1-IHR-DHR-E}
\end{equation}
where $w_2(\infty)$ in (\ref{eq:w2-infty-discounted-IHR-DHR}). 
Otherwise $n_1^* = \infty$ and 
\begin{equation}
w_2(\infty) \ge w_1(\infty) \ge w_1(0), 
\label{eq:w1-w2-result2-IHR-DHR-E}
\end{equation}
where $w_1(\infty)$ is defined in (\ref{eq:w1-infty-discounted-IHR-DHR}) and 
$w_2(\infty)$ in (\ref{eq:w2-infty-discounted-IHR-DHR}). 
\end{lemma}

\paragraph{Proof} 
The results follow directly from the definition of $n_1^*$ combined with 
Lemma~\ref{lem:Whittle-index-discounted-IHR-DHR-lemma-2}(ii).
\hfill $\Box$
\vskip 6pt

\begin{theorem}
\label{thm:Whittle-index-discounted-IHR-DHR}
For the IHR-DHR case, the Whittle index $W_\beta(x)$ related to the relaxed 
optimization problem (\ref{eq:separable-discounted-costs}) is given by the 
following formulas: 
\begin{equation}
W_\beta(2,n_2) = w_2(n_2), 
\quad n_2 \in \{0,1,\ldots\}, 
\label{eq:Whittle-index-2n-discounted-IHR-DHR}
\end{equation}
where $w_2(n_2)$ is defined in (\ref{eq:w2-discounted-IHR-DHR});
\begin{equation}
W_\beta(1,n_1) = w_1(n_1), 
\quad n_1 \in \{0,1,\ldots\}, 
\label{eq:Whittle-index-1n-discounted-IHR-DHR}
\end{equation}
where $w_1(n_1)$ is defined in (\ref{eq:w1-discounted-IHR-DHR}).
\end{theorem}

\paragraph{Proof} 
The proof is slightly different for the two subcases (IHR-DHR-D and DHR-DHR-E). 
For the IHR-DHR-D subcase, the proof is presented in 
Appendix~\ref{app:IHR-DHR-D-proof}. In the DHR-DHR-E subcase, the proof 
is even depending on the parameter $n_1^*$ defined 
in Lemma~\ref{lem:Whittle-index-discounted-IHR-DHR-E-lemma-4}. 
These proofs are presented in Appendices~\ref{app:IHR-DHR-E1-proof} and 
\ref{app:IHR-DHR-E2-proof}.
\hfill $\Box$
\vskip 6pt

Numerical examples of the Whittle index in the IHR-DHR case are given in 
Appendix~\ref{app:whittle-index-examples-discounted-costs}. 
In addition, the special cases IHR-GEO, GEO-DHR, and GEO-GEO 
are discussed in Section~\ref{sec:special-cases}.

\section{Special cases}
\label{sec:special-cases}

In this section, we discuss the Whittle index for the special cases of 
the two-stage jobs where the service time distribution in one of stages, 
say $j$, is {\em geometric} (GEO) with a constant hazard rate (CHR) such as 
\[
\mu_j(n) \equiv \mu_j, 
\quad n \in \{0,1,\ldots\}, 
\]
for some $\mu_j > 0$, while the hazard rate in the other stage is monotonous. 
All the results given in this section are direct corollaries of our 
Theorems~\ref{thm:Whittle-index-discounted-DHR-DHR}-\ref{thm:Whittle-index-discounted-IHR-DHR}.

\subsection{Special case DHR-GEO}
\label{subsec:special-case-DHR-GEO}

Consider the special case DHR-GEO (of cases DHR-DHR and DHR-IHR), where 
the first stage is DHR and 
the second stage geometric with a constant hazard rate 
\[
\mu_2(n) \equiv \mu_2, 
\quad n \in \{0,1,\ldots\}, 
\]
for some $\mu_2 > 0$. In this case, 
\[
\begin{split}
& 
W_\beta(2,n) \equiv h \mu_2 \, \frac{\beta}{1 - \beta}, \\
& 
W_\beta(1,n) = 
h \mu_2 \, 
\frac{\beta \mu_1(n)}
{1 - \beta + \beta(\mu_1(n) + \mu_2)} \, 
\frac{\beta}{1 - \beta}. 
\end{split}
\]
In addition, we have the following ordering among the states: 
\[
\begin{split}
& 
W_\beta(2,\infty) = \ldots = W_\beta(2,1) = W_\beta(2,0) > \\
& \quad 
W_\beta(1,0) \ge W_\beta(1,1) \ge \ldots \ge W_\beta(1,\infty) 
\ge 0, 
\end{split}
\]
where we have defined 
\[
\begin{split}
& 
W_\beta(2,\infty) = 
h \mu_2 \, 
\frac{\beta}{1 - \beta}, \\
& 
W_\beta(1,\infty) = 
h \mu_2 \, 
\frac{\beta \mu_1(\infty)}
{1 - \beta + \beta(\mu_1(\infty) + \mu_2)} \, 
\frac{\beta}{1 - \beta}. 
\end{split}
\]
From the DHR-DHR subcases, this special case belongs to the DHR-DHR-C subcase 
with $n_1^* = -1$.

\subsection{Special case GEO-DHR}
\label{subsec:special-case-GEO-DHR}

Consider the special case GEO-DHR (of cases DHR-DHR and IHR-DHR), where 
the first stage is geometric with a constant hazard rate 
\[
\mu_1(n) \equiv \mu_1, 
\quad n \in \{0,1,\ldots\}, 
\]
for some $\mu_1 > 0$, and the second stage is DHR. 
In this case, 
\[
\begin{split}
& 
W_\beta(2,n) = h \mu_2(n) \, \frac{\beta}{1 - \beta}, \\
& 
W_\beta(1,n) \equiv 
h \mu_1 \, 
\frac{\beta \sum_{i=0}^{\phi_\beta} \beta^i p_2(i)}{1 + \beta \mu_1 \sum_{i=0}^{\phi_\beta} \beta^i \bar p_2(i)} \, 
\frac{\beta}{1 - \beta}, 
\end{split}
\]
where 
\[
\phi_\beta = 
\min\{ n_2 \in \{0,1,\ldots\} \cup \{\infty\} : 
\frac{\beta \mu_1 \sum_{i=0}^{n_2} \beta^i p_2(i)}
{1 + \beta \mu_1 \sum_{i=0}^{n_2} \beta^i \bar p_2(i)} 
> \mu_2(n_2+1) \}.
\]
If $\phi_\beta < \infty$, then 
this special case belongs to the DHR-DHR-B subcase with $n_2^* = \phi_\beta$ 
and to the IHR-DHR-D subcase with $n_2^* = n_2^\circ = \phi_\beta$, and 
we have the following ordering among the states: 
\[
\begin{split}
& 
W_\beta(2,0) \ge W_\beta(2,1) \ge \ldots \ge W_\beta(2,\phi_\beta) \; \ge \\
& \quad 
W_\beta(1,0) = W_\beta(1,1) = \ldots = W_\beta(1,\infty) \; > \\ 
& \quad \quad 
W_\beta(2,\phi_\beta+1) \ge W_\beta(2,\phi_\beta+2) \ge \ldots \ge W_\beta(2,\infty) 
\ge 0, 
\end{split}
\]
where we have defined 
\[
\begin{split}
& 
W_\beta(2,\infty) = 
h \mu_2(\infty) \, 
\frac{\beta}{1 - \beta}, \\
& 
W_\beta(1,\infty) = 
h \mu_1 \, 
\frac{\beta \sum_{i=0}^{\phi_\beta} \beta^i p_2(i)}{1 + \beta \mu_1 \sum_{i=0}^{\phi_\beta} \beta^i \bar p_2(i)} \, 
\frac{\beta}{1 - \beta}. 
\end{split}
\]
But if $\phi_\beta = \infty$, then 
this special case belongs to the DHR-DHR-C subcase with $n_1^* = -1$ and 
to the IHR-DHR-E subcase with $n_1^* = \infty$, and 
we have the following ordering among the states: 
\[
\begin{split}
& 
W_\beta(2,0) \ge W_\beta(2,1) \ge \ldots \ge W_\beta(2,\infty) \; \ge \\
& \quad 
W_\beta(1,0) = W_\beta(1,1) = \ldots = W_\beta(1,\infty) 
\ge 0.
\end{split}
\]

\subsection{Special case IHR-GEO}
\label{subsec:special-case-IHR-GEO}

Consider the special case IHR-GEO (of cases IHR-IHR and IHR-DHR), where 
the first stage is IHR and 
the second stage geometric with a constant hazard rate 
\[
\mu_2(n) \equiv \mu_2, 
\quad n \in \{0,1,\ldots\}, 
\]
for some $\mu_2 > 0$. In this case, 
\[
\begin{split}
& 
W_\beta(2,n) \equiv h \mu_2 \, \frac{\beta}{1 - \beta}, \\
& 
W_\beta(1,n) = 
h \, 
\frac{\frac{1}{1 - \beta} - \sum_{i=0}^{\infty} \beta^i \bar p_1(i|n) 
\Big( 1 + \frac{\beta \mu_1(n+i)}{1 - \beta (1 - \mu_2)} \Big)}
{\sum_{i=0}^{\infty} \beta^i \bar p_1(i|n) 
\Big( 1 + \frac{\beta \mu_1(n+i)}{1 - \beta (1 - \mu_2)} \Big)}. 
\end{split}
\]
In addition, we have the following ordering among the states: 
\[
\begin{split}
& 
W_\beta(2,\infty) = \ldots = W_\beta(2,1) = W_\beta(2,0) > \\
& \quad 
W_\beta(1,\infty) \ge \ldots \ge W_\beta(1,1) \ge W_\beta(1,0) > 0, 
\end{split}
\]
where we have defined 
\[
\begin{split}
& 
W_\beta(2,\infty) = 
h \mu_2 \, 
\frac{\beta}{1 - \beta}, \\
& 
W_\beta(1,\infty) = 
h \mu_2 \, 
\frac{\beta \mu_1(\infty)}
{1 - \beta + \beta(\mu_1(\infty) + \mu_2)} \, 
\frac{\beta}{1 - \beta}. 
\end{split}
\]
From the IHR-DHR subcases it belongs to the IHR-DHR-E subcase with $n_1^* = \infty$.

\subsection{Special case GEO-IHR}
\label{subsec:special-case-GEO-IHR}

Consider the special case GEO-IHR (of cases IHR-IHR and DHR-IHR), where 
the first stage is geometric with a constant hazard rate 
\[
\mu_1(n) \equiv \mu_1, 
\quad n \in \{0,1,\ldots\}, 
\]
for some $\mu_1 > 0$, and the second stage is IHR. 
In this case, 
\[
\begin{split}
& 
W_\beta(2,n) = 
h \, 
\frac{\frac{1}{1 - \beta} - \sum_{i=0}^{\infty} \beta^i \bar p_2(i|n)}
{\sum_{i=0}^{\infty} \beta^i \bar p_2(i|n)}, \\
& 
W_\beta(1,n) \equiv 
h \mu_1 \, 
\frac{\beta \sum_{i=0}^{\infty} \beta^i p_2(i)}{1 + \beta \mu_1 \sum_{i=0}^{\infty} \beta^i \bar p_2(i)} \, 
\frac{\beta}{1 - \beta}.
\end{split}
\]
In addition, we have the following ordering among the states: 
\[
\begin{split}
& 
W_\beta(2,\infty) \ge \ldots \ge W_\beta(2,1) \ge W_\beta(2,0) > \\
& \quad 
W_\beta(1,\infty) = \ldots = W_\beta(1,1) = W_\beta(1,0) > 0, 
\end{split}
\]
where we have defined 
\[
\begin{split}
& 
W_\beta(2,\infty) = 
h \mu_2(\infty) \, 
\frac{\beta}{1 - \beta}, \\
& 
W_\beta(1,\infty) = 
h \mu_1 \, 
\frac{\beta \sum_{i=0}^{\infty} \beta^i p_2(i)}{1 + \beta \mu_1 \sum_{i=0}^{\infty} \beta^i \bar p_2(i)} \, 
\frac{\beta}{1 - \beta}. 
\end{split}
\]

\subsection{Special case GEO-GEO}
\label{subsec:special-case-GEO-GEO}

Consider the special case GEO-GEO (of cases DHR-DHR, IHR-IHR, DHR-IHR, and IHR-DHR), 
where both stages are geometric, 
\[
\mu_1(n) \equiv \mu_1, \quad 
\mu_2(n) \equiv \mu_2, \quad 
n \in \{0,1,\ldots\}, 
\]
for some $\mu_1, \mu_2 > 0$. In this case, 
\[
\begin{split}
& 
W_\beta(2,n) \equiv h \mu_2 \, \frac{\beta}{1 - \beta}, \\
& 
W_\beta(1,n) \equiv 
h \mu_2 \, 
\frac{\beta \mu_1}
{1 - \beta + \beta(\mu_1 + \mu_2)} \, 
\frac{\beta}{1 - \beta}. 
\end{split}
\]
In addition, we have the following ordering among the states: 
\[
\begin{split}
& 
W_\beta(2,0) = W_\beta(2,1) = \ldots = W_\beta(2,\infty) > \\
& \quad 
W_\beta(1,0) = W_\beta(1,1) = \ldots = W_\beta(1,\infty) 
\ge 0.
\end{split}
\]
From the DHR-DHR subcases, this special case belongs to the DHR-DHR-C subcases 
with $n_1^* = -1$, and from the IHR-DHR subcases it belongs to the IHR-DHR-E 
subcases with $n_1^* = \infty$.

\section{Optimal scheduling problem in continuous time with undiscounted costs}
\label{sec:average-cost}

In this section, we move from discounted to undiscounted costs (either average 
or total costs depending on whether we talk about the open or closed version 
of the problem, respectively). In addition, we move from the discrete time 
setup to single server scheduling problems in continuous time. The main results 
are given below in Theorem~\ref{thm:Whittle-index-average}, 
Corollary~\ref{cor:Gittins-index-average}, and 
Theorem~\ref{thm:Gittins-index-average}. 
In addition, we present two conjectures about the characterization of the 
Gittins index for any sequential two-stage and even multistage jobs 
(Conjectures~\ref{con:Gittins-index-average-two-stage} and 
\ref{con:Gittins-index-average-multistage}).

We start by first considering {\em undiscounted} costs in the original 
discrete-time model. Let $W(j,n)$ denote the {\em Whittle index} 
for job~$k$ in state $(j,n)$ related to the undiscounted costs,\footnote{ 
As before, we leave out the index, $k$, from the notation (like $W(j,n)$) 
since we are considering the same job all the time.}
which is derived from the ``discounted cost'' Whittle index $W_\beta(j,n)$ 
as follows: 
\begin{equation}
W(j,n) = \lim_{\beta \to 1} (1 - \beta) W_\beta(j,n), 
\quad j \in \{1,2\}, n \in \{0,1,\ldots\}.
\label{eq:Whittle-index-jn-average}
\end{equation}
Based on Theorems~\ref{thm:Whittle-index-discounted-DHR-DHR}-\ref{thm:Whittle-index-discounted-IHR-DHR}, we get the following results.

\begin{theorem}
\label{thm:Whittle-index-average}
(Discrete time, undiscounted costs) 
\begin{itemize}
\item[(i)]
For the DHR-DHR case, 
\begin{equation}
\begin{split}
& 
W(2,n) = 
h \mu_2(n), \\
& 
W(1,n) = 
h \, 
\sup_{n_2 \ge 0} 
\left( 
\frac{P\{S_2 \le n_2 + 1\}}
{\frac{1}{\mu_1(n)} + E[\min\{S_2, n_2 + 1\}]} 
\right).
\end{split}
\label{eq:Whittle-index-21-average-DHR-DHR}
\end{equation}
 
\item[(ii)]
For the IHR-IHR case, 
\begin{equation}
\begin{split}
& 
W(2,n) = 
\frac{h}
{E[S_2 - n \mid S_2 \ge n + 1]}, \\
& 
W(1,n) = 
\frac{h}
{E[S_1 - n \mid S_1 \ge n + 1] + E[S_2]}.
\end{split}
\label{eq:Whittle-index-21-average-IHR-IHR}
\end{equation}

\item[(iii)]
For the DHR-IHR case, 
\begin{equation}
\begin{split}
& 
W(2,n) = 
\frac{h}
{E[S_2 - n \mid S_2 \ge n + 1]}, \\
& 
W(1,n) = 
\frac{h}
{\frac{1}{\mu_1(n)} + E[S_2]}.
\end{split}
\label{eq:Whittle-index-21-average-DHR-IHR}
\end{equation}

\item[(iv)]
For the IHR-DHR case, 
\begin{equation}
\begin{split}
& 
W(2,n) = 
h \mu_2(n), \\
& 
W(1,n) = 
h \, 
\sup_{n_2 \ge 0} 
\left( 
\frac{P\{S_2 \le n_2 + 1\}}
{E[S_1 - n \mid S_1 \ge n + 1] + E[\min\{S_2, n_2 + 1\}]} 
\right).
\end{split}
\label{eq:Whittle-index-21-average-IHR-DHR}
\end{equation}

\end{itemize}
\end{theorem}

\paragraph{Proof} 
The proof is presented in Appendix~\ref{app:average-proof}.
\hfill $\Box$
\vskip 6pt

Let us now consider the corresponding {\em continuous-time} problem with 
undiscounted costs, which we already introduced in the beginning of 
Section~\ref{sec:intro}. Jobs are still assumed to be sequential two-stage 
jobs so that each job consists of two consecutive continuous-time stages. From 
this on, let $S_j$ denote the random service time of stage $j$ of job $k$ 
taking values in $(0,\infty)$, and let $\mu_j(t)$, $t \in [0,\infty)$, denote 
the corresponding continuous-time {\em hazard rate}, which is assumed to be 
{\em monotonous}, either increasing (IHR) or decreasing (DHR).

In the closed version of the problem, there are $K$ two-stage jobs at time~$0$, 
no new arrivals, and the objective is to minimize the expected total holding 
costs (\ref{eq:optimal-scheduling-problem-closed}). As for the open version, 
there are $K$ classes of two-stage jobs, new jobs arrive according to 
class-specific Poisson processes (so that this is the M/G/1 queueing model), 
and the objective is to minimize the expected average holding costs 
(\ref{eq:optimal-scheduling-problem-open}). For both versions, the scheduling 
problem belongs to the class of multi-armed bandit problems, the optimal 
nonanticipating\footnote{ 
For multistage jobs, a nonanticipating discipline is based on the knowledge of 
the current stage and the service attained in the current stage for each of the 
jobs in the system.}
scheduling discipline is known to be the Gittins index policy, and the Gittins 
index is exactly the same (in both cases) \cite{Git89,Git11}.

On the other hand, as already mentioned in Section~\ref{sec:relaxed-problem}, 
the Whittle index coincides with the Gittins index for multi-armed bandit 
problems \cite{Whi88JAP,Git11}. Thus, we utilize the Whittle index results above 
(Theorem~\ref{thm:Whittle-index-average}) to derive the {\em Gittins index} 
for the corresponding continuous-time scheduling problem.

Let $G(j,a)$ denote the {\em Gittins index} for job~$k$ with current stage $j$ 
and attained service $a$ in the current stage related to undiscounted 
costs. By letting the time slot shrink to $0$, we get from 
Theorem~\ref{thm:Whittle-index-average} the following results for the 
continuous-time model.

\begin{corollary}
\label{cor:Gittins-index-average}
(Continuous time, undiscounted costs)
\begin{itemize}
\item[(i)]
For the DHR-DHR case, 
\begin{equation}
\begin{split}
& 
G(2,a) = 
h \mu_2(a), \\
& 
G(1,a) = 
h \, 
\sup_{\Delta \ge 0} 
\left( 
\frac{P\{S_2 \le \Delta\}}
{\frac{1}{\mu_1(a)} + E[\min\{S_2,\Delta\}]} 
\right).
\end{split}
\label{eq:Gittins-index-21-average-DHR-DHR}
\end{equation}

\item[(ii)]
For the IHR-IHR case, 
\begin{equation}
\begin{split}
& 
G(2,a) = 
\frac{h}
{E[S_2 - a \mid S_2 \ge a]}, \\
& 
G(1,a) = 
\frac{h}
{E[S_1 - a \mid S_1 \ge a] + E[S_2]}.
\end{split}
\label{eq:Gittins-index-21-average-IHR-IHR}
\end{equation}

\item[(iii)]
For the DHR-IHR case, 
\begin{equation}
\begin{split}
& 
G(2,a) = 
\frac{h}
{E[S_2 - a \mid S_2 \ge a]}, \\
& 
G(1,a) = 
\frac{h}
{\frac{1}{\mu_1(a)} + E[S_2]}.
\end{split}
\label{eq:Gittins-index-21-average-DHR-IHR}
\end{equation}

\item[(iv)]
For the IHR-DHR case, 
\begin{equation}
\begin{split}
& 
G(2,a) = 
h \mu_2(a), \\
& 
G(1,a) = 
h \, 
\sup_{\Delta \ge 0} 
\left( 
\frac{P\{S_2 \le \Delta\}}
{E[S_1 - a \mid S_1 \ge a] + E[\min\{S_2,\Delta\}]} 
\right).
\end{split}
\label{eq:Gittins-index-21-average-IHR-DHR}
\end{equation}

\end{itemize}
\end{corollary}

\vskip 6pt
Let then $G_j(a)$ denote the Gittins index for an ordinary single-stage job 
with service time equal to $S_j$. We know from 
\cite{Git89,Git11,Aal09QS,Aal11PEIS} that 
\begin{equation}
G_j(a) = h \sup_{\Delta \ge 0} J_j(a,\Delta), 
\quad a \ge 0, 
\label{eq:Gja}
\end{equation}
where 
\begin{equation}
J_j(a,\Delta) = 
\frac{P\{S_j - a \le \Delta \mid S_j > a\}}
{E[\min\{S_j - a,\Delta\} \mid S_j > a]}, 
\quad 
0 < \Delta < \infty, 
\label{eq:JjaDelta}
\end{equation}
with limits 
\begin{equation}
J_j(a,0) = \mu_j(a), 
\quad 
J_j(a,\infty) = \frac{1}{E[S_j - a \mid S_j > a]}.
\label{eq:JjaDelta-lim}
\end{equation}
In particular, if $S_j$ belongs to DHR, then, by \cite[Proposition~5]{Aal09QS}, 
\begin{equation}
G_j(a) = h J_j(a,0) = h \mu_j(a), 
\label{eq:Gja-DHR}
\end{equation}
and if it belongs to IHR, then, by \cite[Proposition~6]{Aal09QS}, 
\begin{equation}
G_j(a) = h J_j(a,\infty) = \frac{h}{E[S_j - a \mid S_j \ge a]}.
\label{eq:Gja-IHR}
\end{equation}
The results of the previous Corollary can now be summarized as follows.

\begin{theorem}
\label{thm:Gittins-index-average}
(Continuous time, undiscounted costs) \\
For a sequential two-stage job with monotonous hazard rates in both stages, 
\begin{equation}
\begin{split}
& 
G(2,a) = 
G_2(a), \\
& 
G(1,a) = 
h \, 
\sup_{\Delta \ge 0} 
\left( 
\frac{P\{S_2 \le \Delta\}}
{\frac{h}{G_1(a)} + E[\min\{S_2,\Delta\}]} 
\right).
\end{split}
\label{eq:Gittins-index-21-average}
\end{equation}
\end{theorem}

\paragraph{Proof} 
The proof is presented in Appendix~\ref{app:average-continuous-proof}.
\hfill $\Box$
\vskip 6pt

Note that Equation~(\ref{eq:Gittins-index-21-average}) gives a {\em recursive} 
way to compute the Gittins index for a two-stage job when we know the Gittins 
indexes $G_j(a)$ separately for each stage~$j$. Below we conjecture that this 
result can be generalized to sequential two-stage jobs with nonmonotonous 
hazard rate stages. We have not yet proved it but our numerical experiments 
support the claim. In 
Appendix~\ref{app:whittle-index-examples-average-multistage}, we present 
some examples of our numerical experiments.

\begin{conjecture}
\label{con:Gittins-index-average-two-stage}
(Continuous time, undiscounted costs, two stages) \\
For any sequential two-stage job, 
\begin{equation}
\begin{split}
& 
G(2,a) = 
G_2(a), \\
& 
G(1,a) = 
h \, 
\sup_{\Delta \ge 0} 
\left( 
\frac{P\{S_2 \le \Delta\}}
{\frac{h}{G_1(a)} + E[\min\{S_2,\Delta\}]} 
\right).
\end{split}
\label{eq:Gittins-index-21-average-two-stage}
\end{equation}
\end{conjecture}

In fact, our numerical experiments indicate that this recursive method could even 
be generalized to sequential multistage jobs with more than two stages, see 
Appendix~\ref{app:whittle-index-examples-average-multistage}. 

\begin{conjecture}
\label{con:Gittins-index-average-multistage}
(Continuous time, undiscounted costs, multiple stages) \\
For any sequential multistage job with $J$ stages, 
\begin{equation}
\begin{split}
& 
G(J,a) = 
G_J(a), \\
& 
G(j,a) = 
h \, 
\sup_{\Delta \ge 0} 
\left( 
\frac{P\{S_J \le \Delta\}}
{\frac{h}{G(j,a;j,j+1,\ldots,J-1)} + E[\min\{S_J,\Delta\}]} 
\right), \\
& \quad 
j \in {1,2,\ldots,J-1}, 
\end{split}
\label{eq:Gittins-index-J-1-average-multistage}
\end{equation}
where $G(j,a;j,j+1,\ldots,J-1)$ denotes the Gittins index for 
stage $j$ of a sequential multistage job consisting of stages 
$j,j+1,\ldots,J-1$.
\end{conjecture}

\section{Conclusions}
\label{sec:conclusions}
The optimal scheduling problem in single-server queueing systems is a classic 
problem in queueing theory. The Gittins index policy is known to be the 
optimal preemptive nonanticipating policy (both for the open version of the 
problem with Poisson arrivals and the closed version without arrivals) 
minimizing the expected holding costs. While the Gittins index is thoroughly 
characterized for ordinary jobs whose state is described by the attained 
service $a$, it is not the case with jobs that have more complex structure. 
Recently, a class of such jobs, the multistage jobs, were introduced and 
analyzed. The state of a multistage job is described by the pair $(j,a)$, 
where $j$ refers to the current stage and $a$ to the amount of attained 
service in the current stage. It was shown that the computation of Gittins 
index of a multistage job reduces into separable computations for the individual 
stages. The characterization is, however, indirect in the sense that it relies 
on the recursion for an auxiliary function (called SJP function) and not for 
the Gittins index itself.

In this paper, we tried to answer the natural remaining question: Is it possible 
to compute the Gittins index for a multistage job more directly by recursively 
combining the Gittins indexes of its individual stages? According to our results, 
it seems to be possible, at least, for sequential multistage jobs that have 
a fixed (deterministic) sequence of stages. We prove this for sequential 
two-stage jobs that have monotonous hazard rates in both stages, but our 
numerical experiments give an indication that the result could possibly be 
generalized to any sequential multistage jobs.

Our approach, in this paper, is based on the Whittle index originally developed 
in the context of restless bandits. The point is that, for multi-armed bandit 
problems, which our scheduling problem belongs to, the resulting Whittle index 
coincides with the Gittins index. We start with the closed version of the 
corresponding discrete-time single-server scheduling problem, and manage to 
derive the Whittle index for sequential two-stage jobs that have monotonous 
hazard rates in both stages related to minimization of the expected discounted 
holding costs. Even these novel results require quite a lot of work (covering 
the major part of the pages of the paper) and are of independent interest. 
However, from the general point of view, the consequences of these results for 
continuous-time problems with undiscounted (total or average) costs are even 
more interesting revealing the recursive way to compute the Gittins index for 
multistage jobs.

The next step in this research branch is to develop and present a firm proof 
of the two conjectures presented at the end of Section~\ref{sec:average-cost}, 
which are related to the generalization of the recursive rule to any sequential 
multistage jobs.




\newpage

\appendix

\section{Proof of Theorem~\ref{thm:Whittle-index-discounted-DHR-DHR} 
in the DHR-DHR-A subcase}
\label{app:DHR-DHR-A-proof}

\paragraph{Proof} 
We present here the proof of Theorem~\ref{thm:Whittle-index-discounted-DHR-DHR} 
for the DHR-DHR-A subcase. For the other two subcases (DHR-DHR-B and DHR-DHR-C), 
the proof is slightly different and presented 
in Appendices~\ref{app:DHR-DHR-B-proof} and \ref{app:DHR-DHR-C-proof}, 
respectively.

Assume the DHR-DHR-A subcase defined in 
(\ref{eq:DHR-DHR-A1}) and (\ref{eq:DHR-DHR-A2}). 
The main proof is given below in five parts ($1^\circ$--$5^\circ$). 
The idea is to solve the relaxed optimization problem 
(\ref{eq:separable-discounted-costs}) for any $\nu$ by utilizing the 
optimality equations (\ref{eq:opt-eqs-discounted-general}). For the 
proof, we partition the possible values of $\nu$, which is reflected 
by the five parts of the main proof.

For the main proof, we define a sequence of states $x_m$, 
$m \in \{1,2,\ldots\}$, recursively as follows: 
\begin{equation}
\begin{split}
& 
x_1 = (2,0), \\
& 
x_{m+1} = 
\left\{
\begin{array}{ll}
(2,\tilde n_2(m)), & \quad 
\hbox{if $W_\beta(2,\tilde n_2(m)) = 
\max_{x \in {\mathcal S} \setminus \{x_1,\ldots,x_m,*\}} W_\beta(x)$}, \\
(1,\tilde n_1(m)), & \quad 
\hbox{otherwise}, \\
\end{array}
\right.
\end{split}
\label{eq:x-def-DHR-DHR}
\end{equation}
where 
\[
\begin{split}
& 
\tilde n_2(m) = \min\{ n_2 : (2,n_2) \in {\mathcal S} \setminus \{x_1,\ldots,x_m,*\} \}, \\
& 
\tilde n_1(m) = \min\{ n_1 : (1,n_1) \in {\mathcal S} \setminus \{x_1,\ldots,x_m,*\} \}.
\end{split}
\]
We note that, in this DHR-DHR-A subcase, the sequence $(x_m)$ covers all 
the states but $*$, 
\[
\{x_1,x_2,\ldots\} = {\mathcal S} \setminus \{*\}, 
\]
and defines the following ordering among these states: 
\begin{equation}
W_\beta(x_1) \ge W_\beta(x_2) \ge \ldots \ge W_\beta(x_\infty) \ge 0, 
\label{eq:Wx-order-DHR-DHR-A}
\end{equation}
where we have defined 
\begin{equation}
W_\beta(x_\infty) = w_1(\infty) = w_2(\infty) = 
h \mu_2(\infty) \, 
\frac{\beta}{1 - \beta}.
\label{eq:Whittle-index-x-infty-discounted-DHR-DHR-A}
\end{equation}

In addition, for any $m \in \{1,2,\ldots\}$, there are $n_2$ and $n_1$ 
such that $n_2 \in \{0,1,\ldots\}$, $n_1 \in \{-1\} \cup \{0,1,\ldots\}$, and 
\[
\{x_1,\ldots,x_m\} = 
\{(2,0),(2,1),\ldots,(2,n_2)\} \cup \{(1,0),(1,1),\ldots,(1,n_1)\}, 
\]
where the latter part of the right hand side is omitted if $n_1 = -1$. 
If $n_1 \ge 0$, then 
\begin{equation}
\phi(n_1) \le n_2, 
\label{eq:phi-result1-DHR-DHR}
\end{equation}
due to Lemma~\ref{lem:Whittle-index-discounted-DHR-DHR-lemma-1}(i) 
since 
\[
W_\beta(2,\phi(n_1)) \ge W_\beta(1,n_1) > W_\beta(2,n_2+1) 
\]
by (\ref{eq:Whittle-index-order-DHR-DHR}) and (\ref{eq:x-def-DHR-DHR}), 
respectively. Moreover, for any $n_1 \ge -1$, 
\begin{equation}
\phi(n_1+1) \ge n_2, 
\label{eq:phi-result2-DHR-DHR}
\end{equation}
since otherwise, by (\ref{eq:Whittle-index-order-DHR-DHR}) and 
Lemma~\ref{lem:Whittle-index-discounted-DHR-DHR-lemma-1}(i), 
\[
W_\beta(1,n_1+1) > W_\beta(2,\phi(n_1+1)+1) \ge W_\beta(2,n_2), 
\]
which were a contradiction.

\vskip 3pt
$1^\circ$ 
We first assume that $\nu \ge 0$. 
In this case, the optimal decision in state $*$ is not to schedule ($a = 0$), 
the minimum expected discounted cost $V_\beta(*;\nu)$ equals $0$, and the 
optimality equations (\ref{eq:opt-eqs-discounted-general}) read as follows: 
\begin{equation}
\begin{split}
& 
V_\beta(1,n;\nu) = 
h + \min \big\{ \beta V_\beta(1,n;\nu), \\
& \quad 
\nu + \beta \mu_1(n) V_\beta(2,0;\nu) + 
\beta (1 - \mu_1(n)) V_\beta(1,n+1;\nu) \big\}, \\
& 
V_\beta(2,n;\nu) = 
h + \min \big\{ \beta V_\beta(2,n;\nu), \\
& \quad 
\nu + \beta (1 - \mu_2(n)) V_\beta(2,n+1;\nu) \big\}.
\end{split}
\label{eq:opt-eqs-discounted-nu-pos}
\end{equation}
We prove that the policy $\pi$ with activity set 
\[
{\mathcal B}^\pi = \emptyset, 
\]
according to which user~$k$ is not scheduled in any state $x \in {\mathcal S}$, 
is $(\nu,\beta)$-optimal for all 
\[
\nu \in [W_\beta(x_1),\infty), 
\]
where $x_1 = (2,0)$ as defined in (\ref{eq:x-def-DHR-DHR}). 
It remains to prove that policy $\pi$ is optimal for these values of $\nu$ 
in any state $x \in {\mathcal S} \setminus \{*\}$.

We start by first deriving the value function $V_\beta^\pi(x;\nu)$ for 
policy $\pi$ from the so called Howard equations: 
\begin{equation}
V_\beta^\pi(x;\nu) = 
h + \beta V_\beta^\pi(x;\nu), 
\quad x \in {\mathcal S} \setminus \{*\}.
\label{eq:howard-eqs-discounted-DHR-DHR-case-1}
\end{equation}
The unique solution of these linear equations is clearly given by 
\begin{equation}
V_\beta^\pi(x;\nu) = \frac{h}{1 - \beta}, 
\quad x \in {\mathcal S} \setminus \{*\}.
\label{eq:howard-eqs-discounted-DHR-DHR-case-1-solution}
\end{equation}

By (\ref{eq:howard-eqs-discounted-DHR-DHR-case-1-solution}), the following condition 
for optimality of $\pi$ (based on (\ref{eq:opt-eqs-discounted-nu-pos})) in 
state~$x_1 = (2,0)$, 
\[
\beta V_\beta^\pi(2,0;\nu) \le 
\nu + \beta (1 - \mu_2(0)) V_\beta^\pi(2,1;\nu), 
\]
is easily shown to be equivalent with 
\begin{equation}
\nu \ge 
h \mu_2(0) \, \frac{\beta}{1 - \beta}, 
\label{eq:nu-req-discounted-DHR-DHR-case-1}
\end{equation}
where the right hand side equals $W_\beta(2,0)$ given in 
(\ref{eq:Whittle-index-2n-discounted-DHR-DHR}).

Let then $(2,n) \in {\mathcal S} \setminus \{*\}$. Similarly 
by (\ref{eq:howard-eqs-discounted-DHR-DHR-case-1-solution}), the following condition 
for optimality of $\pi$ in state~$(2,n)$, 
\[
\beta V_\beta^\pi(2,n;\nu) \le 
\nu + \beta (1 - \mu_2(n)) V_\beta^\pi(2,n+1;\nu), 
\]
is easily shown to be equivalent with condition 
\[
\nu \ge 
h \mu_2(n) \, \frac{\beta}{1 - \beta}, 
\]
which follows from (\ref{eq:nu-req-discounted-DHR-DHR-case-1}) since 
$\mu_2(0) \ge \mu_2(n)$.

Finally, let $(1,n) \in {\mathcal S} \setminus \{*\}$. Again 
by (\ref{eq:howard-eqs-discounted-DHR-DHR-case-1-solution}), the following condition 
for optimality of $\pi$ in state~$(1,n)$, 
\[
\beta V_\beta^\pi(1,n;\nu) \le 
\nu + \beta \mu_1(n) V_\beta^\pi(2,0;\nu) + 
\beta (1 - \mu_1(n)) V_\beta^\pi(1,n+1;\nu), 
\]
is easily shown to be equivalent with condition $\nu \ge 0$, 
which is satisfied by the assumption made in the beginning of 
this part of the proof. This completes the proof of claim~$1^\circ$.

\vskip 3pt
$2^\circ$ 
We still assume that $\nu \ge 0$ and utilize the optimality equations 
(\ref{eq:opt-eqs-discounted-nu-pos}) given in the beginning of 
part $1^\circ$ of the proof. 
Let $m \in \{1,2,\ldots\}$ and $n_2 \in \{0,1,\ldots\}$ such that 
\[
\{x_1,\ldots,x_m\} = 
\{(2,0),(2,1),\ldots,(2,n_2)\}, 
\]
where $x_m$ is defined in (\ref{eq:x-def-DHR-DHR}). 
We prove that the policy $\pi$ with activity set 
\[
{\mathcal B}^\pi = \{x_1,\ldots,x_m\}, 
\]
according to which user~$k$ is scheduled in states $x_1,\ldots,x_m$, is 
$(\nu,\beta)$-optimal for all 
\[
\nu \in [W_\beta(x_{m+1}),W_\beta(x_m)].
\]
It remains to prove that policy $\pi$ is optimal for these values of $\nu$ 
in any state $x \in {\mathcal S} \setminus \{*\}$. 
This is done below in two parts ($2.1^\circ$ and $2.2^\circ$).

\vskip 3pt
$2.1^\circ$ 
We start by first deriving the value function $V_\beta^\pi(x;\nu)$ for 
policy $\pi$ from the Howard equations: 
\begin{equation}
\begin{split}
& 
V_\beta^\pi(2,n;\nu) = 
h + \nu + \beta (1 - \mu_2(n)) V_\beta^\pi(2,n+1;\nu), 
\quad 
n \in \{0,1,\ldots,n_2\}, \\
& 
V_\beta^\pi(x;\nu) = 
h + \beta V_\beta^\pi(x;\nu), 
\quad x \in {\mathcal S} \setminus \{x_1,\ldots,x_m,*\}.
\end{split}
\label{eq:howard-eqs-discounted-DHR-DHR-case-2}
\end{equation}
The unique solution of these linear equations is given by 
\begin{equation}
\begin{split}
& 
V_\beta^\pi(2,n;\nu) = 
(h + \nu) 
\left( 
\sum_{i = 0}^{n_2-n} \beta^i \bar p_2(i|n) 
\right) \; + \\
& \quad 
h \, \frac{\beta^{n_2-n+1} \bar p_2(n_2-n+1|n)}{1 - \beta}, 
\quad 
n \in \{0,1,\ldots,n_2\}, \\
& 
V_\beta^\pi(x;\nu) = 
\frac{h}{1 - \beta}, 
\quad x \in {\mathcal S} \setminus \{x_1,\ldots,x_m,*\}.
\end{split}
\label{eq:howard-eqs-discounted-DHR-DHR-case-2-solution}
\end{equation}

By (\ref{eq:howard-eqs-discounted-DHR-DHR-case-2-solution}) and some algebraic 
manipulations, the following condition for optimality of $\pi$ in 
state~$x_m = (2,n_2)$, 
\[
V_\beta^\pi(2,n_2;\nu) \le \frac{h}{1 - \beta}, 
\]
can be shown to be equivalent with 
\begin{equation}
\nu \le 
h \mu_2(n_2) \, \frac{\beta}{1 - \beta}, 
\label{eq:nu-req-discounted-DHR-DHR-case-21}
\end{equation}
where the right hand side equals $W_\beta(2,n_2)$ given in 
(\ref{eq:Whittle-index-2n-discounted-DHR-DHR}).

Let then $n \in \{0,1,\ldots,n_2-1\}$. 
Again by (\ref{eq:howard-eqs-discounted-DHR-DHR-case-2-solution}), 
the following condition for optimality of $\pi$ in state~$(2,n)$, 
\[
V_\beta^\pi(2,n;\nu) \le \frac{h}{1 - \beta}, 
\]
can be shown to be equivalent with 
\begin{equation}
\nu \le 
h \, 
\frac{\sum_{i = 0}^{n_2-n} \beta^i p_2(i|n)}
{\sum_{i = 0}^{n_2-n} \beta^i \bar p_2(i|n)} \, 
\frac{\beta}{1 - \beta}.
\label{eq:nu-req-discounted-DHR-DHR-case-21-cont}
\end{equation}
Thus, condition (\ref{eq:nu-req-discounted-DHR-DHR-case-21-cont}) 
follows from (\ref{eq:nu-req-discounted-DHR-DHR-case-21}) 
by Lemma~\ref{lem:Whittle-index-discounted-DHR-DHR-lemma-1}(v).

\vskip 3pt
$2.2^\circ$ 
By (\ref{eq:howard-eqs-discounted-DHR-DHR-case-2-solution}), the following condition 
for optimality of $\pi$ in state~$(2,n_2+1)$, 
\[
\beta V_\beta^\pi(2,n_2+1;\nu) \le 
\nu + \beta (1 - \mu_2(n_2+1)) V_\beta^\pi(2,n_2+2;\nu), 
\]
is easily shown to be equivalent with 
\begin{equation}
\nu \ge 
h \mu_2(n_2+1) \, \frac{\beta}{1 - \beta}, 
\label{eq:nu-req-discounted-DHR-DHR-case-22a}
\end{equation}
where the right hand side equals $W_\beta(2,n_2+1)$ given in 
(\ref{eq:Whittle-index-2n-discounted-DHR-DHR}).

Let then $n \in \{n_2+2,n_2+3,\ldots\}$. 
Again by (\ref{eq:howard-eqs-discounted-DHR-DHR-case-2-solution}), 
the following condition for optimality of $\pi$ in state~$(2,n)$, 
\[
\beta V_\beta^\pi(2,n;\nu) \le 
\nu + \beta (1 - \mu_2(n)) V_\beta^\pi(2,n+1;\nu), 
\]
is easily shown to be equivalent with 
\[
\nu \ge 
h \mu_2(n) \, \frac{\beta}{1 - \beta}, 
\]
which follows from (\ref{eq:nu-req-discounted-DHR-DHR-case-22a}) since 
$\mu_2(n)$ is decreasing.

On the other hand, by (\ref{eq:howard-eqs-discounted-DHR-DHR-case-2-solution}) 
and some algebraic manipulations, the following condition for optimality 
of $\pi$ in state~$(1,0)$, 
\[
\beta V_\beta^\pi(1,0;\nu) \le 
\nu + \beta \mu_1(0) V_\beta^\pi(2,0;\nu) + 
\beta (1 - \mu_1(0)) V_\beta^\pi(1,1;\nu), 
\]
can be shown to be equivalent with 
\begin{equation}
\nu \ge 
h \mu_1(0) \, 
\frac{\beta \sum_{i=0}^{n_2} \beta^i p_2(i)}{1 + \beta \mu_1(0) \sum_{i=0}^{n_2} \beta^i \bar p_2(i)} \, 
\frac{\beta}{1 - \beta}, 
\label{eq:nu-req-discounted-DHR-DHR-case-22b}
\end{equation}
where the right hand side equals $\psi(0,n_2)$ given in 
(\ref{eq:psi-discounted-DHR-DHR}).

Let then $n \in \{1,2,\ldots\}$. 
Again by (\ref{eq:howard-eqs-discounted-DHR-DHR-case-2-solution}) and some 
algebraic manipulations, the following condition for optimality of 
$\pi$ in state~$(1,n)$, 
\[
\beta V_\beta^\pi(1,n;\nu) \le 
\nu + \beta \mu_1(n) V_\beta^\pi(2,0;\nu) + 
\beta (1 - \mu_1(n)) V_\beta^\pi(1,n+1;\nu), 
\]
can be shown to be equivalent with 
\[
\nu \ge 
h \mu_1(n) \, 
\frac{\beta \sum_{i=0}^{n_2} \beta^i p_2(i)}{1 + \beta \mu_1(n) \sum_{i=0}^{n_2} \beta^i \bar p_2(i)} \, 
\frac{\beta}{1 - \beta}, 
\]
which follows from (\ref{eq:nu-req-discounted-DHR-DHR-case-22b}) since 
$\mu_1(n)$ is decreasing.

Finally, by combining (\ref{eq:nu-req-discounted-DHR-DHR-case-22a}) and 
(\ref{eq:nu-req-discounted-DHR-DHR-case-22b}), we get the requirement that 
\begin{equation}
\nu \ge 
\max \{ W_\beta(2,n_2+1), \psi(0,n_2) \}.
\label{eq:nu-req-discounted-DHR-DHR-case-22}
\end{equation}
Now if $W_\beta(2,n_2+1) < \psi(0,n_2)$, then $\phi(0) = n_2$ by 
(\ref{eq:phi-discounted-DHR-DHR}) and (\ref{eq:phi-result2-DHR-DHR}). 
In addition, by (\ref{eq:Whittle-index-1n-discounted-DHR-DHR}), the right 
hand side of (\ref{eq:nu-req-discounted-DHR-DHR-case-22}) equals, under this condition, 
\[
W_\beta(1,0) = \psi(0,n_2) 
\]
and, furthermore, by (\ref{eq:x-def-DHR-DHR}), we have 
\[
x_{m+1} = (1,0) 
\]
so that requirement (\ref{eq:nu-req-discounted-DHR-DHR-case-32}) is equivalent 
with the requirement 
\[
\nu \ge W_\beta(x_{m+1}).
\]
On the other hand, if $W_\beta(2,n_2+1) \ge \psi(0,n_2)$, then, 
$\phi(0) > n_2$ by (\ref{eq:phi-discounted-DHR-DHR}) and 
we have, by (\ref{eq:Whittle-index-order-DHR-DHR}), 
\[
W_\beta(2,n_2+1) \ge W_\beta(2,\phi(0)) \ge W_\beta(1,0) 
\]
and, furthermore, by (\ref{eq:x-def-DHR-DHR}), 
\[
x_{m+1} = (2,n_2+1) 
\]
so that requirement (\ref{eq:nu-req-discounted-DHR-DHR-case-22}) is again equivalent 
with the requirement 
\[
\nu \ge W_\beta(x_{m+1}), 
\]
which completes the proof of claim~$2^\circ$.

\vskip 3pt
$3^\circ$ 
We still assume that $\nu \ge 0$ and utilize the optimality equations 
(\ref{eq:opt-eqs-discounted-nu-pos}). 
Let $m \in \{2,3,\ldots\}$ and $n_1, n_2 \in \{0,1,\ldots\}$ such that 
\[
\{x_1,\ldots,x_m\} = 
\{(2,0),(2,1),\ldots,(2,n_2)\} \cup \{(1,0),(1,1),\ldots,(1,n_1)\}, 
\]
where $x_m$ is defined in (\ref{eq:x-def-DHR-DHR}). 
We prove that the policy $\pi$ with activity set 
\[
{\mathcal B}^\pi = \{x_1,\ldots,x_m\}, 
\]
according to which user~$k$ is scheduled in states $x_1,\ldots,x_m$, is 
$(\nu,\beta)$-optimal for all 
\[
\nu \in [W_\beta(x_{m+1}),W_\beta(x_m)].
\]
It remains to prove that policy $\pi$ is optimal for these values of $\nu$ 
in any state $x \in {\mathcal S} \setminus \{*\}$. 
This is done below in two parts ($3.1^\circ$ and $3.2^\circ$).

\vskip 3pt
$3.1^\circ$ 
We start by first deriving the value function $V_\beta^\pi(x;\nu)$ for 
policy $\pi$ from the Howard equations: 
\begin{equation}
\begin{split}
& 
V_\beta^\pi(1,n;\nu) = 
h + \nu + \beta \mu_1(n) V_\beta^\pi(2,0;\nu) + 
\beta (1 - \mu_1(n)) V_\beta^\pi(1,n+1;\nu), \\
& \quad 
n \in \{0,1,\ldots,n_1\}, \\
& 
V_\beta^\pi(2,n;\nu) = 
h + \nu + \beta (1 - \mu_2(n)) V_\beta^\pi(2,n+1;\nu), \\
& \quad 
n \in \{0,1,\ldots,n_2\}, \\
& 
V_\beta^\pi(x;\nu) = 
h + \beta V_\beta^\pi(x;\nu), 
\quad x \in {\mathcal S} \setminus \{x_1,\ldots,x_m,*\}.
\end{split}
\label{eq:howard-eqs-discounted-DHR-DHR-case-3}
\end{equation}
The unique solution of these linear equations is given by 
\begin{equation}
\begin{split}
& 
V_\beta^\pi(2,n;\nu) = 
(h + \nu) 
\left( 
\sum_{i = 0}^{n_2-n} \beta^i \bar p_2(i|n) 
\right) \; + \\
& \quad 
h \, \frac{\beta^{n_2-n+1} \bar p_2(n_2-n+1|n)}{1 - \beta}, 
\quad 
n \in \{0,1,\ldots,n_2\}, \\
& 
V_\beta^\pi(1,n;\nu) = 
(h + \nu) 
\left( 
\sum_{i = 0}^{n_1-n} \beta^i \bar p_1(i|n)
\left( 
1 + \beta \mu_1(n+i) \frac{V_\beta^\pi(2,0;\nu)}{h + \nu} 
\right) 
\right) \; + \\
& \quad 
h \, \frac{\beta^{n_1-n+1} \bar p_1(n_1-n+1|n)}{1 - \beta}, 
\quad 
n \in \{0,1,\ldots,n_1\}, \\
& 
V_\beta^\pi(x;\nu) = 
\frac{h}{1 - \beta}, 
\quad x \in {\mathcal S} \setminus \{x_1,\ldots,x_m,*\}.
\end{split}
\label{eq:howard-eqs-discounted-DHR-DHR-case-3-solution}
\end{equation}

By (\ref{eq:howard-eqs-discounted-DHR-DHR-case-3-solution}) and some algebraic 
manipulations, the following condition for optimality of $\pi$ in 
state~$(2,n_2)$, 
\[
V_\beta^\pi(2,n_2;\nu) \le \frac{h}{1 - \beta}, 
\]
can be shown to be equivalent with 
\begin{equation}
\nu \le 
h \mu_2(n_2) \, \frac{\beta}{1 - \beta}, 
\label{eq:nu-req-discounted-DHR-DHR-case-31a}
\end{equation}
where the right hand side equals $W_\beta(2,n_2)$ given in 
(\ref{eq:Whittle-index-2n-discounted-DHR-DHR}).

Let then $n \in \{0,1,\ldots,n_2-1\}$. 
Again by (\ref{eq:howard-eqs-discounted-DHR-DHR-case-3-solution}), 
the following condition for optimality of $\pi$ in state~$(2,n)$, 
\[
V_\beta^\pi(2,n;\nu) \le \frac{h}{1 - \beta}, 
\]
can be shown to be equivalent with 
\begin{equation}
\nu \le 
h \, 
\frac{\sum_{i = 0}^{n_2-n} \beta^i p_2(i|n)}
{\sum_{i = 0}^{n_2-n} \beta^i \bar p_2(i|n)} \, 
\frac{\beta}{1 - \beta}.
\label{eq:nu-req-discounted-DHR-DHR-case-31a-cont}
\end{equation}
Thus, condition (\ref{eq:nu-req-discounted-DHR-DHR-case-31a-cont}) 
follows from (\ref{eq:nu-req-discounted-DHR-DHR-case-31a}) 
by Lemma~\ref{lem:Whittle-index-discounted-DHR-DHR-lemma-1}(v).

On the other hand, by (\ref{eq:howard-eqs-discounted-DHR-DHR-case-3-solution}) 
and some algebraic manipulations, the following condition for optimality 
of $\pi$ in state~$(1,n_1)$, 
\[
V_\beta^\pi(1,n_1;\nu) \le \frac{h}{1 - \beta}, 
\]
can be shown to be equivalent with 
\begin{equation}
\nu \le 
h \mu_1(n_1) \, 
\frac{\beta \sum_{i=0}^{n_2} \beta^i p_2(i)}{1 + \beta \mu_1(n_1) \sum_{i=0}^{n_2} \beta^i \bar p_2(i)} \, 
\frac{\beta}{1 - \beta}, 
\label{eq:nu-req-discounted-DHR-DHR-case-31b}
\end{equation}
where the right hand side equals $\psi(n_1,n_2)$ given in 
(\ref{eq:psi-discounted-DHR-DHR}).

Next we show that if conditions 
(\ref{eq:nu-req-discounted-DHR-DHR-case-31a}) and 
(\ref{eq:nu-req-discounted-DHR-DHR-case-31b}) are satisfied, then 
\begin{equation}
V_\beta^\pi(1,0;\nu) \le V_\beta^\pi(1,1;\nu) \le \ldots \le V_\beta^\pi(1,n_1;\nu) \le \frac{h}{1 - \beta}, 
\label{eq:nu-req-discounted-DHR-DHR-case-31b-cont1}
\end{equation}
which implies that policy $\pi$ is optimal in all states 
$\{(1,0),(1,1),\ldots,(1,n_1)\}$. First, as we saw above, condition 
(\ref{eq:nu-req-discounted-DHR-DHR-case-31a}) implies that 
$V_\beta^\pi(2,0;\nu) \le \frac{h}{1 - \beta}$, and condition 
(\ref{eq:nu-req-discounted-DHR-DHR-case-31b}) implies that 
$V_\beta^\pi(1,n_1;\nu) \le \frac{h}{1 - \beta}$. 
Now, by (\ref{eq:howard-eqs-discounted-DHR-DHR-case-3}), we observe that 
\[
\begin{split}
& 
V_\beta^\pi(1,n_1;\nu) - V_\beta^\pi(1,n_1-1;\nu) \; = \\
& \quad 
\beta 
\Big[
\Big( \mu_1(n_1) V_\beta^\pi(2,0;\nu) + (1 - \mu_1(n_1)) \frac{h}{1 - \beta} \Big) \; - \\
& \quad\quad 
\Big( \mu_1(n_1-1) V_\beta^\pi(2,0;\nu) + (1 - \mu_1(n_1-1)) V_\beta^\pi(1,n_1;\nu) \Big) 
\Big] 
\ge 0, 
\end{split}
\]
where the last inequality follows from the facts that $\mu_1(n_1) \le \mu_1(n_1-1)$, 
$V_\beta^\pi(2,0;\nu) \le \frac{h}{1 - \beta}$, and 
$V_\beta^\pi(1,n_1;\nu) \le \frac{h}{1 - \beta}$. 
In addition, again by (\ref{eq:howard-eqs-discounted-DHR-DHR-case-3}), we have 
\[
\begin{split}
& 
V_\beta^\pi(1,n_1;\nu) - V_\beta^\pi(2,0;\nu) \; = \\
& \quad 
h + \nu + \beta (1 - \mu_1(n_1)) \frac{h}{1 - \beta} - 
(1 - \beta \mu_1(n_1)) V_\beta^\pi(2,0;\nu) \; = \\
& \quad 
(1 - \beta) \Big( \frac{h + \nu}{1 - \beta} - V_\beta^\pi(2,0;\nu) \Big) + 
\beta (1 - \mu_1(n_1)) \Big( \frac{h}{1 - \beta} - V_\beta^\pi(2,0;\nu) \Big) 
\ge 0, 
\end{split}
\]
where the last inequality follows from the fact that 
$V_\beta^\pi(2,0;\nu) \le \frac{h}{1 - \beta} \le \frac{h + \nu}{1 - \beta}$. 
So, we have proved that if conditions 
(\ref{eq:nu-req-discounted-DHR-DHR-case-31a}) and 
(\ref{eq:nu-req-discounted-DHR-DHR-case-31b}) are satisfied, then 
\begin{equation}
V_\beta^\pi(2,0;\nu) \le V_\beta^\pi(1,n_1;\nu), \quad 
V_\beta^\pi(1,n_1-1;\nu) \le V_\beta^\pi(1,n_1;\nu).
\label{eq:nu-req-discounted-DHR-DHR-case-31b-cont2}
\end{equation}
Next, again by (\ref{eq:howard-eqs-discounted-DHR-DHR-case-3}), we observe that 
\[
\begin{split}
& 
V_\beta^\pi(1,n_1-1;\nu) - V_\beta^\pi(1,n_1-2;\nu) \; = \\
& \quad 
\beta 
\Big[
\Big( \mu_1(n_1-1) V_\beta^\pi(2,0;\nu) + (1 - \mu_1(n_1-1)) V_\beta^\pi(1,n_1;\nu) \Big) \; - \\
& \quad\quad 
\Big( \mu_1(n_1-2) V_\beta^\pi(2,0;\nu) + (1 - \mu_1(n_1-2)) V_\beta^\pi(1,n_1-1;\nu) \Big) 
\Big] 
\ge 0, 
\end{split}
\]
where the last inequality follows from 
(\ref{eq:nu-req-discounted-DHR-DHR-case-31b-cont2}) and 
the fact that $\mu_1(n_1-1) \le \mu_1(n_1-2)$. 
In addition, again by (\ref{eq:howard-eqs-discounted-DHR-DHR-case-3}), we have 
\[
\begin{split}
& 
V_\beta^\pi(1,n_1-1;\nu) - V_\beta^\pi(2,0;\nu) \; = \\
& \quad 
h + \nu + \beta (1 - \mu_1(n_1-1)) V_\beta^\pi(1,n_1;\nu) - 
(1 - \beta \mu_1(n_1-1)) V_\beta^\pi(2,0;\nu) \; = \\
& \quad 
(1 - \beta) \Big( \frac{h + \nu}{1 - \beta} - V_\beta^\pi(2,0;\nu) \Big) \; + \\
& \quad\quad 
\beta (1 - \mu_1(n_1-1)) \Big( V_\beta^\pi(1,n_1;\nu) - V_\beta^\pi(2,0;\nu) \Big) 
\ge 0, 
\end{split}
\]
where the last inequality follows from the fact that 
$V_\beta^\pi(2,0;\nu) \le V_\beta^\pi(1,n_1;\nu) \le \frac{h}{1 - \beta} \le \frac{h + \nu}{1 - \beta}$. 
So, we have proved that if conditions 
(\ref{eq:nu-req-discounted-DHR-DHR-case-31a}) and 
(\ref{eq:nu-req-discounted-DHR-DHR-case-31b}) are satisfied, then 
\begin{equation}
V_\beta^\pi(2,0;\nu) \le V_\beta^\pi(1,n_1-1;\nu), \quad 
V_\beta^\pi(1,n_1-2;\nu) \le V_\beta^\pi(1,n_1-1;\nu).
\label{eq:nu-req-discounted-DHR-DHR-case-31b-cont3}
\end{equation}
Continuing similarly proves (\ref{eq:nu-req-discounted-DHR-DHR-case-31b-cont1}).

Finally, by combining (\ref{eq:nu-req-discounted-DHR-DHR-case-31a}) and 
(\ref{eq:nu-req-discounted-DHR-DHR-case-31b}), we get the requirement that 
\begin{equation}
\nu \le 
\min \{ W_\beta(2,n_2), \psi(n_1,n_2) \}.
\label{eq:nu-req-discounted-DHR-DHR-case-31}
\end{equation}
Now, if $W_\beta(2,n_2) < \psi(n_1,n_2)$, then the right hand side of 
(\ref{eq:nu-req-discounted-DHR-DHR-case-31}) equals $W_\beta(2,n_2)$ and this 
requirement (\ref{eq:nu-req-discounted-DHR-DHR-case-31}) is satisfied by our 
assumption that $\nu \le W_\beta(x_m)$, since $(2,n_2) \in \{x_1,\ldots,x_m\}$ 
and, by (\ref{eq:Wx-order-DHR-DHR-A}), 
\[
W_\beta(x_m) = \min\{W_\beta(x_1),\ldots,W_\beta(x_m)\} \le W_\beta(2,n_2).
\]
Assume now that 
\[
W_\beta(2,n_2) \ge \psi(n_1,n_2).
\]
Below we show that in this case $\phi(n_1) = n_2$, which implies that 
\[
W_\beta(1,n_1) = \psi(n_1,n_2).
\]
If $\phi(n_1) < n_2$, then we have, by 
(\ref{eq:phi-discounted-DHR-DHR}), 
\[
\psi(n_1,\phi(n_1)) > w_2(\phi(n_1)+1), 
\]
which is, by Lemma~\ref{lem:Whittle-index-discounted-DHR-DHR-lemma-1}(iv), 
equivalent with 
\[
\psi(n_1,\phi(n_1)+1) > w_2(\phi(n_1)+1).
\]
But, by Lemma~\ref{lem:Whittle-index-discounted-DHR-DHR-lemma-1}(i), this implies that 
\[
\psi(n_1,\phi(n_1)+1) > w_2(\phi(n_1)+2), 
\]
which is, again by Lemma~\ref{lem:Whittle-index-discounted-DHR-DHR-lemma-1}(iv), 
equivalent with 
\[
\psi(n_1,\phi(n_1)+2) > w_2(\phi(n_1)+2).
\]
By continuing similarly, we finally end up to the following inequality: 
\[
\psi(n_1,n_2) > w_2(n_2), 
\]
which, however, contradicts our assumption above (i.e., 
$W_\beta(2,n_2) = w_2(n_2) \ge \psi(n_1,n_2)$). 
So, by further taking into account (\ref{eq:phi-result1-DHR-DHR}), we have now 
proved that $\phi(n_1) = n_2$ in this case, which imples that the right hand side 
of (\ref{eq:nu-req-discounted-DHR-DHR-case-31}) equals $W_\beta(1,n_1)$. In addition, 
this requirement (\ref{eq:nu-req-discounted-DHR-DHR-case-31}) is satisfied by our 
assumption that $\nu \le W_\beta(x_m)$, since $(1,n_1) \in \{x_1,\ldots,x_m\}$ 
and, by (\ref{eq:Wx-order-DHR-DHR-A}), 
\[
W_\beta(x_m) = \min\{W_\beta(x_1),\ldots,W_\beta(x_m)\} \le W_\beta(1,n_1).
\]

\vskip 3pt
$3.2^\circ$ 
By (\ref{eq:howard-eqs-discounted-DHR-DHR-case-3-solution}), the following condition 
for optimality of $\pi$ in state~$(2,n_2+1)$, 
\[
\beta V_\beta^\pi(2,n_2+1;\nu) \le 
\nu + \beta (1 - \mu_2(n_2+1)) V_\beta^\pi(2,n_2+2;\nu), 
\]
is easily shown to be equivalent with 
\begin{equation}
\nu \ge 
h \mu_2(n_2+1) \, \frac{\beta}{1 - \beta}, 
\label{eq:nu-req-discounted-DHR-DHR-case-32a}
\end{equation}
where the right hand side equals $W_\beta(2,n_2+1)$ given in 
(\ref{eq:Whittle-index-2n-discounted-DHR-DHR}).

Let then $n \in \{n_2+2,n_2+3,\ldots\}$. 
Again by (\ref{eq:howard-eqs-discounted-DHR-DHR-case-3-solution}), 
the following condition for optimality of $\pi$ in state~$(2,n)$, 
\[
\beta V_\beta^\pi(2,n;\nu) \le 
\nu + \beta (1 - \mu_2(n)) V_\beta^\pi(2,n+1;\nu), 
\]
is easily shown to be equivalent with 
\[
\nu \ge 
h \mu_2(n) \, \frac{\beta}{1 - \beta}, 
\]
which follows from (\ref{eq:nu-req-discounted-DHR-DHR-case-32a}) since 
$\mu_2(n)$ is decreasing.

On the other hand, by (\ref{eq:howard-eqs-discounted-DHR-DHR-case-3-solution}) 
and some algebraic manipulations, the following condition for optimality 
of $\pi$ in state~$(1,n_1+1)$, 
\[
\beta V_\beta^\pi(1,n_1+1;\nu) \le 
\nu + \beta \mu_1(n_1+1) V_\beta^\pi(2,0;\nu) + 
\beta (1 - \mu_1(n_1+1)) V_\beta^\pi(1,n_1+2;\nu), 
\]
can be shown to be equivalent with 
\begin{equation}
\nu \ge 
h \mu_1(n_1+1) \, 
\frac{\beta \sum_{i=0}^{n_2} \beta^i p_2(i)}{1 + \beta \mu_1(n_1+1) \sum_{i=0}^{n_2} \beta^i \bar p_2(i)} \, 
\frac{\beta}{1 - \beta}, 
\label{eq:nu-req-discounted-DHR-DHR-case-32b}
\end{equation}
where the right hand side equals $\psi(n_1+1,n_2)$ given in 
(\ref{eq:psi-discounted-DHR-DHR}).

Let then $n \in \{n_1+2,n_1+3,\ldots\}$. 
Again by (\ref{eq:howard-eqs-discounted-DHR-DHR-case-3-solution}) and some 
algebraic manipulations, the following condition for optimality of 
$\pi$ in state~$(1,n)$, 
\[
\beta V_\beta^\pi(1,n;\nu) \le 
\nu + \beta \mu_1(n) V_\beta^\pi(2,0;\nu) + 
\beta (1 - \mu_1(n)) V_\beta^\pi(1,n+1;\nu), 
\]
can be shown to be equivalent with 
\[
\nu \ge 
h \mu_1(n) \, 
\frac{\beta \sum_{i=0}^{n_2} \beta^i p_2(i)}{1 + \beta \mu_1(n) \sum_{i=0}^{n_2} \beta^i \bar p_2(i)} \, 
\frac{\beta}{1 - \beta}, 
\]
which follows from (\ref{eq:nu-req-discounted-DHR-DHR-case-32b}) since 
$\mu_1(n)$ is decreasing.

Finally, by combining (\ref{eq:nu-req-discounted-DHR-DHR-case-32a}) and 
(\ref{eq:nu-req-discounted-DHR-DHR-case-32b}), we get the requirement that 
\begin{equation}
\nu \ge 
\max \{ W_\beta(2,n_2+1), \psi(n_1+1,n_2) \}.
\label{eq:nu-req-discounted-DHR-DHR-case-32}
\end{equation}
Now if $W_\beta(2,n_2+1) < \psi(n_1+1,n_2)$, then $\phi(n_1+1) = n_2$ by 
(\ref{eq:phi-discounted-DHR-DHR}) and (\ref{eq:phi-result2-DHR-DHR}). 
In addition, by (\ref{eq:Whittle-index-1n-discounted-DHR-DHR}), the right 
hand side of (\ref{eq:nu-req-discounted-DHR-DHR-case-32}) equals, under this condition, 
\[
W_\beta(1,n_1+1) = \psi(n_1+1,n_2) 
\]
and, furthermore, by (\ref{eq:x-def-DHR-DHR}), we have 
\[
x_{m+1} = (1,n_1+1) 
\]
so that requirement (\ref{eq:nu-req-discounted-DHR-DHR-case-32}) is equivalent 
with the requirement 
\[
\nu \ge W_\beta(x_{m+1}).
\]
On the other hand, if $W_\beta(2,n_2+1) \ge \psi(n_1+1,n_2)$, then, 
$\phi(n_1+1) > n_2$ by (\ref{eq:phi-discounted-DHR-DHR}) and 
we have, by (\ref{eq:Whittle-index-order-DHR-DHR}), 
\[
W_\beta(2,n_2+1) \ge W_\beta(2,\phi(n_1+1)) \ge W_\beta(1,n_1+1) 
\]
and, furthermore, by (\ref{eq:x-def-DHR-DHR}), 
\[
x_{m+1} = (2,n_2+1) 
\]
so that requirement (\ref{eq:nu-req-discounted-DHR-DHR-case-32}) is again equivalent 
with the requirement 
\[
\nu \ge W_\beta(x_{m+1}), 
\]
which completes the proof of claim~$3^\circ$.

\vskip 3pt
$4^\circ$ 
We still assume that $\nu \ge 0$ and utilize the optimality equations 
(\ref{eq:opt-eqs-discounted-nu-pos}). 
Now we prove that the policy $\pi$ with activity set 
\[
{\mathcal B}^\pi = {\mathcal S} \setminus \{*\}, 
\]
according to which user~$k$ is scheduled in all the states but $*$, is 
$(\nu,\beta)$-optimal for all 
\[
\nu \in [0,W_\beta(x_\infty)], 
\]
where $W_\beta(x_\infty)$ is defined in (\ref{eq:Whittle-index-x-infty-discounted-DHR-DHR-A}). 
It remains to prove that policy $\pi$ is optimal for these values of $\nu$ 
in any state $x \in {\mathcal S} \setminus \{*\}$.

We start by first deriving the value function $V_\beta^\pi(x;\nu)$ for 
policy $\pi$ from the Howard equations: 
\begin{equation}
\begin{split}
& 
V_\beta^\pi(1,n;\nu) = 
h + \nu + \beta \mu_1(n) V_\beta^\pi(2,0;\nu) + 
\beta (1 - \mu_1(n)) V_\beta^\pi(1,n+1;\nu), \\
& 
V_\beta^\pi(2,n;\nu) = 
h + \nu + \beta (1 - \mu_2(n)) V_\beta^\pi(2,n+1;\nu).
\end{split}
\label{eq:howard-eqs-discounted-DHR-DHR-A-case-4}
\end{equation}
The unique solution of these linear equations is given by 
\begin{equation}
\begin{split}
& 
V_\beta^\pi(2,n;\nu) = 
(h + \nu) 
\left( 
\sum_{i = 0}^\infty \beta^i \bar p_2(i|n) 
\right), \\
& 
V_\beta^\pi(1,n;\nu) = 
(h + \nu) 
\left( 
\sum_{i = 0}^\infty \beta^i \bar p_1(i|n) 
\left( 
1 + \beta \mu_1(n+i) \frac{V_\beta^\pi(2,0;\nu)}{h + \nu} 
\right) 
\right).
\end{split}
\label{eq:howard-eqs-discounted-DHR-DHR-A-case-4-solution}
\end{equation}

Let then $n \in \{0,1,\ldots\}$. 
Since $\bar p_2(i|n)$ is an increasing function of $n$ in the DHR-DHR 
case, we see from (\ref{eq:howard-eqs-discounted-DHR-DHR-A-case-4-solution}) 
that $V_\beta^\pi(2,n;\nu)$, as well, is an increasing function of $n$ 
and approaches 
\[
\begin{split}
& 
\lim_{n \to \infty} V_\beta^\pi(2,n;\nu) = 
(h + \nu) 
\left( 
\sum_{i = 0}^\infty \beta^i (1 - \mu_2(\infty))^i 
\right) \\
& \quad = 
\frac{h + \nu}{1 - \beta (1 - \mu_2(\infty))}.
\end{split}
\]
Thus, the following condition for optimality of $\pi$ in state~$(2,n)$, 
\[
V_\beta^\pi(2,n;\nu) \le \frac{h}{1 - \beta}, 
\]
is satisfied for any $n$ if and only if 
\[
\lim_{n \to \infty} V_\beta^\pi(2,n;\nu) \le \frac{h}{1 - \beta}, 
\]
which is clearly equivalent with condition 
\begin{equation}
\nu \le 
h \mu_2(\infty) \, 
\frac{\beta}{1 - \beta}. 
\label{eq:nu-req-discounted-DHR-DHR-A-case-4a}
\end{equation}
Note that the right hand side equals $w_2(\infty) = W_\beta(x_\infty)$ given in 
(\ref{eq:Whittle-index-x-infty-discounted-DHR-DHR-A}).

Let again $n \in \{0,1,\ldots\}$. 
Since 
\[
\bar p_1(i|n) \mu_1(n+i) = p_1(i|n) = \bar p_1(i|n) - \bar p_1(i+1|n), 
\]
it follows from 
(\ref{eq:howard-eqs-discounted-DHR-DHR-A-case-4-solution}) that 
\begin{equation}
\begin{split}
& 
V_\beta^\pi(1,n;\nu) = 
(h + \nu) 
\Bigg( 
1 + \frac{\beta V_\beta^\pi(2,0;\nu)}{h + \nu} \; + \\
& \quad 
\sum_{i = 1}^\infty \beta^i \bar p_1(i|n) 
\bigg( 
1 - \frac{(1 - \beta) V_\beta^\pi(2,0;\nu)}{h + \nu} 
\bigg) 
\Bigg).
\end{split}
\label{eq:howard-eqs-discounted-DHR-DHR-A-case-4-solution-cont}
\end{equation}
Now, since $\bar p_1(i|n)$ is an increasing function of $n$ in the DHR-DHR 
case and we have above required that 
\[
V_\beta^\pi(2,0;\nu) \le \frac{h}{1 - \beta}, 
\]
we see from (\ref{eq:howard-eqs-discounted-DHR-DHR-A-case-4-solution-cont}) 
that $V_\beta^\pi(1,n;\nu)$, as well, is an increasing function of $n$ 
and approaches, by (\ref{eq:howard-eqs-discounted-DHR-DHR-A-case-4-solution}), 
\[
\begin{split}
& 
\lim_{n \to \infty} V_\beta^\pi(1,n;\nu) = 
(h + \nu) 
\Bigg( 
\sum_{i = 0}^\infty \beta^i (1 - \mu_1(\infty))^i 
\bigg( 
1 + \beta \mu_1(\infty) \frac{V_\beta^\pi(2,0;\nu)}{h + \nu} 
\bigg) 
\Bigg) \\
& \quad = 
(h + \nu) \, 
\frac{1 + \beta \mu_1(\infty) \sum_{i = 0}^\infty \beta^i \bar p_2(i)}
{1 - \beta (1 - \mu_1(\infty))}.
\end{split}
\]
Thus, the following condition for optimality of $\pi$ in state~$(1,n)$, 
\[
V_\beta^\pi(1,n;\nu) \le \frac{h}{1 - \beta}, 
\]
is satisfied for any $n$ if and only if 
\[
\lim_{n \to \infty} V_\beta^\pi(1,n;\nu) \le \frac{h}{1 - \beta}, 
\]
which can be shown to be equivalent with condition 
\begin{equation}
\nu \le 
h \mu_1(\infty) \, 
\frac{\beta \sum_{i=0}^{\infty} \beta^i p_2(i)}
{1 + \beta \mu_1(\infty) \sum_{i=0}^{\infty} \beta^i \bar p_2(i)} \, 
\frac{\beta}{1 - \beta}.
\label{eq:nu-req-discounted-DHR-DHR-A-case-4b}
\end{equation}
Note that the right hand side equals 
$\psi(\infty,\infty) = \lim_{n_1, n_2 \to \infty} \psi(n_1,n_2)$, which 
in this DHR-DHR-A subcase equals $w_1(\infty) = W_\beta(x_\infty)$ given in 
(\ref{eq:Whittle-index-x-infty-discounted-DHR-DHR-A}) by 
Lemma~\ref{lem:Whittle-index-discounted-DHR-DHR-A-lemma-3}(ii).

Finally, by combining (\ref{eq:nu-req-discounted-DHR-DHR-A-case-4a}) and 
(\ref{eq:nu-req-discounted-DHR-DHR-A-case-4b}), we get the requirement that 
\begin{equation}
\nu \le W_\beta(x_\infty), 
\label{eq:nu-req-discounted-DHR-DHR-A-case-4}
\end{equation}
which completes the proof of claim~$4^\circ$.

\vskip 3pt
$5^\circ$ 
Now we assume that $\nu \le 0$. 
In this case, the optimal decision in state~$*$ is to schedule ($a = 1$), 
the minimum expected discounted cost $V_\beta(*;\nu)$ equals $\nu/(1 - \beta)$, 
and the optimality equations (\ref{eq:opt-eqs-discounted-general}) read as follows: 
\begin{equation}
\begin{split}
& 
V_\beta(1,n;\nu) = 
h + \min \big\{ \beta V_\beta(1,n;\nu), \\
& \quad 
\nu + \beta \mu_1(n) V_\beta(2,0;\nu) + 
\beta (1 - \mu_1(n)) V_\beta(1,n+1;\nu) \big\}, \\
& 
V_\beta(2,n;\nu) = 
h + \min \big\{ \beta V_\beta(2,n;\nu), \\
& \quad 
\nu + \beta \mu_2(n) \frac{\nu}{1 - \beta} + 
\beta (1 - \mu_2(n)) V_\beta(2,n+1;\nu) \big\}.
\end{split}
\label{eq:opt-eqs-discounted-nu-neg}
\end{equation}
We prove that the policy $\pi$ with activity set 
\[
{\mathcal B}^\pi = {\mathcal S}, 
\]
according to which user~$k$ is scheduled in all states, is 
$(\nu,\beta)$-optimal for all 
\[
\nu \in (-\infty,0].
\]
It remains to prove that policy $\pi$ is optimal for these values of $\nu$ 
in any state $x \in {\mathcal S} \setminus \{*\}$.

We start by first deriving the value function $V_\beta^\pi(x;\nu)$ for 
policy $\pi$ from the Howard equations: 
\begin{equation}
\begin{split}
& 
V_\beta^\pi(1,n;\nu) = 
h + \nu + \beta \mu_1(n) V_\beta^\pi(2,0;\nu) + 
\beta (1 - \mu_1(n)) V_\beta^\pi(1,n+1;\nu), \\
& 
V_\beta^\pi(2,n;\nu) = 
h + \nu + \beta \mu_2(n) \frac{\nu}{1 - \beta} + 
\beta (1 - \mu_2(n)) V_\beta^\pi(2,n+1;\nu). \\
\end{split}
\label{eq:howard-eqs-discounted-DHR-DHR-A-case-5}
\end{equation}
The unique solution of these linear equations is given by 
\begin{equation}
\begin{split}
& 
V_\beta^\pi(2,n;\nu) = 
h \, 
\left(
\sum_{i = 0}^\infty \beta^i \bar p_2(i|n) 
\right) + 
\frac{\nu}{1 - \beta}, \\
& 
V_\beta^\pi(1,n;\nu) = 
(h + \nu) 
\left( 
\sum_{i = 0}^\infty \beta^i \bar p_1(i|n) 
\right) + 
V_\beta^\pi(2,0;\nu) \beta 
\left( 
\sum_{i = 0}^\infty \beta^i p_1(i|n)
\right).
\end{split}
\label{eq:howard-eqs-discounted-DHR-DHR-A-case-5-solution}
\end{equation}

Let us now define the following auxiliary function: 
\begin{equation}
\begin{split}
& 
\tilde V_\beta^\pi(2,n) = 
h \, 
\left(
\sum_{i = 0}^\infty \beta^i \bar p_2(i|n) 
\right), \\
& 
\tilde V_\beta^\pi(1,n) = 
h \, 
\left( 
\sum_{i = 0}^\infty \beta^i \bar p_1(i|n) 
\right) + 
\tilde V_\beta^\pi(2,0) \beta 
\left( 
\sum_{i = 0}^\infty \beta^i p_1(i|n)
\right), 
\end{split}
\label{eq:discounted-DHR-DHR-A-case-5-auxiliary-function}
\end{equation}
which equals the value function given in 
(\ref{eq:howard-eqs-discounted-DHR-DHR-A-case-4-solution}) 
for $\nu = 0$. Thus, according to part $4^\circ$, we have, 
for any $x \in {\mathcal S} \setminus \{*\}$, 
\[
\tilde V_\beta^\pi(x) \le \frac{h}{1 - \beta}.
\]
Together with the assumption that $\nu \le 0$, we conclude from this that, 
for any $x \in {\mathcal S} \setminus \{*\}$, 
\begin{equation}
V_\beta^\pi(x;\nu) \le \tilde V_\beta^\pi(x) \le \frac{h}{1 - \beta}, 
\label{eq:nu-req-discounted-DHR-DHR-A-case-5}
\end{equation}
which implies that policy $\pi$ is optimal 
for any $x \in {\mathcal S} \setminus \{*\}$. 
This completes the proof of claim~$5^\circ$ and the whole proof of 
Theorem~\ref{thm:Whittle-index-discounted-DHR-DHR} in the DHR-DHR-A subcase.
\hfill $\Box$

\section{Proof of Theorem~\ref{thm:Whittle-index-discounted-DHR-DHR} 
in the DHR-DHR-B subcase}
\label{app:DHR-DHR-B-proof}

\paragraph{Proof} 
We present here the proof of Theorem~\ref{thm:Whittle-index-discounted-DHR-DHR} 
for the DHR-DHR-B subcase. For the other two subcases (DHR-DHR-A and DHR-DHR-C), 
the proof is slightly different and presented 
in Appendices~\ref{app:DHR-DHR-A-proof} and \ref{app:DHR-DHR-C-proof}, 
respectively.

Assume the DHR-DHR-B subcase defined in (\ref{eq:DHR-DHR-B}). 
As in Lemma~\ref{lem:Whittle-index-discounted-DHR-DHR-B-lemma-4}, 
let $n_2^* \in \{0,1,\ldots\}$ denote the smallest $\bar n_2$ satisfying 
condition (\ref{eq:DHR-DHR-B}). 
In addition, let $x_m$, $m \in \{1,2,\ldots\}$, denote the ordered sequence 
of states that is defined in the same way as in the DHR-DHR-A subcase 
(see Appendix~\ref{app:DHR-DHR-A-proof}) 
using the recursive equation (\ref{eq:x-def-DHR-DHR}). However, in this 
DHR-DHR-B subcase, the sequence $(x_m)$ covers only the states 
\[
\{x_1,x_2,\ldots\} = 
\{(2,0),(2,1),\ldots,(2,n_2^*)\} \cup \{(1,0),(1,1),\ldots\}.
\]
Now we have the following ordering among these states: 
\begin{equation}
\begin{split}
& 
W_\beta(x_1) \ge W_\beta(x_2) \ge \ldots \ge W_\beta(1,\infty) \; \ge \\
& \quad 
W_\beta(2,n_2^*+1) \ge W_\beta(2,n_2^*+2) \ge \ldots \ge W_\beta(2,\infty) \ge 0, 
\end{split}
\label{eq:Wx-order-DHR-DHR-B}
\end{equation}
where we have defined 
\begin{equation}
\begin{split}
& 
W_\beta(1,\infty) = w_1(\infty) = \psi(\infty,n_2^*) = 
h \mu_1(\infty) \, 
\frac{\beta \sum_{i=0}^{n_2^*} \beta^i p_2(i)}{1 + \beta \mu_1(\infty) \sum_{i=0}^{n_2^*} \beta^i \bar p_2(i)} \, 
\frac{\beta}{1 - \beta}, \\
& 
W_\beta(2,\infty) = w_2(\infty) = 
h \mu_2(\infty) \, 
\frac{\beta}{1 - \beta}.
\end{split}
\label{eq:Whittle-index-21-infty-discounted-DHR-DHR-B}
\end{equation}

Similarly as in the DHR-DHR-A subcase (see Appendix~\ref{app:DHR-DHR-A-proof}), 
for any $m \in \{1,2,\ldots\}$, there are $n_2$ and $n_1$ such that 
$n_2 \in \{0,1,\ldots,n_2^*\}$, $n_1 \in \{-1\} \cup \{0,1,\ldots\}$, and 
\[
\{x_1,\ldots,x_m\} = 
\{(2,0),(2,1),\ldots,(2,n_2)\} \cup \{(1,0),(1,1),\ldots,(1,n_1)\}, 
\]
where the latter part of the right hand side is omitted if $n_1 = -1$. 
Recall also that these $n_2$ and $n_1$ satisfy results 
(\ref{eq:phi-result1-DHR-DHR}) and (\ref{eq:phi-result2-DHR-DHR}).

The main proof is now given in seven parts ($1^\circ$--$7^\circ$). 
The idea is again to solve the relaxed optimization problem 
(\ref{eq:separable-discounted-costs}) for any $\nu$ by utilizing the 
optimality equations (\ref{eq:opt-eqs-discounted-general}). We partition 
the possible values of $\nu$, which is reflected by the seven parts of 
the main proof. 
However, parts $1^\circ$--$3^\circ$ are exactly the same as in the 
DHR-DHR-A subcase (see Appendix~\ref{app:DHR-DHR-A-proof}). 
Therefore, we omit them here and focus on the remaining parts 
$4^\circ$--$7^\circ$.

\vskip 3pt
$4^\circ$ 
Here we assume that $\nu \ge 0$, and the optimality equations 
(\ref{eq:opt-eqs-discounted-general}) read as given in 
(\ref{eq:opt-eqs-discounted-nu-pos}). 
We prove that the policy $\pi$ with activity set 
\[
{\mathcal B}^\pi = 
\{x_1,x_2,\ldots\} = 
\{(2,0),(2,1),\ldots,(2,n_2^*)\} \cup \{(1,0),(1,1),\ldots\} 
\]
is $(\nu,\beta)$-optimal for all 
\[
\nu \in [W_\beta(2,n_2^*+1),W_\beta(1,\infty)], 
\]
where $W_\beta(1,\infty)$ is defined in 
(\ref{eq:Whittle-index-21-infty-discounted-DHR-DHR-B}). 
It remains to prove that policy $\pi$ is optimal for these values of $\nu$ 
in any state $x \in {\mathcal S} \setminus \{*\}$. 
This is done below in two parts ($4.1^\circ$ and $4.2^\circ$).

\vskip 3pt
$4.1^\circ$ 
We start by first deriving the value function $V_\beta^\pi(x;\nu)$ for 
policy $\pi$ from the Howard equations: 
\begin{equation}
\begin{split}
& 
V_\beta^\pi(1,n;\nu) = 
h + \nu + \beta \mu_1(n) V_\beta^\pi(2,0;\nu) + 
\beta (1 - \mu_1(n)) V_\beta^\pi(1,n+1;\nu), \\
& \quad 
n \in \{0,1,\ldots\}, \\
& 
V_\beta^\pi(2,n;\nu) = 
h + \nu + \beta (1 - \mu_2(n)) V_\beta^\pi(2,n+1;\nu), \\
& \quad 
n \in \{0,1,\ldots,n_2^*\}, \\
& 
V_\beta^\pi(x;\nu) = 
h + \beta V_\beta^\pi(x;\nu), 
\quad x \in \{(2,n_2^*+1),(2,n_2^*+2),\ldots\}.
\end{split}
\label{eq:howard-eqs-discounted-DHR-DHR-B-case-4}
\end{equation}
The unique solution of these linear equations is given by 
\begin{equation}
\begin{split}
& 
V_\beta^\pi(2,n;\nu) = 
(h + \nu) 
\left( 
\sum_{i = 0}^{n_2^*-n} \beta^i \bar p_2(i|n) 
\right) + 
h \, \frac{\beta^{n_2^*-n+1} \bar p_2(n_2^*-n+1|n)}{1 - \beta}, \\
& \quad 
n \in \{0,1,\ldots,n_2^*\}, \\
& 
V_\beta^\pi(1,n;\nu) = 
(h + \nu) 
\left( 
\sum_{i = 0}^\infty \beta^i \bar p_1(i|n) 
\left( 
1 + \beta \mu_1(n+i) \frac{V_\beta^\pi(2,0;\nu)}{h + \nu} 
\right) 
\right), \\
& \quad 
n \in \{0,1,\ldots\},\\
& 
V_\beta^\pi(x;\nu) = 
\frac{h}{1 - \beta}, 
\quad x \in \{(2,n_2^*+1),(2,n_2^*+2),\ldots\}.
\end{split}
\label{eq:howard-eqs-discounted-DHR-DHR-B-case-4-solution}
\end{equation}

By (\ref{eq:howard-eqs-discounted-DHR-DHR-B-case-4-solution}) and some algebraic 
manipulations, the following condition for optimality of $\pi$ in 
state~$(2,n_2^*)$, 
\[
V_\beta^\pi(2,n_2^*;\nu) \le \frac{h}{1 - \beta}, 
\]
can be shown to be equivalent with 
\begin{equation}
\nu \le 
h \mu_2(n_2^*) \, \frac{\beta}{1 - \beta}, 
\label{eq:nu-req-discounted-DHR-DHR-B-case-41a}
\end{equation}
where the right hand side equals $W_\beta(2,n_2^*)$ given in 
(\ref{eq:Whittle-index-2n-discounted-DHR-DHR}). 
Note that (\ref{eq:nu-req-discounted-DHR-DHR-B-case-41a}) follows from the 
requirement that $\nu \le W_\beta(1,\infty)$ since 
$W_\beta(1,\infty) \le W_\beta(2,n_2^*)$ by (\ref{eq:Wx-order-DHR-DHR-B}).

Let then $n \in \{0,1,\ldots,n_2^*-1\}$. 
Again by (\ref{eq:howard-eqs-discounted-DHR-DHR-B-case-4-solution}), 
the following condition for optimality of $\pi$ in state~$(2,n)$, 
\[
V_\beta^\pi(2,n;\nu) \le \frac{h}{1 - \beta}, 
\]
can be shown to be equivalent with 
\begin{equation}
\nu \le 
h \, 
\frac{\sum_{i = 0}^{n_2^*-n} \beta^i p_2(i|n)}
{\sum_{i = 0}^{n_2^*-n} \beta^i \bar p_2(i|n)} \, 
\frac{\beta}{1 - \beta}.
\label{eq:nu-req-discounted-DHR-DHR-B-case-41a-cont}
\end{equation}
Thus, condition (\ref{eq:nu-req-discounted-DHR-DHR-B-case-41a-cont}) 
follows from (\ref{eq:nu-req-discounted-DHR-DHR-B-case-41a}) 
by Lemma~\ref{lem:Whittle-index-discounted-DHR-DHR-lemma-1}(v).

Let then $n \in \{0,1,\ldots\}$. 
Since 
\[
\bar p_1(i|n) \mu_1(n+i) = p_1(i|n) = \bar p_1(i|n) - \bar p_1(i+1|n), 
\]
it follows from 
(\ref{eq:howard-eqs-discounted-DHR-DHR-B-case-4-solution}) that 
\begin{equation}
\begin{split}
& 
V_\beta^\pi(1,n;\nu) = 
(h + \nu) 
\Bigg( 
1 + \frac{\beta V_\beta^\pi(2,0;\nu)}{h + \nu} \; + \\
& \quad 
\sum_{i = 1}^\infty \beta^i \bar p_1(i|n) 
\bigg( 
1 - \frac{(1 - \beta) V_\beta^\pi(2,0;\nu)}{h + \nu} 
\bigg) 
\Bigg).
\end{split}
\label{eq:howard-eqs-discounted-DHR-DHR-B-case-4-solution-cont}
\end{equation}
Now, since $\bar p_1(i|n)$ is an increasing function of $n$ in the DHR-DHR 
case and we have above required that 
\[
V_\beta^\pi(2,0;\nu) \le \frac{h}{1 - \beta}, 
\]
we see from (\ref{eq:howard-eqs-discounted-DHR-DHR-B-case-4-solution-cont}) 
that $V_\beta^\pi(1,n;\nu)$, as well, is an increasing function of $n$ 
and approaches, by~(\ref{eq:howard-eqs-discounted-DHR-DHR-B-case-4-solution}), 
\[
\begin{split}
& 
\lim_{n \to \infty} V_\beta^\pi(1,n;\nu) = 
(h + \nu) 
\Bigg( 
\sum_{i = 0}^\infty \beta^i (1 - \mu_1(\infty))^i 
\bigg( 
1 + \beta \mu_1(\infty) \frac{V_\beta^\pi(2,0;\nu)}{h + \nu} 
\bigg) 
\Bigg) \\
& \quad \; = 
\frac{h + \nu}{1 - \beta (1 - \mu_1(\infty))} 
\bigg( 
1 + \beta \mu_1(\infty) \sum_{j = 0}^{n_2^*} \beta^j \bar p_2(j) + 
\beta \mu_1(\infty) \frac{h}{h + \nu} \frac{\beta^{n_2^*+1} \bar p_2(n_2^*+1)}{1 - \beta}
\bigg) \\
& \quad = 
(h + \nu) 
\frac{1 + \beta \mu_1(\infty) \sum_{i = 0}^{n_2^*} \beta^i \bar p_2(i)}
{1 - \beta (1 - \mu_1(\infty))} + 
\frac{h}{1 - \beta (1 - \mu_1(\infty))} \frac{\beta^{n_2^*+1} \bar p_2(n_2^*+1)}{1 - \beta}.
\end{split}
\]
Thus, the following condition for optimality of $\pi$ in state~$(1,n)$, 
\[
V_\beta^\pi(1,n;\nu) \le \frac{h}{1 - \beta}, 
\]
is satisfied for any $n$ if and only if 
\[
\lim_{n \to \infty} V_\beta^\pi(1,n;\nu) \le \frac{h}{1 - \beta}, 
\]
which can be shown to be equivalent with condition 
\begin{equation}
\nu \le 
h \mu_1(\infty) \, 
\frac{\beta \sum_{i=0}^{n_2^*} \beta^i p_2(i)}
{1 + \beta \mu_1(\infty) \sum_{i=0}^{n_2^*} \beta^i \bar p_2(i)} \, 
\frac{\beta}{1 - \beta}.
\label{eq:nu-req-discounted-DHR-DHR-B-case-41b}
\end{equation}
Note that the right hand side equals $W_\beta(1,\infty)$ given in 
(\ref{eq:Whittle-index-21-infty-discounted-DHR-DHR-B}).

\vskip 3pt
$4.2^\circ$ 
By (\ref{eq:howard-eqs-discounted-DHR-DHR-B-case-4-solution}), the following 
condition for optimality of $\pi$ in state~$(2,n_2^*+1)$, 
\[
\beta V_\beta^\pi(2,n_2^*+1;\nu) \le 
\nu + \beta (1 - \mu_2(n_2^*+1)) V_\beta^\pi(2,n_2^*+2;\nu), 
\]
is easily shown to be equivalent with 
\begin{equation}
\nu \ge 
h \mu_2(n_2^*+1) \, \frac{\beta}{1 - \beta}, 
\label{eq:nu-req-discounted-DHR-DHR-B-case-42}
\end{equation}
where the right hand side equals $W_\beta(2,n_2^*+1)$ 
given in (\ref{eq:Whittle-index-2n-discounted-DHR-DHR}).

Let then $n \in \{n_2^*+2,n_2^*+3,\ldots\}$. 
Again by (\ref{eq:howard-eqs-discounted-DHR-DHR-B-case-4-solution}), 
the following condition for optimality of $\pi$ in state~$(2,n)$, 
\[
\beta V_\beta^\pi(2,n;\nu) \le 
\nu + \beta (1 - \mu_2(n)) V_\beta^\pi(2,n+1;\nu), 
\]
is easily shown to be equivalent with 
\[
\nu \ge 
h \mu_2(n) \, \frac{\beta}{1 - \beta}, 
\]
which follows from (\ref{eq:nu-req-discounted-DHR-DHR-B-case-42}) since 
$\mu_2(n)$ is decreasing. 
This completes the proof of claim~$4^\circ$.

\vskip 3pt
$5^\circ$ 
We still assume that $\nu \ge 0$ and utilize the optimality equations 
(\ref{eq:opt-eqs-discounted-nu-pos}). 
Let $m \in \{1,2,\ldots\}$. 
We prove that the policy $\pi$ with activity set 
\[
\begin{split}
& 
{\mathcal B}^\pi = \{x_1,x_2,\ldots\} \cup \{(2,n_2^*+1),\ldots,(2,n_2^*+m)\} \; = \\
& \quad 
\{(2,0),(2,1),\ldots,(2,n_2^*+m)\} \cup \{(1,0),(1,1),\ldots\} 
\end{split}
\]
is $(\nu,\beta)$-optimal for all 
\[
\nu \in [W_\beta(2,n_2^*+m+1),W_\beta(2,n_2^*+m)].
\]
It remains to prove that policy $\pi$ is optimal for these values of $\nu$ 
in any state $x \in {\mathcal S} \setminus \{*\}$. 
This is done below in two parts ($5.1^\circ$ and $5.2^\circ$).

\vskip 3pt
$5.1^\circ$ 
We start by first deriving the value function $V_\beta^\pi(x;\nu)$ for 
policy $\pi$ from the Howard equations: 
\begin{equation}
\begin{split}
& 
V_\beta^\pi(1,n;\nu) = 
h + \nu + \beta \mu_1(n) V_\beta^\pi(2,0;\nu) + 
\beta (1 - \mu_1(n)) V_\beta^\pi(1,n+1;\nu), \\
& \quad 
n \in \{0,1,\ldots\}, \\
& 
V_\beta^\pi(2,n;\nu) = 
h + \nu + \beta (1 - \mu_2(n)) V_\beta^\pi(2,n+1;\nu), \\
& \quad 
n \in \{0,1,\ldots,n_2^*+m\}, \\
& 
V_\beta^\pi(x;\nu) = 
h + \beta V_\beta^\pi(x;\nu), 
\quad x \in \{(2,n_2^*+m+1),(2,n_2^*+m+2),\ldots\}.
\end{split}
\label{eq:howard-eqs-discounted-DHR-DHR-B-case-5}
\end{equation}
The unique solution of these linear equations is given by 
\begin{equation}
\begin{split}
& 
V_\beta^\pi(2,n;\nu) = 
(h + \nu) 
\left( 
\sum_{i = 0}^{n_2^*+m-n} \beta^i \bar p_2(i|n) 
\right) + 
h \, \frac{\beta^{n_2^*+m-n+1} \bar p_2(n_2^*+m-n+1|n)}{1 - \beta}, \\
& \quad 
n \in \{0,1,\ldots,n_2^*+m\}, \\
& 
V_\beta^\pi(1,n;\nu) = 
(h + \nu) 
\left( 
\sum_{i = 0}^\infty \beta^i \bar p_1(i|n) 
\left( 
1 + \beta \mu_1(n+i) \frac{V_\beta^\pi(2,0;\nu)}{h + \nu} 
\right) 
\right), \\
& \quad 
n \in \{0,1,\ldots\},\\
& 
V_\beta^\pi(x;\nu) = 
\frac{h}{1 - \beta}, 
\quad x \in \{(2,n_2^*+m+1),(2,n_2^*+m+2),\ldots\}.
\end{split}
\label{eq:howard-eqs-discounted-DHR-DHR-B-case-5-solution}
\end{equation}

By (\ref{eq:howard-eqs-discounted-DHR-DHR-B-case-5-solution}) and some algebraic 
manipulations, the following condition for optimality of $\pi$ in 
state~$(2,n_2^*+m)$, 
\[
V_\beta^\pi(2,n_2^*+m;\nu) \le \frac{h}{1 - \beta}, 
\]
can be shown to be equivalent with 
\begin{equation}
\nu \le 
h \mu_2(n_2^*+m) \, \frac{\beta}{1 - \beta}, 
\label{eq:nu-req-discounted-DHR-DHR-B-case-51a}
\end{equation}
where the right hand side equals $W_\beta(2,n_2^*+m)$ 
given in (\ref{eq:Whittle-index-2n-discounted-DHR-DHR}).

Let then $n \in \{0,1,\ldots,n_2^*+m-1\}$. 
Again by (\ref{eq:howard-eqs-discounted-DHR-DHR-B-case-5-solution}), 
the following condition for optimality of $\pi$ in state~$(2,n)$, 
\[
V_\beta^\pi(2,n;\nu) \le \frac{h}{1 - \beta}, 
\]
can be shown to be equivalent with 
\begin{equation}
\nu \le 
h \, 
\frac{\sum_{i = 0}^{n_2^*+m-n} \beta^i p_2(i|n)}
{\sum_{i = 0}^{n_2^*+m-n} \beta^i \bar p_2(i|n)} \, 
\frac{\beta}{1 - \beta}.
\label{eq:nu-req-discounted-DHR-DHR-B-case-51a-cont}
\end{equation}
Thus, condition (\ref{eq:nu-req-discounted-DHR-DHR-B-case-51a-cont}) 
follows from (\ref{eq:nu-req-discounted-DHR-DHR-B-case-51a}) 
by Lemma~\ref{lem:Whittle-index-discounted-DHR-DHR-lemma-1}(v).

Let then $n \in \{0,1,\ldots\}$. 
Since 
\[
\bar p_1(i|n) \mu_1(n+i) = p_1(i|n) = \bar p_1(i|n) - \bar p_1(i+1|n), 
\]
it follows from 
(\ref{eq:howard-eqs-discounted-DHR-DHR-B-case-5-solution}) that 
\begin{equation}
\begin{split}
& 
V_\beta^\pi(1,n;\nu) = 
(h + \nu) 
\Bigg( 
1 + \frac{\beta V_\beta^\pi(2,0;\nu)}{h + \nu} \; + \\
& \quad 
\sum_{i = 1}^\infty \beta^i \bar p_1(i|n) 
\bigg( 
1 - \frac{(1 - \beta) V_\beta^\pi(2,0;\nu)}{h + \nu} 
\bigg) 
\Bigg).
\end{split}
\label{eq:howard-eqs-discounted-DHR-DHR-B-case-5-solution-cont}
\end{equation}
Now, since $\bar p_1(i|n)$ is an increasing function of $n$ in the DHR-DHR 
case and we have above required that 
\[
V_\beta^\pi(2,0;\nu) \le \frac{h}{1 - \beta}, 
\]
we see from (\ref{eq:howard-eqs-discounted-DHR-DHR-B-case-5-solution-cont}) 
that $V_\beta^\pi(1,n;\nu)$, as well, is an increasing function of $n$ 
and approaches, by~(\ref{eq:howard-eqs-discounted-DHR-DHR-B-case-5-solution}), 
\[
\begin{split}
& 
\lim_{n \to \infty} V_\beta^\pi(1,n;\nu) = 
(h + \nu) 
\Bigg( 
\sum_{i = 0}^\infty \beta^i (1 - \mu_1(\infty))^i 
\bigg( 
1 + \beta \mu_1(\infty) \frac{V_\beta^\pi(2,0;\nu)}{h + \nu} 
\bigg) 
\Bigg) \\
& \quad \; = 
\frac{h + \nu}{1 - \beta (1 - \mu_1(\infty))} 
\bigg( 
1 + \beta \mu_1(\infty) \sum_{j = 0}^{n_2^*+m} \beta^j \bar p_2(j) + 
\beta \mu_1(\infty) 
\frac{h}{h + \nu} \frac{\beta^{n_2^*+m+1} \bar p_2(n_2^*+m+1)}{1 - \beta}
\bigg) \\
& \quad \; = 
(h + \nu) 
\frac{1 + \beta \mu_1(\infty) \sum_{i = 0}^{n_2^*+m} \beta^i \bar p_2(i)}
{1 - \beta (1 - \mu_1(\infty))} + 
\frac{h}{1 - \beta (1 - \mu_1(\infty))} 
\frac{\beta^{n_2^*+m+1} \bar p_2(n_2^*+m+1)}{1 - \beta}.
\end{split}
\]
Thus, the following condition for optimality of $\pi$ in state~$(1,n)$, 
\[
V_\beta^\pi(1,n;\nu) \le \frac{h}{1 - \beta}, 
\]
is satisfied for any $n$ if and only if 
\[
\lim_{n \to \infty} V_\beta^\pi(1,n;\nu) \le \frac{h}{1 - \beta}, 
\]
which can be shown to be equivalent with condition 
\begin{equation}
\nu \le 
h \mu_1(\infty) \, 
\frac{\beta \sum_{i=0}^{n_2^*+m} \beta^i p_2(i)}
{1 + \beta \mu_1(\infty) \sum_{i=0}^{n_2^*+m} \beta^i \bar p_2(i)} \, 
\frac{\beta}{1 - \beta}.
\label{eq:nu-req-discounted-DHR-DHR-B-case-51b}
\end{equation}
Note that the right hand side equals 
$\psi(\infty,n_2^*+m) = \lim_{n_1 \to \infty} \psi(n_1,n_2^*+m)$ given 
in (\ref{eq:psi-infty-n2-discounted-DHR-DHR}).

Next we prove that $\psi(\infty,n_2^*+m) \ge w_2(n_2^*+m)$. 
Whenever $n_1$ is sufficiently large, $\phi(n_1) = n_2^*$ by 
Lemma~\ref{lem:Whittle-index-discounted-DHR-DHR-B-lemma-4}(i), 
which implies, by (\ref{eq:Whittle-index-order-DHR-DHR}), that 
\[
\psi(n_1,n_2^*) = \psi(n_1,\phi(n_1)) = w_1(n_1) > w_2(\phi(n_1)+1) = w_2(n_2^*+1).
\]
However, by Lemma~\ref{lem:Whittle-index-discounted-DHR-DHR-lemma-1}(iv), 
this is equivalent with 
\[
\psi(n_1,n_2^*+1) > w_2(n_2^*+1).
\]
But, by Lemma~\ref{lem:Whittle-index-discounted-DHR-DHR-lemma-1}(i), this implies that 
\[
\psi(n_1,n_2^*+1) > w_2(n_2^*+2), 
\]
which is, again by Lemma~\ref{lem:Whittle-index-discounted-DHR-DHR-lemma-1}(iv), 
equivalent with 
\[
\psi(n_1,n_2^*+2) > w_2(n_2^*+2).
\]
By continuing similarly, we finally end up to the following inequality: 
\[
\psi(n_1,n_2^*+m) > w_2(n_2^*+m).
\]
Since this is true for any $n_1$ sufficiently large, we conclude that 
\[
\psi(\infty,n_2^*+m) = \lim_{n_1 \to \infty} \psi(n_1,n_2^*+m) \ge w_2(n_2^*+m), 
\]
which, in turn, proves that requirement 
(\ref{eq:nu-req-discounted-DHR-DHR-B-case-51b}) follows from 
(\ref{eq:nu-req-discounted-DHR-DHR-B-case-51a}).

\vskip 3pt
$5.2^\circ$ 
By (\ref{eq:howard-eqs-discounted-DHR-DHR-B-case-5-solution}), the following condition 
for optimality of $\pi$ in state~$(2,n_2^*+m+1)$, 
\[
\beta V_\beta^\pi(2,n_2^*+m+1;\nu) \le 
\nu + \beta (1 - \mu_2(n_2^*+m+1)) V_\beta^\pi(2,n_2^*+m+2;\nu), 
\]
is easily shown to be equivalent with 
\begin{equation}
\nu \ge 
h \mu_2(n_2^*+m+1) \, \frac{\beta}{1 - \beta}, 
\label{eq:nu-req-discounted-DHR-DHR-B-case-52}
\end{equation}
where the right hand side equals $W_\beta(2,n_2^*+m+1)$ 
given in (\ref{eq:Whittle-index-2n-discounted-DHR-DHR}).

Let then $n \in \{n_2^*+m+2,n_2^*+m+3,\ldots\}$. 
Again by (\ref{eq:howard-eqs-discounted-DHR-DHR-B-case-5-solution}), 
the following condition for optimality of $\pi$ in state~$(2,n)$, 
\[
\beta V_\beta^\pi(2,n;\nu) \le 
\nu + \beta (1 - \mu_2(n)) V_\beta^\pi(2,n+1;\nu), 
\]
is easily shown to be equivalent with 
\[
\nu \ge 
h \mu_2(n) \, \frac{\beta}{1 - \beta}, 
\]
which follows from (\ref{eq:nu-req-discounted-DHR-DHR-B-case-52}) since 
$\mu_2(n)$ is decreasing. 
This completes the proof of claim~$5^\circ$.

\vskip 3pt
$6^\circ$ 
We still assume that $\nu \ge 0$ and utilize the optimality equations 
(\ref{eq:opt-eqs-discounted-nu-pos}). 
Now we prove that the policy $\pi$ with activity set 
\[
{\mathcal B}^\pi = {\mathcal S} \setminus \{*\} 
\]
is $(\nu,\beta)$-optimal for all 
\[
\nu \in [0,W_\beta(2,\infty)], 
\]
where $W_\beta(2,\infty)$ is defined in (\ref{eq:Whittle-index-21-infty-discounted-DHR-DHR-B}). 
It remains to prove that policy $\pi$ is optimal for these values of $\nu$ 
in any state $x \in {\mathcal S} \setminus \{*\}$.

We start by first deriving the value function $V_\beta^\pi(x;\nu)$ for 
policy $\pi$ from the Howard equations: 
\begin{equation}
\begin{split}
& 
V_\beta^\pi(1,n;\nu) = 
h + \nu + \beta \mu_1(n) V_\beta^\pi(2,0;\nu) + 
\beta (1 - \mu_1(n)) V_\beta^\pi(1,n+1;\nu), \\
& 
V_\beta^\pi(2,n;\nu) = 
h + \nu + \beta (1 - \mu_2(n)) V_\beta^\pi(2,n+1;\nu).
\end{split}
\label{eq:howard-eqs-discounted-DHR-DHR-B-case-6}
\end{equation}
The unique solution of these linear equations is given by 
\begin{equation}
\begin{split}
& 
V_\beta^\pi(2,n;\nu) = 
(h + \nu) 
\left( 
\sum_{i = 0}^\infty \beta^i \bar p_2(i|n) 
\right), \\
& 
V_\beta^\pi(1,n;\nu) = 
(h + \nu) 
\left( 
\sum_{i = 0}^\infty \beta^i \bar p_1(i|n) 
\left( 
1 + \beta \mu_1(n+i) \frac{V_\beta^\pi(2,0;\nu)}{h + \nu} 
\right) 
\right).
\end{split}
\label{eq:howard-eqs-discounted-DHR-DHR-B-case-6-solution}
\end{equation}

Let then $n \in \{0,1,\ldots\}$. 
Since $\bar p_2(i|n)$ is an increasing function of $n$ in the DHR-DHR 
case, we see from (\ref{eq:howard-eqs-discounted-DHR-DHR-B-case-6-solution}) 
that $V_\beta^\pi(2,n;\nu)$, as well, is an increasing function of $n$ 
and approaches 
\[
\begin{split}
& 
\lim_{n \to \infty} V_\beta^\pi(2,n;\nu) = 
(h + \nu) 
\left( 
\sum_{i = 0}^\infty \beta^i (1 - \mu_2(\infty))^i 
\right) \\
& \quad = 
\frac{h + \nu}{1 - \beta (1 - \mu_2(\infty))}.
\end{split}
\]
Thus, the following condition for optimality of $\pi$ in state~$(2,n)$, 
\[
V_\beta^\pi(2,n;\nu) \le \frac{h}{1 - \beta}, 
\]
is satisfied for any $n$ if and only if 
\[
\lim_{n \to \infty} V_\beta^\pi(2,n;\nu) \le \frac{h}{1 - \beta}, 
\]
which is clearly equivalent with condition 
\begin{equation}
\nu \le 
h \mu_2(\infty) \, 
\frac{\beta}{1 - \beta}. 
\label{eq:nu-req-discounted-DHR-DHR-B-case-6a}
\end{equation}
Note that the right hand side equals $W_\beta(2,\infty)$ given in 
(\ref{eq:Whittle-index-21-infty-discounted-DHR-DHR-B}).

Let then $n \in \{0,1,\ldots\}$. 
Since 
\[
\bar p_1(i|n) \mu_1(n+i) = p_1(i|n) = \bar p_1(i|n) - \bar p_1(i+1|n), 
\]
it follows from (\ref{eq:howard-eqs-discounted-DHR-DHR-B-case-6-solution}) that 
\begin{equation}
\begin{split}
& 
V_\beta^\pi(1,n;\nu) = 
(h + \nu) 
\Bigg( 
1 + \frac{\beta V_\beta^\pi(2,0;\nu)}{h + \nu} \; + \\
& \quad 
\sum_{i = 1}^\infty \beta^i \bar p_1(i|n) 
\bigg( 
1 - \frac{(1 - \beta) V_\beta^\pi(2,0;\nu)}{h + \nu} 
\bigg) 
\Bigg).
\end{split}
\label{eq:howard-eqs-discounted-DHR-DHR-B-case-6-solution-cont}
\end{equation}
Now, since $\bar p_1(i|n)$ is an increasing function of $n$ in the DHR-DHR 
case and we have above required that 
\[
V_\beta^\pi(2,0;\nu) \le \frac{h}{1 - \beta}, 
\]
we see from (\ref{eq:howard-eqs-discounted-DHR-DHR-B-case-6-solution-cont}) 
that $V_\beta^\pi(1,n;\nu)$, as well, is an increasing function of $n$ 
and approaches, by (\ref{eq:howard-eqs-discounted-DHR-DHR-B-case-6-solution}), 
\[
\begin{split}
& 
\lim_{n \to \infty} V_\beta^\pi(1,n;\nu) = 
(h + \nu) 
\Bigg( 
\sum_{i = 0}^\infty \beta^i (1 - \mu_1(\infty))^i 
\bigg( 
1 + \beta \mu_1(\infty) \frac{V_\beta^\pi(2,0;\nu)}{h + \nu} 
\bigg) 
\Bigg) \\
& \quad = 
(h + \nu) \, 
\frac{1 + \beta \mu_1(\infty) \sum_{i = 0}^\infty \beta^i \bar p_2(i)}
{1 - \beta (1 - \mu_1(\infty))}.
\end{split}
\]
Thus, the following condition for optimality of $\pi$ in state~$(1,n)$, 
\[
V_\beta^\pi(1,n;\nu) \le \frac{h}{1 - \beta}, 
\]
is satisfied for any $n$ if and only if 
\[
\lim_{n \to \infty} V_\beta^\pi(1,n;\nu) \le \frac{h}{1 - \beta}, 
\]
which can be shown to be equivalent with condition 
\begin{equation}
\nu \le 
h \mu_1(\infty) \, 
\frac{\beta \sum_{i=0}^{\infty} \beta^i p_2(i)}
{1 + \beta \mu_1(\infty) \sum_{i=0}^{\infty} \beta^i \bar p_2(i)} \, 
\frac{\beta}{1 - \beta}.
\label{eq:nu-req-discounted-DHR-DHR-B-case-6b}
\end{equation}
Note that the right hand side equals 
$\psi(\infty,\infty) = \lim_{n_1, n_2 \to \infty} \psi(n_1,n_2)$.

Next we prove that $\psi(\infty,\infty) \ge w_2(\infty)$. 
Whenever $n_1$ is sufficiently large, $\phi(n_1) = n_2^*$ by 
Lemma~\ref{lem:Whittle-index-discounted-DHR-DHR-B-lemma-4}(i), 
which implies, by (\ref{eq:Whittle-index-order-DHR-DHR}), that 
\[
\psi(n_1,n_2^*) = \psi(n_1,\phi(n_1)) = w_1(n_1) > w_2(\phi(n_1)+1) = w_2(n_2^*+1).
\]
However, by Lemma~\ref{lem:Whittle-index-discounted-DHR-DHR-lemma-1}(iv), 
this is equivalent with 
\[
\psi(n_1,n_2^*+1) > w_2(n_2^*+1).
\]
But, by Lemma~\ref{lem:Whittle-index-discounted-DHR-DHR-lemma-1}(i), this implies that 
\[
\psi(n_1,n_2^*+1) > w_2(n_2^*+2), 
\]
which is, again by Lemma~\ref{lem:Whittle-index-discounted-DHR-DHR-lemma-1}(iv), 
equivalent with 
\[
\psi(n_1,n_2^*+2) > w_2(n_2^*+2).
\]
By continuing similarly, we finally end up to the following inequality: 
for any $n_2 \ge n_2^*+1$, 
\[
\psi(n_1,n_2) > w_2(n_2).
\]
Thus, 
\[
\psi(n_1,\infty) = \lim_{n_2 \to \infty} \psi(n_1,n_2) \ge 
\lim_{n_2 \to \infty} w_2(n_2) = w_2(\infty).
\]
Since this is true for any $n_1$ sufficiently large, we conclude that 
\[
\psi(\infty,\infty) = \lim_{n_1 \to \infty} \psi(n_1,\infty) \ge w_2(\infty), 
\]
which, in turn, proves that requirement 
(\ref{eq:nu-req-discounted-DHR-DHR-B-case-6b}) follows from 
(\ref{eq:nu-req-discounted-DHR-DHR-B-case-6a}). 
This completes the proof of claim~$6^\circ$.

\vskip 3pt
$7^\circ$ 
Finally, we assume that $\nu \le 0$. 
However, the claim that the policy $\pi$ with activity set 
\[
{\mathcal B}^\pi = {\mathcal S} 
\]
is $(\nu,\beta)$-optimal for all 
\[
\nu \in (-\infty,0] 
\]
can be proved similarly as the corresponding claim~$5^\circ$ in the DHR-DHR-A 
subcase (see Appendix~\ref{app:DHR-DHR-A-proof}). 
Therefore we may omit the proof here.
\hfill $\Box$

\section{Proof of Theorem~\ref{thm:Whittle-index-discounted-DHR-DHR} 
in the DHR-DHR-C subcase}
\label{app:DHR-DHR-C-proof}

\paragraph{Proof} 
We present here the proof of Theorem~\ref{thm:Whittle-index-discounted-DHR-DHR} 
for the DHR-DHR-C subcase. For the other two subcases (DHR-DHR-A and DHR-DHR-B), 
the proof is slightly different and presented 
in Appendices~\ref{app:DHR-DHR-A-proof} and \ref{app:DHR-DHR-B-proof}, 
respectively.

Assume the DHR-DHR-C subcase defined in (\ref{eq:DHR-DHR-C}). As in 
Lemma~\ref{lem:Whittle-index-discounted-DHR-DHR-C-lemma-5}, 
let $n_1^* \in \{-1\} \cup \{0,1,\ldots\}$ denote the smallest $\bar n_1$ 
satisfying condition (\ref{eq:DHR-DHR-C}). 
In addition, let $x_m$, $m \in \{1,2,\ldots\}$, denote the ordered sequence 
of states that is defined in the same way as in the DHR-DHR-A subcase 
(see Appendix~\ref{app:DHR-DHR-A-proof}) 
using the recursive equation (\ref{eq:x-def-DHR-DHR}). However, in this 
DHR-DHR-C subcase, the sequence $(x_m)$ covers only the states 
\[
\{x_1,x_2,\ldots\} = 
\{(2,0),(2,1),\ldots\} \cup \{(1,0),(1,1),\ldots,(1,n_1^*)\}.
\]
Now we have the following ordering among these states: 
\begin{equation}
\begin{split}
& 
W_\beta(x_1) \ge W_\beta(x_2) \ge \ldots \ge W_\beta(2,\infty) \; \ge \\
& \quad 
W_\beta(1,n_1^*+1) \ge W_\beta(1,n_1^*+2) \ge \ldots \ge W_\beta(1,\infty) \ge 0, 
\end{split}
\label{eq:Wx-order-DHR-DHR-C}
\end{equation}
where we have defined 
\begin{equation}
\begin{split}
& 
W_\beta(2,\infty) = w_2(\infty) = 
h \mu_2(\infty) \, 
\frac{\beta}{1 - \beta}, \\
& 
W_\beta(1,\infty) = w_1(\infty) = \psi(\infty,\infty) = 
h \mu_1(\infty) \, 
\frac{\beta \sum_{i=0}^{\infty} \beta^i p_2(i)}{1 + \beta \mu_1(\infty) \sum_{i=0}^{\infty} \beta^i \bar p_2(i)} \, 
\frac{\beta}{1 - \beta}.
\end{split}
\label{eq:Whittle-index-21-infty-discounted-DHR-DHR-C}
\end{equation}

Similarly as in the DHR-DHR-A and DHR-DHR-B subcases (see 
Appendices~\ref{app:DHR-DHR-A-proof} and \ref{app:DHR-DHR-B-proof}, 
respectively), for any $m \in \{1,2,\ldots\}$, there are $n_2$ and 
$n_1$ such that $n_2 \in \{0,1,\ldots\}$, $n_1 \in \{-1\} \cup 
\{0,1,\ldots,n_1^*\}$, and 
\[
\{x_1,\ldots,x_m\} = 
\{(2,0),(2,1),\ldots,(2,n_2)\} \cup \{(1,0),(1,1),\ldots,(1,n_1)\}, 
\]
where the latter part of the right hand side is omitted if $n_1 = -1$. 
Recall also that these $n_2$ and $n_1$ satisfy results 
(\ref{eq:phi-result1-DHR-DHR}) and (\ref{eq:phi-result2-DHR-DHR}).

The main proof is now given in seven parts ($1^\circ$--$7^\circ$). 
The idea is again to solve the relaxed optimization problem 
(\ref{eq:separable-discounted-costs}) for any $\nu$ by utilizing the 
optimality equations (\ref{eq:opt-eqs-discounted-general}). We partition 
the possible values of $\nu$, which is reflected by the seven parts of 
the main proof. 
However, parts $1^\circ$--$3^\circ$ are exactly the same as in the 
DHR-DHR-A subcase (see Appendix~\ref{app:DHR-DHR-A-proof}).\footnote{
In fact, part $3^\circ$ is not even needed for the special case where 
$n_1^* = -1$.}
Therefore, we omit them here and focus 
on the remaining parts $4^\circ$--$7^\circ$.

\vskip 3pt
$4^\circ$ 
Here we assume that $\nu \ge 0$, and the optimality equations 
(\ref{eq:opt-eqs-discounted-general}) read as given in 
(\ref{eq:opt-eqs-discounted-nu-pos}). 
We prove that the policy $\pi$ with activity set 
\[
{\mathcal B}^\pi = 
\{x_1,x_2,\ldots\} = 
\{(2,0),(2,1),\ldots\} \cup \{(1,0),(1,1),\ldots,(1,n_1^*)\} 
\]
is $(\nu,\beta)$-optimal for all 
\[
\nu \in [W_\beta(1,n_1^*+1),W_\beta(2,\infty)], 
\]
where $W_\beta(2,\infty)$ is defined in 
(\ref{eq:Whittle-index-21-infty-discounted-DHR-DHR-C}). 
It remains to prove that policy $\pi$ is optimal for these values of $\nu$ 
in any state $x \in {\mathcal S} \setminus \{*\}$. 
This is done below in two parts ($4.1^\circ$ and $4.2^\circ$).

\vskip 3pt
$4.1^\circ$ 
We start by first deriving the value function $V_\beta^\pi(x;\nu)$ for 
policy $\pi$ from the Howard equations: 
\begin{equation}
\begin{split}
& 
V_\beta^\pi(1,n;\nu) = 
h + \nu + \beta \mu_1(n) V_\beta^\pi(2,0;\nu) + 
\beta (1 - \mu_1(n)) V_\beta^\pi(1,n+1;\nu), \\
& \quad 
n \in \{0,1,\ldots,n_1^*\}, \\
& 
V_\beta^\pi(2,n;\nu) = 
h + \nu + \beta (1 - \mu_2(n)) V_\beta^\pi(2,n+1;\nu), \\
& \quad 
n \in \{0,1,\ldots\}, \\
& 
V_\beta^\pi(x;\nu) = 
h + \beta V_\beta^\pi(x;\nu), 
\quad x \in \{(1,n_1^*+1),(1,n_1^*+2),\ldots\}.
\end{split}
\label{eq:howard-eqs-discounted-DHR-DHR-C-case-4}
\end{equation}
The unique solution of these linear equations is given by 
\begin{equation}
\begin{split}
& 
V_\beta^\pi(2,n;\nu) = 
(h + \nu) 
\left( 
\sum_{i = 0}^{\infty} \beta^i \bar p_2(i|n) 
\right), 
\quad 
n \in \{0,1,\ldots\}, \\
& 
V_\beta^\pi(1,n;\nu) = 
(h + \nu) 
\left( 
\sum_{i = 0}^{n_1^*-n} \beta^i \bar p_1(i|n) 
\left( 
1 + \beta \mu_1(n+i) \frac{V_\beta^\pi(2,0;\nu)}{h + \nu} 
\right) 
\right) \; + \\
& \quad 
h \, \frac{\beta^{n_1^*-n+1} \bar p_1(n_1^*-n+1|n)}{1 - \beta}, 
\quad 
n \in \{0,1,\ldots,n_1^*\}, \\
& 
V_\beta^\pi(x;\nu) = 
\frac{h}{1 - \beta}, 
\quad x \in \{(1,n_1^*+1),(1,n_1^*+2),\ldots\}.
\end{split}
\label{eq:howard-eqs-discounted-DHR-DHR-C-case-4-solution}
\end{equation}

Let then $n \in \{0,1,\ldots\}$. 
Since $\bar p_2(i|n)$ is an increasing function of $n$ in the DHR-DHR 
case, we see from (\ref{eq:howard-eqs-discounted-DHR-DHR-C-case-4-solution}) 
that $V_\beta^\pi(2,n;\nu)$, as well, is an increasing function of $n$ 
and approaches 
\[
\begin{split}
& 
\lim_{n \to \infty} V_\beta^\pi(2,n;\nu) = 
(h + \nu) 
\left( 
\sum_{i = 0}^\infty \beta^i (1 - \mu_2(\infty))^i 
\right) \\
& \quad = 
\frac{h + \nu}{1 - \beta (1 - \mu_2(\infty))}.
\end{split}
\]
Thus, the following condition for optimality of $\pi$ in state~$(2,n)$, 
\[
V_\beta^\pi(2,n;\nu) \le \frac{h}{1 - \beta}, 
\]
is satisfied for any $n$ if and only if 
\[
\lim_{n \to \infty} V_\beta^\pi(2,n;\nu) \le \frac{h}{1 - \beta}, 
\]
which is clearly equivalent with condition 
\begin{equation}
\nu \le 
h \mu_2(\infty) \, 
\frac{\beta}{1 - \beta}. 
\label{eq:nu-req-discounted-DHR-DHR-C-case-41a}
\end{equation}
Note that the right hand side equals $W_\beta(2,\infty)$ given in 
(\ref{eq:Whittle-index-21-infty-discounted-DHR-DHR-C}).

If $n_1^* = -1$, then part $4.1^\circ$ of the proof is complete here, but 
otherwise we still have to continue. So below we assume (until the end of 
part $4.1^\circ$) that $n_1^* \ge 0$.

By (\ref{eq:howard-eqs-discounted-DHR-DHR-C-case-4-solution}) and some 
algebraic manipulations, the following condition for optimality of $\pi$ 
in state~$(1,n_1^*)$, 
\[
V_\beta^\pi(1,n_1^*;\nu) \le \frac{h}{1 - \beta}, 
\]
can be shown to be equivalent with 
\begin{equation}
\nu \le 
h \mu_1(n_1^*) \, 
\frac{\beta \sum_{i=0}^{\infty} \beta^i p_2(i)}{1 + \beta \mu_1(n_1^*) \sum_{i=0}^{\infty} \beta^i \bar p_2(i)} \, 
\frac{\beta}{1 - \beta}, 
\label{eq:nu-req-discounted-DHR-DHR-C-case-41b}
\end{equation}
where the right hand side equals 
$\psi(n_1^*,\infty) = \lim_{n_2 \to \infty} \psi(n_1^*,n_2)$. 

Next we prove that $\psi(n_1^*,\infty) \ge w_2(\infty)$. First, by definition, 
\[
w_1(n_1^*) = \psi(n_1^*,\phi(n_1^*)) > w_2(\phi(n_1^*)+1).
\]
However, by Lemma~\ref{lem:Whittle-index-discounted-DHR-DHR-lemma-1}(iv), 
this is equivalent with 
\[
\psi(n_1,\phi(n_1^*)+1) > w_2(n_2^*+1).
\]
But, by Lemma~\ref{lem:Whittle-index-discounted-DHR-DHR-lemma-1}(i), this implies that 
\[
\psi(n_1^*,\phi(n_1^*)+1) > w_2(\phi(n_1^*)+2), 
\]
which is, again by Lemma~\ref{lem:Whittle-index-discounted-DHR-DHR-lemma-1}(iv), 
equivalent with 
\[
\psi(n_1^*,\phi(n_1^*)+2) > w_2(\phi(n_1^*)+2).
\]
By continuing similarly, we finally end up to the following inequality: 
for any $n_2 \ge \phi(n_1^*)+1$, 
\[
\psi(n_1^*,n_2) > w_2(n_2).
\]
Thus, we conclude that 
\[
\psi(n_1^*,\infty) = \lim_{n_2 \to \infty} \psi(n_1^*,n_2) \ge 
\lim_{n_2 \to \infty} w_2(n_2) = w_2(\infty), 
\]
which, in turn, proves that requirement 
(\ref{eq:nu-req-discounted-DHR-DHR-C-case-41b}) follows from 
(\ref{eq:nu-req-discounted-DHR-DHR-C-case-41a}).

Now we show that if conditions 
(\ref{eq:nu-req-discounted-DHR-DHR-C-case-41a}) and 
(\ref{eq:nu-req-discounted-DHR-DHR-C-case-41b}) are satisfied, then 
\begin{equation}
V_\beta^\pi(1,0;\nu) \le V_\beta^\pi(1,1;\nu) \le \ldots \le V_\beta^\pi(1,n_1^*;\nu) \le \frac{h}{1 - \beta}, 
\label{eq:nu-req-discounted-DHR-DHR-C-case-41b-cont1}
\end{equation}
which implies that policy $\pi$ is optimal in all states 
$\{(1,0),(1,1),\ldots,(1,n_1^*)\}$. First, as we saw above, condition 
(\ref{eq:nu-req-discounted-DHR-DHR-C-case-41a}) implies that 
$V_\beta^\pi(2,0;\nu) \le \frac{h}{1 - \beta}$, and condition 
(\ref{eq:nu-req-discounted-DHR-DHR-C-case-41b}) implies that 
$V_\beta^\pi(1,n_1^*;\nu) \le \frac{h}{1 - \beta}$. 
Now, by (\ref{eq:howard-eqs-discounted-DHR-DHR-C-case-4}), we observe that 
\[
\begin{split}
& 
V_\beta^\pi(1,n_1^*;\nu) - V_\beta^\pi(1,n_1^*-1;\nu) \; = \\
& \quad 
\beta 
\Big[
\Big( \mu_1(n_1^*) V_\beta^\pi(2,0;\nu) + (1 - \mu_1(n_1^*)) \frac{h}{1 - \beta} \Big) \; - \\
& \quad\quad 
\Big( \mu_1(n_1^*-1) V_\beta^\pi(2,0;\nu) + (1 - \mu_1(n_1^*-1)) V_\beta^\pi(1,n_1^*;\nu) \Big) 
\Big] 
\ge 0, 
\end{split}
\]
where the last inequality follows from the facts that $\mu_1(n_1^*) \le \mu_1(n_1^*-1)$, 
$V_\beta^\pi(2,0;\nu) \le \frac{h}{1 - \beta}$, and 
$V_\beta^\pi(1,n_1^*;\nu) \le \frac{h}{1 - \beta}$. 
In addition, again by (\ref{eq:howard-eqs-discounted-DHR-DHR-C-case-4}), we have 
\[
\begin{split}
& 
V_\beta^\pi(1,n_1^*;\nu) - V_\beta^\pi(2,0;\nu) \; = \\
& \quad 
h + \nu + \beta (1 - \mu_1(n_1^*)) \frac{h}{1 - \beta} - 
(1 - \beta \mu_1(n_1^*)) V_\beta^\pi(2,0;\nu) \; = \\
& \quad 
(1 - \beta) \Big( \frac{h + \nu}{1 - \beta} - V_\beta^\pi(2,0;\nu) \Big) + 
\beta (1 - \mu_1(n_1^*)) \Big( \frac{h}{1 - \beta} - V_\beta^\pi(2,0;\nu) \Big) 
\ge 0, 
\end{split}
\]
where the last inequality follows from the fact that 
$V_\beta^\pi(2,0;\nu) \le \frac{h}{1 - \beta} \le \frac{h + \nu}{1 - \beta}$. 
So, we have proved that if conditions 
(\ref{eq:nu-req-discounted-DHR-DHR-C-case-41a}) and 
(\ref{eq:nu-req-discounted-DHR-DHR-C-case-41b}) are satisfied, then 
\begin{equation}
V_\beta^\pi(2,0;\nu) \le V_\beta^\pi(1,n_1^*;\nu), \quad 
V_\beta^\pi(1,n_1^*-1;\nu) \le V_\beta^\pi(1,n_1^*;\nu).
\label{eq:nu-req-discounted-DHR-DHR-C-case-41b-cont2}
\end{equation}
Next, again by (\ref{eq:howard-eqs-discounted-DHR-DHR-C-case-4}), we observe that 
\[
\begin{split}
& 
V_\beta^\pi(1,n_1^*-1;\nu) - V_\beta^\pi(1,n_1^*-2;\nu) \; = \\
& \quad 
\beta 
\Big[
\Big( \mu_1(n_1^*-1) V_\beta^\pi(2,0;\nu) + (1 - \mu_1(n_1^*-1)) V_\beta^\pi(1,n_1^*;\nu) \Big) \; - \\
& \quad\quad 
\Big( \mu_1(n_1^*-2) V_\beta^\pi(2,0;\nu) + (1 - \mu_1(n_1^*-2)) V_\beta^\pi(1,n_1^*-1;\nu) \Big) 
\Big] 
\ge 0, 
\end{split}
\]
where the last inequality follows from 
(\ref{eq:nu-req-discounted-DHR-DHR-C-case-41b-cont2}) and 
the fact that $\mu_1(n_1^*-1) \le \mu_1(n_1^*-2)$. 
In addition, again by (\ref{eq:howard-eqs-discounted-DHR-DHR-C-case-4}), we have 
\[
\begin{split}
& 
V_\beta^\pi(1,n_1^*-1;\nu) - V_\beta^\pi(2,0;\nu) \; = \\
& \quad 
h + \nu + \beta (1 - \mu_1(n_1^*-1)) V_\beta^\pi(1,n_1^*;\nu) - 
(1 - \beta \mu_1(n_1^*-1)) V_\beta^\pi(2,0;\nu) \; = \\
& \quad 
(1 - \beta) \Big( \frac{h + \nu}{1 - \beta} - V_\beta^\pi(2,0;\nu) \Big) \; + \\
& \quad\quad 
\beta (1 - \mu_1(n_1^*-1)) \Big( V_\beta^\pi(1,n_1^*;\nu) - V_\beta^\pi(2,0;\nu) \Big) 
\ge 0, 
\end{split}
\]
where the last inequality follows from the fact that 
$V_\beta^\pi(2,0;\nu) \le V_\beta^\pi(1,n_1^*;\nu) \le \frac{h}{1 - \beta} \le \frac{h + \nu}{1 - \beta}$. 
So, we have proved that if conditions 
(\ref{eq:nu-req-discounted-DHR-DHR-C-case-41a}) and 
(\ref{eq:nu-req-discounted-DHR-DHR-C-case-41b}) are satisfied, then 
\begin{equation}
V_\beta^\pi(2,0;\nu) \le V_\beta^\pi(1,n_1^*-1;\nu), \quad 
V_\beta^\pi(1,n_1^*-2;\nu) \le V_\beta^\pi(1,n_1^*-1;\nu).
\label{eq:nu-req-discounted-DHR-DHR-C-case-41b-cont3}
\end{equation}
Continuing similarly proves claim (\ref{eq:nu-req-discounted-DHR-DHR-C-case-41b-cont1}).

\vskip 3pt
$4.2^\circ$ 
By (\ref{eq:howard-eqs-discounted-DHR-DHR-C-case-4-solution}) 
and some algebraic manipulations, the following condition for optimality 
of $\pi$ in state $(1,n_1^*+1)$, 
\[
\beta V_\beta^\pi(1,n_1^*+1;\nu) \le 
\nu + \beta \mu_1(n_1^*+1) V_\beta^\pi(2,0;\nu) + 
\beta (1 - \mu_1(n_1^*+1)) V_\beta^\pi(1,n_1^*+2;\nu), 
\]
can be shown to be equivalent with 
\begin{equation}
\nu \ge 
h \mu_1(n_1^*+1) \, 
\frac{\beta \sum_{i=0}^{\infty} \beta^i p_2(i)}{1 + \beta \mu_1(n_1^*+1) \sum_{i=0}^{\infty} \beta^i \bar p_2(i)} \, 
\frac{\beta}{1 - \beta}, 
\label{eq:nu-req-discounted-DHR-DHR-C-case-42}
\end{equation}
where the right hand side equals 
$W_\beta(1,n_1^*+1) = \psi(n_1^*+1,\infty) = \lim_{n_2 \to \infty} \psi(n_1^*+1,n_2)$.

Let then $n \in \{n_1^*+2,n_1^*+3,\ldots\}$. 
Again by (\ref{eq:howard-eqs-discounted-DHR-DHR-C-case-4-solution}) and some 
algebraic manipulations, the following condition for optimality of 
$\pi$ in state~$(1,n)$, 
\[
\beta V_\beta^\pi(1,n;\nu) \le 
\nu + \beta \mu_1(n) V_\beta^\pi(2,0;\nu) + 
\beta (1 - \mu_1(n)) V_\beta^\pi(1,n+1;\nu), 
\]
can be shown to be equivalent with 
\[
\nu \ge 
h \mu_1(n) \, 
\frac{\beta \sum_{i=0}^{\infty} \beta^i p_2(i)}{1 + \beta \mu_1(n) \sum_{i=0}^{\infty} \beta^i \bar p_2(i)} \, 
\frac{\beta}{1 - \beta}, 
\]
which follows from (\ref{eq:nu-req-discounted-DHR-DHR-C-case-42}) since 
$\mu_1(n)$ is decreasing. 
This completes the proof of claim~$4^\circ$.

\vskip 3pt
$5^\circ$ 
We still assume that $\nu \ge 0$ and utilize the optimality equations 
(\ref{eq:opt-eqs-discounted-nu-pos}). 
Let $m \in \{1,2,\ldots\}$. 
We prove that the policy $\pi$ with activity set 
\[
\begin{split}
& 
{\mathcal B}^\pi = \{x_1,x_2,\ldots\} \cup \{(1,n_1^*+1),\ldots,(1,n_1^*+m)\} \; = \\
& \quad 
\{(2,0),(2,1),\ldots\} \cup \{(1,0),(1,1),\ldots,(1,n_1^*+m)\} 
\end{split}
\]
is $(\nu,\beta)$-optimal for all 
\[
\nu \in [W_\beta(1,n_1^*+m+1),W_\beta(1,n_1^*+m)].
\]
It remains to prove that policy $\pi$ is optimal for these values of $\nu$ 
in any state $x \in {\mathcal S} \setminus \{*\}$. 
This is done below in two parts ($5.1^\circ$ and $5.2^\circ$).

\vskip 3pt
$5.1^\circ$ 
We start by first deriving the value function $V_\beta^\pi(x;\nu)$ for 
policy $\pi$ from the Howard equations: 
\begin{equation}
\begin{split}
& 
V_\beta^\pi(1,n;\nu) = 
h + \nu + \beta \mu_1(n) V_\beta^\pi(2,0;\nu) + 
\beta (1 - \mu_1(n)) V_\beta^\pi(1,n+1;\nu), \\
& \quad 
n \in \{0,1,\ldots,n_1^*+m\}, \\
& 
V_\beta^\pi(2,n;\nu) = 
h + \nu + \beta (1 - \mu_2(n)) V_\beta^\pi(2,n+1;\nu), \\
& \quad 
n \in \{0,1,\ldots\}, \\
& 
V_\beta^\pi(x;\nu) = 
h + \beta V_\beta^\pi(x;\nu), 
\quad x \in \{(1,n_1^*+m+1),(1,n_1^*+m+2),\ldots\}.
\end{split}
\label{eq:howard-eqs-discounted-DHR-DHR-C-case-5}
\end{equation}
The unique solution of these linear equations is given by 
\begin{equation}
\begin{split}
& 
V_\beta^\pi(2,n;\nu) = 
(h + \nu) 
\left( 
\sum_{i = 0}^{\infty} \beta^i \bar p_2(i|n) 
\right), 
\quad 
n \in \{0,1,\ldots\}, \\
& 
V_\beta^\pi(1,n;\nu) = 
(h + \nu) 
\left( 
\sum_{i = 0}^{n_1^*+m-n} \beta^i \bar p_1(i|n) 
\left( 
1 + \beta \mu_1(n+i) \frac{V_\beta^\pi(2,0;\nu)}{h + \nu} 
\right) 
\right) \; + \\
& \quad 
h \, \frac{\beta^{n_1^*+m-n+1} \bar p_1(n_1^*+m-n+1|n)}{1 - \beta}, 
\quad 
n \in \{0,1,\ldots,n_1^*+m\}, \\
& 
V_\beta^\pi(x;\nu) = 
\frac{h}{1 - \beta}, 
\quad x \in \{(1,n_1^*+m+1),(1,n_1^*+m+2),\ldots\}.
\end{split}
\label{eq:howard-eqs-discounted-DHR-DHR-C-case-5-solution}
\end{equation}

Let then $n \in \{0,1,\ldots\}$. 
Since $\bar p_2(i|n)$ is an increasing function of $n$ in the DHR-DHR 
case, we see from (\ref{eq:howard-eqs-discounted-DHR-DHR-C-case-5-solution}) 
that $V_\beta^\pi(2,n;\nu)$, as well, is an increasing function of $n$ 
and approaches 
\[
\begin{split}
& 
\lim_{n \to \infty} V_\beta^\pi(2,n;\nu) = 
(h + \nu) 
\left( 
\sum_{i = 0}^\infty \beta^i (1 - \mu_2(\infty))^i 
\right) \\
& \quad = 
\frac{h + \nu}{1 - \beta (1 - \mu_2(\infty))}.
\end{split}
\]
Thus, the following condition for optimality of $\pi$ in state~$(2,n)$, 
\[
V_\beta^\pi(2,n;\nu) \le \frac{h}{1 - \beta}, 
\]
is satisfied for any $n$ if and only if 
\[
\lim_{n \to \infty} V_\beta^\pi(2,n;\nu) \le \frac{h}{1 - \beta}, 
\]
which is clearly equivalent with condition 
\begin{equation}
\nu \le 
h \mu_2(\infty) \, 
\frac{\beta}{1 - \beta}. 
\label{eq:nu-req-discounted-DHR-DHR-C-case-51a}
\end{equation}
Note that the right hand side equals $W_\beta(2,\infty)$ given in 
(\ref{eq:Whittle-index-21-infty-discounted-DHR-DHR-C}).

On the other hand, by (\ref{eq:howard-eqs-discounted-DHR-DHR-C-case-5-solution}) 
and some algebraic manipulations, the following condition for optimality 
of $\pi$ in state $(1,n_1^*+m)$, 
\[
V_\beta^\pi(1,n_1^*+m;\nu) \le \frac{h}{1 - \beta}, 
\]
can be shown to be equivalent with 
\begin{equation}
\nu \le 
h \mu_1(n_1^*+m) \, 
\frac{\beta \sum_{i=0}^{\infty} \beta^i p_2(i)}{1 + \beta \mu_1(n_1^*+m) \sum_{i=0}^{\infty} \beta^i \bar p_2(i)} \, 
\frac{\beta}{1 - \beta}, 
\label{eq:nu-req-discounted-DHR-DHR-C-case-51b}
\end{equation}
where the right hand side equals 
$W_\beta(1,n_1^*+m) = \psi(n_1^*+m,\infty) = \lim_{n_2 \to \infty} \psi(n_1^*+m,n_2)$. 
Note that (\ref{eq:nu-req-discounted-DHR-DHR-C-case-51a}) follows from 
(\ref{eq:nu-req-discounted-DHR-DHR-C-case-51b}) since 
$W_\beta(2,\infty) \ge W_\beta(1,n_1^*+m)$ by (\ref{eq:Wx-order-DHR-DHR-C}).

Now we show that if conditions 
(\ref{eq:nu-req-discounted-DHR-DHR-C-case-51a}) and 
(\ref{eq:nu-req-discounted-DHR-DHR-C-case-51b}) are satisfied, then 
\begin{equation}
V_\beta^\pi(1,0;\nu) \le V_\beta^\pi(1,1;\nu) \le \ldots \le V_\beta^\pi(1,n_1^*+m;\nu) \le \frac{h}{1 - \beta}, 
\label{eq:nu-req-discounted-DHR-DHR-C-case-51b-cont1}
\end{equation}
which implies that policy $\pi$ is optimal in all states 
$\{(1,0),(1,1),\ldots,(1,n_1^*+m)\}$. First, as we saw above, condition 
(\ref{eq:nu-req-discounted-DHR-DHR-C-case-51a}) implies that 
$V_\beta^\pi(2,0;\nu) \le \frac{h}{1 - \beta}$, and condition 
(\ref{eq:nu-req-discounted-DHR-DHR-C-case-51b}) implies that 
$V_\beta^\pi(1,n_1^*+m;\nu) \le \frac{h}{1 - \beta}$. 
Now, by (\ref{eq:howard-eqs-discounted-DHR-DHR-C-case-5}), we observe that 
\[
\begin{split}
& 
V_\beta^\pi(1,n_1^*+m;\nu) - V_\beta^\pi(1,n_1^*+m-1;\nu) \; = \\
& \quad 
\beta 
\Big[
\Big( \mu_1(n_1^*+m) V_\beta^\pi(2,0;\nu) + (1 - \mu_1(n_1^*+m)) \frac{h}{1 - \beta} \Big) \; - \\
& \quad\quad 
\Big( \mu_1(n_1^*+m-1) V_\beta^\pi(2,0;\nu) + (1 - \mu_1(n_1^*+m-1)) V_\beta^\pi(1,n_1^*+m;\nu) \Big) 
\Big] 
\ge 0, 
\end{split}
\]
where the last inequality follows from the facts that $\mu_1(n_1^*+m) \le \mu_1(n_1^*+m-1)$, 
$V_\beta^\pi(2,0;\nu) \le \frac{h}{1 - \beta}$, and 
$V_\beta^\pi(1,n_1^*+m;\nu) \le \frac{h}{1 - \beta}$. 
In addition, again by (\ref{eq:howard-eqs-discounted-DHR-DHR-C-case-5}), we have 
\[
\begin{split}
& 
V_\beta^\pi(1,n_1^*+m;\nu) - V_\beta^\pi(2,0;\nu) \; = \\
& \quad 
h + \nu + \beta (1 - \mu_1(n_1^*+m)) \frac{h}{1 - \beta} - 
(1 - \beta \mu_1(n_1^*+m)) V_\beta^\pi(2,0;\nu) \; = \\
& \quad 
(1 - \beta) \Big( \frac{h + \nu}{1 - \beta} - V_\beta^\pi(2,0;\nu) \Big) + 
\beta (1 - \mu_1(n_1^*+m)) \Big( \frac{h}{1 - \beta} - V_\beta^\pi(2,0;\nu) \Big) 
\ge 0, 
\end{split}
\]
where the last inequality follows from the fact that 
$V_\beta^\pi(2,0;\nu) \le \frac{h}{1 - \beta} \le \frac{h + \nu}{1 - \beta}$. 
So, we have proved that if conditions 
(\ref{eq:nu-req-discounted-DHR-DHR-C-case-51a}) and 
(\ref{eq:nu-req-discounted-DHR-DHR-C-case-51b}) are satisfied, then 
\begin{equation}
V_\beta^\pi(2,0;\nu) \le V_\beta^\pi(1,n_1^*+m;\nu), \quad 
V_\beta^\pi(1,n_1^*+m-1;\nu) \le V_\beta^\pi(1,n_1^*+m;\nu).
\label{eq:nu-req-discounted-DHR-DHR-C-case-51b-cont2}
\end{equation}
Next, again by (\ref{eq:howard-eqs-discounted-DHR-DHR-C-case-5}), we observe that 
\[
\begin{split}
& 
V_\beta^\pi(1,n_1^*+m-1;\nu) - V_\beta^\pi(1,n_1^*+m-2;\nu) \; = \\
& \quad 
\beta 
\Big[
\Big( \mu_1(n_1^*+m-1) V_\beta^\pi(2,0;\nu) + (1 - \mu_1(n_1^*+m-1)) V_\beta^\pi(1,n_1^*+m;\nu) \Big) \; - \\
& \quad\quad 
\Big( \mu_1(n_1^*+m-2) V_\beta^\pi(2,0;\nu) + (1 - \mu_1(n_1^*+m-2)) V_\beta^\pi(1,n_1^*+m-1;\nu) \Big) 
\Big] 
\ge 0, 
\end{split}
\]
where the last inequality follows from 
(\ref{eq:nu-req-discounted-DHR-DHR-C-case-51b-cont2}) and 
the fact that $\mu_1(n_1^*+m-1) \le \mu_1(n_1^*+m-2)$. 
In addition, again by (\ref{eq:howard-eqs-discounted-DHR-DHR-C-case-5}), we have 
\[
\begin{split}
& 
V_\beta^\pi(1,n_1^*+m-1;\nu) - V_\beta^\pi(2,0;\nu) \; = \\
& \quad 
h + \nu + \beta (1 - \mu_1(n_1^*+m-1)) V_\beta^\pi(1,n_1^*+m;\nu) - 
(1 - \beta \mu_1(n_1^*+m-1)) V_\beta^\pi(2,0;\nu) \; = \\
& \quad 
(1 - \beta) \Big( \frac{h + \nu}{1 - \beta} - V_\beta^\pi(2,0;\nu) \Big) \; + \\
& \quad\quad 
\beta (1 - \mu_1(n_1^*+m-1)) \Big( V_\beta^\pi(1,n_1^*+m;\nu) - V_\beta^\pi(2,0;\nu) \Big) 
\ge 0, 
\end{split}
\]
where the last inequality follows from the fact that 
$V_\beta^\pi(2,0;\nu) \le V_\beta^\pi(1,n_1^*+m;\nu) \le \frac{h}{1 - \beta} \le \frac{h + \nu}{1 - \beta}$. 
So, we have proved that if conditions 
(\ref{eq:nu-req-discounted-DHR-DHR-C-case-51a}) and 
(\ref{eq:nu-req-discounted-DHR-DHR-C-case-51b}) are satisfied, then 
\begin{equation}
V_\beta^\pi(2,0;\nu) \le V_\beta^\pi(1,n_1^*+m-1;\nu), \quad 
V_\beta^\pi(1,n_1^*+m-2;\nu) \le V_\beta^\pi(1,n_1^*+m-1;\nu).
\label{eq:nu-req-discounted-DHR-DHR-C-case-51b-cont3}
\end{equation}
Continuing similarly proves claim (\ref{eq:nu-req-discounted-DHR-DHR-C-case-51b-cont1}).

\vskip 3pt
$5.2^\circ$ 
By (\ref{eq:howard-eqs-discounted-DHR-DHR-C-case-5-solution}) 
and some algebraic manipulations, the following condition for optimality 
of $\pi$ in state $(1,n_1^*+m+1)$, 
\[
\begin{split}
& 
\beta V_\beta^\pi(1,n_1^*+m+1;\nu) \; \le \\
& \quad 
\nu + \beta \mu_1(n_1^*+m+1) V_\beta^\pi(2,0;\nu) + 
\beta (1 - \mu_1(n_1^*+m+1)) V_\beta^\pi(1,n_1^*+m+2;\nu), 
\end{split}
\]
can be shown to be equivalent with 
\begin{equation}
\nu \ge 
h \mu_1(n_1^*+m+1) \, 
\frac{\beta \sum_{i=0}^{\infty} \beta^i p_2(i)}{1 + \beta \mu_1(n_1^*+m+1) \sum_{i=0}^{\infty} \beta^i \bar p_2(i)} \, 
\frac{\beta}{1 - \beta}, 
\label{eq:nu-req-discounted-DHR-DHR-C-case-52}
\end{equation}
where the right hand side equals 
$W_\beta(1,n_1^*+m+1) = \psi(n_1^*+m+1,\infty) = \lim_{n_2 \to \infty} 
\psi(n_1^*+m+1,n_2)$.

Let then $n \in \{n_1^*+m+2,n_1^*+m+3,\ldots\}$. 
Again by (\ref{eq:howard-eqs-discounted-DHR-DHR-C-case-5-solution}) and some 
algebraic manipulations, the following condition for optimality of 
$\pi$ in state~$(1,n)$, 
\[
\beta V_\beta^\pi(1,n;\nu) \le 
\nu + \beta \mu_1(n) V_\beta^\pi(2,0;\nu) + 
\beta (1 - \mu_1(n)) V_\beta^\pi(1,n+1;\nu), 
\]
can be shown to be equivalent with 
\[
\nu \ge 
h \mu_1(n) \, 
\frac{\beta \sum_{i=0}^{\infty} \beta^i p_2(i)}{1 + \beta \mu_1(n) \sum_{i=0}^{\infty} \beta^i \bar p_2(i)} \, 
\frac{\beta}{1 - \beta}, 
\]
which follows from (\ref{eq:nu-req-discounted-DHR-DHR-C-case-52}) since 
$\mu_1(n)$ is decreasing. 
This completes the proof of claim~$5^\circ$.

\vskip 3pt
$6^\circ$ 
We still assume that $\nu \ge 0$ and utilize the optimality equations 
(\ref{eq:opt-eqs-discounted-nu-pos}). 
Now we prove that the policy $\pi$ with activity set 
\[
{\mathcal B}^\pi = {\mathcal S} \setminus \{*\}, 
\]
is $(\nu,\beta)$-optimal for all 
\[
\nu \in [0,W_\beta(1,\infty)], 
\]
where $W_\beta(1,\infty)$ is defined in (\ref{eq:Whittle-index-21-infty-discounted-DHR-DHR-C}). 
It remains to prove that policy $\pi$ is optimal for these values of $\nu$ 
in any state $x \in {\mathcal S} \setminus \{*\}$.

We start by first deriving the value function $V_\beta^\pi(x;\nu)$ for 
policy $\pi$ from the Howard equations: 
\begin{equation}
\begin{split}
& 
V_\beta^\pi(1,n;\nu) = 
h + \nu + \beta \mu_1(n) V_\beta^\pi(2,0;\nu) + 
\beta (1 - \mu_1(n)) V_\beta^\pi(1,n+1;\nu), \\
& 
V_\beta^\pi(2,n;\nu) = 
h + \nu + \beta (1 - \mu_2(n)) V_\beta^\pi(2,n+1;\nu).
\end{split}
\label{eq:howard-eqs-discounted-DHR-DHR-C-case-6}
\end{equation}
The unique solution of these linear equations is given by 
\begin{equation}
\begin{split}
& 
V_\beta^\pi(2,n;\nu) = 
(h + \nu) 
\left( 
\sum_{i = 0}^\infty \beta^i \bar p_2(i|n) 
\right), \\
& 
V_\beta^\pi(1,n;\nu) = 
(h + \nu) 
\left( 
\sum_{i = 0}^\infty \beta^i \bar p_1(i|n) 
\left( 
1 + \beta \mu_1(n+i) \frac{V_\beta^\pi(2,0;\nu)}{h + \nu} 
\right) 
\right).
\end{split}
\label{eq:howard-eqs-discounted-DHR-DHR-C-case-6-solution}
\end{equation}

Let then $n \in \{0,1,\ldots\}$. 
Since $\bar p_2(i|n)$ is an increasing function of $n$ in the DHR-DHR 
case, we see from (\ref{eq:howard-eqs-discounted-DHR-DHR-C-case-6-solution}) 
that $V_\beta^\pi(2,n;\nu)$, as well, is an increasing function of $n$ 
and approaches 
\[
\begin{split}
& 
\lim_{n \to \infty} V_\beta^\pi(2,n;\nu) = 
(h + \nu) 
\left( 
\sum_{i = 0}^\infty \beta^i (1 - \mu_2(\infty))^i 
\right) \\
& \quad = 
\frac{h + \nu}{1 - \beta (1 - \mu_2(\infty))}.
\end{split}
\]
Thus, the following condition for optimality of $\pi$ in state~$(2,n)$, 
\[
V_\beta^\pi(2,n;\nu) \le \frac{h}{1 - \beta}, 
\]
is satisfied for any $n$ if and only if 
\[
\lim_{n \to \infty} V_\beta^\pi(2,n;\nu) \le \frac{h}{1 - \beta}, 
\]
which is clearly equivalent with condition 
\begin{equation}
\nu \le 
h \mu_2(\infty) \, 
\frac{\beta}{1 - \beta}. 
\label{eq:nu-req-discounted-DHR-DHR-C-case-6a}
\end{equation}
Note that the right hand side equals $W_\beta(2,\infty)$ given in 
(\ref{eq:Whittle-index-21-infty-discounted-DHR-DHR-C}).

Let again $n \in \{0,1,\ldots\}$. 
Since 
\[
\bar p_1(i|n) \mu_1(n+i) = p_1(i|n) = \bar p_1(i|n) - \bar p_1(i+1|n), 
\]
it follows from 
(\ref{eq:howard-eqs-discounted-DHR-DHR-C-case-6-solution}) that 
\begin{equation}
\begin{split}
& 
V_\beta^\pi(1,n;\nu) = 
(h + \nu) 
\Bigg( 
1 + \frac{\beta V_\beta^\pi(2,0;\nu)}{h + \nu} \; + \\
& \quad 
\sum_{i = 1}^\infty \beta^i \bar p_1(i|n) 
\bigg( 
1 - \frac{(1 - \beta) V_\beta^\pi(2,0;\nu)}{h + \nu} 
\bigg) 
\Bigg).
\end{split}
\label{eq:howard-eqs-discounted-DHR-DHR-C-case-6-solution-cont}
\end{equation}
Now, since $\bar p_1(i|n)$ is an increasing function of $n$ in the DHR-DHR 
case and we have above required that 
\[
V_\beta^\pi(2,0;\nu) \le \frac{h}{1 - \beta}, 
\]
we see from (\ref{eq:howard-eqs-discounted-DHR-DHR-C-case-6-solution-cont}) 
that $V_\beta^\pi(1,n;\nu)$, as well, is an increasing function of $n$ 
and approaches, by (\ref{eq:howard-eqs-discounted-DHR-DHR-C-case-6-solution}), 
\[
\begin{split}
& 
\lim_{n \to \infty} V_\beta^\pi(1,n;\nu) = 
(h + \nu) 
\Bigg( 
\sum_{i = 0}^\infty \beta^i (1 - \mu_1(\infty))^i 
\bigg( 
1 + \beta \mu_1(\infty) \frac{V_\beta^\pi(2,0;\nu)}{h + \nu} 
\bigg) 
\Bigg) \\
& \quad = 
(h + \nu) \, 
\frac{1 + \beta \mu_1(\infty) \sum_{i = 0}^\infty \beta^i \bar p_2(i)}
{1 - \beta (1 - \mu_1(\infty))}.
\end{split}
\]
Thus, the following condition for optimality of $\pi$ in state~$(1,n)$, 
\[
V_\beta^\pi(1,n;\nu) \le \frac{h}{1 - \beta}, 
\]
is satisfied for any $n$ if and only if 
\[
\lim_{n \to \infty} V_\beta^\pi(1,n;\nu) \le \frac{h}{1 - \beta}, 
\]
which can be shown to be equivalent with condition 
\begin{equation}
\nu \le 
h \mu_1(\infty) \, 
\frac{\beta \sum_{i=0}^{\infty} \beta^i p_2(i)}
{1 + \beta \mu_1(\infty) \sum_{i=0}^{\infty} \beta^i \bar p_2(i)} \, 
\frac{\beta}{1 - \beta}.
\label{eq:nu-req-discounted-DHR-DHR-C-case-6b}
\end{equation}
Note that the right hand side equals 
$\psi(\infty,\infty) = \lim_{n_1, n_2 \to \infty} \psi(n_1,n_2)$, which 
in this DHR-DHR-C subcase equals $W_\beta(1,\infty)$ given in 
(\ref{eq:Whittle-index-21-infty-discounted-DHR-DHR-C}). 
Note also that (\ref{eq:nu-req-discounted-DHR-DHR-C-case-6a}) follows from 
(\ref{eq:nu-req-discounted-DHR-DHR-C-case-6b}) since 
$W_\beta(2,\infty) \ge W_\beta(1,\infty)$ by (\ref{eq:Wx-order-DHR-DHR-C}).
This completes the proof of claim~$6^\circ$.

\vskip 3pt
$7^\circ$ 
Finally, we assume that $\nu \le 0$. 
However, the claim that the policy $\pi$ with activity set 
\[
{\mathcal B}^\pi = {\mathcal S} 
\]
is $(\nu,\beta)$-optimal for all 
\[
\nu \in (-\infty,0] 
\]
can be proved similarly as the corresponding claim~$5^\circ$ in the DHR-DHR-A 
subcase (see Appendix~\ref{app:DHR-DHR-A-proof}). 
Therefore we may omit the proof here.
\hfill $\Box$

\section{Proof of Theorem~\ref{thm:Whittle-index-discounted-IHR-IHR} 
(the IHR-IHR case)}
\label{app:IHR-IHR-proof}

\paragraph{Proof} 
Assume the IHR-IHR case. 
Note that, by Lemmas~\ref{lem:Whittle-index-discounted-IHR-IHR-lemma-1} and 
\ref{lem:Whittle-index-discounted-IHR-IHR-lemma-2}, we have the following 
ordering among the states: 
\begin{equation}
\begin{split}
& 
W_\beta(2,\infty) \ge \ldots \ge W_\beta(2,1) \ge W_\beta(2,0) > \\
& \quad 
W_\beta(1,\infty) \ge \ldots \ge W_\beta(1,1) \ge W_\beta(1,0) > 0, 
\end{split}
\label{eq:Wx-order-IHR-IHR}
\end{equation}
where we have defined 
\begin{equation}
\begin{split}
& 
W_\beta(2,\infty) = w_2(\infty) = 
h \mu_2(\infty) \, 
\frac{\beta}{1 - \beta}, \\
& 
W_\beta(1,\infty) = w_1(\infty) = 
h \mu_1(\infty) \, 
\frac{\beta \sum_{i=0}^{\infty} \beta^i p_2(i)}
{1 + \beta \mu_1(\infty) \sum_{i=0}^{\infty} \beta^i \bar p_2(i)} \, 
\frac{\beta}{1 - \beta}.
\end{split}
\label{eq:Whittle-index-21-infty-discounted-IHR-IHR}
\end{equation}

The main proof is now given in six parts ($1^\circ$--$6^\circ$). 
The idea is again to solve the relaxed optimization problem 
(\ref{eq:separable-discounted-costs}) for any $\nu$ by utilizing the 
optimality equations (\ref{eq:opt-eqs-discounted-general}). We partition 
the possible values of $\nu$, which is reflected by the six parts of 
the main proof.

\vskip 3pt
$1^\circ$ 
We first assume that $\nu \ge 0$. 
In this case, the optimal decision in state $*$ is not to schedule ($a = 0$), 
the minimum expected discounted cost $V_\beta(*;\nu)$ equals $0$, and the 
optimality equations (\ref{eq:opt-eqs-discounted-general}) read as given 
in (\ref{eq:opt-eqs-discounted-nu-pos}). 
We prove that the policy $\pi$ with activity set 
\[
{\mathcal B}^\pi = \emptyset, 
\]
according to which user~$k$ is not scheduled in any state $x \in {\mathcal S}$, 
is $(\nu,\beta)$-optimal for all 
\[
\nu \in [W_\beta(2,\infty),\infty), 
\]
where $W_\beta(2,\infty)$ as defined in 
(\ref{eq:Whittle-index-21-infty-discounted-IHR-IHR}). 
It remains to prove that policy $\pi$ is optimal for these values of $\nu$ 
in any state $x \in {\mathcal S} \setminus \{*\}$.

We start by first deriving the value function $V_\beta^\pi(x;\nu)$ for 
policy $\pi$ from the Howard equations: 
\begin{equation}
V_\beta^\pi(x;\nu) = 
h + \beta V_\beta^\pi(x;\nu), 
\quad x \in {\mathcal S} \setminus \{*\}.
\label{eq:howard-eqs-discounted-IHR-IHR-case-1}
\end{equation}
The unique solution of these linear equations is clearly given by 
\begin{equation}
V_\beta^\pi(x;\nu) = \frac{h}{1 - \beta}, 
\quad x \in {\mathcal S} \setminus \{*\}.
\label{eq:howard-eqs-discounted-IHR-IHR-case-1-solution}
\end{equation}

Let $n \in \{0,1,\ldots\}$. 
By (\ref{eq:howard-eqs-discounted-IHR-IHR-case-1-solution}), the following 
condition for optimality of $\pi$ (based on (\ref{eq:opt-eqs-discounted-nu-pos})) 
in state~$(2,n)$, 
\[
\beta V_\beta^\pi(2,n;\nu) \le 
\nu + \beta (1 - \mu_2(n)) V_\beta^\pi(2,n+1;\nu), 
\]
is easily shown to be equivalent with 
\[
\nu \ge 
h \mu_2(n) \, \frac{\beta}{1 - \beta}.
\]
Since $\mu_2(n)$ is increasing, we conclude that policy $\pi$ is optimal 
in any state~$(2,n)$ if and only if 
\begin{equation}
\nu \ge 
\lim_{n \to \infty} 
h \mu_2(n) \, \frac{\beta}{1 - \beta} = 
h \mu_2(\infty) \, \frac{\beta}{1 - \beta}, 
\label{eq:nu-req-discounted-IHR-IHR-case-1}
\end{equation}
where the right hand side equals $W_\beta(2,\infty)$ given in 
(\ref{eq:Whittle-index-21-infty-discounted-IHR-IHR}).

Let $n \in \{0,1,\ldots\}$. 
Again by (\ref{eq:howard-eqs-discounted-IHR-IHR-case-1-solution}), 
the following condition for optimality of $\pi$ in state~$(1,n)$, 
\[
\beta V_\beta^\pi(1,n;\nu) \le 
\nu + \beta \mu_1(n) V_\beta^\pi(2,0;\nu) + 
\beta (1 - \mu_1(n)) V_\beta^\pi(1,n+1;\nu), 
\]
is easily shown to be equivalent with condition $\nu \ge 0$, 
which is satisfied by the assumption made in the beginning of 
this part of the proof. This completes the proof of claim~$1^\circ$.

\vskip 3pt
$2^\circ$ 
We still assume that $\nu \ge 0$ and utilize the optimality equations 
(\ref{eq:opt-eqs-discounted-nu-pos}). 
Let $m \in \{1,2,\ldots\}$. 
We prove that the policy $\pi$ with activity set 
\[
{\mathcal B}^\pi = \{(2,m),(2,m+1),\ldots\}, 
\]
according to which user~$k$ is scheduled in states $(2,m),(2,m+1),\ldots$, is 
$(\nu,\beta)$-optimal for all 
\[
\nu \in [W_\beta(2,m-1),W_\beta(2,m)].
\]
It remains to prove that policy $\pi$ is optimal for these values of $\nu$ 
in any state $x \in {\mathcal S} \setminus \{*\}$. 
This is done below in two parts ($2.1^\circ$ and $2.2^\circ$).

\vskip 3pt
$2.1^\circ$ 
We start by first deriving the value function $V_\beta^\pi(x;\nu)$ for 
policy $\pi$ from the Howard equations: 
\begin{equation}
\begin{split}
& 
V_\beta^\pi(2,n;\nu) = 
h + \nu + \beta (1 - \mu_2(n)) V_\beta^\pi(2,n+1;\nu), 
\quad n \in \{m,m+1,\ldots\}, \\
& 
V_\beta^\pi(x;\nu) = 
h + \beta V_\beta^\pi(x;\nu), 
\quad x \in {\mathcal S} \setminus \{(2,m),(2,m+1),\ldots,*\}.
\end{split}
\label{eq:howard-eqs-discounted-IHR-IHR-case-2}
\end{equation}
The unique solution of these linear equations is given by 
\begin{equation}
\begin{split}
& 
V_\beta^\pi(2,n;\nu) = 
(h + \nu) 
\left( 
\sum_{i = 0}^{\infty} \beta^i \bar p_2(i|n) 
\right), 
\quad 
n \in \{m,m+1,\ldots\}, \\
& 
V_\beta^\pi(x;\nu) = 
\frac{h}{1 - \beta}, 
\quad x \in {\mathcal S} \setminus\{(2,m),(2,m+1),\ldots,*\}.
\end{split}
\label{eq:howard-eqs-discounted-IHR-IHR-case-2-solution}
\end{equation}

By (\ref{eq:howard-eqs-discounted-IHR-IHR-case-2-solution}) and some algebraic 
manipulations, the following condition for optimality of $\pi$ in state~$(2,m)$, 
\[
V_\beta^\pi(2,m;\nu) \le \frac{h}{1 - \beta}, 
\]
can be shown to be equivalent with 
\begin{equation}
\nu \le 
h \, 
\frac{\frac{1}{1 - \beta} - \sum_{i=0}^{\infty} \beta^i \bar p_2(i|m)}
{\sum_{i=0}^{\infty} \beta^i \bar p_2(i|m)}, 
\label{eq:nu-req-discounted-IHR-IHR-case-21}
\end{equation}
where the right hand side equals $W_\beta(2,m)$ given in 
(\ref{eq:Whittle-index-2n-discounted-IHR-IHR}).

Let then $n \in \{m+1,m+2,\ldots\}$. 
Since $\bar p_2(i|n)$ is a decreasing function of $n$ in the IHR-IHR 
case, we see from (\ref{eq:howard-eqs-discounted-IHR-IHR-case-2-solution}) 
that $V_\beta^\pi(2,n;\nu)$ is decreasing with respect to $n$. 
Therefore, the following condition for optimality of $\pi$ in state~$(2,n)$, 
\[
V_\beta^\pi(2,n;\nu) \le \frac{h}{1 - \beta}, 
\]
follows immediately from (\ref{eq:nu-req-discounted-IHR-IHR-case-21}).

\vskip 3pt
$2.2^\circ$ 
By (\ref{eq:howard-eqs-discounted-IHR-IHR-case-2-solution}) and some algebraic 
manipulations, the following condition for optimality of $\pi$ in state~$(2,m-1)$, 
\[
\beta V_\beta^\pi(2,m-1;\nu) \le 
\nu + \beta (1 - \mu_2(m-1)) V_\beta^\pi(2,m;\nu), 
\]
can be shown to be equivalent with 
\begin{equation}
\nu \ge 
h \, 
\frac{\frac{1}{1 - \beta} - \sum_{i=0}^{\infty} \beta^i \bar p_2(i|m-1)}
{\sum_{i=0}^{\infty} \beta^i \bar p_2(i|m-1)}, 
\label{eq:nu-req-discounted-IHR-IHR-case-22}
\end{equation}
where the right hand side equals $W_\beta(2,m-1)$ given in 
(\ref{eq:Whittle-index-2n-discounted-IHR-IHR}).

Let then $n \in \{0,1,\ldots,m-2\}$. 
By (\ref{eq:howard-eqs-discounted-IHR-IHR-case-2-solution}), the following 
condition for optimality of $\pi$ in state~$(2,n)$, 
\[
\beta V_\beta^\pi(2,n;\nu) \le 
\nu + \beta (1 - \mu_2(m)) V_\beta^\pi(2,n+1;\nu), 
\]
is easily shown to be equivalent with 
\[
\nu \ge 
h \mu_2(n) \frac{\beta}{1 - \beta}.
\]
By Lemma~\ref{lem:Whittle-index-discounted-IHR-IHR-lemma-1}, we conclude 
that this condition follows from (\ref{eq:nu-req-discounted-IHR-IHR-case-22}).

Finally, let $n \in \{0,1,\ldots\}$. 
Again by (\ref{eq:howard-eqs-discounted-IHR-IHR-case-2-solution}), 
the following condition for optimality of $\pi$ in state~$(1,n)$, 
\[
\beta V_\beta^\pi(1,n;\nu) \le 
\nu + \beta \mu_1(n) V_\beta^\pi(2,0;\nu) + 
\beta (1 - \mu_1(n)) V_\beta^\pi(1,n+1;\nu), 
\]
is easily shown to be equivalent with condition $\nu \ge 0$, 
which is satisfied by the assumption made in the beginning of 
this part of the proof. This completes the proof of claim~$2^\circ$.

\vskip 3pt
$3^\circ$ 
We still assume that $\nu \ge 0$ and utilize the optimality equations 
(\ref{eq:opt-eqs-discounted-nu-pos}). 
Now we prove that the policy $\pi$ with activity set 
\[
{\mathcal B}^\pi = \{(2,0),(2,1),\ldots\}, 
\]
according to which user~$k$ is scheduled in states $(2,0),(2,1),\ldots$, is 
$(\nu,\beta)$-optimal for all 
\[
\nu \in [W_\beta(1,\infty),W_\beta(2,0)].
\]
It remains to prove that policy $\pi$ is optimal for these values of $\nu$ 
in any state $x \in {\mathcal S} \setminus \{*\}$. 
This is done below in two parts ($3.1^\circ$ and $3.2^\circ$).

\vskip 3pt
$3.1^\circ$ 
We start by first deriving the value function $V_\beta^\pi(x;\nu)$ for 
policy $\pi$ from the Howard equations: 
\begin{equation}
\begin{split}
& 
V_\beta^\pi(2,n;\nu) = 
h + \nu + \beta (1 - \mu_2(n)) V_\beta^\pi(2,n+1;\nu), 
\quad 
n \in \{0,1,\ldots\}, \\
& 
V_\beta^\pi(x;\nu) = 
h + \beta V_\beta^\pi(x;\nu), 
\quad x \in {\mathcal S} \setminus \{(2,0),(2,1),\ldots,*\}.
\end{split}
\label{eq:howard-eqs-discounted-IHR-IHR-case-3}
\end{equation}
The unique solution of these linear equations is given by 
\begin{equation}
\begin{split}
& 
V_\beta^\pi(2,n;\nu) = 
(h + \nu) 
\left( 
\sum_{i = 0}^{\infty} \beta^i \bar p_2(i|n) 
\right), 
\quad 
n \in \{0,1,\ldots\}, \\
& 
V_\beta^\pi(x;\nu) = 
\frac{h}{1 - \beta}, 
\quad x \in {\mathcal S} \setminus\{(2,0),(2,1),\ldots,*\}.
\end{split}
\label{eq:howard-eqs-discounted-IHR-IHR-case-3-solution}
\end{equation}

By (\ref{eq:howard-eqs-discounted-IHR-IHR-case-3-solution}) and some algebraic 
manipulations, the following condition for optimality of $\pi$ in state~$(2,0)$, 
\[
V_\beta^\pi(2,0;\nu) \le \frac{h}{1 - \beta}, 
\]
can be shown to be equivalent with 
\begin{equation}
\nu \le 
h \, 
\frac{\frac{1}{1 - \beta} - \sum_{i=0}^{\infty} \beta^i \bar p_2(i)}
{\sum_{i=0}^{\infty} \beta^i \bar p_2(i)}, 
\label{eq:nu-req-discounted-IHR-IHR-case-31}
\end{equation}
where the right hand side equals $W_\beta(2,0)$ given in 
(\ref{eq:Whittle-index-2n-discounted-IHR-IHR}).

Let then $n \in \{1,2,\ldots\}$. 
Since $\bar p_2(i|n)$ is a decreasing function of $n$ in the IHR-IHR 
case, we see from (\ref{eq:howard-eqs-discounted-IHR-IHR-case-3-solution}) 
that $V_\beta^\pi(2,n;\nu)$ is decreasing with respect to $n$. 
Therefore, the following condition for optimality of $\pi$ in state~$(2,n)$, 
\[
V_\beta^\pi(2,n;\nu) \le \frac{h}{1 - \beta}, 
\]
follows immediately from (\ref{eq:nu-req-discounted-IHR-IHR-case-31}).

\vskip 3pt
$3.2^\circ$ 
Let $n \in \{0,1,\ldots\}$. 
By (\ref{eq:howard-eqs-discounted-IHR-IHR-case-3-solution}) and some 
algebraic manipulations, the following condition for optimality of $\pi$ 
in state~$(1,n)$, 
\[
\beta V_\beta^\pi(1,n;\nu) \le 
\nu + \beta \mu_1(n) V_\beta^\pi(2,0;\nu) + 
\beta (1 - \mu_1(n)) V_\beta^\pi(1,n+1;\nu), 
\]
can be shown to be equivalent with 
\[
\begin{split}
& 
\nu \ge 
h \beta \mu_1(n) \, 
\frac{\frac{1}{1 - \beta} - \sum_{i=0}^{\infty} \beta^i \bar p_2(i|n)}
{1 + \beta \mu_1(n) \sum_{i=0}^{\infty} \beta^i \bar p_2(i|n)} \; = \\
& \quad 
h \mu_1(n) \, 
\frac{\beta \sum_{i=0}^{\infty} \beta^i p_2(i)}
{1 + \beta \mu_1(n) \sum_{i=0}^{\infty} \beta^i \bar p_2(i)} \, 
\frac{\beta}{1 - \beta}.
\end{split}
\]
Since $\mu_1(n)$ is increasing, we conclude that policy $\pi$ is optimal 
in any state~$(1,n)$ if and only if 
\begin{equation}
\nu \ge 
h \mu_1(\infty) \, 
\frac{\beta \sum_{i=0}^{\infty} \beta^i p_2(i)}
{1 + \beta \mu_1(\infty) \sum_{i=0}^{\infty} \beta^i \bar p_2(i)} \, 
\frac{\beta}{1 - \beta}, 
\label{eq:nu-req-discounted-IHR-IHR-case-32}
\end{equation}
where the right hand side equals $W_\beta(1,\infty)$ given in 
(\ref{eq:Whittle-index-21-infty-discounted-IHR-IHR}). 
This completes the proof of claim~$3^\circ$.

\vskip 3pt
$4^\circ$ 
We still assume that $\nu \ge 0$ and utilize the optimality equations 
(\ref{eq:opt-eqs-discounted-nu-pos}). 
Let $m \in \{1,2,\ldots\}$. 
We prove that the policy $\pi$ with activity set 
\[
{\mathcal B}^\pi = \{(2,0),(2,1),\ldots\} \cup \{(1,m),(1,m+1),\ldots\} 
\]
is $(\nu,\beta)$-optimal for all 
\[
\nu \in [W_\beta(1,m-1),W_\beta(1,m)].
\]
It remains to prove that policy $\pi$ is optimal for these values of $\nu$ 
in any state $x \in {\mathcal S} \setminus \{*\}$. 
This is done below in two parts ($4.1^\circ$ and $4.2^\circ$).

\vskip 3pt
$4.1^\circ$ 
We start by first deriving the value function $V_\beta^\pi(x;\nu)$ for 
policy $\pi$ from the Howard equations: 
\begin{equation}
\begin{split}
& 
V_\beta^\pi(1,n;\nu) = 
h + \nu + \beta \mu_1(n) V_\beta^\pi(2,0;\nu) \; + \\
& \quad 
\beta (1 - \mu_1(n)) V_\beta^\pi(1,n+1;\nu), 
\quad 
n \in \{m,m+1,\ldots\}, \\
& 
V_\beta^\pi(2,n;\nu) = 
h + \nu + \beta (1 - \mu_2(n)) V_\beta^\pi(2,n+1;\nu), \\
& \quad 
n \in \{0,1,\ldots\}, \\
& 
V_\beta^\pi(x;\nu) = 
h + \beta V_\beta^\pi(x;\nu), \\
& \quad 
x \in {\mathcal S} \setminus \{(1,m),(1,m+1),\ldots,(2,0),(2,1),\ldots,*\}.
\end{split}
\label{eq:howard-eqs-discounted-IHR-IHR-case-4}
\end{equation}
The unique solution of these linear equations is given by 
\begin{equation}
\begin{split}
& 
V_\beta^\pi(2,n;\nu) = 
(h + \nu) 
\left( 
\sum_{i = 0}^{\infty} \beta^i \bar p_2(i|n) 
\right), 
\quad 
n \in \{0,1,\ldots\}, \\
& 
V_\beta^\pi(1,n;\nu) = 
(h + \nu) 
\left( 
\sum_{i = 0}^{\infty} \beta^i \bar p_1(i|n) 
\left( 
1 + \beta \mu_1(n+i) \frac{V_\beta^\pi(2,0;\nu)}{h + \nu} 
\right) 
\right), \\
& \quad 
n \in \{m,m+1,\ldots\}, \\
& 
V_\beta^\pi(x;\nu) = 
\frac{h}{1 - \beta}, \\
& \quad 
x \in {\mathcal S} \setminus \{(1,m),(1,m+1),\ldots,(2,0),(2,1),\ldots,*\}.
\end{split}
\label{eq:howard-eqs-discounted-IHR-IHR-case-4-solution}
\end{equation}

By (\ref{eq:howard-eqs-discounted-IHR-IHR-case-4-solution}) and some algebraic 
manipulations, the following condition for optimality of $\pi$ in state~$(1,m)$, 
\[
V_\beta^\pi(1,m;\nu) \le \frac{h}{1 - \beta}, 
\]
can be shown to be equivalent with 
\begin{equation}
\nu \le 
h \, 
\frac{\frac{1}{1 - \beta} - \sum_{i=0}^{\infty} \beta^i \bar p_1(i|m) 
\Big( 1 + \beta \mu_1(m+i) \sum_{j=0}^{\infty} \beta^{j} \bar p_2(j) \Big)}
{\sum_{i=0}^{\infty} \beta^i \bar p_1(i|m) 
\Big( 1 + \beta \mu_1(m+i) \sum_{j=0}^{\infty} \beta^{j} \bar p_2(j) \Big)}, 
\label{eq:nu-req-discounted-IHR-IHR-case-41}
\end{equation}
where the right hand side equals $W_\beta(1,m)$ given in 
(\ref{eq:Whittle-index-1n-discounted-IHR-IHR}).

Let then $n \in \{m+1,m+2,\ldots\}$. 
Similarly as above, by (\ref{eq:howard-eqs-discounted-IHR-IHR-case-4-solution}) 
and some algebraic manipulations, the following condition for optimality of 
$\pi$ in state~$(1,n)$, 
\[
V_\beta^\pi(1,n;\nu) \le \frac{h}{1 - \beta}, 
\]
can be shown to be equivalent with 
\[
\nu \le 
h \, 
\frac{\frac{1}{1 - \beta} - \sum_{i=0}^{\infty} \beta^i \bar p_1(i|n) 
\Big( 1 + \beta \mu_1(n+i) \sum_{j=0}^{\infty} \beta^{j} \bar p_2(j) \Big)}
{\sum_{i=0}^{\infty} \beta^i \bar p_1(i|n) 
\Big( 1 + \beta \mu_1(n+i) \sum_{j=0}^{\infty} \beta^{j} \bar p_2(j) \Big)}, 
\]
where the right hand side equals $W_\beta(1,n)$. Thus, we conclude that this 
condition follows from (\ref{eq:nu-req-discounted-IHR-IHR-case-41}) 
by (\ref{eq:Wx-order-IHR-IHR}).

Let then $n \in \{0,1,\ldots\}$. 
Again by (\ref{eq:howard-eqs-discounted-IHR-IHR-case-4-solution}) 
and some algebraic manipulations, the following condition for optimality of 
$\pi$ in state~$(2,n)$, 
\[
V_\beta^\pi(2,n;\nu) \le \frac{h}{1 - \beta}, 
\]
can be shown to be equivalent with 
\[
\nu \le 
h \, 
\frac{\frac{1}{1 - \beta} - \sum_{i=0}^{\infty} \beta^i \bar p_2(i|n)}
{\sum_{i=0}^{\infty} \beta^i \bar p_2(i|n)}, 
\]
where the right hand side equals $W_\beta(2,n)$ given in 
(\ref{eq:Whittle-index-2n-discounted-IHR-IHR}). Thus, we conclude that this 
condition follows from (\ref{eq:nu-req-discounted-IHR-IHR-case-41}) by 
(\ref{eq:Wx-order-IHR-IHR}).

\vskip 3pt
$4.2^\circ$ 
By (\ref{eq:howard-eqs-discounted-IHR-IHR-case-4-solution}) and some algebraic 
manipulations, the following condition for optimality of $\pi$ in state~$(1,m-1)$, 
\[
\beta V_\beta^\pi(1,m-1;\nu) \le 
\nu + \beta \mu_1(m-1) V_\beta^\pi(2,0;\nu) + 
\beta (1 - \mu_1(m-1)) V_\beta^\pi(1,m;\nu), 
\]
can be shown to be equivalent with 
\begin{equation}
\nu \ge 
h \, 
\frac{\frac{1}{1 - \beta} - \sum_{i=0}^{\infty} \beta^i \bar p_1(i|m-1) 
\Big( 1 + \beta \mu_1(m-1+i) \sum_{j=0}^{\infty} \beta^{j} \bar p_2(j) \Big)}
{\sum_{i=0}^{\infty} \beta^i \bar p_1(i|m-1) 
\Big( 1 + \beta \mu_1(m-1+i) \sum_{j=0}^{\infty} \beta^{j} \bar p_2(j) \Big)}, 
\label{eq:nu-req-discounted-IHR-IHR-case-42}
\end{equation}
where the right hand side equals $W_\beta(1,m-1)$ given in 
(\ref{eq:Whittle-index-1n-discounted-IHR-IHR}).

Let $n \in \{0,1,\ldots,m-2\}$. 
By (\ref{eq:howard-eqs-discounted-IHR-IHR-case-4-solution}) and some 
algebraic manipulations, the following condition for optimality of $\pi$ 
in state~$(1,n)$, 
\[
\beta V_\beta^\pi(1,n;\nu) \le 
\nu + \beta \mu_1(n) V_\beta^\pi(2,0;\nu) + 
\beta (1 - \mu_1(n)) V_\beta^\pi(1,n+1;\nu), 
\]
can be shown to be equivalent with 
\[
\begin{split}
& 
\nu \ge 
h \beta \mu_1(n) \, 
\frac{\frac{1}{1 - \beta} - \sum_{i=0}^{\infty} \beta^i \bar p_2(i|n)}
{1 + \beta \mu_1(n) \sum_{i=0}^{\infty} \beta^i \bar p_2(i|n)} \; = \\
& \quad 
h \mu_1(n) \, 
\frac{\beta \sum_{i=0}^{\infty} \beta^i p_2(i)}
{1 + \beta \mu_1(n) \sum_{i=0}^{\infty} \beta^i \bar p_2(i)} \, 
\frac{\beta}{1 - \beta}.
\end{split}
\]
By Lemma~\ref{lem:Whittle-index-discounted-IHR-IHR-lemma-2}, we conclude 
that this condition follows from (\ref{eq:nu-req-discounted-IHR-IHR-case-42}), 
which completes the proof of claim~$4^\circ$.

\vskip 3pt
$5^\circ$ 
We still assume that $\nu \ge 0$ and utilize the optimality equations 
(\ref{eq:opt-eqs-discounted-nu-pos}). 
Now we prove that the policy $\pi$ with activity set 
\[
{\mathcal B}^\pi = {\mathcal S} \setminus \{*\}, 
\]
according to which user~$k$ is scheduled in all the states but $*$, is 
$(\nu,\beta)$-optimal for all 
\[
\nu \in [0,W_\beta(1,0)].
\]
It remains to prove that policy $\pi$ is optimal for these values of $\nu$ 
in any state $x \in {\mathcal S} \setminus \{*\}$. 

We start by first deriving the value function $V_\beta^\pi(x;\nu)$ for 
policy $\pi$ from the Howard equations: 
\begin{equation}
\begin{split}
& 
V_\beta^\pi(1,n;\nu) = 
h + \nu + \beta \mu_1(n) V_\beta^\pi(2,0;\nu) + 
\beta (1 - \mu_1(n)) V_\beta^\pi(1,n+1;\nu), \\
& 
V_\beta^\pi(2,n;\nu) = 
h + \nu + \beta (1 - \mu_2(n)) V_\beta^\pi(2,n+1;\nu).
\end{split}
\label{eq:howard-eqs-discounted-IHR-IHR-case-5}
\end{equation}
The unique solution of these linear equations is given by 
\begin{equation}
\begin{split}
& 
V_\beta^\pi(2,n;\nu) = 
(h + \nu) 
\left( 
\sum_{i = 0}^{\infty} \beta^i \bar p_2(i|n) 
\right), \\
& 
V_\beta^\pi(1,n;\nu) = 
(h + \nu) 
\left( 
\sum_{i = 0}^{\infty} \beta^i \bar p_1(i|n) 
\left( 
1 + \beta \mu_1(n+i) \frac{V_\beta^\pi(2,0;\nu)}{h + \nu} 
\right) 
\right).
\end{split}
\label{eq:howard-eqs-discounted-IHR-IHR-case-5-solution}
\end{equation}

By (\ref{eq:howard-eqs-discounted-IHR-IHR-case-5-solution}) and some algebraic 
manipulations, the following condition for optimality of $\pi$ in state~$(1,0)$, 
\[
V_\beta^\pi(1,0;\nu) \le \frac{h}{1 - \beta}, 
\]
can be shown to be equivalent with 
\begin{equation}
\nu \le 
h \, 
\frac{\frac{1}{1 - \beta} - \sum_{i=0}^{\infty} \beta^i \bar p_1(i) 
\Big( 1 + \beta \mu_1(i) \sum_{j=0}^{\infty} \beta^{j} \bar p_2(j) \Big)}
{\sum_{i=0}^{\infty} \beta^i \bar p_1(i) 
\Big( 1 + \beta \mu_1(i) \sum_{j=0}^{\infty} \beta^{j} \bar p_2(j) \Big)}, 
\label{eq:nu-req-discounted-IHR-IHR-case-5}
\end{equation}
where the right hand side equals $W_\beta(1,0)$ given in 
(\ref{eq:Whittle-index-1n-discounted-IHR-IHR}).

Let then $n \in \{1,2,\ldots\}$. 
Similarly as above, by (\ref{eq:howard-eqs-discounted-IHR-IHR-case-5-solution}) 
and some algebraic manipulations, the following condition for optimality of 
$\pi$ in state~$(1,n)$, 
\[
V_\beta^\pi(1,n;\nu) \le \frac{h}{1 - \beta}, 
\]
can be shown to be equivalent with 
\[
\nu \le 
h \, 
\frac{\frac{1}{1 - \beta} - \sum_{i=0}^{\infty} \beta^i \bar p_1(i|n) 
\Big( 1 + \beta \mu_1(n+i) \sum_{j=0}^{\infty} \beta^{j} \bar p_2(j) \Big)}
{\sum_{i=0}^{\infty} \beta^i \bar p_1(i|n) 
\Big( 1 + \beta \mu_1(n+i) \sum_{j=0}^{\infty} \beta^{j} \bar p_2(j) \Big)}, 
\]
where the right hand side equals $W_\beta(1,n)$. Thus, we conclude that this 
condition follows from (\ref{eq:nu-req-discounted-IHR-IHR-case-5}) 
by (\ref{eq:Wx-order-IHR-IHR}).

Let then $n \in \{0,1,\ldots\}$. 
Again by (\ref{eq:howard-eqs-discounted-IHR-IHR-case-5-solution}) 
and some algebraic manipulations, the following condition for optimality of 
$\pi$ in state~$(2,n)$, 
\[
V_\beta^\pi(2,n;\nu) \le \frac{h}{1 - \beta}, 
\]
can be shown to be equivalent with 
\[
\nu \le 
h \, 
\frac{\frac{1}{1 - \beta} - \sum_{i=0}^{\infty} \beta^i \bar p_2(i|n)}
{\sum_{i=0}^{\infty} \beta^i \bar p_2(i|n)}, 
\]
where the right hand side equals $W_\beta(2,n)$ given in 
(\ref{eq:Whittle-index-2n-discounted-IHR-IHR}). Thus, we conclude that this 
condition follows from (\ref{eq:nu-req-discounted-IHR-IHR-case-5}) by 
(\ref{eq:Wx-order-IHR-IHR}), which completes the proof of claim~$5^\circ$.

\vskip 3pt
$6^\circ$ 
Finally, we assume that $\nu \le 0$. 
However, the claim that the policy $\pi$ with activity set 
\[
{\mathcal B}^\pi = {\mathcal S} 
\]
is $(\nu,\beta)$-optimal for all 
\[
\nu \in (-\infty,0] 
\]
can be proved similarly as the corresponding claim~$5^\circ$ in the proof 
of Theorem~\ref{thm:Whittle-index-discounted-DHR-DHR} 
(the DHR-DHR-A subcase, see Appendix~\ref{app:DHR-DHR-A-proof}). 
Therefore we may omit the proof here.
\hfill $\Box$

\section{Proof of Theorem~\ref{thm:Whittle-index-discounted-DHR-IHR} 
(the DHR-IHR case)}
\label{app:DHR-IHR-proof}

\paragraph{Proof} 
Assume the DHR-IHR case. 
Note that, by Lemmas~\ref{lem:Whittle-index-discounted-DHR-IHR-lemma-1} and 
\ref{lem:Whittle-index-discounted-DHR-IHR-lemma-2}, we have the following 
ordering among the states: 
\begin{equation}
\begin{split}
& 
W_\beta(2,\infty) \ge \ldots \ge W_\beta(2,1) \ge W_\beta(2,0) > \\
& \quad 
W_\beta(1,0) \ge W_\beta(1,1) \ge \ldots \ge W_\beta(1,\infty) 
\ge 0, 
\end{split}
\label{eq:Wx-order-DHR-IHR}
\end{equation}
where we have defined 
\begin{equation}
\begin{split}
& 
W_\beta(2,\infty) = w_2(\infty) = 
h \mu_2(\infty) \, 
\frac{\beta}{1 - \beta}, \\
& 
W_\beta(1,\infty) = w_1(\infty) = 
h \mu_1(\infty) \, 
\frac{\beta \sum_{i=0}^{\infty} \beta^i p_2(i)}
{1 + \beta \mu_1(\infty) \sum_{i=0}^{\infty} \beta^i \bar p_2(i)} \, 
\frac{\beta}{1 - \beta}.
\end{split}
\label{eq:Whittle-index-21-infty-discounted-DHR-IHR}
\end{equation}

The main proof is now given in six parts ($1^\circ$--$6^\circ$). 
The idea is again to solve the relaxed optimization problem 
(\ref{eq:separable-discounted-costs}) for any $\nu$ by utilizing the 
optimality equations (\ref{eq:opt-eqs-discounted-general}). We partition 
the possible values of $\nu$, which is reflected by the six parts of 
the main proof. 
However, parts $1^\circ$--$2^\circ$ are exactly the same as in 
the proof of Theorem~\ref{thm:Whittle-index-discounted-IHR-IHR} 
(case IHR-IHR). Therefore, we omit them here and focus on the 
remaining parts $3^\circ$--$6^\circ$.

\vskip 3pt
$3^\circ$
Here we assume that $\nu \ge 0$, and the optimality equations 
(\ref{eq:opt-eqs-discounted-general}) read as given in 
(\ref{eq:opt-eqs-discounted-nu-pos}). 
We prove that the policy $\pi$ with activity set 
\[
{\mathcal B}^\pi = \{(2,0),(2,1),\ldots\}, 
\]
according to which user~$k$ is scheduled in states $(2,0),(2,1),\ldots$, is 
$(\nu,\beta)$-optimal for all 
\[
\nu \in [W_\beta(1,0),W_\beta(2,0)].
\]
It remains to prove that policy $\pi$ is optimal for these values of $\nu$ 
in any state $x \in {\mathcal S} \setminus \{*\}$. 
This is done below in two parts ($3.1^\circ$ and $3.2^\circ$).

\vskip 3pt
$3.1^\circ$ 
We start by first deriving the value function $V_\beta^\pi(x;\nu)$ for 
policy $\pi$ from the Howard equations: 
\begin{equation}
\begin{split}
& 
V_\beta^\pi(2,n;\nu) = 
h + \nu + \beta (1 - \mu_2(n)) V_\beta^\pi(2,n+1;\nu), 
\quad 
n \in \{0,1,\ldots\}, \\
& 
V_\beta^\pi(x;\nu) = 
h + \beta V_\beta^\pi(x;\nu), 
\quad x \in {\mathcal S} \setminus \{(2,0),(2,1),\ldots,*\}.
\end{split}
\label{eq:howard-eqs-discounted-DHR-IHR-case-3}
\end{equation}
The unique solution of these linear equations is given by 
\begin{equation}
\begin{split}
& 
V_\beta^\pi(2,n;\nu) = 
(h + \nu) 
\left( 
\sum_{i = 0}^{\infty} \beta^i \bar p_2(i|n) 
\right), 
\quad 
n \in \{0,1,\ldots\}, \\
& 
V_\beta^\pi(x;\nu) = 
\frac{h}{1 - \beta}, 
\quad x \in {\mathcal S} \setminus\{(2,0),(2,1),\ldots,*\}.
\end{split}
\label{eq:howard-eqs-discounted-DHR-IHR-case-3-solution}
\end{equation}

By (\ref{eq:howard-eqs-discounted-DHR-IHR-case-3-solution}) and some algebraic 
manipulations, the following condition for optimality of $\pi$ in state~$(2,0)$, 
\[
V_\beta^\pi(2,0;\nu) \le \frac{h}{1 - \beta}, 
\]
can be shown to be equivalent with 
\begin{equation}
\nu \le 
h \, 
\frac{\frac{1}{1 - \beta} - \sum_{i=0}^{\infty} \beta^i \bar p_2(i)}
{\sum_{i=0}^{\infty} \beta^i \bar p_2(i)}, 
\label{eq:nu-req-discounted-DHR-IHR-case-31}
\end{equation}
where the right hand side equals $W_\beta(2,0)$ given in 
(\ref{eq:Whittle-index-2n-discounted-DHR-IHR}).

Let then $n \in \{1,2,\ldots\}$. 
Since $\bar p_2(i|n)$ is a decreasing function of $n$ in the DHR-IHR 
case, we see from (\ref{eq:howard-eqs-discounted-DHR-IHR-case-3-solution}) 
that $V_\beta^\pi(2,n;\nu)$ is decreasing with respect to $n$. 
Therefore, the following condition for optimality of $\pi$ in state~$(2,n)$, 
\[
V_\beta^\pi(2,n;\nu) \le \frac{h}{1 - \beta}, 
\]
follows immediately from (\ref{eq:nu-req-discounted-DHR-IHR-case-31}).

\vskip 3pt
$3.2^\circ$ 
Let $n \in \{0,1,\ldots\}$. 
By (\ref{eq:howard-eqs-discounted-DHR-IHR-case-3-solution}) and some 
algebraic manipulations, the following condition for optimality of $\pi$ 
in state~$(1,n)$, 
\[
\beta V_\beta^\pi(1,n;\nu) \le 
\nu + \beta \mu_1(n) V_\beta^\pi(2,0;\nu) + 
\beta (1 - \mu_1(n)) V_\beta^\pi(1,n+1;\nu), 
\]
can be shown to be equivalent with 
\[
\begin{split}
& 
\nu \ge 
h \beta \mu_1(n) \, 
\frac{\frac{1}{1 - \beta} - \sum_{i=0}^{\infty} \beta^i \bar p_2(i|n)}
{1 + \beta \mu_1(n) \sum_{i=0}^{\infty} \beta^i \bar p_2(i|n)} \; = \\
& \quad 
h \mu_1(n) \, 
\frac{\beta \sum_{i=0}^{\infty} \beta^i p_2(i)}
{1 + \beta \mu_1(n) \sum_{i=0}^{\infty} \beta^i \bar p_2(i)} \, 
\frac{\beta}{1 - \beta}.
\end{split}
\]
Since $\mu_1(n)$ is decreasing, we conclude that policy $\pi$ is optimal 
in any state~$(1,n)$ if and only if 
\begin{equation}
\nu \ge 
h \mu_1(0) \, 
\frac{\beta \sum_{i=0}^{\infty} \beta^i p_2(i)}
{1 + \beta \mu_1(0) \sum_{i=0}^{\infty} \beta^i \bar p_2(i)} \, 
\frac{\beta}{1 - \beta}, 
\label{eq:nu-req-discounted-DHR-IHR-case-32}
\end{equation}
where the right hand side equals $W_\beta(1,0)$ given in 
(\ref{eq:Whittle-index-1n-discounted-DHR-IHR}). 
This completes the proof of claim~$3^\circ$.

\vskip 3pt
$4^\circ$ 
We still assume that $\nu \ge 0$ and utilize the optimality equations 
(\ref{eq:opt-eqs-discounted-nu-pos}). 
Let $m \in \{0,1,\ldots\}$. 
We prove that the policy $\pi$ with activity set 
\[
{\mathcal B}^\pi = \{(2,0),(2,1),\ldots\} \cup \{(1,0),(1,1),\ldots,(1,m)\} 
\]
is $(\nu,\beta)$-optimal for all 
\[
\nu \in [W_\beta(1,m+1),W_\beta(1,m)].
\]
It remains to prove that policy $\pi$ is optimal for these values of $\nu$ 
in any state $x \in {\mathcal S} \setminus \{*\}$. 
This is done below in two parts ($4.1^\circ$ and $4.2^\circ$).

\vskip 3pt
$4.1^\circ$ 
We start by first deriving the value function $V_\beta^\pi(x;\nu)$ for 
policy $\pi$ from the Howard equations: 
\begin{equation}
\begin{split}
& 
V_\beta^\pi(1,n;\nu) = 
h + \nu + \beta \mu_1(n) V_\beta^\pi(2,0;\nu) \; + \\
& \quad 
\beta (1 - \mu_1(n)) V_\beta^\pi(1,n+1;\nu), 
\quad 
n \in \{0,1,\ldots,m\}, \\
& 
V_\beta^\pi(2,n;\nu) = 
h + \nu + \beta (1 - \mu_2(n)) V_\beta^\pi(2,n+1;\nu), \\
& \quad 
n \in \{0,1,\ldots\}, \\
& 
V_\beta^\pi(x;\nu) = 
h + \beta V_\beta^\pi(x;\nu), \\
& \quad 
x \in {\mathcal S} \setminus \{(2,0),(2,1),\ldots,(1,0),(1,1),\ldots,(1,m),*\}.
\end{split}
\label{eq:howard-eqs-discounted-DHR-IHR-case-4}
\end{equation}
The unique solution of these linear equations is given by 
\begin{equation}
\begin{split}
& 
V_\beta^\pi(2,n;\nu) = 
(h + \nu) 
\left( 
\sum_{i = 0}^{\infty} \beta^i \bar p_2(i|n) 
\right), 
\quad 
n \in \{0,1,\ldots\}, \\
& 
V_\beta^\pi(1,n;\nu) = 
(h + \nu) 
\left( 
\sum_{i = 0}^{m-n} \beta^i \bar p_1(i|n) 
\left( 
1 + \beta \mu_1(n+i) \frac{V_\beta^\pi(2,0;\nu)}{h + \nu} 
\right) 
\right) \; + \\
& \quad 
h \, \frac{\beta^{m-n+1} \bar p_1(m-n+1|n)}{1 - \beta}, 
\quad 
n \in \{0,1,\ldots,m\}, \\
& 
V_\beta^\pi(x;\nu) = 
\frac{h}{1 - \beta}, 
\quad 
x \in {\mathcal S} \setminus \{(2,0),(2,1),\ldots,(1,0),(1,1),\ldots,(1,m),*\}.
\end{split}
\label{eq:howard-eqs-discounted-DHR-IHR-case-4-solution}
\end{equation}

By (\ref{eq:howard-eqs-discounted-DHR-IHR-case-4-solution}) 
and some algebraic manipulations, the following condition for optimality 
of $\pi$ in state~$(1,m)$, 
\[
V_\beta^\pi(1,m;\nu) \le \frac{h}{1 - \beta}, 
\]
can be shown to be equivalent with 
\begin{equation}
\nu \le 
h \mu_1(m) \, 
\frac{\beta \sum_{i=0}^{\infty} \beta^i p_2(i)}
{1 + \beta \mu_1(m) \sum_{i=0}^{\infty} \beta^i \bar p_2(i)} \, 
\frac{\beta}{1 - \beta}, 
\label{eq:nu-req-discounted-DHR-IHR-case-41}
\end{equation}
where the right hand side equals $W_\beta(1,m)$ given in 
(\ref{eq:Whittle-index-1n-discounted-DHR-IHR}).

Let then $n \in \{0,1,\ldots\}$. 
Again by (\ref{eq:howard-eqs-discounted-DHR-IHR-case-4-solution}) 
and some algebraic manipulations, the following condition for optimality of 
$\pi$ in state~$(2,n)$, 
\[
V_\beta^\pi(2,n;\nu) \le \frac{h}{1 - \beta}, 
\]
can be shown to be equivalent with 
\[
\nu \le 
h \, 
\frac{\frac{1}{1 - \beta} - \sum_{i=0}^{\infty} \beta^i \bar p_2(i|n)}
{\sum_{i=0}^{\infty} \beta^i \bar p_2(i|n)}, 
\]
where the right hand side equals $W_\beta(2,n)$ given in 
(\ref{eq:Whittle-index-2n-discounted-DHR-IHR}). Thus, we conclude that 
this condition follows from (\ref{eq:nu-req-discounted-DHR-IHR-case-41}) 
by (\ref{eq:Wx-order-DHR-IHR}).

Next we show that if condition 
(\ref{eq:nu-req-discounted-DHR-IHR-case-41}) is satisfied, then 
\begin{equation}
V_\beta^\pi(1,0;\nu) \le V_\beta^\pi(1,1;\nu) \le \ldots \le V_\beta^\pi(1,m;\nu) \le \frac{h}{1 - \beta}, 
\label{eq:nu-req-discounted-DHR-IHR-case-41-cont1}
\end{equation}
which implies that policy $\pi$ is optimal in all states 
$\{(1,0),(1,1),\ldots,(1,m)\}$. First, as we saw above, condition 
(\ref{eq:nu-req-discounted-DHR-IHR-case-41}) implies that 
$V_\beta^\pi(2,0;\nu) \le \frac{h}{1 - \beta}$ and 
$V_\beta^\pi(1,m;\nu) \le \frac{h}{1 - \beta}$. 
Now, by (\ref{eq:howard-eqs-discounted-DHR-IHR-case-4}), we observe that 
\[
\begin{split}
& 
V_\beta^\pi(1,m;\nu) - V_\beta^\pi(1,m-1;\nu) \; = \\
& \quad 
\beta 
\Big[
\Big( \mu_1(m) V_\beta^\pi(2,0;\nu) + (1 - \mu_1(m)) \frac{h}{1 - \beta} \Big) \; - \\
& \quad\quad 
\Big( \mu_1(m-1) V_\beta^\pi(2,0;\nu) + (1 - \mu_1(m-1)) V_\beta^\pi(1,m;\nu) \Big) 
\Big] 
\ge 0, 
\end{split}
\]
where the last inequality follows from the facts that $\mu_1(m) \le \mu_1(m-1)$, 
$V_\beta^\pi(2,0;\nu) \le \frac{h}{1 - \beta}$, and 
$V_\beta^\pi(1,m;\nu) \le \frac{h}{1 - \beta}$. 
In addition, again by (\ref{eq:howard-eqs-discounted-DHR-IHR-case-4}), we have 
\[
\begin{split}
& 
V_\beta^\pi(1,m;\nu) - V_\beta^\pi(2,0;\nu) \; = \\
& \quad 
h + \nu + \beta (1 - \mu_1(m)) \frac{h}{1 - \beta} - 
(1 - \beta \mu_1(m)) V_\beta^\pi(2,0;\nu) \; = \\
& \quad 
(1 - \beta) \Big( \frac{h + \nu}{1 - \beta} - V_\beta^\pi(2,0;\nu) \Big) + 
\beta (1 - \mu_1(m)) \Big( \frac{h}{1 - \beta} - V_\beta^\pi(2,0;\nu) \Big) 
\ge 0, 
\end{split}
\]
where the last inequality follows from the fact that 
$V_\beta^\pi(2,0;\nu) \le \frac{h}{1 - \beta} \le \frac{h + \nu}{1 - \beta}$. 
So, we have proved that if condition 
(\ref{eq:nu-req-discounted-DHR-IHR-case-41}) is satisfied, then 
\begin{equation}
V_\beta^\pi(2,0;\nu) \le V_\beta^\pi(1,m;\nu), \quad 
V_\beta^\pi(1,m-1;\nu) \le V_\beta^\pi(1,m;\nu).
\label{eq:nu-req-discounted-DHR-IHR-case-41-cont2}
\end{equation}
Next, again by (\ref{eq:howard-eqs-discounted-DHR-IHR-case-4}), we observe that 
\[
\begin{split}
& 
V_\beta^\pi(1,m-1;\nu) - V_\beta^\pi(1,m-2;\nu) \; = \\
& \quad 
\beta 
\Big[
\Big( \mu_1(m-1) V_\beta^\pi(2,0;\nu) + (1 - \mu_1(m-1)) V_\beta^\pi(1,m;\nu) \Big) \; - \\
& \quad\quad 
\Big( \mu_1(m-2) V_\beta^\pi(2,0;\nu) + (1 - \mu_1(m-2)) V_\beta^\pi(1,m-1;\nu) \Big) 
\Big] 
\ge 0, 
\end{split}
\]
where the last inequality follows from 
(\ref{eq:nu-req-discounted-DHR-IHR-case-41-cont2}) and 
the fact that $\mu_1(m-1) \le \mu_1(m-2)$. 
In addition, again by (\ref{eq:howard-eqs-discounted-DHR-IHR-case-4}), we have 
\[
\begin{split}
& 
V_\beta^\pi(1,m-1;\nu) - V_\beta^\pi(2,0;\nu) \; = \\
& \quad 
h + \nu + \beta (1 - \mu_1(m-1)) V_\beta^\pi(1,m;\nu) - 
(1 - \beta \mu_1(m-1)) V_\beta^\pi(2,0;\nu) \; = \\
& \quad 
(1 - \beta) \Big( \frac{h + \nu}{1 - \beta} - V_\beta^\pi(2,0;\nu) \Big) \; + \\
& \quad\quad 
\beta (1 - \mu_1(m-1)) \Big( V_\beta^\pi(1,m;\nu) - V_\beta^\pi(2,0;\nu) \Big) 
\ge 0,
\end{split}
\]
where the last inequality follows from the fact that 
$V_\beta^\pi(2,0;\nu) \le V_\beta^\pi(1,m;\nu) \le \frac{h}{1 - \beta} \le \frac{h + \nu}{1 - \beta}$. 
So, we have proved that if condition 
(\ref{eq:nu-req-discounted-DHR-IHR-case-41}) is satisfied, then 
\begin{equation}
V_\beta^\pi(2,0;\nu) \le V_\beta^\pi(1,m-1;\nu), \quad 
V_\beta^\pi(1,m-2;\nu) \le V_\beta^\pi(1,m-1;\nu).
\label{eq:nu-req-discounted-DHR-IHR-case-41-cont3}
\end{equation}
Continuing similarly proves claim (\ref{eq:nu-req-discounted-DHR-IHR-case-41-cont1}).

\vskip 3pt
$4.2^\circ$ 
By (\ref{eq:howard-eqs-discounted-DHR-IHR-case-4-solution}) 
and some algebraic manipulations, the following condition for optimality 
of $\pi$ in state~$(1,m+1)$, 
\[
\beta V_\beta^\pi(1,m+1;\nu) \le 
\nu + \beta \mu_1(m+1) V_\beta^\pi(2,0;\nu) + 
\beta (1 - \mu_1(m+1)) V_\beta^\pi(1,m+2;\nu), 
\]
can be shown to be equivalent with 
\begin{equation}
\nu \ge 
h \mu_1(m+1) \, 
\frac{\beta \sum_{i=0}^{\infty} \beta^i p_2(i)}
{1 + \beta \mu_1(m+1) \sum_{i=0}^{\infty} \beta^i \bar p_2(i)} \, 
\frac{\beta}{1 - \beta}, 
\label{eq:nu-req-discounted-DHR-IHR-case-42}
\end{equation}
where the right hand side equals $W_\beta(1,m+1)$ given in 
(\ref{eq:Whittle-index-1n-discounted-DHR-IHR}).

Let then $n \in \{m+2,m+3,\ldots\}$. 
Again by (\ref{eq:howard-eqs-discounted-DHR-IHR-case-4-solution}) and some 
algebraic manipulations, the following condition for optimality of 
$\pi$ in state~$(1,n)$, 
\[
\beta V_\beta^\pi(1,n;\nu) \le 
\nu + \beta \mu_1(n) V_\beta^\pi(2,0;\nu) + 
\beta (1 - \mu_1(n)) V_\beta^\pi(1,n+1;\nu), 
\]
can be shown to be equivalent with 
\[
\nu \ge 
h \mu_1(n) \, 
\frac{\beta \sum_{i=0}^{\infty} \beta^i p_2(i)}
{1 + \beta \mu_1(n) \sum_{i=0}^{\infty} \beta^i \bar p_2(i)} \, 
\frac{\beta}{1 - \beta}, 
\]
which follows from (\ref{eq:nu-req-discounted-DHR-IHR-case-42}) since 
$\mu_1(n)$ is decreasing. 
This completes the proof of claim~$4^\circ$.

\vskip 3pt
$5^\circ$ 
We still assume that $\nu \ge 0$ and utilize the optimality equations 
(\ref{eq:opt-eqs-discounted-nu-pos}). 
Now we prove that the policy $\pi$ with activity set 
\[
{\mathcal B}^\pi = {\mathcal S} \setminus \{*\}, 
\]
is $(\nu,\beta)$-optimal for all 
\[
\nu \in [0,W_\beta(1,\infty)], 
\]
where $W_\beta(1,\infty)$ is defined in 
(\ref{eq:Whittle-index-21-infty-discounted-DHR-IHR}). 
It remains to prove that policy $\pi$ is optimal for these values of $\nu$ 
in any state $x \in {\mathcal S} \setminus \{*\}$.

We start by first deriving the value function $V_\beta^\pi(x;\nu)$ for 
policy $\pi$ from the Howard equations: 
\begin{equation}
\begin{split}
& 
V_\beta^\pi(1,n;\nu) = 
h + \nu + \beta \mu_1(n) V_\beta^\pi(2,0;\nu) + 
\beta (1 - \mu_1(n)) V_\beta^\pi(1,n+1;\nu), \\
& 
V_\beta^\pi(2,n;\nu) = 
h + \nu + \beta (1 - \mu_2(n)) V_\beta^\pi(2,n+1;\nu).
\end{split}
\label{eq:howard-eqs-discounted-DHR-IHR-case-5}
\end{equation}
The unique solution of these linear equations is given by 
\begin{equation}
\begin{split}
& 
V_\beta^\pi(2,n;\nu) = 
(h + \nu) 
\left( 
\sum_{i = 0}^\infty \beta^i \bar p_2(i|n) 
\right), \\
& 
V_\beta^\pi(1,n;\nu) = 
(h + \nu) 
\left( 
\sum_{i = 0}^\infty \beta^i \bar p_1(i|n) 
\left( 
1 + \beta \mu_1(n+i) \frac{V_\beta^\pi(2,0;\nu)}{h + \nu} 
\right) 
\right).
\end{split}
\label{eq:howard-eqs-discounted-DHR-IHR-case-5-solution}
\end{equation}

Let $n \in \{0,1,\ldots\}$. 
By (\ref{eq:howard-eqs-discounted-DHR-IHR-case-5-solution}) 
and some algebraic manipulations, the following condition for optimality of 
$\pi$ in state~$(2,n)$, 
\[
V_\beta^\pi(2,n;\nu) \le \frac{h}{1 - \beta}, 
\]
can be shown to be equivalent with 
\begin{equation}
\nu \le 
h \, 
\frac{\frac{1}{1 - \beta} - \sum_{i=0}^{\infty} \beta^i \bar p_2(i|n)}
{\sum_{i=0}^{\infty} \beta^i \bar p_2(i|n)}, 
\label{eq:nu-req-discounted-DHR-IHR-case-5a}
\end{equation}
where the right hand side equals $W_\beta(2,n)$ given in 
(\ref{eq:Whittle-index-2n-discounted-DHR-IHR}). 

Let again $n \in \{0,1,\ldots\}$. 
Since 
\[
\bar p_1(i|n) \mu_1(n+i) = p_1(i|n) = \bar p_1(i|n) - \bar p_1(i+1|n), 
\]
it follows from 
(\ref{eq:howard-eqs-discounted-DHR-IHR-case-5-solution}) that 
\begin{equation}
\begin{split}
& 
V_\beta^\pi(1,n;\nu) = 
(h + \nu) 
\Bigg( 
1 + \frac{\beta V_\beta^\pi(2,0;\nu)}{h + \nu} \; + \\
& \quad 
\sum_{i = 1}^\infty \beta^i \bar p_1(i|n) 
\bigg( 
1 - \frac{(1 - \beta) V_\beta^\pi(2,0;\nu)}{h + \nu} 
\bigg) 
\Bigg).
\end{split}
\label{eq:howard-eqs-discounted-DHR-IHR-case-5-solution-cont}
\end{equation}
Now, since $\bar p_1(i|n)$ is an increasing function of $n$ in the DHR-IHR 
case and we have above required that 
\[
V_\beta^\pi(2,0;\nu) \le \frac{h}{1 - \beta}, 
\]
we see from (\ref{eq:howard-eqs-discounted-DHR-IHR-case-5-solution-cont}) 
that $V_\beta^\pi(1,n;\nu)$, as well, is an increasing function of $n$ 
and approaches, by (\ref{eq:howard-eqs-discounted-DHR-IHR-case-5-solution}), 
\[
\begin{split}
& 
\lim_{n \to \infty} V_\beta^\pi(1,n;\nu) = 
(h + \nu) 
\Bigg( 
\sum_{i = 0}^\infty \beta^i (1 - \mu_1(\infty))^i 
\bigg( 
1 + \beta \mu_1(\infty) \frac{V_\beta^\pi(2,0;\nu)}{h + \nu} 
\bigg) 
\Bigg) \\
& \quad = 
(h + \nu) \, 
\frac{1 + \beta \mu_1(\infty) \sum_{i = 0}^\infty \beta^i \bar p_2(i)}
{1 - \beta (1 - \mu_1(\infty))}.
\end{split}
\]
Thus, the following condition for optimality of $\pi$ in state~$(1,n)$, 
\[
V_\beta^\pi(1,n;\nu) \le \frac{h}{1 - \beta}, 
\]
is satisfied for any $n$ if and only if 
\[
\lim_{n \to \infty} V_\beta^\pi(1,n;\nu) \le \frac{h}{1 - \beta}, 
\]
which can be shown to be equivalent with condition 
\begin{equation}
\nu \le 
h \mu_1(\infty) \, 
\frac{\beta \sum_{i=0}^{\infty} \beta^i p_2(i)}
{1 + \beta \mu_1(\infty) \sum_{i=0}^{\infty} \beta^i \bar p_2(i)} \, 
\frac{\beta}{1 - \beta}.
\label{eq:nu-req-discounted-DHR-IHR-case-5b}
\end{equation}
Note that the right hand side equals $W_\beta(1,\infty)$ given in 
(\ref{eq:Whittle-index-21-infty-discounted-DHR-IHR}). 
Note also that (\ref{eq:nu-req-discounted-DHR-IHR-case-5a}) follows from 
(\ref{eq:nu-req-discounted-DHR-IHR-case-5b}) by 
(\ref{eq:Wx-order-DHR-IHR}). 
This completes the proof of claim~$5^\circ$.

\vskip 3pt
$6^\circ$ 
Finally, we assume that $\nu \le 0$. 
However, the claim that the policy $\pi$ with activity set 
\[
{\mathcal B}^\pi = {\mathcal S} 
\]
is $(\nu,\beta)$-optimal for all 
\[
\nu \in (-\infty,0] 
\]
can be proved similarly as the corresponding claim~$5^\circ$ in the proof 
of Theorem~\ref{thm:Whittle-index-discounted-DHR-DHR} 
(the DHR-DHR-A subcase, see Appendix~\ref{app:DHR-DHR-A-proof}). 
Therefore we may omit the proof here.
\hfill $\Box$

\section{Proof of Theorem~\ref{thm:Whittle-index-discounted-IHR-DHR} 
in the IHR-DHR-D subcase}
\label{app:IHR-DHR-D-proof}

\paragraph{Proof} 
We present here the proof for the IHR-DHR-D subcase. 
In the DHR-DHR-E subcase, the proof is slightly different and 
even depending on the parameter $n_1^*$ defined in 
Lemma~\ref{lem:Whittle-index-discounted-IHR-DHR-E-lemma-4}. 
These proofs are presented in Appendices~\ref{app:IHR-DHR-E1-proof} and 
\ref{app:IHR-DHR-E2-proof}.

Assume the IHR-DHR-D subcase defined in (\ref{eq:IHR-DHR-D}). 
As in Lemma~\ref{lem:Whittle-index-discounted-IHR-DHR-D-lemma-3}, 
let $n_2^*$ denote the smallest $\bar n_2 \in \{0,1,\ldots\}$ satisfying 
condition (\ref{eq:DHR-DHR-B}). In addition, let us define 
$n_2^\circ \in \{0,1,\ldots,n_2^*\}$ as follows: 
\begin{equation}
n_2^\circ = \phi(\infty) = \max \{ n_2 : w_2(n_2) \ge w_1(\infty) \}, 
\label{eq:n-2-circ-IHR-DHR-D}
\end{equation}
where $w_1(\infty)$ is defined in (\ref{eq:w1-infty-discounted-IHR-DHR}). 
Furthermore, we denote here 
\begin{equation}
{\mathcal S}^* = {\mathcal S} \setminus \{(2,n_2^*+1),(2,n_2^*+2),\ldots\}.
\label{eq:mathcal-S-circ-IHR-DHR-D}
\end{equation}

For the main proof, we define a sequence of states $y_m$, 
$m \in \{1,2,\ldots\}$, recursively as follows: 
\begin{equation}
\begin{split}
& 
y_1 = (1,0), \\
& 
y_{m+1} = 
\left\{
\begin{array}{ll}
(1,\tilde n_1(m)), & \quad 
\hbox{if $W_\beta(1,\tilde n_1(m)) = 
\min_{x \in {\mathcal S}^* \setminus \{y_1,\ldots,y_m,*\}} W_\beta(x)$}, \\
(2,\tilde n_2(m)), & \quad 
\hbox{otherwise}, \\
\end{array}
\right.
\end{split}
\label{eq:y-def-IHR-DHR-D}
\end{equation}
where 
\[
\begin{split}
& 
\tilde n_1(m) = \min\{ n_1 : (1,n_1) \in {\mathcal S}^* \setminus \{y_1,\ldots,y_m,*\} \}, \\
& 
\tilde n_2(m) = \max\{ n_2 : (2,n_2) \in {\mathcal S}^* \setminus \{y_1,\ldots,y_m,*\} \}.
\end{split}
\]
Note that, in this IHR-DHR-D subcase, the sequence $(y_m)$ covers the states 
\[
\{y_1,y_2,\ldots\} = 
\{(2,n_2^\circ+1),(2,n_2^\circ+2),\ldots,(2,n_2^*)\} \cup \{(1,0),(1,1),\ldots\}, 
\]
where the former part of the right hand side is omitted if $n_2^\circ = n_2^*$. 
Note also that, by 
Lemmas~\ref{lem:Whittle-index-discounted-IHR-DHR-lemma-1}, 
\ref{lem:Whittle-index-discounted-IHR-DHR-lemma-2}, and 
\ref{lem:Whittle-index-discounted-IHR-DHR-D-lemma-3}, 
we have the following ordering among the states: 
\begin{equation}
\begin{split}
& 
W_\beta(2,0) \ge W_\beta(2,1) \ge \ldots \ge W_\beta(2,n_2^\circ) \ge \\
& \quad 
W_\beta(1,\infty) \ge \ldots \ge W_\beta(y_2) \ge W_\beta(y_1) = W_\beta(1,0) > \\ 
& \quad \quad 
W_\beta(2,n_2^*+1) \ge W_\beta(2,n_2^*+2) \ge \ldots \ge W_\beta(2,\infty) 
\ge 0, 
\end{split}
\label{eq:Wx-order-IHR-DHR-D}
\end{equation}
where we have defined 
\begin{equation}
\begin{split}
& 
W_\beta(1,\infty) = w_1(\infty) = 
h \mu_1(\infty) \, 
\frac{\beta \sum_{i=0}^{n_2^\circ} \beta^i p_2(i)}
{1 + \beta \mu_1(\infty) \sum_{i=0}^{n_2^\circ} \beta^i \bar p_2(i)} \, 
\frac{\beta}{1 - \beta}, \\
& 
W_\beta(2,\infty) = w_2(\infty) = 
h \mu_2(\infty) \, 
\frac{\beta}{1 - \beta}.
\end{split}
\label{eq:Whittle-index-21-infty-discounted-IHR-DHR-D}
\end{equation}

In addition, for any $m \in \{1,2,\ldots\}$, there are $n_2$ 
and $n_1$ such that $n_2 \in \{n_2^\circ,n_2^\circ+1,\ldots,n_2^*\}$, 
$n_1 \in \{0,1,\ldots\}$, and 
\[
\begin{split}
& 
\{(2,0),(2,1),\ldots,(2,n_2^\circ)\} \cup \{y_m,y_{m+1},\ldots\} \; = \\
& \quad 
\{(2,0),(2,1),\ldots,(2,n_2)\} \cup \{(1,n_1),(1,n_1+1),\ldots\}.
\end{split}
\]
Now 
\begin{equation}
\phi(n_1) \le n_2, 
\label{eq:phi-result1-IHR-DHR-D}
\end{equation}
due to Lemma~\ref{lem:Whittle-index-discounted-IHR-DHR-lemma-1}(i) 
since 
\[
W_\beta(2,\phi(n_1)) \ge W_\beta(1,n_1) > W_\beta(2,n_2+1) 
\]
by (\ref{eq:Whittle-index-order-IHR-DHR}) and (\ref{eq:y-def-IHR-DHR-D}), 
respectively. Moreover, if $n_1 \ge 1$, then 
\begin{equation}
\phi(n_1-1) \ge n_2, 
\label{eq:phi-result2-IHR-DHR-D}
\end{equation}
since otherwise, by (\ref{eq:Whittle-index-order-IHR-DHR}) and 
Lemma~\ref{lem:Whittle-index-discounted-IHR-DHR-lemma-1}(i), 
\[
W_\beta(1,n_1-1) > W_\beta(2,\phi(n_1-1)+1) \ge W_\beta(2,n_2), 
\]
which were a contradiction.

The main proof is now given in eight parts ($1^\circ$--$8^\circ$). 
The idea is again to solve the relaxed optimization problem 
(\ref{eq:separable-discounted-costs}) for any $\nu$ by utilizing the 
optimality equations (\ref{eq:opt-eqs-discounted-general}). We partition 
the possible values of $\nu$, which is reflected by the eight parts of 
the main proof. 
However, part $1^\circ$ is exactly the same as in 
the proof of Theorem~\ref{thm:Whittle-index-discounted-DHR-DHR} 
(the DHR-DHR-A subcase, see Appendix~\ref{app:DHR-DHR-A-proof}). 
Therefore, we omit it here and focus on the remaining parts 
$2^\circ$--$8^\circ$.

\vskip 3pt
$2^\circ$
Here we assume that $\nu \ge 0$, and the optimality equations 
(\ref{eq:opt-eqs-discounted-general}) read as given in 
(\ref{eq:opt-eqs-discounted-nu-pos}). 
Let $n_2 \in \{0,1,\ldots,n_2^\circ-1\}$. 
We prove that the policy $\pi$ with activity set 
\[
{\mathcal B}^\pi = \{(2,0),\ldots,(2,n_2)\}, 
\]
according to which user~$k$ is scheduled in states $(2,0),\ldots,(2,n_2)$, 
is $(\nu,\beta)$-optimal for all 
\[
\nu \in [W_\beta(2,n_2+1),W_\beta(2,n_2)].
\]
It remains to prove that policy $\pi$ is optimal for these values of $\nu$ 
in any state $x \in {\mathcal S} \setminus \{*\}$. 
This is done below in two parts ($2.1^\circ$ and $2.2^\circ$).

\vskip 3pt
$2.1^\circ$ 
We start by first deriving the value function $V_\beta^\pi(x;\nu)$ for 
policy $\pi$ from the Howard equations: 
\begin{equation}
\begin{split}
& 
V_\beta^\pi(2,n;\nu) = 
h + \nu + \beta (1 - \mu_2(n)) V_\beta^\pi(2,n+1;\nu), 
\quad n \in \{0,1,\ldots,n_2\}, \\
& 
V_\beta^\pi(x;\nu) = 
h + \beta V_\beta^\pi(x;\nu), 
\quad x \in {\mathcal S} \setminus \{(2,0),\ldots,(2,n_2),*\}.
\end{split}
\label{eq:howard-eqs-discounted-IHR-DHR-case-2}
\end{equation}
The unique solution of these linear equations is given by 
\begin{equation}
\begin{split}
& 
V_\beta^\pi(2,n;\nu) = 
(h + \nu) 
\left( 
\sum_{i = 0}^{n_2-n} \beta^i \bar p_2(i|n) 
\right) \; + \\
& \quad 
h \, \frac{\beta^{n_2-n+1} \bar p_2(n_2-n+1|n)}{1 - \beta}, 
\quad 
n \in \{0,1,\ldots,n_2\}, \\
& 
V_\beta^\pi(x;\nu) = 
\frac{h}{1 - \beta}, 
\quad x \in {\mathcal S} \setminus \{(2,0),\ldots,(2,n_2),*\}.
\end{split}
\label{eq:howard-eqs-discounted-IHR-DHR-case-2-solution}
\end{equation}

By (\ref{eq:howard-eqs-discounted-IHR-DHR-case-2-solution}) and some algebraic 
manipulations, the following condition for optimality of $\pi$ in 
state~$(2,n_2)$, 
\[
V_\beta^\pi(2,n_2;\nu) \le \frac{h}{1 - \beta}, 
\]
can be shown to be equivalent with 
\begin{equation}
\nu \le 
h \mu_2(n_2) \, \frac{\beta}{1 - \beta}, 
\label{eq:nu-req-discounted-IHR-DHR-case-21}
\end{equation}
where the right hand side equals $W_\beta(2,n_2)$ given in 
(\ref{eq:Whittle-index-2n-discounted-IHR-DHR}).

Let then $n \in \{0,1,\ldots,n_2-1\}$. 
Again by (\ref{eq:howard-eqs-discounted-IHR-DHR-case-2-solution}), 
the following condition for optimality of $\pi$ in state~$(2,n)$, 
\[
V_\beta^\pi(2,n;\nu) \le \frac{h}{1 - \beta}, 
\]
can be shown to be equivalent with 
\begin{equation}
\nu \le 
h \, 
\frac{\sum_{i = 0}^{n_2-n} \beta^i p_2(i|n)}
{\sum_{i = 0}^{n_2-n} \beta^i \bar p_2(i|n)} \, 
\frac{\beta}{1 - \beta}.
\label{eq:nu-req-discounted-IHR-DHR-case-21-cont}
\end{equation}
Thus, condition (\ref{eq:nu-req-discounted-IHR-DHR-case-21-cont}) 
follows from (\ref{eq:nu-req-discounted-IHR-DHR-case-21}) 
by Lemma~\ref{lem:Whittle-index-discounted-IHR-DHR-lemma-1}(vi).

\vskip 3pt
$2.2^\circ$ 
By (\ref{eq:howard-eqs-discounted-IHR-DHR-case-2-solution}), the following condition 
for optimality of $\pi$ in state~$(2,n_2+1)$, 
\[
\beta V_\beta^\pi(2,n_2+1;\nu) \le 
\nu + \beta (1 - \mu_2(n_2+1)) V_\beta^\pi(2,n_2+2;\nu), 
\]
is easily shown to be equivalent with 
\begin{equation}
\nu \ge 
h \mu_2(n_2+1) \, \frac{\beta}{1 - \beta}, 
\label{eq:nu-req-discounted-IHR-DHR-case-22a}
\end{equation}
where the right hand side equals $W_\beta(2,n_2+1)$ given in 
(\ref{eq:Whittle-index-2n-discounted-IHR-DHR}).

Let then $n \in \{n_2+2,n_2+3,\ldots\}$. 
Again by (\ref{eq:howard-eqs-discounted-IHR-DHR-case-2-solution}), 
the following condition for optimality of $\pi$ in state~$(2,n)$, 
\[
\beta V_\beta^\pi(2,n;\nu) \le 
\nu + \beta (1 - \mu_2(n)) V_\beta^\pi(2,n+1;\nu), 
\]
is easily shown to be equivalent with 
\[
\nu \ge 
h \mu_2(n) \, \frac{\beta}{1 - \beta}, 
\]
which follows from (\ref{eq:nu-req-discounted-IHR-DHR-case-22a}) since 
$\mu_2(n)$ is decreasing.

Let then $n \in \{0,1,\ldots\}$. 
By (\ref{eq:howard-eqs-discounted-IHR-DHR-case-2-solution}) and some algebraic 
manipulations, the following condition for optimality of 
$\pi$ in state~$(1,n)$, 
\[
\beta V_\beta^\pi(1,n;\nu) \le 
\nu + \beta \mu_1(n) V_\beta^\pi(2,0;\nu) + 
\beta (1 - \mu_1(n)) V_\beta^\pi(1,n+1;\nu), 
\]
can be shown to be equivalent with 
\[
\nu \ge 
h \mu_1(n) \, 
\frac{\beta \sum_{i=0}^{n_2} \beta^i p_2(i)}{1 + \beta \mu_1(n) \sum_{i=0}^{n_2} \beta^i \bar p_2(i)} \, 
\frac{\beta}{1 - \beta}.
\]
Since $\mu_1(n)$ is increasing, we conclude that policy $\pi$ is optimal 
in any state~$(1,n)$ if and only if 
\begin{equation}
\nu \ge 
h \mu_1(\infty) \, 
\frac{\beta \sum_{i=0}^{n_2} \beta^i p_2(i)}
{1 + \beta \mu_1(\infty) \sum_{i=0}^{n_2} \beta^i \bar p_2(i)} \, 
\frac{\beta}{1 - \beta}.
\label{eq:nu-req-discounted-IHR-DHR-case-22b}
\end{equation}
However, by (\ref{eq:n-2-circ-IHR-DHR-D}) and taking into account the fact 
that $n_2 < n_2^\circ$, we have 
\[
\begin{split}
& 
\mu_2(n_2+1) \ge 
\mu_2(n_2^\circ) \; \ge \\
& \quad 
\frac{\beta \mu_1(\infty) \sum_{i=0}^{n_2^\circ} \beta^i p_2(i)}
{1 + \beta \mu_1(\infty) \sum_{i=0}^{n_2^\circ} \beta^i \bar p_2(i)} \ge 
\frac{\beta \mu_1(\infty) \sum_{i=0}^{n_2} \beta^i p_2(i)}
{1 + \beta \mu_1(\infty) \sum_{i=0}^{n_2} \beta^i \bar p_2(i)}, 
\end{split}
\]
which implies that condition (\ref{eq:nu-req-discounted-IHR-DHR-case-22b}) 
follows from (\ref{eq:nu-req-discounted-IHR-DHR-case-22a}). 
This completes the proof of claim~$2^\circ$.

\vskip 3pt
$3^\circ$ 
We still assume that $\nu \ge 0$ and utilize the optimality equations 
(\ref{eq:opt-eqs-discounted-nu-pos}). 
Now we prove that the policy $\pi$ with activity set 
\[
{\mathcal B}^\pi = \{(2,0),\ldots,(2,n_2^\circ)\}, 
\]
according to which user~$k$ is scheduled in states $(2,0),\ldots,(2,n_2^\circ)$, 
is $(\nu,\beta)$-optimal for all 
\[
\nu \in [W_\beta(1,\infty),W_\beta(2,n_2^\circ)].
\]
It remains to prove that policy $\pi$ is optimal for these values of $\nu$ 
in any state $x \in {\mathcal S} \setminus \{*\}$. 
This is done below in two parts ($3.1^\circ$ and $3.2^\circ$).

\vskip 3pt
$3.1^\circ$ 
We start by first deriving the value function $V_\beta^\pi(x;\nu)$ for 
policy $\pi$ from the Howard equations: 
\begin{equation}
\begin{split}
& 
V_\beta^\pi(2,n;\nu) = 
h + \nu + \beta (1 - \mu_2(n)) V_\beta^\pi(2,n+1;\nu), 
\quad n \in \{0,1,\ldots,n_2^\circ\}, \\
& 
V_\beta^\pi(x;\nu) = 
h + \beta V_\beta^\pi(x;\nu), 
\quad x \in {\mathcal S} \setminus \{(2,0),\ldots,(2,n_2^\circ),*\}.
\end{split}
\label{eq:howard-eqs-discounted-IHR-DHR-case-3}
\end{equation}
The unique solution of these linear equations is given by 
\begin{equation}
\begin{split}
& 
V_\beta^\pi(2,n;\nu) = 
(h + \nu) 
\left( 
\sum_{i = 0}^{n_2^\circ-n} \beta^i \bar p_2(i|n) 
\right) \; + \\
& \quad 
h \, \frac{\beta^{n_2^\circ-n+1} \bar p_2(n_2^\circ-n+1|n)}{1 - \beta}, 
\quad 
n \in \{0,1,\ldots,n_2^\circ\}, \\
& 
V_\beta^\pi(x;\nu) = 
\frac{h}{1 - \beta}, 
\quad x \in {\mathcal S} \setminus \{(2,0),\ldots,(2,n_2^\circ),*\}.
\end{split}
\label{eq:howard-eqs-discounted-IHR-DHR-case-3-solution}
\end{equation}

By (\ref{eq:howard-eqs-discounted-IHR-DHR-case-3-solution}) and some algebraic 
manipulations, the following condition for optimality of $\pi$ in 
state~$(2,n_2^\circ)$, 
\[
V_\beta^\pi(2,n_2^\circ;\nu) \le \frac{h}{1 - \beta}, 
\]
can be shown to be equivalent with 
\begin{equation}
\nu \le 
h \mu_2(n_2^\circ) \, \frac{\beta}{1 - \beta}, 
\label{eq:nu-req-discounted-IHR-DHR-case-31}
\end{equation}
where the right hand side equals $W_\beta(2,n_2^\circ)$ given in 
(\ref{eq:Whittle-index-2n-discounted-IHR-DHR}).

Let then $n \in \{0,1,\ldots,n_2^\circ-1\}$. 
Again by (\ref{eq:howard-eqs-discounted-IHR-DHR-case-3-solution}), 
the following condition for optimality of $\pi$ in state~$(2,n)$, 
\[
V_\beta^\pi(2,n;\nu) \le \frac{h}{1 - \beta}, 
\]
can be shown to be equivalent with 
\begin{equation}
\nu \le 
h \, 
\frac{\sum_{i = 0}^{n_2^\circ-n} \beta^i p_2(i|n)}
{\sum_{i = 0}^{n_2^\circ-n} \beta^i \bar p_2(i|n)} \, 
\frac{\beta}{1 - \beta}.
\label{eq:nu-req-discounted-IHR-DHR-case-31-cont}
\end{equation}
Thus, condition (\ref{eq:nu-req-discounted-IHR-DHR-case-31-cont}) 
follows from (\ref{eq:nu-req-discounted-IHR-DHR-case-31}) 
by Lemma~\ref{lem:Whittle-index-discounted-IHR-DHR-lemma-1}(vi).

\vskip 3pt
$3.2^\circ$ 
By (\ref{eq:howard-eqs-discounted-IHR-DHR-case-3-solution}), the following condition 
for optimality of $\pi$ in state~$(2,n_2^\circ+1)$, 
\[
\beta V_\beta^\pi(2,n_2^\circ+1;\nu) \le 
\nu + \beta (1 - \mu_2(n_2^\circ+1)) V_\beta^\pi(2,n_2^\circ+2;\nu), 
\]
is easily shown to be equivalent with 
\begin{equation}
\nu \ge 
h \mu_2(n_2^\circ+1) \, \frac{\beta}{1 - \beta}, 
\label{eq:nu-req-discounted-IHR-DHR-case-32a}
\end{equation}
where the right hand side equals $W_\beta(2,n_2^\circ+1)$ given in 
(\ref{eq:Whittle-index-2n-discounted-IHR-DHR}).

Let then $n \in \{n_2^\circ+2,n_2^\circ+3,\ldots\}$. 
Again by (\ref{eq:howard-eqs-discounted-IHR-DHR-case-3-solution}), 
the following condition for optimality of $\pi$ in state~$(2,n)$, 
\[
\beta V_\beta^\pi(2,n;\nu) \le 
\nu + \beta (1 - \mu_2(n)) V_\beta^\pi(2,n+1;\nu), 
\]
is easily shown to be equivalent with 
\[
\nu \ge 
h \mu_2(n) \, \frac{\beta}{1 - \beta}, 
\]
which follows from (\ref{eq:nu-req-discounted-IHR-DHR-case-32a}) since 
$\mu_2(n)$ is decreasing.

Let then $n \in \{0,1,\ldots\}$. 
By (\ref{eq:howard-eqs-discounted-IHR-DHR-case-3-solution}) and some algebraic 
manipulations, the following condition for optimality of 
$\pi$ in state~$(1,n)$, 
\[
\beta V_\beta^\pi(1,n;\nu) \le 
\nu + \beta \mu_1(n) V_\beta^\pi(2,0;\nu) + 
\beta (1 - \mu_1(n)) V_\beta^\pi(1,n+1;\nu), 
\]
can be shown to be equivalent with 
\[
\nu \ge 
h \mu_1(n) \, 
\frac{\beta \sum_{i=0}^{n_2^\circ} \beta^i p_2(i)}
{1 + \beta \mu_1(n) \sum_{i=0}^{n_2^\circ} \beta^i \bar p_2(i)} \, 
\frac{\beta}{1 - \beta}.
\]
Since $\mu_1(n)$ is increasing, we conclude that policy $\pi$ is optimal 
in any state~$(1,n)$ if and only if 
\begin{equation}
\nu \ge 
h \mu_1(\infty) \, 
\frac{\beta \sum_{i=0}^{n_2^\circ} \beta^i p_2(i)}
{1 + \beta \mu_1(\infty) \sum_{i=0}^{n_2^\circ} \beta^i \bar p_2(i)} \, 
\frac{\beta}{1 - \beta}, 
\label{eq:nu-req-discounted-IHR-DHR-case-32b}
\end{equation}
where the right hand side equals $W_\beta(1,\infty)$ given in 
(\ref{eq:Whittle-index-21-infty-discounted-IHR-DHR-D}).

Finally, by (\ref{eq:Wx-order-IHR-DHR-D}), we have 
$W_\beta(1,\infty) \ge W_\beta(2,n_2^\circ+1)$, which implies that 
condition (\ref{eq:nu-req-discounted-IHR-DHR-case-32a}) follows from 
(\ref{eq:nu-req-discounted-IHR-DHR-case-32b}). 
This completes the proof of claim~$3^\circ$.

\vskip 3pt
$4^\circ$ 
We still assume that $\nu \ge 0$ and utilize the optimality equations 
(\ref{eq:opt-eqs-discounted-nu-pos}). 
Let $m \in \{2,3,\ldots\}$, $n_2 \in \{n_2^\circ,n_2^\circ+1,\ldots,n_2^*\}$, 
and $n_1 \in \{1,2,\ldots\}$ such that 
\[
\begin{split}
& 
\{(2,0),(2,1),\ldots,(2,n_2^\circ)\} \cup \{y_m,y_{m+1},\ldots\} \; = \\
& \quad 
\{(2,0),(2,1),\ldots,(2,n_2)\} \cup \{(1,n_1),(1,n_1+1),\ldots\}, 
\end{split}
\]
where $y_m$ is defined in (\ref{eq:y-def-IHR-DHR-D}). 
We prove that the policy $\pi$ with activity set 
\[
{\mathcal B}^\pi = 
\{(2,0),(2,1),\ldots,(2,n_2^\circ)\} \cup \{y_m,y_{m+1},\ldots\} 
\]
is $(\nu,\beta)$-optimal for all 
\[
\nu \in [W_\beta(y_{m-1}),W_\beta(y_m)].
\]
It remains to prove that policy $\pi$ is optimal for these values of $\nu$ 
in any state $x \in {\mathcal S} \setminus \{*\}$. 
This is done below in two parts ($4.1^\circ$ and $4.2^\circ$).

\vskip 3pt
$4.1^\circ$ 
We start by first deriving the value function $V_\beta^\pi(x;\nu)$ for 
policy $\pi$ from the Howard equations: 
\begin{equation}
\begin{split}
& 
V_\beta^\pi(1,n;\nu) = 
h + \nu + \beta \mu_1(n) V_\beta^\pi(2,0;\nu) \; + \\
& \quad 
\beta (1 - \mu_1(n)) V_\beta^\pi(1,n+1;\nu), 
\quad 
n \in \{n_1,n_1+1,\ldots\}, \\
& 
V_\beta^\pi(2,n;\nu) = 
h + \nu + \beta (1 - \mu_2(n)) V_\beta^\pi(2,n+1;\nu), \\
& \quad 
n \in \{0,1,\ldots,n_2\}, \\
& 
V_\beta^\pi(x;\nu) = 
h + \beta V_\beta^\pi(x;\nu), \\
& \quad 
x \in {\mathcal S} \setminus 
\{(2,0),(2,1),\ldots,(2,n_2),(1,n_1),(1,n_1+1),\ldots,*\}.
\end{split}
\label{eq:howard-eqs-discounted-IHR-DHR-D-case-4}
\end{equation}
The unique solution of these linear equations is given by 
\begin{equation}
\begin{split}
& 
V_\beta^\pi(2,n;\nu) = 
(h + \nu) 
\left( 
\sum_{i = 0}^{n_2-n} \beta^i \bar p_2(i|n) 
\right) \; + \\
& \quad 
h \, \frac{\beta^{n_2-n+1} \bar p_2(n_2-n+1|n)}{1 - \beta}, 
\quad 
n \in \{0,1,\ldots,n_2\}, \\
& 
V_\beta^\pi(1,n;\nu) = 
(h + \nu) 
\left( 
\sum_{i = 0}^\infty \beta^i \bar p_1(i|n) 
\left( 
1 + \beta \mu_1(n+i) \frac{V_\beta^\pi(2,0;\nu)}{h + \nu} 
\right) 
\right), \\
& \quad 
n \in \{n_1,n_1+1,\ldots\},\\
& 
V_\beta^\pi(x;\nu) = 
\frac{h}{1 - \beta}, \\
& \quad 
x \in {\mathcal S} \setminus 
\{(2,0),(2,1),\ldots,(2,n_2),(1,n_1),(1,n_1+1),\ldots,*\}.
\end{split}
\label{eq:howard-eqs-discounted-IHR-DHR-D-case-4-solution}
\end{equation}

By (\ref{eq:howard-eqs-discounted-IHR-DHR-D-case-4-solution}) and some algebraic 
manipulations, the following condition for optimality of $\pi$ in 
state~$(2,n_2)$, 
\[
V_\beta^\pi(2,n_2;\nu) \le \frac{h}{1 - \beta}, 
\]
can be shown to be equivalent with 
\begin{equation}
\nu \le 
h \mu_2(n_2) \, \frac{\beta}{1 - \beta}, 
\label{eq:nu-req-discounted-IHR-DHR-D-case-41a}
\end{equation}
where the right hand side equals $W_\beta(2,n_2)$ given in 
(\ref{eq:Whittle-index-2n-discounted-IHR-DHR}).

Let then $n \in \{0,1,\ldots,n_2-1\}$. 
Again by (\ref{eq:howard-eqs-discounted-IHR-DHR-D-case-4-solution}), 
the following condition for optimality of $\pi$ in state~$(2,n)$, 
\[
V_\beta^\pi(2,n;\nu) \le \frac{h}{1 - \beta}, 
\]
can be shown to be equivalent with 
\begin{equation}
\nu \le 
h \, 
\frac{\sum_{i = 0}^{n_2-n} \beta^i p_2(i|n)}
{\sum_{i = 0}^{n_2-n} \beta^i \bar p_2(i|n)} \, 
\frac{\beta}{1 - \beta}.
\label{eq:nu-req-discounted-IHR-DHR-D-case-41a-cont}
\end{equation}
Thus, condition (\ref{eq:nu-req-discounted-IHR-DHR-D-case-41a-cont}) 
follows from (\ref{eq:nu-req-discounted-IHR-DHR-D-case-41a}) 
by Lemma~\ref{lem:Whittle-index-discounted-IHR-DHR-lemma-1}(vi).

On the other hand, by (\ref{eq:howard-eqs-discounted-IHR-DHR-D-case-4-solution}) 
and some algebraic manipulations, the following condition for optimality 
of $\pi$ in state~$(1,n_1)$, 
\[
V_\beta^\pi(1,n_1;\nu) \le \frac{h}{1 - \beta}, 
\]
can be shown to be equivalent with 
\begin{equation}
\nu \le 
h \, 
\frac{\frac{1}{1 - \beta} - \sum_{i=0}^{\infty} \beta^i \bar p_1(i|n_1) 
\left( 1 + \mu_1(n_1+i) \Big( 1 - \beta \sum_{j=0}^{n_2} \beta^{j} p_2(j) \Big) \, \frac{\beta}{1 - \beta} \right)}
{\sum_{i=0}^{\infty} \beta^i \bar p_1(i|n_1) \Big( 1 + \mu_1(n_1+i) \beta \sum_{j=0}^{n_2} \beta^{j} \bar p_2(j) \Big)}, 
\label{eq:nu-req-discounted-IHR-DHR-D-case-41b}
\end{equation}
where the right hand side equals $\psi(n_1,n_2)$ given in 
(\ref{eq:psi-discounted-IHR-DHR}).

Let then $n \in \{n_1,n_1+1,\ldots\}$. 
Since 
\[
\bar p_1(i|n) \mu_1(n+i) = p_1(i|n) = \bar p_1(i|n) - \bar p_1(i+1|n), 
\]
it follows from 
(\ref{eq:howard-eqs-discounted-IHR-DHR-D-case-4-solution}) that 
\begin{equation}
\begin{split}
& 
V_\beta^\pi(1,n;\nu) = 
(h + \nu) 
\Bigg( 
1 + \frac{\beta V_\beta^\pi(2,0;\nu)}{h + \nu} \; + \\
& \quad 
\sum_{i = 1}^\infty \beta^i \bar p_1(i|n) 
\bigg( 
1 - \frac{(1 - \beta) V_\beta^\pi(2,0;\nu)}{h + \nu} 
\bigg) 
\Bigg).
\end{split}
\label{eq:howard-eqs-discounted-IHR-DHR-D-case-4-solution-cont}
\end{equation}
Now, since $\bar p_1(i|n)$ is an decreasing function of $n$ in the IHR-DHR 
case and we have above required that 
\[
V_\beta^\pi(2,0;\nu) \le \frac{h}{1 - \beta}, 
\]
we see from (\ref{eq:howard-eqs-discounted-IHR-DHR-D-case-4-solution-cont}) 
that $V_\beta^\pi(1,n;\nu)$, as well, is a decreasing function of $n$. 
Thus, condition (\ref{eq:nu-req-discounted-IHR-DHR-D-case-41b}) implies that 
the following condition for optimality of $\pi$ in state~$(1,n)$, 
\[
V_\beta^\pi(1,n;\nu) \le \frac{h}{1 - \beta}, 
\]
is satisfied for any $n \in \{n_1,n_1+1,\ldots\}$.

Finally, by combining (\ref{eq:nu-req-discounted-IHR-DHR-D-case-41a}) and 
(\ref{eq:nu-req-discounted-IHR-DHR-D-case-41b}), we get the requirement that 
\begin{equation}
\nu \le 
\min \{ W_\beta(2,n_2), \psi(n_1,n_2) \}.
\label{eq:nu-req-discounted-IHR-DHR-D-case-41}
\end{equation}
Now, if $W_\beta(2,n_2) < \psi(n_1,n_2)$, then the right hand side of 
(\ref{eq:nu-req-discounted-IHR-DHR-D-case-41}) equals $W_\beta(2,n_2)$ and this 
requirement (\ref{eq:nu-req-discounted-IHR-DHR-D-case-41}) is satisfied by our 
assumption that $\nu \le W_\beta(y_m)$, since 
$(2,n_2) \in \{(2,0),(2,1),\ldots,(2,n_2^\circ),y_m,y_{m+1},\ldots\}$ 
and, by (\ref{eq:Wx-order-IHR-DHR-D}), 
\[
\begin{split}
& 
W_\beta(y_m) = 
\min\{W_\beta(2,0),W_\beta(2,1),\ldots,W_\beta(2,n_2^\circ),W_\beta(y_m),W_\beta(y_{m+1}),\ldots\} \\
& \quad 
\le \; 
W_\beta(2,n_2).
\end{split}
\]
Assume now that 
\[
W_\beta(2,n_2) \ge \psi(n_1,n_2).
\]
Below we show that in this case $\phi(n_1) = n_2$, which implies that 
\[
W_\beta(1,n_1) = \psi(n_1,n_2).
\]
If $\phi(n_1) < n_2$, then we have, by 
(\ref{eq:phi-discounted-IHR-DHR}), 
\[
\psi(n_1,\phi(n_1)) > w_2(\phi(n_1)+1), 
\]
which is, by Lemma~\ref{lem:Whittle-index-discounted-IHR-DHR-lemma-1}(iv), 
equivalent with 
\[
\psi(n_1,\phi(n_1)+1) > w_2(\phi(n_1)+1).
\]
But, by Lemma~\ref{lem:Whittle-index-discounted-IHR-DHR-lemma-1}(i), 
this implies that 
\[
\psi(n_1,\phi(n_1)+1) > w_2(\phi(n_1)+2), 
\]
which is, again by Lemma~\ref{lem:Whittle-index-discounted-IHR-DHR-lemma-1}(iv), 
equivalent with 
\[
\psi(n_1,\phi(n_1)+2) > w_2(\phi(n_1)+2).
\]
By continuing similarly, we finally end up to the following inequality: 
\[
\psi(n_1,n_2) > w_2(n_2), 
\]
which, however, contradicts our assumption above (i.e., 
$W_\beta(2,n_2) = w_2(n_2) \ge \psi(n_1,n_2)$). 
So, by further taking into account (\ref{eq:phi-result1-IHR-DHR-D}), we have now 
proved that $\phi(n_1) = n_2$ in this case, which imples that the right hand side 
of (\ref{eq:nu-req-discounted-IHR-DHR-D-case-41}) equals $W_\beta(1,n_1)$. 
In addition, this requirement (\ref{eq:nu-req-discounted-IHR-DHR-D-case-41}) 
is satisfied by our assumption that $\nu \le W_\beta(y_m)$, since 
$(1,n_1) \in \{y_m,y_{m+1},\ldots\}$ and, by (\ref{eq:Wx-order-IHR-DHR-D}), 
\[
W_\beta(y_m) = 
\min\{W_\beta(y_m),W_\beta(y_{m+1}),\ldots\} \le 
W_\beta(1,n_1).
\]

\vskip 3pt
$4.2^\circ$ 
By (\ref{eq:howard-eqs-discounted-IHR-DHR-D-case-4-solution}), the following condition 
for optimality of $\pi$ in state~$(2,n_2+1)$, 
\[
\beta V_\beta^\pi(2,n_2+1;\nu) \le 
\nu + \beta (1 - \mu_2(n_2+1)) V_\beta^\pi(2,n_2+2;\nu), 
\]
is easily shown to be equivalent with 
\begin{equation}
\nu \ge 
h \mu_2(n_2+1) \, \frac{\beta}{1 - \beta}, 
\label{eq:nu-req-discounted-IHR-DHR-D-case-42a}
\end{equation}
where the right hand side equals $W_\beta(2,n_2+1)$ given in 
(\ref{eq:Whittle-index-2n-discounted-IHR-DHR}).

Let then $n \in \{n_2+2,n_2+3,\ldots\}$. 
Again by (\ref{eq:howard-eqs-discounted-IHR-DHR-D-case-4-solution}), 
the following condition for optimality of $\pi$ in state~$(2,n)$, 
\[
\beta V_\beta^\pi(2,n;\nu) \le 
\nu + \beta (1 - \mu_2(n)) V_\beta^\pi(2,n+1;\nu), 
\]
is easily shown to be equivalent with 
\[
\nu \ge 
h \mu_2(n) \, \frac{\beta}{1 - \beta}, 
\]
which follows from (\ref{eq:nu-req-discounted-IHR-DHR-D-case-42a}) since 
$\mu_2(n)$ is decreasing.

On the other hand, by (\ref{eq:howard-eqs-discounted-IHR-DHR-D-case-4-solution}) 
and some algebraic manipulations, the following condition for optimality of 
$\pi$ in state~$(1,n_1-1)$, 
\[
\beta V_\beta^\pi(1,n_1-1;\nu) \le 
\nu + \beta \mu_1(n_1-1) V_\beta^\pi(2,0;\nu) + 
\beta (1 - \mu_1(n_1-1)) V_\beta^\pi(1,n_1;\nu), 
\]
can be shown to be equivalent with 
\begin{equation}
\nu \ge 
h \, 
\frac{\frac{1}{1 - \beta} - \sum_{i=0}^{\infty} \beta^i \bar p_1(i|n_1-1) 
\left( 1 + \mu_1(n_1-1+i) \Big( 1 - \beta \sum_{j=0}^{n_2} \beta^{j} p_2(j) \Big) \, \frac{\beta}{1 - \beta} \right)}
{\sum_{i=0}^{\infty} \beta^i \bar p_1(i|n_1-1) \Big( 1 + \mu_1(n_1-1+i) \beta \sum_{j=0}^{n_2} \beta^{j} \bar p_2(j) \Big)},
\label{eq:nu-req-discounted-IHR-DHR-D-case-42b}
\end{equation}
where the right hand side equals $\psi(n_1-1,n_2)$ given in 
(\ref{eq:psi-discounted-IHR-DHR}).

Let then $n \in \{0,1,\ldots,n_1-2\}$. 
By (\ref{eq:howard-eqs-discounted-IHR-DHR-D-case-4-solution}) and some 
algebraic manipulations, the following condition for optimality of $\pi$ 
in state~$(1,n)$, 
\[
\beta V_\beta^\pi(1,n;\nu) \le 
\nu + \beta \mu_1(n) V_\beta^\pi(2,0;\nu) + 
\beta (1 - \mu_1(n)) V_\beta^\pi(1,n+1;\nu), 
\]
can be shown to be equivalent with 
\[
\begin{split}
& 
\nu \ge 
h \mu_1(n) \, 
\frac{\beta \sum_{i=0}^{n_2} \beta^i p_2(i)}
{1 + \beta \mu_1(n) \sum_{i=0}^{n_2} \beta^i \bar p_2(i)} \, 
\frac{\beta}{1 - \beta}.
\end{split}
\]
By Lemma~\ref{lem:Whittle-index-discounted-IHR-DHR-lemma-1}(vii), we conclude 
that this condition follows from (\ref{eq:nu-req-discounted-IHR-DHR-D-case-42b}).

Finally, by combining (\ref{eq:nu-req-discounted-IHR-DHR-D-case-42a}) and 
(\ref{eq:nu-req-discounted-IHR-DHR-D-case-42b}), we get the requirement that 
\begin{equation}
\nu \ge 
\max \{ W_\beta(2,n_2+1), \psi(n_1-1,n_2) \}.
\label{eq:nu-req-discounted-IHR-DHR-D-case-42}
\end{equation}
Now if $W_\beta(2,n_2+1) < \psi(n_1-1,n_2)$, then $\phi(n_1-1) = n_2$ by 
(\ref{eq:phi-discounted-IHR-DHR}) and (\ref{eq:phi-result2-IHR-DHR-D}). 
In addition, by (\ref{eq:Whittle-index-1n-discounted-IHR-DHR}), the right 
hand side of (\ref{eq:nu-req-discounted-IHR-DHR-D-case-42}) equals, under this condition, 
\[
W_\beta(1,n_1-1) = \psi(n_1-1,n_2) 
\]
and, furthermore, by (\ref{eq:y-def-IHR-DHR-D}), we have 
\[
y_{m-1} = (1,n_1-1) 
\]
so that requirement (\ref{eq:nu-req-discounted-IHR-DHR-D-case-42}) is equivalent 
with the requirement 
\[
\nu \ge W_\beta(y_{m-1}).
\]
On the other hand, if $W_\beta(2,n_2+1) \ge \psi(n_1-1,n_2)$, then, 
$\phi(n_1-1) > n_2$ by (\ref{eq:phi-discounted-IHR-DHR}) and 
we have, by (\ref{eq:Whittle-index-order-IHR-DHR}), 
\[
W_\beta(2,n_2+1) \ge W_\beta(2,\phi(n_1-1)) \ge W_\beta(1,n_1-1) 
\]
and, furthermore, by (\ref{eq:y-def-IHR-DHR-D}), 
\[
y_{m-1} = (2,n_2+1) 
\]
so that requirement (\ref{eq:nu-req-discounted-IHR-DHR-D-case-42}) is again 
equivalent with the requirement 
\[
\nu \ge W_\beta(y_{m-1}), 
\]
which completes the proof of claim~$4^\circ$.

\vskip 3pt
$5^\circ$ 
We still assume that $\nu \ge 0$ and utilize the optimality equations 
(\ref{eq:opt-eqs-discounted-nu-pos}). 
Now we prove that the policy $\pi$ with activity set 
\[
\begin{split}
& 
{\mathcal B}^\pi = 
\{(2,0),(2,1),\ldots,(2,n_2^\circ)\} \cup \{y_1,y_2,\ldots\} \; = \\
& \quad 
\{(2,0),(2,1),\ldots,(2,n_2^*)\} \cup \{(1,0),(1,1),\ldots\} 
\end{split}
\]
is $(\nu,\beta)$-optimal for all 
\[
\nu \in [W_\beta(2,n_2^*+1),W_\beta(y_1)].
\]
It remains to prove that policy $\pi$ is optimal for these values of $\nu$ 
in any state $x \in {\mathcal S} \setminus \{*\}$. 
This is done below in two parts ($5.1^\circ$ and $5.2^\circ$).

\vskip 3pt
$5.1^\circ$ 
We start by first deriving the value function $V_\beta^\pi(x;\nu)$ for 
policy $\pi$ from the Howard equations: 
\begin{equation}
\begin{split}
& 
V_\beta^\pi(1,n;\nu) = 
h + \nu + \beta \mu_1(n) V_\beta^\pi(2,0;\nu) \; + \\
& \quad 
\beta (1 - \mu_1(n)) V_\beta^\pi(1,n+1;\nu), 
\quad 
n \in \{0,1,\ldots\}, \\
& 
V_\beta^\pi(2,n;\nu) = 
h + \nu + \beta (1 - \mu_2(n)) V_\beta^\pi(2,n+1;\nu), \\
& \quad 
n \in \{0,1,\ldots,n_2^*\}, \\
& 
V_\beta^\pi(x;\nu) = 
h + \beta V_\beta^\pi(x;\nu), \\
& \quad 
x \in {\mathcal S} \setminus 
\{(2,0),(2,1),\ldots,(2,n_2^*),(1,0),(1,1),\ldots,*\}.
\end{split}
\label{eq:howard-eqs-discounted-IHR-DHR-D-case-5}
\end{equation}
The unique solution of these linear equations is given by 
\begin{equation}
\begin{split}
& 
V_\beta^\pi(2,n;\nu) = 
(h + \nu) 
\left( 
\sum_{i = 0}^{n_2^*-n} \beta^i \bar p_2(i|n) 
\right) \; + \\
& \quad 
h \, \frac{\beta^{n_2^*-n+1} \bar p_2(n_2^*-n+1|n)}{1 - \beta}, 
\quad 
n \in \{0,1,\ldots,n_2^*\}, \\
& 
V_\beta^\pi(1,n;\nu) = 
(h + \nu) 
\left( 
\sum_{i = 0}^\infty \beta^i \bar p_1(i|n) 
\left( 
1 + \beta \mu_1(n+i) \frac{V_\beta^\pi(2,0;\nu)}{h + \nu} 
\right) 
\right), \\
& \quad 
n \in \{0,1,\ldots\},\\
& 
V_\beta^\pi(x;\nu) = 
\frac{h}{1 - \beta}, \\
& \quad 
x \in {\mathcal S} \setminus 
\{(2,0),(2,1),\ldots,(2,n_2^*),(1,0),(1,1),\ldots,*\}.
\end{split}
\label{eq:howard-eqs-discounted-IHR-DHR-D-case-5-solution}
\end{equation}

By (\ref{eq:howard-eqs-discounted-IHR-DHR-D-case-5-solution}) and some algebraic 
manipulations, the following condition for optimality of $\pi$ in 
state~$(2,n_2^*)$, 
\[
V_\beta^\pi(2,n_2^*;\nu) \le \frac{h}{1 - \beta}, 
\]
can be shown to be equivalent with 
\begin{equation}
\nu \le 
h \mu_2(n_2^*) \, \frac{\beta}{1 - \beta}, 
\label{eq:nu-req-discounted-IHR-DHR-D-case-51a}
\end{equation}
where the right hand side equals $W_\beta(2,n_2^*)$ given in 
(\ref{eq:Whittle-index-2n-discounted-IHR-DHR}). 
Note that (\ref{eq:nu-req-discounted-IHR-DHR-D-case-51a}) follows from the 
requirement that $\nu \le W_\beta(y_1)$ since 
$W_\beta(y_1) \le W_\beta(2,n_2^*)$ by (\ref{eq:Wx-order-IHR-DHR-D}).

Let then $n \in \{0,1,\ldots,n_2^*-1\}$. 
Again by (\ref{eq:howard-eqs-discounted-IHR-DHR-D-case-5-solution}), 
the following condition for optimality of $\pi$ in state~$(2,n)$, 
\[
V_\beta^\pi(2,n;\nu) \le \frac{h}{1 - \beta}, 
\]
can be shown to be equivalent with 
\begin{equation}
\nu \le 
h \, 
\frac{\sum_{i = 0}^{n_2^*-n} \beta^i p_2(i|n)}
{\sum_{i = 0}^{n_2^*-n} \beta^i \bar p_2(i|n)} \, 
\frac{\beta}{1 - \beta}.
\label{eq:nu-req-discounted-IHR-DHR-D-case-51a-cont}
\end{equation}
Thus, condition (\ref{eq:nu-req-discounted-IHR-DHR-D-case-51a-cont}) 
follows from (\ref{eq:nu-req-discounted-IHR-DHR-D-case-51a}) 
by Lemma~\ref{lem:Whittle-index-discounted-IHR-DHR-lemma-1}(vi).

On the other hand, by (\ref{eq:howard-eqs-discounted-IHR-DHR-D-case-5-solution}) 
and some algebraic manipulations, the following condition for optimality 
of $\pi$ in state~$y_1 = (1,0)$, 
\[
V_\beta^\pi(1,0;\nu) \le \frac{h}{1 - \beta}, 
\]
can be shown to be equivalent with 
\begin{equation}
\nu \le 
h \, 
\frac{\frac{1}{1 - \beta} - \sum_{i=0}^{\infty} \beta^i \bar p_1(i) 
\left( 1 + \mu_1(i) \Big( 1 - \beta \sum_{j=0}^{n_2^*} \beta^{j} p_2(j) \Big) \, \frac{\beta}{1 - \beta} \right)}
{\sum_{i=0}^{\infty} \beta^i \bar p_1(i) \Big( 1 + \mu_1(i) \beta \sum_{j=0}^{n_2^*} \beta^{j} \bar p_2(j) \Big)}, 
\label{eq:nu-req-discounted-IHR-DHR-D-case-51b}
\end{equation}
where the right hand side equals $\psi(0,n_2^*)$ given in 
(\ref{eq:psi-discounted-IHR-DHR}). Moreover, it follows from 
the definition of $n_2^*$ and (\ref{eq:phi-discounted-IHR-DHR}) 
that $\phi(0) = n_2^*$, which implies that the right hand side of 
(\ref{eq:nu-req-discounted-IHR-DHR-D-case-51b}) equals 
$W_\beta(y_1) = W_\beta(1,0)$ given in 
(\ref{eq:Whittle-index-1n-discounted-IHR-DHR}).

Let then $n \in \{0,1,\ldots\}$. 
Since 
\[
\bar p_1(i|n) \mu_1(n+i) = p_1(i|n) = \bar p_1(i|n) - \bar p_1(i+1|n), 
\]
it follows from 
(\ref{eq:howard-eqs-discounted-IHR-DHR-D-case-5-solution}) that 
\begin{equation}
\begin{split}
& 
V_\beta^\pi(1,n;\nu) = 
(h + \nu) 
\Bigg( 
1 + \frac{\beta V_\beta^\pi(2,0;\nu)}{h + \nu} \; + \\
& \quad 
\sum_{i = 1}^\infty \beta^i \bar p_1(i|n) 
\bigg( 
1 - \frac{(1 - \beta) V_\beta^\pi(2,0;\nu)}{h + \nu} 
\bigg) 
\Bigg).
\end{split}
\label{eq:howard-eqs-discounted-IHR-DHR-D-case-5-solution-cont}
\end{equation}
Now, since $\bar p_1(i|n)$ is an decreasing function of $n$ in the IHR-DHR 
case and we have above required that 
\[
V_\beta^\pi(2,0;\nu) \le \frac{h}{1 - \beta}, 
\]
we see from (\ref{eq:howard-eqs-discounted-IHR-DHR-D-case-5-solution-cont}) 
that $V_\beta^\pi(1,n;\nu)$, as well, is a decreasing function of $n$. 
Thus, condition (\ref{eq:nu-req-discounted-IHR-DHR-D-case-51b}) implies that 
the following condition for optimality of $\pi$ in state~$(1,n)$, 
\[
V_\beta^\pi(1,n;\nu) \le \frac{h}{1 - \beta}, 
\]
is satisfied for any $n \in \{0,1,\ldots\}$.

\vskip 3pt
$5.2^\circ$ 
By (\ref{eq:howard-eqs-discounted-IHR-DHR-D-case-5-solution}), 
the following condition for optimality of $\pi$ in state~$(2,n_2^*+1)$, 
\[
\beta V_\beta^\pi(2,n_2^*+1;\nu) \le 
\nu + \beta (1 - \mu_2(n_2^*+1)) V_\beta^\pi(2,n_2^*+2;\nu), 
\]
is easily shown to be equivalent with 
\begin{equation}
\nu \ge 
h \mu_2(n_2^*+1) \, \frac{\beta}{1 - \beta}, 
\label{eq:nu-req-discounted-IHR-DHR-D-case-52}
\end{equation}
where the right hand side equals $W_\beta(2,n_2^*+1)$ given in 
(\ref{eq:Whittle-index-2n-discounted-IHR-DHR}).

Let then $n \in \{n_2^*+2,n_2^*+3,\ldots\}$. 
Again by (\ref{eq:howard-eqs-discounted-IHR-DHR-D-case-5-solution}), 
the following condition for optimality of $\pi$ in state~$(2,n)$, 
\[
\beta V_\beta^\pi(2,n;\nu) \le 
\nu + \beta (1 - \mu_2(n)) V_\beta^\pi(2,n+1;\nu), 
\]
is easily shown to be equivalent with 
\[
\nu \ge 
h \mu_2(n) \, \frac{\beta}{1 - \beta}, 
\]
which follows from (\ref{eq:nu-req-discounted-IHR-DHR-D-case-52}) since 
$\mu_2(n)$ is decreasing. 
This completes the proof of claim~$5^\circ$.

\vskip 3pt
$6^\circ$ 
We still assume that $\nu \ge 0$ and utilize the optimality equations 
(\ref{eq:opt-eqs-discounted-nu-pos}). 
Let $n_2 \in \{n_2^*+1,n_2^*+2,\ldots\}$. 
We prove that the policy $\pi$ with activity set 
\[
{\mathcal B}^\pi = 
\{(2,0),(2,1),\ldots,(2,n_2)\} \cup \{(1,0),(1,1),\ldots\} 
\]
is $(\nu,\beta)$-optimal for all 
\[
\nu \in [W_\beta(2,n_2+1),W_\beta(2,n_2)].
\]
It remains to prove that policy $\pi$ is optimal for these values of $\nu$ 
in any state $x \in {\mathcal S} \setminus \{*\}$. 
This is done below in two parts ($6.1^\circ$ and $6.2^\circ$).

\vskip 3pt
$6.1^\circ$ 
We start by first deriving the value function $V_\beta^\pi(x;\nu)$ for 
policy $\pi$ from the Howard equations: 
\begin{equation}
\begin{split}
& 
V_\beta^\pi(1,n;\nu) = 
h + \nu + \beta \mu_1(n) V_\beta^\pi(2,0;\nu) \; + \\
& \quad 
\beta (1 - \mu_1(n)) V_\beta^\pi(1,n+1;\nu), 
\quad 
n \in \{0,1,\ldots\}, \\
& 
V_\beta^\pi(2,n;\nu) = 
h + \nu + \beta (1 - \mu_2(n)) V_\beta^\pi(2,n+1;\nu), \\
& \quad 
n \in \{0,1,\ldots,n_2\}, \\
& 
V_\beta^\pi(x;\nu) = 
h + \beta V_\beta^\pi(x;\nu), \\
& \quad 
x \in {\mathcal S} \setminus 
\{(2,0),(2,1),\ldots,(2,n_2),(1,0),(1,1),\ldots,*\}.
\end{split}
\label{eq:howard-eqs-discounted-IHR-DHR-D-case-6}
\end{equation}
The unique solution of these linear equations is given by 
\begin{equation}
\begin{split}
& 
V_\beta^\pi(2,n;\nu) = 
(h + \nu) 
\left( 
\sum_{i = 0}^{n_2-n} \beta^i \bar p_2(i|n) 
\right) \; + \\
& \quad 
h \, \frac{\beta^{n_2-n+1} \bar p_2(n_2-n+1|n)}{1 - \beta}, 
\quad 
n \in \{0,1,\ldots,n_2\}, \\
& 
V_\beta^\pi(1,n;\nu) = 
(h + \nu) 
\left( 
\sum_{i = 0}^\infty \beta^i \bar p_1(i|n) 
\left( 
1 + \beta \mu_1(n+i) \frac{V_\beta^\pi(2,0;\nu)}{h + \nu} 
\right) 
\right), \\
& \quad 
n \in \{0,1,\ldots\},\\
& 
V_\beta^\pi(x;\nu) = 
\frac{h}{1 - \beta}, \\
& \quad 
x \in {\mathcal S} \setminus 
\{(2,0),(2,1),\ldots,(2,n_2),(1,0),(1,1),\ldots,*\}.
\end{split}
\label{eq:howard-eqs-discounted-IHR-DHR-D-case-6-solution}
\end{equation}

By (\ref{eq:howard-eqs-discounted-IHR-DHR-D-case-6-solution}) and some algebraic 
manipulations, the following condition for optimality of $\pi$ in 
state~$(2,n_2)$, 
\[
V_\beta^\pi(2,n_2;\nu) \le \frac{h}{1 - \beta}, 
\]
can be shown to be equivalent with 
\begin{equation}
\nu \le 
h \mu_2(n_2) \, \frac{\beta}{1 - \beta}, 
\label{eq:nu-req-discounted-IHR-DHR-D-case-61a}
\end{equation}
where the right hand side equals $W_\beta(2,n_2)$ given in 
(\ref{eq:Whittle-index-2n-discounted-IHR-DHR}). 

Let then $n \in \{0,1,\ldots,n_2-1\}$. 
Again by (\ref{eq:howard-eqs-discounted-IHR-DHR-D-case-6-solution}), 
the following condition for optimality of $\pi$ in state~$(2,n)$, 
\[
V_\beta^\pi(2,n;\nu) \le \frac{h}{1 - \beta}, 
\]
can be shown to be equivalent with 
\begin{equation}
\nu \le 
h \, 
\frac{\sum_{i = 0}^{n_2-n} \beta^i p_2(i|n)}
{\sum_{i = 0}^{n_2-n} \beta^i \bar p_2(i|n)} \, 
\frac{\beta}{1 - \beta}.
\label{eq:nu-req-discounted-IHR-DHR-D-case-61a-cont}
\end{equation}
Thus, condition (\ref{eq:nu-req-discounted-IHR-DHR-D-case-61a-cont}) 
follows from (\ref{eq:nu-req-discounted-IHR-DHR-D-case-61a}) 
by Lemma~\ref{lem:Whittle-index-discounted-IHR-DHR-lemma-1}(vi).

On the other hand, by (\ref{eq:howard-eqs-discounted-IHR-DHR-D-case-6-solution}) 
and some algebraic manipulations, the following condition for optimality 
of $\pi$ in state~$(1,0)$, 
\[
V_\beta^\pi(1,0;\nu) \le \frac{h}{1 - \beta}, 
\]
can be shown to be equivalent with 
\begin{equation}
\nu \le 
h \, 
\frac{\frac{1}{1 - \beta} - \sum_{i=0}^{\infty} \beta^i \bar p_1(i) 
\left( 1 + \mu_1(i) \Big( 1 - \beta \sum_{j=0}^{n_2} \beta^{j} p_2(j) \Big) \, \frac{\beta}{1 - \beta} \right)}
{\sum_{i=0}^{\infty} \beta^i \bar p_1(i) \Big( 1 + \mu_1(i) \beta \sum_{j=0}^{n_2} \beta^{j} \bar p_2(j) \Big)}, 
\label{eq:nu-req-discounted-IHR-DHR-D-case-61b}
\end{equation}
where the right hand side equals $\psi(0,n_2)$ given in 
(\ref{eq:psi-discounted-IHR-DHR}).

Next we prove that $\psi(0,n_2) > w_2(n_2)$. 
First, it follows from the definition of $n_2^*$ and 
(\ref{eq:phi-discounted-IHR-DHR}) that $\phi(0) = n_2^*$, 
which implies, by (\ref{eq:Whittle-index-order-IHR-DHR}), that 
\[
\psi(0,n_2^*) = \psi(0,\phi(0)) = w_1(0) > w_2(\phi(0)+1) = w_2(n_2^*+1).
\]
However, by Lemma~\ref{lem:Whittle-index-discounted-IHR-DHR-lemma-1}(iv), 
this is equivalent with 
\[
\psi(0,n_2^*+1) > w_2(n_2^*+1).
\]
But, by Lemma~\ref{lem:Whittle-index-discounted-IHR-DHR-lemma-1}(i), 
this implies that 
\[
\psi(0,n_2^*+1) > w_2(n_2^*+2), 
\]
which is, again by Lemma~\ref{lem:Whittle-index-discounted-IHR-DHR-lemma-1}(iv), 
equivalent with 
\[
\psi(0,n_2^*+2) > w_2(n_2^*+2).
\]
By continuing similarly, we finally end up to the following inequality: 
\[
\psi(0,n_2) > w_2(n_2), 
\] 
which, in turn, proves that requirement 
(\ref{eq:nu-req-discounted-IHR-DHR-D-case-61b}) follows from 
(\ref{eq:nu-req-discounted-IHR-DHR-D-case-61a}). 

Let then $n \in \{0,1,\ldots\}$. 
Since 
\[
\bar p_1(i|n) \mu_1(n+i) = p_1(i|n) = \bar p_1(i|n) - \bar p_1(i+1|n), 
\]
it follows from 
(\ref{eq:howard-eqs-discounted-IHR-DHR-D-case-6-solution}) that 
\begin{equation}
\begin{split}
& 
V_\beta^\pi(1,n;\nu) = 
(h + \nu) 
\Bigg( 
1 + \frac{\beta V_\beta^\pi(2,0;\nu)}{h + \nu} \; + \\
& \quad 
\sum_{i = 1}^\infty \beta^i \bar p_1(i|n) 
\bigg( 
1 - \frac{(1 - \beta) V_\beta^\pi(2,0;\nu)}{h + \nu} 
\bigg) 
\Bigg).
\end{split}
\label{eq:howard-eqs-discounted-IHR-DHR-D-case-6-solution-cont}
\end{equation}
Now, since $\bar p_1(i|n)$ is an decreasing function of $n$ in the IHR-DHR 
case and we have above required that 
\[
V_\beta^\pi(2,0;\nu) \le \frac{h}{1 - \beta}, 
\]
we see from (\ref{eq:howard-eqs-discounted-IHR-DHR-D-case-6-solution-cont}) 
that $V_\beta^\pi(1,n;\nu)$, as well, is a decreasing function of $n$. 
Thus, condition (\ref{eq:nu-req-discounted-IHR-DHR-D-case-61b}) implies that 
the following condition for optimality of $\pi$ in state~$(1,n)$, 
\[
V_\beta^\pi(1,n;\nu) \le \frac{h}{1 - \beta}, 
\]
is satisfied for any $n \in \{0,1,\ldots\}$.

\vskip 3pt
$6.2^\circ$ 
By (\ref{eq:howard-eqs-discounted-IHR-DHR-D-case-6-solution}), 
the following condition for optimality of $\pi$ in state~$(2,n_2+1)$, 
\[
\beta V_\beta^\pi(2,n_2+1;\nu) \le 
\nu + \beta (1 - \mu_2(n_2+1)) V_\beta^\pi(2,n_2+2;\nu), 
\]
is easily shown to be equivalent with 
\begin{equation}
\nu \ge 
h \mu_2(n_2+1) \, \frac{\beta}{1 - \beta}, 
\label{eq:nu-req-discounted-IHR-DHR-D-case-62}
\end{equation}
where the right hand side equals $W_\beta(2,n_2+1)$ given in 
(\ref{eq:Whittle-index-2n-discounted-IHR-DHR}).

Let then $n \in \{n_2+2,n_2+3,\ldots\}$. 
Again by (\ref{eq:howard-eqs-discounted-IHR-DHR-D-case-6-solution}), 
the following condition for optimality of $\pi$ in state~$(2,n)$, 
\[
\beta V_\beta^\pi(2,n;\nu) \le 
\nu + \beta (1 - \mu_2(n)) V_\beta^\pi(2,n+1;\nu), 
\]
is easily shown to be equivalent with 
\[
\nu \ge 
h \mu_2(n) \, \frac{\beta}{1 - \beta}, 
\]
which follows from (\ref{eq:nu-req-discounted-IHR-DHR-D-case-62}) since 
$\mu_2(n)$ is decreasing. 
This completes the proof of claim~$6^\circ$.

\vskip 3pt
$7^\circ$ 
We still assume that $\nu \ge 0$ and utilize the optimality equations 
(\ref{eq:opt-eqs-discounted-nu-pos}). 
Now we prove that the policy $\pi$ with activity set 
\[
{\mathcal B}^\pi = {\mathcal S} \setminus \{*\} 
\]
is $(\nu,\beta)$-optimal for all 
\[
\nu \in [0,W_\beta(2,\infty)], 
\]
where $W_\beta(2,\infty)$ is defined in (\ref{eq:Whittle-index-21-infty-discounted-IHR-DHR-D}). 
It remains to prove that policy $\pi$ is optimal for these values of $\nu$ 
in any state $x \in {\mathcal S} \setminus \{*\}$.

We start by first deriving the value function $V_\beta^\pi(x;\nu)$ for 
policy $\pi$ from the Howard equations: 
\begin{equation}
\begin{split}
& 
V_\beta^\pi(1,n;\nu) = 
h + \nu + \beta \mu_1(n) V_\beta^\pi(2,0;\nu) + 
\beta (1 - \mu_1(n)) V_\beta^\pi(1,n+1;\nu), \\
& 
V_\beta^\pi(2,n;\nu) = 
h + \nu + \beta (1 - \mu_2(n)) V_\beta^\pi(2,n+1;\nu).
\end{split}
\label{eq:howard-eqs-discounted-IHR-DHR-D-case-7}
\end{equation}
The unique solution of these linear equations is given by 
\begin{equation}
\begin{split}
& 
V_\beta^\pi(2,n;\nu) = 
(h + \nu) 
\left( 
\sum_{i = 0}^\infty \beta^i \bar p_2(i|n) 
\right), \\
& 
V_\beta^\pi(1,n;\nu) = 
(h + \nu) 
\left( 
\sum_{i = 0}^\infty \beta^i \bar p_1(i|n) 
\left( 
1 + \beta \mu_1(n+i) \frac{V_\beta^\pi(2,0;\nu)}{h + \nu} 
\right) 
\right).
\end{split}
\label{eq:howard-eqs-discounted-IHR-DHR-D-case-7-solution}
\end{equation}

Let then $n \in \{0,1,\ldots\}$. 
Since $\bar p_2(i|n)$ is an increasing function of $n$ in the IHR-DHR 
case, we see from (\ref{eq:howard-eqs-discounted-IHR-DHR-D-case-7-solution}) 
that $V_\beta^\pi(2,n;\nu)$, as well, is an increasing function of $n$ 
and approaches 
\[
\begin{split}
& 
\lim_{n \to \infty} V_\beta^\pi(2,n;\nu) = 
(h + \nu) 
\left( 
\sum_{i = 0}^\infty \beta^i (1 - \mu_2(\infty))^i 
\right) \\
& \quad = 
\frac{h + \nu}{1 - \beta (1 - \mu_2(\infty))}.
\end{split}
\]
Thus, the following condition for optimality of $\pi$ in state~$(2,n)$, 
\[
V_\beta^\pi(2,n;\nu) \le \frac{h}{1 - \beta}, 
\]
is satisfied for any $n$ if and only if 
\[
\lim_{n \to \infty} V_\beta^\pi(2,n;\nu) \le \frac{h}{1 - \beta}, 
\]
which is clearly equivalent with condition 
\begin{equation}
\nu \le 
h \mu_2(\infty) \, 
\frac{\beta}{1 - \beta}. 
\label{eq:nu-req-discounted-IHR-DHR-D-case-7a}
\end{equation}
Note that the right hand side equals $w_2(\infty) = W_\beta(2,\infty)$ given in 
(\ref{eq:Whittle-index-21-infty-discounted-IHR-DHR-D}).

On the other hand, by (\ref{eq:howard-eqs-discounted-IHR-DHR-D-case-7-solution}) 
and some algebraic manipulations, the following condition for optimality 
of $\pi$ in state~$(1,0)$, 
\[
V_\beta^\pi(1,0;\nu) \le \frac{h}{1 - \beta}, 
\]
can be shown to be equivalent with 
\begin{equation}
\nu \le 
h \, 
\frac{\frac{1}{1 - \beta} - \sum_{i=0}^{\infty} \beta^i \bar p_1(i) 
\left( 1 + \mu_1(i) \Big( 1 - \beta \sum_{j=0}^{\infty} \beta^{j} p_2(j) \Big) \, \frac{\beta}{1 - \beta} \right)}
{\sum_{i=0}^{\infty} \beta^i \bar p_1(i) \Big( 1 + \mu_1(i) \beta \sum_{j=0}^{\infty} \beta^{j} \bar p_2(j) \Big)}, 
\label{eq:nu-req-discounted-IHR-DHR-D-case-7b}
\end{equation}
where the right hand side equals $\psi(0,\infty)$ given in 
(\ref{eq:psi-n1-infty-discounted-IHR-DHR}).

Next we prove that $\psi(0,\infty) \ge w_2(\infty)$. 
First, it follows from the definition of $n_2^*$ and 
(\ref{eq:phi-discounted-IHR-DHR}) that $\phi(0) = n_2^*$, 
which implies, by (\ref{eq:Whittle-index-order-IHR-DHR}), that 
\[
\psi(0,n_2^*) = \psi(0,\phi(0)) = w_1(0) > w_2(\phi(0)+1) = w_2(n_2^*+1).
\]
However, by Lemma~\ref{lem:Whittle-index-discounted-IHR-DHR-lemma-1}(iv), 
this is equivalent with 
\[
\psi(0,n_2^*+1) > w_2(n_2^*+1).
\]
But, by Lemma~\ref{lem:Whittle-index-discounted-IHR-DHR-lemma-1}(i), 
this implies that 
\[
\psi(0,n_2^*+1) > w_2(n_2^*+2), 
\]
which is, again by Lemma~\ref{lem:Whittle-index-discounted-IHR-DHR-lemma-1}(iv), 
equivalent with 
\[
\psi(0,n_2^*+2) > w_2(n_2^*+2).
\]
By continuing similarly, we finally end up (in the limit) to the following 
inequality: 
\[
\psi(0,\infty) \ge w_2(\infty), 
\] 
which, in turn, proves that requirement 
(\ref{eq:nu-req-discounted-IHR-DHR-D-case-7b}) follows from 
(\ref{eq:nu-req-discounted-IHR-DHR-D-case-7a}). 

Let then $n \in \{0,1,\ldots\}$. 
Since 
\[
\bar p_1(i|n) \mu_1(n+i) = p_1(i|n) = \bar p_1(i|n) - \bar p_1(i+1|n), 
\]
it follows from 
(\ref{eq:howard-eqs-discounted-IHR-DHR-D-case-7-solution}) that 
\begin{equation}
\begin{split}
& 
V_\beta^\pi(1,n;\nu) = 
(h + \nu) 
\Bigg( 
1 + \frac{\beta V_\beta^\pi(2,0;\nu)}{h + \nu} \; + \\
& \quad 
\sum_{i = 1}^\infty \beta^i \bar p_1(i|n) 
\bigg( 
1 - \frac{(1 - \beta) V_\beta^\pi(2,0;\nu)}{h + \nu} 
\bigg) 
\Bigg).
\end{split}
\label{eq:howard-eqs-discounted-IHR-DHR-D-case-7-solution-cont}
\end{equation}
Now, since $\bar p_1(i|n)$ is an decreasing function of $n$ in the IHR-DHR 
case and we have above required that 
\[
V_\beta^\pi(2,0;\nu) \le \frac{h}{1 - \beta}, 
\]
we see from (\ref{eq:howard-eqs-discounted-IHR-DHR-D-case-7-solution-cont}) 
that $V_\beta^\pi(1,n;\nu)$, as well, is a decreasing function of $n$. 
Thus, condition (\ref{eq:nu-req-discounted-IHR-DHR-D-case-7b}) implies that 
the following condition for optimality of $\pi$ in state~$(1,n)$, 
\[
V_\beta^\pi(1,n;\nu) \le \frac{h}{1 - \beta}, 
\]
is satisfied for any $n \in \{0,1,\ldots\}$. 
This completes the proof of claim~$7^\circ$.

\vskip 3pt
$8^\circ$ 
Finally, we assume that $\nu \le 0$. 
However, the claim that the policy $\pi$ with activity set 
\[
{\mathcal B}^\pi = {\mathcal S} 
\]
is $(\nu,\beta)$-optimal for all 
\[
\nu \in (-\infty,0] 
\]
can be proved similarly as the corresponding claim~$5^\circ$ in the proof 
of Theorem~\ref{thm:Whittle-index-discounted-DHR-DHR} 
(the DHR-DHR-A subcase, see Appendix~\ref{app:DHR-DHR-A-proof}). 
Therefore we may omit the proof here.
\hfill $\Box$

\section{Proof of Theorem~\ref{thm:Whittle-index-discounted-IHR-DHR} 
in the IHR-DHR-E subcase with $n_1^* < \infty$}
\label{app:IHR-DHR-E1-proof}

\paragraph{Proof} 
Assume the IHR-DHR-E subcase defined in (\ref{eq:IHR-DHR-E}). 
As in Lemma~\ref{lem:Whittle-index-discounted-IHR-DHR-E-lemma-4}, 
let $n_1^*$ denote the greatest $\bar n_1$ satisfying (\ref{eq:IHR-DHR-E}). 
In this proof we assume that $n_1^* < \infty$. 
Under the assumption that $n_1^* = \infty$, the proof is slightly different 
and presented in Appendix~\ref{app:IHR-DHR-E2-proof}.

Similarly as in the IHR-DHR-D subcase, 
define $n_2^\circ \in \{0,1,\ldots\}$ as follows: 
\begin{equation}
n_2^\circ = \phi(\infty) = \max \{ n_2 : w_2(n_2) \ge w_1(\infty) \}, 
\label{eq:n-2-circ-IHR-DHR-E1}
\end{equation}
where $w_2(n_2)$ is defined in (\ref{eq:w2-discounted-IHR-DHR}) and 
$w_1(\infty)$ in (\ref{eq:w1-infty-discounted-IHR-DHR}). 
In this proof, we define $n_2^* \in \{0,1,\ldots\}$ as follows: 
\begin{equation}
n_2^* = \phi(n_1^*).
\label{eq:n-2-*-IHR-DHR-E1}
\end{equation}
Furthermore, we denote 
\begin{equation}
{\mathcal S}^* = {\mathcal S} \setminus \{(2,n_2^*+1),(2,n_2^*+2),\ldots\}.
\label{eq:mathcal-S-circ-IHR-DHR-E1}
\end{equation}
Now we define a sequence of states $z_m$, $m \in \{1,2,\ldots\}$, 
recursively as follows: 
\begin{equation}
\begin{split}
& 
z_1 = (1,n_1^*), \\
& 
z_{m+1} = 
\left\{
\begin{array}{ll}
(1,\tilde n_1(m)), & \quad 
\hbox{if $W_\beta(1,\tilde n_1(m)) = 
\min_{x \in {\mathcal S}^* \setminus \{z_1,\ldots,z_m,*\}} W_\beta(x)$}, \\
(2,\tilde n_2(m)), & \quad 
\hbox{otherwise}, \\
\end{array}
\right.
\end{split}
\label{eq:z-def-IHR-DHR-E1}
\end{equation}
where 
\[
\begin{split}
& 
\tilde n_1(m) = \min\{ n_1 : (1,n_1) \in {\mathcal S}^* \setminus \{z_1,\ldots,z_m,*\} \}, \\
& 
\tilde n_2(m) = \max\{ n_2 : (2,n_2) \in {\mathcal S}^* \setminus \{z_1,\ldots,z_m,*\} \}.
\end{split}
\]
Note that, in this IHR-DHR-E subcase with $n_1^* < \infty$, 
the sequence $(z_m)$ covers the states 
\[
\{z_1,z_2,\ldots\} = 
\{(2,n_2^\circ+1),(2,n_2^\circ+2),\ldots,(2,n_2^*)\} \cup \{(1,n_1^*),(1,n_1^*+1),\ldots\}, 
\]
where the former part of the right hand side is omitted if $n_2^\circ = n_2^*$. 
Note also that, by 
Lemmas~\ref{lem:Whittle-index-discounted-IHR-DHR-lemma-1}, 
\ref{lem:Whittle-index-discounted-IHR-DHR-lemma-2}, and 
\ref{lem:Whittle-index-discounted-IHR-DHR-E-lemma-4}, 
we have the following ordering among the states: 
\begin{equation}
\begin{split}
& 
W_\beta(2,0) \ge W_\beta(2,1) \ge \ldots \ge W_\beta(2,n_2^\circ) \ge \\
& \quad 
W_\beta(1,\infty) \ge \ldots \ge W_\beta(z_2) \ge W_\beta(z_1) = W_\beta(1,n_1^*) > \\ 
& \quad \quad 
W_\beta(2,n_2^*+1) \ge W_\beta(2,n_2^*+2) \ge \ldots \ge W_\beta(2,\infty) \ge \\
& \quad \quad \quad 
W_\beta(1,n_1^*-1) \ge \ldots \ge W_\beta(1,1) \ge W_\beta(1,0) 
\ge 0, 
\end{split}
\label{eq:Wx-order-IHR-DHR-E1}
\end{equation}
where we have defined 
\begin{equation}
\begin{split}
& 
W_\beta(1,\infty) = w_1(\infty) = 
h \mu_1(\infty) \, 
\frac{\beta \sum_{i=0}^{n_2^\circ} \beta^i p_2(i)}
{1 + \beta \mu_1(\infty) \sum_{i=0}^{n_2^\circ} \beta^i \bar p_2(i)} \, 
\frac{\beta}{1 - \beta}, \\
& 
W_\beta(2,\infty) = w_2(\infty) = 
h \mu_2(\infty) \, 
\frac{\beta}{1 - \beta}.
\end{split}
\label{eq:Whittle-index-21-infty-discounted-IHR-DHR-E1}
\end{equation}

In addition, for any $m \in \{1,2,\ldots\}$, there are $n_2$ 
and $n_1$ such that $n_2 \in \{n_2^\circ,n_2^\circ+1,\ldots,n_2^*\}$, 
$n_1 \in \{n_1^*,n_1^*+1,\ldots\}$, and 
\[
\begin{split}
& 
\{(2,0),(2,1),\ldots,(2,n_2^\circ)\} \cup \{z_m,z_{m+1},\ldots\} \; = \\
& \quad 
\{(2,0),(2,1),\ldots,(2,n_2)\} \cup \{(1,n_1),(1,n_1+1),\ldots\}.
\end{split}
\]
Now 
\begin{equation}
\phi(n_1) \le n_2, 
\label{eq:phi-result1-IHR-DHR-E1}
\end{equation}
due to Lemma~\ref{lem:Whittle-index-discounted-IHR-DHR-lemma-1}(i) 
since 
\[
W_\beta(2,\phi(n_1)) \ge W_\beta(1,n_1) > W_\beta(2,n_2+1) 
\]
by (\ref{eq:Whittle-index-order-IHR-DHR}) and (\ref{eq:z-def-IHR-DHR-E1}), 
respectively. Moreover, 
\begin{equation}
\phi(n_1-1) \ge n_2, 
\label{eq:phi-result2-IHR-DHR-E1}
\end{equation}
since otherwise, by (\ref{eq:Whittle-index-order-IHR-DHR}) and 
Lemma~\ref{lem:Whittle-index-discounted-IHR-DHR-lemma-1}(i), 
\[
W_\beta(1,n_1-1) > W_\beta(2,\phi(n_1-1)+1) \ge W_\beta(2,n_2), 
\]
which were a contradiction.

The main proof is now given in ten parts ($1^\circ$--$10^\circ$). 
The idea is again to solve the relaxed optimization problem 
(\ref{eq:separable-discounted-costs}) for any $\nu$ by utilizing the 
optimality equations (\ref{eq:opt-eqs-discounted-general}). We partition 
the possible values of $\nu$, which is reflected by the ten parts of 
the main proof. 
However, part $1^\circ$ is exactly the same as in 
the proof of Theorem~\ref{thm:Whittle-index-discounted-DHR-DHR} 
(the DHR-DHR-A subcase, see Appendix~\ref{app:DHR-DHR-A-proof}). 
In addition, parts $2^\circ$ and $3^\circ$ are exactly 
the same as in the subcase IHR-DHR-D. Therefore, we omit those parts 
here and focus on the remaining parts $4^\circ$--$10^\circ$.

\vskip 3pt
$4^\circ$ 
Here we assume that $\nu \ge 0$, and the optimality equations 
(\ref{eq:opt-eqs-discounted-general}) read as given in 
(\ref{eq:opt-eqs-discounted-nu-pos}). 
Let $m \in \{2,3,\ldots\}$, $n_2 \in \{n_2^\circ,n_2^\circ+1,\ldots,n_2^*\}$, 
and $n_1 \in \{1,2,\ldots\}$ such that 
\[
\begin{split}
& 
\{(2,0),(2,1),\ldots,(2,n_2^\circ)\} \cup \{z_m,z_{m+1},\ldots\} \; = \\
& \quad 
\{(2,0),(2,1),\ldots,(2,n_2)\} \cup \{(1,n_1),(1,n_1+1),\ldots\}, 
\end{split}
\]
where $z_m$ is defined in (\ref{eq:z-def-IHR-DHR-E1}). 
We prove that the policy $\pi$ with activity set 
\[
{\mathcal B}^\pi = 
\{(2,0),(2,1),\ldots,(2,n_2^\circ)\} \cup \{z_m,z_{m+1},\ldots\} 
\]
is $(\nu,\beta)$-optimal for all 
\[
\nu \in [W_\beta(z_{m-1}),W_\beta(z_m)].
\]
It remains to prove that policy $\pi$ is optimal for these values of $\nu$ 
in any state $x \in {\mathcal S} \setminus \{*\}$. 
This is done below in two parts ($4.1^\circ$ and $4.2^\circ$).

\vskip 3pt
$4.1^\circ$ 
We start by first deriving the value function $V_\beta^\pi(x;\nu)$ for 
policy $\pi$ from the Howard equations: 
\begin{equation}
\begin{split}
& 
V_\beta^\pi(1,n;\nu) = 
h + \nu + \beta \mu_1(n) V_\beta^\pi(2,0;\nu) \; + \\
& \quad 
\beta (1 - \mu_1(n)) V_\beta^\pi(1,n+1;\nu), 
\quad 
n \in \{n_1,n_1+1,\ldots\}, \\
& 
V_\beta^\pi(2,n;\nu) = 
h + \nu + \beta (1 - \mu_2(n)) V_\beta^\pi(2,n+1;\nu), \\
& \quad 
n \in \{0,1,\ldots,n_2\}, \\
& 
V_\beta^\pi(x;\nu) = 
h + \beta V_\beta^\pi(x;\nu), \\
& \quad 
x \in {\mathcal S} \setminus 
\{(2,0),(2,1),\ldots,(2,n_2),(1,n_1),(1,n_1+1),\ldots,*\}.
\end{split}
\label{eq:howard-eqs-discounted-IHR-DHR-E1-case-4}
\end{equation}
The unique solution of these linear equations is given by 
\begin{equation}
\begin{split}
& 
V_\beta^\pi(2,n;\nu) = 
(h + \nu) 
\left( 
\sum_{i = 0}^{n_2-n} \beta^i \bar p_2(i|n) 
\right) \; + \\
& \quad 
h \, \frac{\beta^{n_2-n+1} \bar p_2(n_2-n+1|n)}{1 - \beta}, 
\quad 
n \in \{0,1,\ldots,n_2\}, \\
& 
V_\beta^\pi(1,n;\nu) = 
(h + \nu) 
\left( 
\sum_{i = 0}^\infty \beta^i \bar p_1(i|n) 
\left( 
1 + \beta \mu_1(n+i) \frac{V_\beta^\pi(2,0;\nu)}{h + \nu} 
\right) 
\right), \\
& \quad 
n \in \{n_1,n_1+1,\ldots\},\\
& 
V_\beta^\pi(x;\nu) = 
\frac{h}{1 - \beta}, 
\quad 
x \in {\mathcal S} \setminus 
\{(2,0),(2,1),\ldots,(2,n_2),(1,n_1),(1,n_1+1),\ldots,*\}.
\end{split}
\label{eq:howard-eqs-discounted-IHR-DHR-E1-case-4-solution}
\end{equation}

By (\ref{eq:howard-eqs-discounted-IHR-DHR-E1-case-4-solution}) and some algebraic 
manipulations, the following condition for optimality of $\pi$ in 
state~$(2,n_2)$, 
\[
V_\beta^\pi(2,n_2;\nu) \le \frac{h}{1 - \beta}, 
\]
can be shown to be equivalent with 
\begin{equation}
\nu \le 
h \mu_2(n_2) \, \frac{\beta}{1 - \beta}, 
\label{eq:nu-req-discounted-IHR-DHR-E1-case-41a}
\end{equation}
where the right hand side equals $W_\beta(2,n_2)$ given in 
(\ref{eq:Whittle-index-2n-discounted-IHR-DHR}).

Let then $n \in \{0,1,\ldots,n_2-1\}$. 
Again by (\ref{eq:howard-eqs-discounted-IHR-DHR-E1-case-4-solution}), 
the following condition for optimality of $\pi$ in state~$(2,n)$, 
\[
V_\beta^\pi(2,n;\nu) \le \frac{h}{1 - \beta}, 
\]
can be shown to be equivalent with 
\begin{equation}
\nu \le 
h \, 
\frac{\sum_{i = 0}^{n_2-n} \beta^i p_2(i|n)}
{\sum_{i = 0}^{n_2-n} \beta^i \bar p_2(i|n)} \, 
\frac{\beta}{1 - \beta}.
\label{eq:nu-req-discounted-IHR-DHR-E1-case-41a-cont}
\end{equation}
Thus, condition (\ref{eq:nu-req-discounted-IHR-DHR-E1-case-41a-cont}) 
follows from (\ref{eq:nu-req-discounted-IHR-DHR-E1-case-41a}) 
by Lemma~\ref{lem:Whittle-index-discounted-IHR-DHR-lemma-1}(vi).

On the other hand, by (\ref{eq:howard-eqs-discounted-IHR-DHR-E1-case-4-solution}) 
and some algebraic manipulations, the following condition for optimality 
of $\pi$ in state~$(1,n_1)$, 
\[
V_\beta^\pi(1,n_1;\nu) \le \frac{h}{1 - \beta}, 
\]
can be shown to be equivalent with 
\begin{equation}
\nu \le 
h \, 
\frac{\frac{1}{1 - \beta} - \sum_{i=0}^{\infty} \beta^i \bar p_1(i|n_1) 
\left( 1 + \mu_1(n_1+i) \Big( 1 - \beta \sum_{j=0}^{n_2} \beta^{j} p_2(j) \Big) \, \frac{\beta}{1 - \beta} \right)}
{\sum_{i=0}^{\infty} \beta^i \bar p_1(i|n_1) \Big( 1 + \mu_1(n_1+i) \beta \sum_{j=0}^{n_2} \beta^{j} \bar p_2(j) \Big)}, 
\label{eq:nu-req-discounted-IHR-DHR-E1-case-41b}
\end{equation}
where the right hand side equals $\psi(n_1,n_2)$ given in 
(\ref{eq:psi-discounted-IHR-DHR}).

Let then $n \in \{n_1,n_1+1,\ldots\}$. 
Since 
\[
\bar p_1(i|n) \mu_1(n+i) = p_1(i|n) = \bar p_1(i|n) - \bar p_1(i+1|n), 
\]
it follows from 
(\ref{eq:howard-eqs-discounted-IHR-DHR-E1-case-4-solution}) that 
\begin{equation}
\begin{split}
& 
V_\beta^\pi(1,n;\nu) = 
(h + \nu) 
\Bigg( 
1 + \frac{\beta V_\beta^\pi(2,0;\nu)}{h + \nu} \; + \\
& \quad 
\sum_{i = 1}^\infty \beta^i \bar p_1(i|n) 
\bigg( 
1 - \frac{(1 - \beta) V_\beta^\pi(2,0;\nu)}{h + \nu} 
\bigg) 
\Bigg).
\end{split}
\label{eq:howard-eqs-discounted-IHR-DHR-E1-case-4-solution-cont}
\end{equation}
Now, since $\bar p_1(i|n)$ is an decreasing function of $n$ in the IHR-DHR 
case and we have above required that 
\[
V_\beta^\pi(2,0;\nu) \le \frac{h}{1 - \beta}, 
\]
we see from (\ref{eq:howard-eqs-discounted-IHR-DHR-E1-case-4-solution-cont}) 
that $V_\beta^\pi(1,n;\nu)$, as well, is a decreasing function of $n$. 
Thus, condition (\ref{eq:nu-req-discounted-IHR-DHR-E1-case-41b}) implies that 
the following condition for optimality of $\pi$ in state~$(1,n)$, 
\[
V_\beta^\pi(1,n;\nu) \le \frac{h}{1 - \beta}, 
\]
is satisfied for any $n \in \{n_1,n_1+1,\ldots\}$.

Finally, by combining (\ref{eq:nu-req-discounted-IHR-DHR-E1-case-41a}) and 
(\ref{eq:nu-req-discounted-IHR-DHR-E1-case-41b}), we get the requirement that 
\begin{equation}
\nu \le 
\min \{ W_\beta(2,n_2), \psi(n_1,n_2) \}.
\label{eq:nu-req-discounted-IHR-DHR-E1-case-41}
\end{equation}
Now, if $W_\beta(2,n_2) < \psi(n_1,n_2)$, then the right hand side of 
(\ref{eq:nu-req-discounted-IHR-DHR-E1-case-41}) equals $W_\beta(2,n_2)$ and this 
requirement (\ref{eq:nu-req-discounted-IHR-DHR-E1-case-41}) is satisfied by our 
assumption that $\nu \le W_\beta(z_m)$, since 
$(2,n_2) \in \{(2,0),(2,1),\ldots,(2,n_2^\circ),z_m,z_{m+1},\ldots\}$ 
and, by (\ref{eq:Wx-order-IHR-DHR-E1}), 
\[
\begin{split}
& 
W_\beta(z_m) = 
\min\{W_\beta(2,0),W_\beta(2,1),\ldots,W_\beta(2,n_2^\circ),W_\beta(z_m),W_\beta(z_{m+1}),\ldots\} \\
& \quad 
\le \; 
W_\beta(2,n_2).
\end{split}
\]
Assume now that 
\[
W_\beta(2,n_2) \ge \psi(n_1,n_2).
\]
Below we show that in this case $\phi(n_1) = n_2$, which implies that 
\[
W_\beta(1,n_1) = \psi(n_1,n_2).
\]
If $\phi(n_1) < n_2$, then we have, by 
(\ref{eq:phi-discounted-IHR-DHR}), 
\[
\psi(n_1,\phi(n_1)) > w_2(\phi(n_1)+1), 
\]
which is, by Lemma~\ref{lem:Whittle-index-discounted-IHR-DHR-lemma-1}(iv), 
equivalent with 
\[
\psi(n_1,\phi(n_1)+1) > w_2(\phi(n_1)+1).
\]
But, by Lemma~\ref{lem:Whittle-index-discounted-IHR-DHR-lemma-1}(i), 
this implies that 
\[
\psi(n_1,\phi(n_1)+1) > w_2(\phi(n_1)+2), 
\]
which is, again by Lemma~\ref{lem:Whittle-index-discounted-IHR-DHR-lemma-1}(iv), 
equivalent with 
\[
\psi(n_1,\phi(n_1)+2) > w_2(\phi(n_1)+2).
\]
By continuing similarly, we finally end up to the following inequality: 
\[
\psi(n_1,n_2) > w_2(n_2), 
\]
which, however, contradicts our assumption above (i.e., 
$W_\beta(2,n_2) = w_2(n_2) \ge \psi(n_1,n_2)$). 
So, by further taking into account (\ref{eq:phi-result1-IHR-DHR-E1}), we have now 
proved that $\phi(n_1) = n_2$ in this case, which imples that the right hand side 
of (\ref{eq:nu-req-discounted-IHR-DHR-E1-case-41}) equals $W_\beta(1,n_1)$. 
In addition, this requirement (\ref{eq:nu-req-discounted-IHR-DHR-E1-case-41}) 
is satisfied by our assumption that $\nu \le W_\beta(z_m)$, since 
$(1,n_1) \in \{z_m,z_{m+1},\ldots\}$ and, by (\ref{eq:Wx-order-IHR-DHR-E1}), 
\[
W_\beta(z_m) = 
\min\{W_\beta(z_m),W_\beta(z_{m+1}),\ldots\} \le 
W_\beta(1,n_1).
\]

\vskip 3pt
$4.2^\circ$ 
By (\ref{eq:howard-eqs-discounted-IHR-DHR-E1-case-4-solution}), the following 
condition for optimality of $\pi$ in state~$(2,n_2+1)$, 
\[
\beta V_\beta^\pi(2,n_2+1;\nu) \le 
\nu + \beta (1 - \mu_2(n_2+1)) V_\beta^\pi(2,n_2+2;\nu), 
\]
is easily shown to be equivalent with 
\begin{equation}
\nu \ge 
h \mu_2(n_2+1) \, \frac{\beta}{1 - \beta}, 
\label{eq:nu-req-discounted-IHR-DHR-E1-case-42a}
\end{equation}
where the right hand side equals $W_\beta(2,n_2+1)$ given in 
(\ref{eq:Whittle-index-2n-discounted-IHR-DHR}).

Let then $n \in \{n_2+2,n_2+3,\ldots\}$. 
Again by (\ref{eq:howard-eqs-discounted-IHR-DHR-E1-case-4-solution}), 
the following condition for optimality of $\pi$ in state~$(2,n)$, 
\[
\beta V_\beta^\pi(2,n;\nu) \le 
\nu + \beta (1 - \mu_2(n)) V_\beta^\pi(2,n+1;\nu), 
\]
is easily shown to be equivalent with 
\[
\nu \ge 
h \mu_2(n) \, \frac{\beta}{1 - \beta}, 
\]
which follows from (\ref{eq:nu-req-discounted-IHR-DHR-E1-case-42a}) since 
$\mu_2(n)$ is decreasing.

On the other hand, by (\ref{eq:howard-eqs-discounted-IHR-DHR-E1-case-4-solution}) 
and some algebraic manipulations, the following condition for optimality of 
$\pi$ in state~$(1,n_1-1)$, 
\[
\beta V_\beta^\pi(1,n_1-1;\nu) \le 
\nu + \beta \mu_1(n_1-1) V_\beta^\pi(2,0;\nu) + 
\beta (1 - \mu_1(n_1-1)) V_\beta^\pi(1,n_1;\nu), 
\]
can be shown to be equivalent with 
\begin{equation}
\nu \ge 
h \, 
\frac{\frac{1}{1 - \beta} - \sum_{i=0}^{\infty} \beta^i \bar p_1(i|n_1-1) 
\left( 1 + \mu_1(n_1-1+i) \Big( 1 - \beta \sum_{j=0}^{n_2} \beta^{j} p_2(j) \Big) \, \frac{\beta}{1 - \beta} \right)}
{\sum_{i=0}^{\infty} \beta^i \bar p_1(i|n_1-1) \Big( 1 + \mu_1(n_1-1+i) \beta \sum_{j=0}^{n_2} \beta^{j} \bar p_2(j) \Big)},
\label{eq:nu-req-discounted-IHR-DHR-E1-case-42b}
\end{equation}
where the right hand side equals $\psi(n_1-1,n_2)$ given in 
(\ref{eq:psi-discounted-IHR-DHR}).

Let then $n \in \{0,1,\ldots,n_1-2\}$. 
By (\ref{eq:howard-eqs-discounted-IHR-DHR-E1-case-4-solution}) and some 
algebraic manipulations, the following condition for optimality of $\pi$ 
in state~$(1,n)$, 
\[
\beta V_\beta^\pi(1,n;\nu) \le 
\nu + \beta \mu_1(n) V_\beta^\pi(2,0;\nu) + 
\beta (1 - \mu_1(n)) V_\beta^\pi(1,n+1;\nu), 
\]
can be shown to be equivalent with 
\[
\begin{split}
& 
\nu \ge 
h \mu_1(n) \, 
\frac{\beta \sum_{i=0}^{n_2} \beta^i p_2(i)}
{1 + \beta \mu_1(n) \sum_{i=0}^{n_2} \beta^i \bar p_2(i)} \, 
\frac{\beta}{1 - \beta}.
\end{split}
\]
By Lemma~\ref{lem:Whittle-index-discounted-IHR-DHR-lemma-1}(vii), we conclude 
that this condition follows from (\ref{eq:nu-req-discounted-IHR-DHR-E1-case-42b}).

Finally, by combining (\ref{eq:nu-req-discounted-IHR-DHR-E1-case-42a}) and 
(\ref{eq:nu-req-discounted-IHR-DHR-E1-case-42b}), we get the requirement that 
\begin{equation}
\nu \ge 
\max \{ W_\beta(2,n_2+1), \psi(n_1-1,n_2) \}.
\label{eq:nu-req-discounted-IHR-DHR-E1-case-42}
\end{equation}
Now if $W_\beta(2,n_2+1) < \psi(n_1-1,n_2)$, then $\phi(n_1-1) = n_2$ by 
(\ref{eq:phi-discounted-IHR-DHR}) and (\ref{eq:phi-result2-IHR-DHR-E1}). 
In addition, by (\ref{eq:Whittle-index-1n-discounted-IHR-DHR}), the right 
hand side of (\ref{eq:nu-req-discounted-IHR-DHR-E1-case-42}) equals, under this condition, 
\[
W_\beta(1,n_1-1) = \psi(n_1-1,n_2) 
\]
and, furthermore, by (\ref{eq:z-def-IHR-DHR-E1}), we have 
\[
z_{m-1} = (1,n_1-1) 
\]
so that requirement (\ref{eq:nu-req-discounted-IHR-DHR-E1-case-42}) is equivalent 
with the requirement 
\[
\nu \ge W_\beta(z_{m-1}).
\]

On the other hand, if $W_\beta(2,n_2+1) \ge \psi(n_1-1,n_2)$, then, 
$\phi(n_1-1) > n_2$ by (\ref{eq:phi-discounted-IHR-DHR}) and 
we have, by (\ref{eq:Whittle-index-order-IHR-DHR}), 
\[
W_\beta(2,n_2+1) \ge W_\beta(2,\phi(n_1-1)) \ge W_\beta(1,n_1-1) 
\]
and, furthermore, by (\ref{eq:z-def-IHR-DHR-E1}), 
\[
z_{m-1} = (2,n_2+1) 
\]
so that requirement (\ref{eq:nu-req-discounted-IHR-DHR-E1-case-42}) is again 
equivalent with the requirement 
\[
\nu \ge W_\beta(z_{m-1}), 
\]
which completes the proof of claim~$4^\circ$.

\vskip 3pt
$5^\circ$ 
We still assume that $\nu \ge 0$ and utilize the optimality equations 
(\ref{eq:opt-eqs-discounted-nu-pos}). 
Now we prove that the policy $\pi$ with activity set 
\[
\begin{split}
& 
{\mathcal B}^\pi = 
\{(2,0),(2,1),\ldots,(2,n_2^\circ)\} \cup \{z_1,z_2,\ldots\} \; = \\
& \quad 
\{(2,0),(2,1),\ldots,(2,n_2^*)\} \cup \{(1,n_1^*),(1,n_1^*+1),\ldots\} 
\end{split}
\]
is $(\nu,\beta)$-optimal for all 
\[
\nu \in [W_\beta(2,n_2^*+1),W_\beta(z_1)].
\]
It remains to prove that policy $\pi$ is optimal for these values of $\nu$ 
in any state $x \in {\mathcal S} \setminus \{*\}$. 
This is done below in two parts ($5.1^\circ$ and $5.2^\circ$).

\vskip 3pt
$5.1^\circ$ 
We start by first deriving the value function $V_\beta^\pi(x;\nu)$ for 
policy $\pi$ from the Howard equations: 
\begin{equation}
\begin{split}
& 
V_\beta^\pi(1,n;\nu) = 
h + \nu + \beta \mu_1(n) V_\beta^\pi(2,0;\nu) \; + \\
& \quad 
\beta (1 - \mu_1(n)) V_\beta^\pi(1,n+1;\nu), 
\quad 
n \in \{n_1^*,n_1^*+1,\ldots\}, \\
& 
V_\beta^\pi(2,n;\nu) = 
h + \nu + \beta (1 - \mu_2(n)) V_\beta^\pi(2,n+1;\nu), \\
& \quad 
n \in \{0,1,\ldots,n_2^*\}, \\
& 
V_\beta^\pi(x;\nu) = 
h + \beta V_\beta^\pi(x;\nu), \\
& \quad 
x \in {\mathcal S} \setminus 
\{(2,0),(2,1),\ldots,(2,n_2^*),(1,n_1^*),(1,n_1^*+1),\ldots,*\}.
\end{split}
\label{eq:howard-eqs-discounted-IHR-DHR-E1-case-5}
\end{equation}
The unique solution of these linear equations is given by 
\begin{equation}
\begin{split}
& 
V_\beta^\pi(2,n;\nu) = 
(h + \nu) 
\left( 
\sum_{i = 0}^{n_2^*-n} \beta^i \bar p_2(i|n) 
\right) \; + \\
& \quad 
h \, \frac{\beta^{n_2^*-n+1} \bar p_2(n_2^*-n+1|n)}{1 - \beta}, 
\quad 
n \in \{0,1,\ldots,n_2^*\}, \\
& 
V_\beta^\pi(1,n;\nu) = 
(h + \nu) 
\left( 
\sum_{i = 0}^\infty \beta^i \bar p_1(i|n) 
\left( 
1 + \beta \mu_1(n+i) \frac{V_\beta^\pi(2,0;\nu)}{h + \nu} 
\right) 
\right), \\
& \quad 
n \in \{n_1^*,n_1^*+1,\ldots\},\\
& 
V_\beta^\pi(x;\nu) = 
\frac{h}{1 - \beta}, 
\quad 
x \in {\mathcal S} \setminus 
\{(2,0),(2,1),\ldots,(2,n_2^*),(1,n_1^*),(1,n_1^*+1),\ldots,*\}.
\end{split}
\label{eq:howard-eqs-discounted-IHR-DHR-E1-case-5-solution}
\end{equation}

By (\ref{eq:howard-eqs-discounted-IHR-DHR-E1-case-5-solution}) and some algebraic 
manipulations, the following condition for optimality of $\pi$ in 
state~$(2,n_2^*)$, 
\[
V_\beta^\pi(2,n_2^*;\nu) \le \frac{h}{1 - \beta}, 
\]
can be shown to be equivalent with 
\begin{equation}
\nu \le 
h \mu_2(n_2^*) \, \frac{\beta}{1 - \beta}, 
\label{eq:nu-req-discounted-IHR-DHR-E1-case-51a}
\end{equation}
where the right hand side equals $W_\beta(2,n_2^*)$ given in 
(\ref{eq:Whittle-index-2n-discounted-IHR-DHR}). 
Note that (\ref{eq:nu-req-discounted-IHR-DHR-E1-case-51a}) follows from the 
requirement that $\nu \le W_\beta(z_1)$ since 
$W_\beta(z_1) \le W_\beta(2,n_2^*)$ by (\ref{eq:Wx-order-IHR-DHR-E1}).

Let then $n \in \{0,1,\ldots,n_2^*-1\}$. 
Again by (\ref{eq:howard-eqs-discounted-IHR-DHR-E1-case-5-solution}), 
the following condition for optimality of $\pi$ in state~$(2,n)$, 
\[
V_\beta^\pi(2,n;\nu) \le \frac{h}{1 - \beta}, 
\]
can be shown to be equivalent with 
\begin{equation}
\nu \le 
h \, 
\frac{\sum_{i = 0}^{n_2^*-n} \beta^i p_2(i|n)}
{\sum_{i = 0}^{n_2^*-n} \beta^i \bar p_2(i|n)} \, 
\frac{\beta}{1 - \beta}.
\label{eq:nu-req-discounted-IHR-DHR-E1-case-51a-cont}
\end{equation}
Thus, condition (\ref{eq:nu-req-discounted-IHR-DHR-E1-case-51a-cont}) 
follows from (\ref{eq:nu-req-discounted-IHR-DHR-E1-case-51a}) 
by Lemma~\ref{lem:Whittle-index-discounted-IHR-DHR-lemma-1}(vi).

On the other hand, by (\ref{eq:howard-eqs-discounted-IHR-DHR-E1-case-5-solution}) 
and some algebraic manipulations, the following condition for optimality 
of $\pi$ in state~$z_1 = (1,n_1^*)$, 
\[
V_\beta^\pi(1,n_1^*;\nu) \le \frac{h}{1 - \beta}, 
\]
can be shown to be equivalent with 
\begin{equation}
\nu \le 
h \, 
\frac{\frac{1}{1 - \beta} - \sum_{i=0}^{\infty} \beta^i \bar p_1(i|n_1^*) 
\left( 1 + \mu_1(n_1^*+i) \Big( 1 - \beta \sum_{j=0}^{n_2^*} \beta^{j} p_2(j) \Big) \, \frac{\beta}{1 - \beta} \right)}
{\sum_{i=0}^{\infty} \beta^i \bar p_1(i|n_1^*) \Big( 1 + \mu_1(n_1^*+i) \beta \sum_{j=0}^{n_2^*} \beta^{j} \bar p_2(j) \Big)}, 
\label{eq:nu-req-discounted-IHR-DHR-E1-case-51b}
\end{equation}
where the right hand side equals $\psi(n_1^*,n_2^*)$ given in 
(\ref{eq:psi-discounted-IHR-DHR}). Moreover, $\phi(n_1^*) = n_2^*$ 
by (\ref{eq:n-2-*-IHR-DHR-E1}), which implies that the right hand side of 
(\ref{eq:nu-req-discounted-IHR-DHR-E1-case-51b}) equals 
$W_\beta(z_1) = W_\beta(1,n_1^*)$ given in 
(\ref{eq:Whittle-index-1n-discounted-IHR-DHR}).

Let then $n \in \{n_1^*,n_1^*+1,\ldots\}$. 
Since 
\[
\bar p_1(i|n) \mu_1(n+i) = p_1(i|n) = \bar p_1(i|n) - \bar p_1(i+1|n), 
\]
it follows from 
(\ref{eq:howard-eqs-discounted-IHR-DHR-E1-case-5-solution}) that 
\begin{equation}
\begin{split}
& 
V_\beta^\pi(1,n;\nu) = 
(h + \nu) 
\Bigg( 
1 + \frac{\beta V_\beta^\pi(2,0;\nu)}{h + \nu} \; + \\
& \quad 
\sum_{i = 1}^\infty \beta^i \bar p_1(i|n) 
\bigg( 
1 - \frac{(1 - \beta) V_\beta^\pi(2,0;\nu)}{h + \nu} 
\bigg) 
\Bigg).
\end{split}
\label{eq:howard-eqs-discounted-IHR-DHR-E1-case-5-solution-cont}
\end{equation}
Now, since $\bar p_1(i|n)$ is an decreasing function of $n$ in the IHR-DHR 
case and we have above required that 
\[
V_\beta^\pi(2,0;\nu) \le \frac{h}{1 - \beta}, 
\]
we see from (\ref{eq:howard-eqs-discounted-IHR-DHR-E1-case-5-solution-cont}) 
that $V_\beta^\pi(1,n;\nu)$, as well, is a decreasing function of $n$. 
Thus, condition (\ref{eq:nu-req-discounted-IHR-DHR-E1-case-51b}) implies that 
the following condition for optimality of $\pi$ in state~$(1,n)$, 
\[
V_\beta^\pi(1,n;\nu) \le \frac{h}{1 - \beta}, 
\]
is satisfied for any $n \in \{n_1^*,n_1^*+1,\ldots\}$.

\vskip 3pt
$5.2^\circ$ 
By (\ref{eq:howard-eqs-discounted-IHR-DHR-E1-case-5-solution}), 
the following condition for optimality of $\pi$ in state~$(2,n_2^*+1)$, 
\[
\beta V_\beta^\pi(2,n_2^*+1;\nu) \le 
\nu + \beta (1 - \mu_2(n_2^*+1)) V_\beta^\pi(2,n_2^*+2;\nu), 
\]
is easily shown to be equivalent with 
\begin{equation}
\nu \ge 
h \mu_2(n_2^*+1) \, \frac{\beta}{1 - \beta}, 
\label{eq:nu-req-discounted-IHR-DHR-E1-case-52a}
\end{equation}
where the right hand side equals $W_\beta(2,n_2^*+1)$ given in 
(\ref{eq:Whittle-index-2n-discounted-IHR-DHR}).

Let then $n \in \{n_2^*+2,n_2^*+3,\ldots\}$. 
Again by (\ref{eq:howard-eqs-discounted-IHR-DHR-E1-case-5-solution}), 
the following condition for optimality of $\pi$ in state~$(2,n)$, 
\[
\beta V_\beta^\pi(2,n;\nu) \le 
\nu + \beta (1 - \mu_2(n)) V_\beta^\pi(2,n+1;\nu), 
\]
is easily shown to be equivalent with 
\[
\nu \ge 
h \mu_2(n) \, \frac{\beta}{1 - \beta}, 
\]
which follows from (\ref{eq:nu-req-discounted-IHR-DHR-E1-case-52a}) since 
$\mu_2(n)$ is decreasing.

In addition, by (\ref{eq:howard-eqs-discounted-IHR-DHR-E1-case-5-solution}) 
and some algebraic manipulations, the following condition for optimality 
of $\pi$ in state~$(1,n_1^*-1)$, 
\[
\beta V_\beta^\pi(1,n_1^*-1;\nu) \le 
\nu + \beta \mu_1(n_1^*-1) V_\beta^\pi(2,0;\nu) + \beta (1 - \mu_1(n_1^*-1)) V_\beta^\pi(1,n_1^*;\nu), 
\]
can be shown to be equivalent with 
\begin{equation}
\nu \ge 
h \, 
\frac{\frac{1}{1 - \beta} - \sum_{i=0}^{\infty} \beta^i \bar p_1(i|n_1^*-1) 
\left( 1 + \mu_1(n_1^*-1+i) \Big( 1 - \beta \sum_{j=0}^{n_2^*} \beta^{j} p_2(j) \Big) \, \frac{\beta}{1 - \beta} \right)}
{\sum_{i=0}^{\infty} \beta^i \bar p_1(i|n_1^*-1) \Big( 1 + \mu_1(n_1^*-1+i) \beta \sum_{j=0}^{n_2^*} \beta^{j} \bar p_2(j) \Big)}, 
\label{eq:nu-req-discounted-IHR-DHR-E1-case-52b}
\end{equation}
where the right hand side equals $\psi(n_1^*-1,n_2^*)$ given in 
(\ref{eq:psi-discounted-IHR-DHR}). However, this inequality 
(\ref{eq:nu-req-discounted-IHR-DHR-E1-case-52b}) follows from 
requirement (\ref{eq:nu-req-discounted-IHR-DHR-E1-case-52a}), since 
\[
\psi(n_1^*-1,n_2^*) \le w_2(n_2^*+1) = W_\beta(2,n_2^*+1), 
\]
which is due to the fact that $\phi(n_1^*-1) = \infty$.

Finally, let $n \in \{0,1,\ldots,n_1^*-2\}$. 
By (\ref{eq:howard-eqs-discounted-IHR-DHR-E1-case-5-solution}) 
and some algebraic manipulations, the following condition for optimality 
of $\pi$ in state~$(1,n)$, 
\[
\beta V_\beta^\pi(1,n;\nu) \le 
\nu + \beta \mu_1(n) V_\beta^\pi(2,0;\nu) + \beta (1 - \mu_1(n)) V_\beta^\pi(1,n+1;\nu), 
\]
can be shown to be equivalent with 
\begin{equation}
\nu \ge 
h \mu_1(n) \, 
\frac{\beta \sum_{i=0}^{n_2^*} \beta^i p_2(i)}
{1 + \beta \mu_1(n) \sum_{i=0}^{n_2^*} \beta^i \bar p_2(i)} \, 
\frac{\beta}{1 - \beta}.
\label{eq:nu-req-discounted-IHR-DHR-E1-case-52b-cont}
\end{equation}
However, this inequality 
(\ref{eq:nu-req-discounted-IHR-DHR-E1-case-52b-cont}) follows again from 
requirement (\ref{eq:nu-req-discounted-IHR-DHR-E1-case-52a}), since 
\[
\psi(n,n_2^*) \ge 
h \mu_1(n) \, 
\frac{\beta \sum_{i=0}^{n_2^*} \beta^i p_2(i)}
{1 + \beta \mu_1(n) \sum_{i=0}^{n_2^*} \beta^i \bar p_2(i)} \, 
\frac{\beta}{1 - \beta} 
\]
by Lemma~\ref{lem:Whittle-index-discounted-IHR-DHR-lemma-1}(vii) and 
\[
\psi(n,n_2^*) \le w_2(n_2^*+1) = W_\beta(2,n_2^*+1), 
\]
which is due to the fact that $\phi(n) = \infty$. 
This completes the proof of claim~$5^\circ$.

\vskip 3pt
$6^\circ$ 
We still assume that $\nu \ge 0$ and utilize the optimality equations 
(\ref{eq:opt-eqs-discounted-nu-pos}). 
Let $n_2 \in \{n_2^*+1,n_2^*+2,\ldots\}$. 
We prove that the policy $\pi$ with activity set 
\[
{\mathcal B}^\pi = 
\{(2,0),(2,1),\ldots,(2,n_2)\} \cup \{(1,n_1^*),(1,n_1^*+1),\ldots\} 
\]
is $(\nu,\beta)$-optimal for all 
\[
\nu \in [W_\beta(2,n_2+1),W_\beta(2,n_2)].
\]
It remains to prove that policy $\pi$ is optimal for these values of $\nu$ 
in any state $x \in {\mathcal S} \setminus \{*\}$. 
This is done below in two parts ($6.1^\circ$ and $6.2^\circ$).

\vskip 3pt
$6.1^\circ$ 
We start by first deriving the value function $V_\beta^\pi(x;\nu)$ for 
policy $\pi$ from the Howard equations: 
\begin{equation}
\begin{split}
& 
V_\beta^\pi(1,n;\nu) = 
h + \nu + \beta \mu_1(n) V_\beta^\pi(2,0;\nu) \; + \\
& \quad 
\beta (1 - \mu_1(n)) V_\beta^\pi(1,n+1;\nu), 
\quad 
n \in \{n_1^*,n_1^*+1,\ldots\}, \\
& 
V_\beta^\pi(2,n;\nu) = 
h + \nu + \beta (1 - \mu_2(n)) V_\beta^\pi(2,n+1;\nu), \\
& \quad 
n \in \{0,1,\ldots,n_2\}, \\
& 
V_\beta^\pi(x;\nu) = 
h + \beta V_\beta^\pi(x;\nu), \\
& \quad 
x \in {\mathcal S} \setminus 
\{(2,0),(2,1),\ldots,(2,n_2),(1,n_1^*),(1,n_1^*+1),\ldots,*\}.
\end{split}
\label{eq:howard-eqs-discounted-IHR-DHR-E1-case-6}
\end{equation}
The unique solution of these linear equations is given by 
\begin{equation}
\begin{split}
& 
V_\beta^\pi(2,n;\nu) = 
(h + \nu) 
\left( 
\sum_{i = 0}^{n_2-n} \beta^i \bar p_2(i|n) 
\right) \; + \\
& \quad 
h \, \frac{\beta^{n_2-n+1} \bar p_2(n_2-n+1|n)}{1 - \beta}, 
\quad 
n \in \{0,1,\ldots,n_2\}, \\
& 
V_\beta^\pi(1,n;\nu) = 
(h + \nu) 
\left( 
\sum_{i = 0}^\infty \beta^i \bar p_1(i|n) 
\left( 
1 + \beta \mu_1(n+i) \frac{V_\beta^\pi(2,0;\nu)}{h + \nu} 
\right) 
\right), \\
& \quad 
n \in \{n_1^*,n_1^*+1,\ldots\},\\
& 
V_\beta^\pi(x;\nu) = 
\frac{h}{1 - \beta}, 
\quad 
x \in {\mathcal S} \setminus 
\{(2,0),(2,1),\ldots,(2,n_2),(1,n_1^*),(1,n_1^*+1),\ldots,*\}.
\end{split}
\label{eq:howard-eqs-discounted-IHR-DHR-E1-case-6-solution}
\end{equation}

By (\ref{eq:howard-eqs-discounted-IHR-DHR-E1-case-6-solution}) and some algebraic 
manipulations, the following condition for optimality of $\pi$ in 
state~$(2,n_2)$, 
\[
V_\beta^\pi(2,n_2;\nu) \le \frac{h}{1 - \beta}, 
\]
can be shown to be equivalent with 
\begin{equation}
\nu \le 
h \mu_2(n_2) \, \frac{\beta}{1 - \beta}, 
\label{eq:nu-req-discounted-IHR-DHR-E1-case-61a}
\end{equation}
where the right hand side equals $W_\beta(2,n_2)$ given in 
(\ref{eq:Whittle-index-2n-discounted-IHR-DHR}). 

Let then $n \in \{0,1,\ldots,n_2-1\}$. 
Again by (\ref{eq:howard-eqs-discounted-IHR-DHR-E1-case-6-solution}), 
the following condition for optimality of $\pi$ in state~$(2,n)$, 
\[
V_\beta^\pi(2,n;\nu) \le \frac{h}{1 - \beta}, 
\]
can be shown to be equivalent with 
\begin{equation}
\nu \le 
h \, 
\frac{\sum_{i = 0}^{n_2-n} \beta^i p_2(i|n)}
{\sum_{i = 0}^{n_2-n} \beta^i \bar p_2(i|n)} \, 
\frac{\beta}{1 - \beta}.
\label{eq:nu-req-discounted-IHR-DHR-E1-case-61a-cont}
\end{equation}
Thus, condition (\ref{eq:nu-req-discounted-IHR-DHR-E1-case-61a-cont}) 
follows from (\ref{eq:nu-req-discounted-IHR-DHR-E1-case-61a}) 
by Lemma~\ref{lem:Whittle-index-discounted-IHR-DHR-lemma-1}(vi).

On the other hand, by (\ref{eq:howard-eqs-discounted-IHR-DHR-E1-case-6-solution}) 
and some algebraic manipulations, the following condition for optimality 
of $\pi$ in state~$(1,n_1^*)$, 
\[
V_\beta^\pi(1,n_1^*;\nu) \le \frac{h}{1 - \beta}, 
\]
can be shown to be equivalent with 
\begin{equation}
\nu \le 
h \, 
\frac{\frac{1}{1 - \beta} - \sum_{i=0}^{\infty} \beta^i \bar p_1(i|n_1^*) 
\left( 1 + \mu_1(n_1^*+i) \Big( 1 - \beta \sum_{j=0}^{n_2} \beta^{j} p_2(j) \Big) \, \frac{\beta}{1 - \beta} \right)}
{\sum_{i=0}^{\infty} \beta^i \bar p_1(i|n_1^*) \Big( 1 + \mu_1(n_1^*+i) \beta \sum_{j=0}^{n_2} \beta^{j} \bar p_2(j) \Big)}, 
\label{eq:nu-req-discounted-IHR-DHR-E1-case-61b}
\end{equation}
where the right hand side equals $\psi(n_1^*,n_2)$ given in 
(\ref{eq:psi-discounted-IHR-DHR}).

Next we prove that $\psi(n_1^*,n_2) > w_2(n_2)$. 
First, $\phi(n_1^*) = n_2^*$ by (\ref{eq:n-2-*-IHR-DHR-E1}), 
which implies, by (\ref{eq:Whittle-index-order-IHR-DHR}), that 
\[
\psi(n_1^*,n_2^*) = \psi(n_1^*,\phi(n_1^*)) = w_1(n_1^*) > w_2(\phi(n_1^*)+1) = w_2(n_2^*+1).
\]
However, by Lemma~\ref{lem:Whittle-index-discounted-IHR-DHR-lemma-1}(iv), 
this is equivalent with 
\[
\psi(n_1^*,n_2^*+1) > w_2(n_2^*+1).
\]
But, by Lemma~\ref{lem:Whittle-index-discounted-IHR-DHR-lemma-1}(i), 
this implies that 
\[
\psi(n_1^*,n_2^*+1) > w_2(n_2^*+2), 
\]
which is, again by Lemma~\ref{lem:Whittle-index-discounted-IHR-DHR-lemma-1}(iv), 
equivalent with 
\[
\psi(n_1^*,n_2^*+2) > w_2(n_2^*+2).
\]
By continuing similarly, we finally end up to the following inequality: 
\[
\psi(n_1^*,n_2) > w_2(n_2), 
\] 
which, in turn, proves that requirement 
(\ref{eq:nu-req-discounted-IHR-DHR-E1-case-61b}) follows from 
(\ref{eq:nu-req-discounted-IHR-DHR-E1-case-61a}). 

Let then $n \in \{n_1^*,n_1^*+1,\ldots\}$. 
Since 
\[
\bar p_1(i|n) \mu_1(n+i) = p_1(i|n) = \bar p_1(i|n) - \bar p_1(i+1|n), 
\]
it follows from 
(\ref{eq:howard-eqs-discounted-IHR-DHR-E1-case-6-solution}) that 
\begin{equation}
\begin{split}
& 
V_\beta^\pi(1,n;\nu) = 
(h + \nu) 
\Bigg( 
1 + \frac{\beta V_\beta^\pi(2,0;\nu)}{h + \nu} \; + \\
& \quad 
\sum_{i = 1}^\infty \beta^i \bar p_1(i|n) 
\bigg( 
1 - \frac{(1 - \beta) V_\beta^\pi(2,0;\nu)}{h + \nu} 
\bigg) 
\Bigg).
\end{split}
\label{eq:howard-eqs-discounted-IHR-DHR-E1-case-6-solution-cont}
\end{equation}
Now, since $\bar p_1(i|n)$ is an decreasing function of $n$ in the IHR-DHR 
case and we have above required that 
\[
V_\beta^\pi(2,0;\nu) \le \frac{h}{1 - \beta}, 
\]
we see from (\ref{eq:howard-eqs-discounted-IHR-DHR-E1-case-6-solution-cont}) 
that $V_\beta^\pi(1,n;\nu)$, as well, is a decreasing function of $n$. 
Thus, condition (\ref{eq:nu-req-discounted-IHR-DHR-E1-case-61b}) implies that 
the following condition for optimality of $\pi$ in state~$(1,n)$, 
\[
V_\beta^\pi(1,n;\nu) \le \frac{h}{1 - \beta}, 
\]
is satisfied for any $n \in \{n_1^*,n_1^*+1,\ldots\}$.

\vskip 3pt
$6.2^\circ$ 
By (\ref{eq:howard-eqs-discounted-IHR-DHR-E1-case-6-solution}), 
the following condition for optimality of $\pi$ in state~$(2,n_2+1)$, 
\[
\beta V_\beta^\pi(2,n_2+1;\nu) \le 
\nu + \beta (1 - \mu_2(n_2+1)) V_\beta^\pi(2,n_2+2;\nu), 
\]
is easily shown to be equivalent with 
\begin{equation}
\nu \ge 
h \mu_2(n_2+1) \, \frac{\beta}{1 - \beta}, 
\label{eq:nu-req-discounted-IHR-DHR-E1-case-62a}
\end{equation}
where the right hand side equals $W_\beta(2,n_2+1)$ given in 
(\ref{eq:Whittle-index-2n-discounted-IHR-DHR}).

Let then $n \in \{n_2+2,n_2+3,\ldots\}$. 
Again by (\ref{eq:howard-eqs-discounted-IHR-DHR-E1-case-6-solution}), 
the following condition for optimality of $\pi$ in state~$(2,n)$, 
\[
\beta V_\beta^\pi(2,n;\nu) \le 
\nu + \beta (1 - \mu_2(n)) V_\beta^\pi(2,n+1;\nu), 
\]
is easily shown to be equivalent with 
\[
\nu \ge 
h \mu_2(n) \, \frac{\beta}{1 - \beta}, 
\]
which follows from (\ref{eq:nu-req-discounted-IHR-DHR-E1-case-62a}) since 
$\mu_2(n)$ is decreasing.

In addition, by (\ref{eq:howard-eqs-discounted-IHR-DHR-E1-case-6-solution}) 
and some algebraic manipulations, the following condition for optimality 
of $\pi$ in state~$(1,n_1^*-1)$, 
\[
\beta V_\beta^\pi(1,n_1^*-1;\nu) \le 
\nu + \beta \mu_1(n_1^*-1) V_\beta^\pi(2,0;\nu) + \beta (1 - \mu_1(n_1^*-1)) V_\beta^\pi(1,n_1^*;\nu), 
\]
can be shown to be equivalent with 
\begin{equation}
\nu \ge 
h \, 
\frac{\frac{1}{1 - \beta} - \sum_{i=0}^{\infty} \beta^i \bar p_1(i|n_1^*-1) 
\left( 1 + \mu_1(n_1^*-1+i) \Big( 1 - \beta \sum_{j=0}^{n_2} \beta^{j} p_2(j) \Big) \, \frac{\beta}{1 - \beta} \right)}
{\sum_{i=0}^{\infty} \beta^i \bar p_1(i|n_1^*-1) \Big( 1 + \mu_1(n_1^*-1+i) \beta \sum_{j=0}^{n_2} \beta^{j} \bar p_2(j) \Big)}, 
\label{eq:nu-req-discounted-IHR-DHR-E1-case-62b}
\end{equation}
where the right hand side equals $\psi(n_1^*-1,n_2)$ given in 
(\ref{eq:psi-discounted-IHR-DHR}). However, this inequality 
(\ref{eq:nu-req-discounted-IHR-DHR-E1-case-62b}) follows from 
requirement (\ref{eq:nu-req-discounted-IHR-DHR-E1-case-62a}), since 
\[
\psi(n_1^*-1,n_2) \le w_2(n_2+1) = W_\beta(2,n_2+1), 
\]
which is due to the fact that $\phi(n_1^*-1) = \infty$.

Finally, let $n \in \{0,1,\ldots,n_1^*-2\}$. 
By (\ref{eq:howard-eqs-discounted-IHR-DHR-E1-case-6-solution}) 
and some algebraic manipulations, the following condition for optimality 
of $\pi$ in state~$(1,n)$, 
\[
\beta V_\beta^\pi(1,n;\nu) \le 
\nu + \beta \mu_1(n) V_\beta^\pi(2,0;\nu) + \beta (1 - \mu_1(n)) V_\beta^\pi(1,n+1;\nu), 
\]
can be shown to be equivalent with 
\begin{equation}
\nu \ge 
h \mu_1(n) \, 
\frac{\beta \sum_{i=0}^{n_2} \beta^i p_2(i)}
{1 + \beta \mu_1(n) \sum_{i=0}^{n_2} \beta^i \bar p_2(i)} \, 
\frac{\beta}{1 - \beta}.
\label{eq:nu-req-discounted-IHR-DHR-E1-case-62b-cont}
\end{equation}
However, this inequality 
(\ref{eq:nu-req-discounted-IHR-DHR-E1-case-62b-cont}) follows again from 
requirement (\ref{eq:nu-req-discounted-IHR-DHR-E1-case-62a}), since 
\[
\psi(n,n_2) \ge 
h \mu_1(n) \, 
\frac{\beta \sum_{i=0}^{n_2} \beta^i p_2(i)}
{1 + \beta \mu_1(n) \sum_{i=0}^{n_2} \beta^i \bar p_2(i)} \, 
\frac{\beta}{1 - \beta} 
\]
by Lemma~\ref{lem:Whittle-index-discounted-IHR-DHR-lemma-1}(vii) and 
\[
\psi(n,n_2) \le w_2(n_2+1) = W_\beta(2,n_2+1), 
\]
which is due to the fact that $\phi(n) = \infty$. 
This completes the proof of claim~$6^\circ$.

\vskip 3pt
$7^\circ$ 
We still assume that $\nu \ge 0$ and utilize the optimality equations 
(\ref{eq:opt-eqs-discounted-nu-pos}). 
Now we prove that the policy $\pi$ with activity set 
\[
{\mathcal B}^\pi = 
\{(2,0),(2,1),\ldots\} \cup \{(1,n_1^*),(1,n_1^*+1),\ldots\} 
\]
is $(\nu,\beta)$-optimal for all 
\[
\nu \in [W_\beta(1,n_1^*-1),W_\beta(2,\infty)], 
\]
where $W_\beta(2,\infty)$ is defined in (\ref{eq:Whittle-index-21-infty-discounted-IHR-DHR-E1}). 
It remains to prove that policy $\pi$ is optimal for these values of $\nu$ 
in any state $x \in {\mathcal S} \setminus \{*\}$. 
This is done below in two parts ($7.1^\circ$ and $7.2^\circ$).

\vskip 3pt
$7.1^\circ$ 
We start by first deriving the value function $V_\beta^\pi(x;\nu)$ for 
policy $\pi$ from the Howard equations: 
\begin{equation}
\begin{split}
& 
V_\beta^\pi(1,n;\nu) = 
h + \nu + \beta \mu_1(n) V_\beta^\pi(2,0;\nu) \; + \\
& \quad 
\beta (1 - \mu_1(n)) V_\beta^\pi(1,n+1;\nu), 
\quad 
n \in \{n_1^*,n_1^*+1,\ldots\}, \\
& 
V_\beta^\pi(2,n;\nu) = 
h + \nu + \beta (1 - \mu_2(n)) V_\beta^\pi(2,n+1;\nu), \\
& \quad 
n \in \{0,1,\ldots\}, \\
& 
V_\beta^\pi(x;\nu) = 
h + \beta V_\beta^\pi(x;\nu), \\
& \quad 
x \in {\mathcal S} \setminus 
\{(2,0),(2,1),\ldots,(1,n_1^*),(1,n_1^*+1),\ldots,*\}.
\end{split}
\label{eq:howard-eqs-discounted-IHR-DHR-E1-case-7}
\end{equation}
The unique solution of these linear equations is given by 
\begin{equation}
\begin{split}
& 
V_\beta^\pi(2,n;\nu) = 
(h + \nu) 
\left( 
\sum_{i = 0}^{\infty} \beta^i \bar p_2(i|n) 
\right), 
\quad 
n \in \{0,1,\ldots\}, \\
& 
V_\beta^\pi(1,n;\nu) = 
(h + \nu) 
\left( 
\sum_{i = 0}^\infty \beta^i \bar p_1(i|n) 
\left( 
1 + \beta \mu_1(n+i) \frac{V_\beta^\pi(2,0;\nu)}{h + \nu} 
\right) 
\right), \\
& \quad 
n \in \{n_1^*,n_1^*+1,\ldots\},\\
& 
V_\beta^\pi(x;\nu) = 
\frac{h}{1 - \beta}, 
\quad 
x \in {\mathcal S} \setminus 
\{(2,0),(2,1),\ldots,(1,n_1^*),(1,n_1^*+1),\ldots,*\}.
\end{split}
\label{eq:howard-eqs-discounted-IHR-DHR-E1-case-7-solution}
\end{equation}

Let then $n \in \{0,1,\ldots\}$. 
Since $\bar p_2(i|n)$ is an increasing function of $n$ in the IHR-DHR 
case, we see from (\ref{eq:howard-eqs-discounted-IHR-DHR-E1-case-7-solution}) 
that $V_\beta^\pi(2,n;\nu)$, as well, is an increasing function of $n$ 
and approaches 
\[
\begin{split}
& 
\lim_{n \to \infty} V_\beta^\pi(2,n;\nu) = 
(h + \nu) 
\left( 
\sum_{i = 0}^\infty \beta^i (1 - \mu_2(\infty))^i 
\right) \\
& \quad = 
\frac{h + \nu}{1 - \beta (1 - \mu_2(\infty))}.
\end{split}
\]
Thus, the following condition for optimality of $\pi$ in state~$(2,n)$, 
\[
V_\beta^\pi(2,n;\nu) \le \frac{h}{1 - \beta}, 
\]
is satisfied for any $n$ if and only if 
\[
\lim_{n \to \infty} V_\beta^\pi(2,n;\nu) \le \frac{h}{1 - \beta}, 
\]
which is clearly equivalent with condition 
\begin{equation}
\nu \le 
h \mu_2(\infty) \, 
\frac{\beta}{1 - \beta}. 
\label{eq:nu-req-discounted-IHR-DHR-E1-case-71a}
\end{equation}
Note that the right hand side equals $w_2(\infty) = W_\beta(2,\infty)$ given in 
(\ref{eq:Whittle-index-21-infty-discounted-IHR-DHR-E1}).

On the other hand, by (\ref{eq:howard-eqs-discounted-IHR-DHR-E1-case-7-solution}) 
and some algebraic manipulations, the following condition for optimality 
of $\pi$ in state~$(1,n_1^*)$, 
\[
V_\beta^\pi(1,n_1^*;\nu) \le \frac{h}{1 - \beta}, 
\]
can be shown to be equivalent with 
\begin{equation}
\nu \le 
h \, 
\frac{\frac{1}{1 - \beta} - \sum_{i=0}^{\infty} \beta^i \bar p_1(i|n_1^*) 
\left( 1 + \mu_1(n_1^*+i) \Big( 1 - \beta \sum_{j=0}^{\infty} \beta^{j} p_2(j) \Big) \, \frac{\beta}{1 - \beta} \right)}
{\sum_{i=0}^{\infty} \beta^i \bar p_1(i|n_1^*) \Big( 1 + \mu_1(n_1^*+i) \beta \sum_{j=0}^{\infty} \beta^{j} \bar p_2(j) \Big)}, 
\label{eq:nu-req-discounted-IHR-DHR-E1-case-71b}
\end{equation}
where the right hand side equals $\psi(n_1^*,\infty)$ given in 
(\ref{eq:psi-n1-infty-discounted-IHR-DHR}).

Next we prove that $\psi(n_1^*,\infty) \ge w_2(\infty)$. 
First, $\phi(n_1^*) = n_2^*$ by (\ref{eq:n-2-*-IHR-DHR-E1}), 
which implies, by (\ref{eq:Whittle-index-order-IHR-DHR}), that 
\[
\psi(n_1^*,n_2^*) = \psi(n_1^*,\phi(n_1^*)) = w_1(n_1^*) > w_2(\phi(n_1^*)+1) = w_2(n_2^*+1).
\]
However, by Lemma~\ref{lem:Whittle-index-discounted-IHR-DHR-lemma-1}(iv), 
this is equivalent with 
\[
\psi(n_1^*,n_2^*+1) > w_2(n_2^*+1).
\]
But, by Lemma~\ref{lem:Whittle-index-discounted-IHR-DHR-lemma-1}(i), 
this implies that 
\[
\psi(n_1^*,n_2^*+1) > w_2(n_2^*+2), 
\]
which is, again by Lemma~\ref{lem:Whittle-index-discounted-IHR-DHR-lemma-1}(iv), 
equivalent with 
\[
\psi(n_1^*,n_2^*+2) > w_2(n_2^*+2).
\]
By continuing similarly, we finally end up (in the limit) to the following 
inequality: 
\[
\psi(n_1^*,\infty) \ge w_2(\infty), 
\] 
which, in turn, proves that requirement 
(\ref{eq:nu-req-discounted-IHR-DHR-E1-case-71b}) follows from 
(\ref{eq:nu-req-discounted-IHR-DHR-E1-case-71a}). 

Let then $n \in \{n_1^*,n_1^*+1,\ldots\}$. 
Since 
\[
\bar p_1(i|n) \mu_1(n+i) = p_1(i|n) = \bar p_1(i|n) - \bar p_1(i+1|n), 
\]
it follows from 
(\ref{eq:howard-eqs-discounted-IHR-DHR-E1-case-7-solution}) that 
\begin{equation}
\begin{split}
& 
V_\beta^\pi(1,n;\nu) = 
(h + \nu) 
\Bigg( 
1 + \frac{\beta V_\beta^\pi(2,0;\nu)}{h + \nu} \; + \\
& \quad 
\sum_{i = 1}^\infty \beta^i \bar p_1(i|n) 
\bigg( 
1 - \frac{(1 - \beta) V_\beta^\pi(2,0;\nu)}{h + \nu} 
\bigg) 
\Bigg).
\end{split}
\label{eq:howard-eqs-discounted-IHR-DHR-E1-case-7-solution-cont}
\end{equation}
Now, since $\bar p_1(i|n)$ is an decreasing function of $n$ in the IHR-DHR 
case and we have above required that 
\[
V_\beta^\pi(2,0;\nu) \le \frac{h}{1 - \beta}, 
\]
we see from (\ref{eq:howard-eqs-discounted-IHR-DHR-E1-case-7-solution-cont}) 
that $V_\beta^\pi(1,n;\nu)$, as well, is a decreasing function of $n$. 
Thus, condition (\ref{eq:nu-req-discounted-IHR-DHR-E1-case-71b}) implies that 
the following condition for optimality of $\pi$ in state~$(1,n)$, 
\[
V_\beta^\pi(1,n;\nu) \le \frac{h}{1 - \beta}, 
\]
is satisfied for any $n \in \{n_1^*,n_1^*+1,\ldots\}$.

\vskip 3pt
$7.2^\circ$ 
By (\ref{eq:howard-eqs-discounted-IHR-DHR-E1-case-7-solution}) 
and some algebraic manipulations, the following condition for optimality 
of $\pi$ in state~$(1,n_1^*-1)$, 
\[
\beta V_\beta^\pi(1,n_1^*-1;\nu) \le 
\nu + \beta \mu_1(n_1^*-1) V_\beta^\pi(2,0;\nu) + \beta (1 - \mu_1(n_1^*-1)) V_\beta^\pi(1,n_1^*;\nu), 
\]
can be shown to be equivalent with 
\begin{equation}
\nu \ge 
h \, 
\frac{\frac{1}{1 - \beta} - \sum_{i=0}^{\infty} \beta^i \bar p_1(i|n_1^*-1) 
\left( 1 + \mu_1(n_1^*-1+i) \Big( 1 - \beta \sum_{j=0}^{\infty} \beta^{j} p_2(j) \Big) \, \frac{\beta}{1 - \beta} \right)}
{\sum_{i=0}^{\infty} \beta^i \bar p_1(i|n_1^*-1) \Big( 1 + \mu_1(n_1^*-1+i) \beta \sum_{j=0}^{\infty} \beta^{j} \bar p_2(j) \Big)}, 
\label{eq:nu-req-discounted-IHR-DHR-E1-case-72}
\end{equation}
where the right hand side equals $W_\beta(1,n_1^*-1)$ given in 
(\ref{eq:Whittle-index-1n-discounted-IHR-DHR}), which is due to the fact 
that $\phi(n_1^*-1) = \infty$.

Let then $n \in \{0,1,\ldots,n_1^*-2\}$. 
By (\ref{eq:howard-eqs-discounted-IHR-DHR-E1-case-7-solution}) 
and some algebraic manipulations, the following condition for optimality 
of $\pi$ in state~$(1,n)$, 
\[
\beta V_\beta^\pi(1,n;\nu) \le 
\nu + \beta \mu_1(n) V_\beta^\pi(2,0;\nu) + \beta (1 - \mu_1(n)) V_\beta^\pi(1,n+1;\nu), 
\]
can be shown to be equivalent with 
\begin{equation}
\nu \ge 
h \mu_1(n) \, 
\frac{\beta \sum_{i=0}^{\infty} \beta^i p_2(i)}
{1 + \beta \mu_1(n) \sum_{i=0}^{\infty} \beta^i \bar p_2(i)} \, 
\frac{\beta}{1 - \beta}.
\label{eq:nu-req-discounted-IHR-DHR-E1-case-72-cont}
\end{equation}
However, this inequality 
(\ref{eq:nu-req-discounted-IHR-DHR-E1-case-72-cont}) follows from 
(\ref{eq:nu-req-discounted-IHR-DHR-E1-case-72}), since 
\[
W_\beta(1,n_1^*-1) \ge 
W_\beta(1,n) = 
\psi(n,\infty) \ge 
h \mu_1(n) \, 
\frac{\beta \sum_{i=0}^{\infty} \beta^i p_2(i)}
{1 + \beta \mu_1(n) \sum_{i=0}^{\infty} \beta^i \bar p_2(i)} \, 
\frac{\beta}{1 - \beta} 
\]
by (\ref{eq:Wx-order-IHR-DHR-E1}), the fact that $\phi(n) = \infty$, 
and Lemma~\ref{lem:Whittle-index-discounted-IHR-DHR-lemma-1}(vii). 
This completes the proof of claim~$7^\circ$.

\vskip 3pt
$8^\circ$ 
We still assume that $\nu \ge 0$ and utilize the optimality equations 
(\ref{eq:opt-eqs-discounted-nu-pos}). 
Let $n_1 \in \{1,2,\ldots,n_1^*-1\}$. 
We prove that the policy $\pi$ with activity set 
\[
{\mathcal B}^\pi = 
\{(2,0),(2,1),\ldots\} \cup \{(1,n_1),(1,n_1+1),\ldots\} 
\]
is $(\nu,\beta)$-optimal for all 
\[
\nu \in [W_\beta(1,n_1-1),W_\beta(1,n_1)].
\]
It remains to prove that policy $\pi$ is optimal for these values of $\nu$ 
in any state $x \in {\mathcal S} \setminus \{*\}$. 
This is done below in two parts ($8.1^\circ$ and $8.2^\circ$).

\vskip 3pt
$8.1^\circ$ 
We start by first deriving the value function $V_\beta^\pi(x;\nu)$ for 
policy $\pi$ from the Howard equations: 
\begin{equation}
\begin{split}
& 
V_\beta^\pi(1,n;\nu) = 
h + \nu + \beta \mu_1(n) V_\beta^\pi(2,0;\nu) \; + \\
& \quad 
\beta (1 - \mu_1(n)) V_\beta^\pi(1,n+1;\nu), 
\quad 
n \in \{n_1,n_1+1,\ldots\}, \\
& 
V_\beta^\pi(2,n;\nu) = 
h + \nu + \beta (1 - \mu_2(n)) V_\beta^\pi(2,n+1;\nu), \\
& \quad 
n \in \{0,1,\ldots\}, \\
& 
V_\beta^\pi(x;\nu) = 
h + \beta V_\beta^\pi(x;\nu), \\
& \quad 
x \in {\mathcal S} \setminus 
\{(2,0),(2,1),\ldots,(1,n_1),(1,n_1+1),\ldots,*\}.
\end{split}
\label{eq:howard-eqs-discounted-IHR-DHR-E1-case-8}
\end{equation}
The unique solution of these linear equations is given by 
\begin{equation}
\begin{split}
& 
V_\beta^\pi(2,n;\nu) = 
(h + \nu) 
\left( 
\sum_{i = 0}^{\infty} \beta^i \bar p_2(i|n) 
\right), 
\quad 
n \in \{0,1,\ldots\}, \\
& 
V_\beta^\pi(1,n;\nu) = 
(h + \nu) 
\left( 
\sum_{i = 0}^\infty \beta^i \bar p_1(i|n) 
\left( 
1 + \beta \mu_1(n+i) \frac{V_\beta^\pi(2,0;\nu)}{h + \nu} 
\right) 
\right), \\
& \quad 
n \in \{n_1,n_1+1,\ldots\},\\
& 
V_\beta^\pi(x;\nu) = 
\frac{h}{1 - \beta}, 
\quad 
x \in {\mathcal S} \setminus 
\{(2,0),(2,1),\ldots,(1,n_1),(1,n_1+1),\ldots,*\}.
\end{split}
\label{eq:howard-eqs-discounted-IHR-DHR-E1-case-8-solution}
\end{equation}

Let then $n \in \{0,1,\ldots\}$. 
Since $\bar p_2(i|n)$ is an increasing function of $n$ in the IHR-DHR 
case, we see from (\ref{eq:howard-eqs-discounted-IHR-DHR-E1-case-8-solution}) 
that $V_\beta^\pi(2,n;\nu)$, as well, is an increasing function of $n$ 
and approaches 
\[
\begin{split}
& 
\lim_{n \to \infty} V_\beta^\pi(2,n;\nu) = 
(h + \nu) 
\left( 
\sum_{i = 0}^\infty \beta^i (1 - \mu_2(\infty))^i 
\right) \\
& \quad = 
\frac{h + \nu}{1 - \beta (1 - \mu_2(\infty))}.
\end{split}
\]
Thus, the following condition for optimality of $\pi$ in state~$(2,n)$, 
\[
V_\beta^\pi(2,n;\nu) \le \frac{h}{1 - \beta}, 
\]
is satisfied for any $n$ if and only if 
\[
\lim_{n \to \infty} V_\beta^\pi(2,n;\nu) \le \frac{h}{1 - \beta}, 
\]
which is clearly equivalent with condition 
\begin{equation}
\nu \le 
h \mu_2(\infty) \, 
\frac{\beta}{1 - \beta}. 
\label{eq:nu-req-discounted-IHR-DHR-E1-case-81a}
\end{equation}
Note that the right hand side equals $w_2(\infty) = W_\beta(2,\infty)$ given in 
(\ref{eq:Whittle-index-21-infty-discounted-IHR-DHR-E1}). 
Note also that (\ref{eq:nu-req-discounted-IHR-DHR-E1-case-81a}) follows 
from the requirement that $\nu \le W_\beta(1,n_1)$ since 
$W_\beta(1,n_1) \le W_\beta(2,\infty)$ by (\ref{eq:Wx-order-IHR-DHR-E1}).

On the other hand, by (\ref{eq:howard-eqs-discounted-IHR-DHR-E1-case-8-solution}) 
and some algebraic manipulations, the following condition for optimality 
of $\pi$ in state~$(1,n_1)$, 
\[
V_\beta^\pi(1,n_1;\nu) \le \frac{h}{1 - \beta}, 
\]
can be shown to be equivalent with 
\begin{equation}
\nu \le 
h \, 
\frac{\frac{1}{1 - \beta} - \sum_{i=0}^{\infty} \beta^i \bar p_1(i|n_1) 
\left( 1 + \mu_1(n_1+i) \Big( 1 - \beta \sum_{j=0}^{\infty} \beta^{j} p_2(j) \Big) \, \frac{\beta}{1 - \beta} \right)}
{\sum_{i=0}^{\infty} \beta^i \bar p_1(i|n_1) \Big( 1 + \mu_1(n_1+i) \beta \sum_{j=0}^{\infty} \beta^{j} \bar p_2(j) \Big)}, 
\label{eq:nu-req-discounted-IHR-DHR-E1-case-81b}
\end{equation}
where the right hand side equals $W_\beta(1,n_1)$ given in 
(\ref{eq:Whittle-index-1n-discounted-IHR-DHR}), which is due to the fact 
that $\phi(n_1) = \infty$. 

Let then $n \in \{n_1,n_1+1,\ldots\}$. 
Since 
\[
\bar p_1(i|n) \mu_1(n+i) = p_1(i|n) = \bar p_1(i|n) - \bar p_1(i+1|n), 
\]
it follows from 
(\ref{eq:howard-eqs-discounted-IHR-DHR-E1-case-8-solution}) that 
\begin{equation}
\begin{split}
& 
V_\beta^\pi(1,n;\nu) = 
(h + \nu) 
\Bigg( 
1 + \frac{\beta V_\beta^\pi(2,0;\nu)}{h + \nu} \; + \\
& \quad 
\sum_{i = 1}^\infty \beta^i \bar p_1(i|n) 
\bigg( 
1 - \frac{(1 - \beta) V_\beta^\pi(2,0;\nu)}{h + \nu} 
\bigg) 
\Bigg).
\end{split}
\label{eq:howard-eqs-discounted-IHR-DHR-E1-case-8-solution-cont}
\end{equation}
Now, since $\bar p_1(i|n)$ is an decreasing function of $n$ in the IHR-DHR 
case and we have above required that 
\[
V_\beta^\pi(2,0;\nu) \le \frac{h}{1 - \beta}, 
\]
we see from (\ref{eq:howard-eqs-discounted-IHR-DHR-E1-case-8-solution-cont}) 
that $V_\beta^\pi(1,n;\nu)$, as well, is a decreasing function of $n$. 
Thus, condition (\ref{eq:nu-req-discounted-IHR-DHR-E1-case-81b}) implies that 
the following condition for optimality of $\pi$ in state~$(1,n)$, 
\[
V_\beta^\pi(1,n;\nu) \le \frac{h}{1 - \beta}, 
\]
is satisfied for any $n \in \{n_1,n_1+1,\ldots\}$.

\vskip 3pt
$8.2^\circ$ 
By (\ref{eq:howard-eqs-discounted-IHR-DHR-E1-case-8-solution}) 
and some algebraic manipulations, the following condition for optimality 
of $\pi$ in state~$(1,n_1-1)$, 
\[
\beta V_\beta^\pi(1,n_1-1;\nu) \le 
\nu + \beta \mu_1(n_1-1) V_\beta^\pi(2,0;\nu) + \beta (1 - \mu_1(n_1-1)) V_\beta^\pi(1,n_1;\nu), 
\]
can be shown to be equivalent with 
\begin{equation}
\nu \ge 
h \, 
\frac{\frac{1}{1 - \beta} - \sum_{i=0}^{\infty} \beta^i \bar p_1(i|n_1-1) 
\left( 1 + \mu_1(n_1-1+i) \Big( 1 - \beta \sum_{j=0}^{\infty} \beta^{j} p_2(j) \Big) \, \frac{\beta}{1 - \beta} \right)}
{\sum_{i=0}^{\infty} \beta^i \bar p_1(i|n_1-1) \Big( 1 + \mu_1(n_1-1+i) \beta \sum_{j=0}^{\infty} \beta^{j} \bar p_2(j) \Big)}, 
\label{eq:nu-req-discounted-IHR-DHR-E1-case-82}
\end{equation}
where the right hand side equals $W_\beta(1,n_1-1)$ given in 
(\ref{eq:Whittle-index-1n-discounted-IHR-DHR}), which is due to the fact 
that $\phi(n_1-1) = \infty$.

Let then $n \in \{0,1,\ldots,n_1-2\}$. 
By (\ref{eq:howard-eqs-discounted-IHR-DHR-E1-case-8-solution}) 
and some algebraic manipulations, the following condition for optimality 
of $\pi$ in state~$(1,n)$, 
\[
\beta V_\beta^\pi(1,n;\nu) \le 
\nu + \beta \mu_1(n) V_\beta^\pi(2,0;\nu) + \beta (1 - \mu_1(n)) V_\beta^\pi(1,n+1;\nu), 
\]
can be shown to be equivalent with 
\begin{equation}
\nu \ge 
h \mu_1(n) \, 
\frac{\beta \sum_{i=0}^{\infty} \beta^i p_2(i)}
{1 + \beta \mu_1(n) \sum_{i=0}^{\infty} \beta^i \bar p_2(i)} \, 
\frac{\beta}{1 - \beta}.
\label{eq:nu-req-discounted-IHR-DHR-E1-case-82-cont}
\end{equation}
However, this inequality 
(\ref{eq:nu-req-discounted-IHR-DHR-E1-case-82-cont}) follows from 
(\ref{eq:nu-req-discounted-IHR-DHR-E1-case-82}), since 
\[
W_\beta(1,n_1-1) \ge 
W_\beta(1,n) = 
\psi(n,\infty) \ge 
h \mu_1(n) \, 
\frac{\beta \sum_{i=0}^{\infty} \beta^i p_2(i)}
{1 + \beta \mu_1(n) \sum_{i=0}^{\infty} \beta^i \bar p_2(i)} \, 
\frac{\beta}{1 - \beta} 
\]
by (\ref{eq:Wx-order-IHR-DHR-E1}), the fact that $\phi(n) = \infty$, 
and Lemma~\ref{lem:Whittle-index-discounted-IHR-DHR-lemma-1}(vii). 
This completes the proof of claim~$8^\circ$.

\vskip 3pt
$9^\circ$ 
We still assume that $\nu \ge 0$ and utilize the optimality equations 
(\ref{eq:opt-eqs-discounted-nu-pos}). 
Now we prove that the policy $\pi$ with activity set 
\[
{\mathcal B}^\pi = {\mathcal S} \setminus \{*\} 
\]
is $(\nu,\beta)$-optimal for all 
\[
\nu \in [0,W_\beta(1,0)].
\]
It remains to prove that policy $\pi$ is optimal for these values of $\nu$ 
in any state $x \in {\mathcal S} \setminus \{*\}$.

We start by first deriving the value function $V_\beta^\pi(x;\nu)$ for 
policy $\pi$ from the Howard equations: 
\begin{equation}
\begin{split}
& 
V_\beta^\pi(1,n;\nu) = 
h + \nu + \beta \mu_1(n) V_\beta^\pi(2,0;\nu) + 
\beta (1 - \mu_1(n)) V_\beta^\pi(1,n+1;\nu), \\
& 
V_\beta^\pi(2,n;\nu) = 
h + \nu + \beta (1 - \mu_2(n)) V_\beta^\pi(2,n+1;\nu).
\end{split}
\label{eq:howard-eqs-discounted-IHR-DHR-E1-case-9}
\end{equation}
The unique solution of these linear equations is given by 
\begin{equation}
\begin{split}
& 
V_\beta^\pi(2,n;\nu) = 
(h + \nu) 
\left( 
\sum_{i = 0}^{\infty} \beta^i \bar p_2(i|n) 
\right), \\
& 
V_\beta^\pi(1,n;\nu) = 
(h + \nu) 
\left( 
\sum_{i = 0}^{\infty} \beta^i \bar p_1(i|n) 
\left( 
1 + \beta \mu_1(n+i) \frac{V_\beta^\pi(2,0;\nu)}{h + \nu} 
\right) 
\right).
\end{split}
\label{eq:howard-eqs-discounted-IHR-DHR-E1-case-9-solution}
\end{equation}

Let then $n \in \{0,1,\ldots\}$. 
Since $\bar p_2(i|n)$ is an increasing function of $n$ in the IHR-DHR 
case, we see from (\ref{eq:howard-eqs-discounted-IHR-DHR-E1-case-9-solution}) 
that $V_\beta^\pi(2,n;\nu)$, as well, is an increasing function of $n$ 
and approaches 
\[
\begin{split}
& 
\lim_{n \to \infty} V_\beta^\pi(2,n;\nu) = 
(h + \nu) 
\left( 
\sum_{i = 0}^\infty \beta^i (1 - \mu_2(\infty))^i 
\right) \\
& \quad = 
\frac{h + \nu}{1 - \beta (1 - \mu_2(\infty))}.
\end{split}
\]
Thus, the following condition for optimality of $\pi$ in state~$(2,n)$, 
\[
V_\beta^\pi(2,n;\nu) \le \frac{h}{1 - \beta}, 
\]
is satisfied for any $n$ if and only if 
\[
\lim_{n \to \infty} V_\beta^\pi(2,n;\nu) \le \frac{h}{1 - \beta}, 
\]
which is clearly equivalent with condition 
\begin{equation}
\nu \le 
h \mu_2(\infty) \, 
\frac{\beta}{1 - \beta}. 
\label{eq:nu-req-discounted-IHR-DHR-E1-case-9a}
\end{equation}
Note that the right hand side equals $w_2(\infty) = W_\beta(2,\infty)$ given in 
(\ref{eq:Whittle-index-21-infty-discounted-IHR-DHR-E1}). 
Note also that (\ref{eq:nu-req-discounted-IHR-DHR-E1-case-9a}) follows 
from the requirement that $\nu \le W_\beta(1,0)$ since 
$W_\beta(1,0) \le W_\beta(2,\infty)$ by (\ref{eq:Wx-order-IHR-DHR-E1}).

On the other hand, by (\ref{eq:howard-eqs-discounted-IHR-DHR-E1-case-9-solution}) 
and some algebraic manipulations, the following condition for optimality 
of $\pi$ in state~$(1,0)$, 
\[
V_\beta^\pi(1,0;\nu) \le \frac{h}{1 - \beta}, 
\]
can be shown to be equivalent with 
\begin{equation}
\nu \le 
h \, 
\frac{\frac{1}{1 - \beta} - \sum_{i=0}^{\infty} \beta^i \bar p_1(i) 
\left( 1 + \mu_1(i) \Big( 1 - \beta \sum_{j=0}^{\infty} \beta^{j} p_2(j) \Big) \, \frac{\beta}{1 - \beta} \right)}
{\sum_{i=0}^{\infty} \beta^i \bar p_1(i) \Big( 1 + \mu_1(i) \beta \sum_{j=0}^{\infty} \beta^{j} \bar p_2(j) \Big)}, 
\label{eq:nu-req-discounted-IHR-DHR-E1-case-9b}
\end{equation}
where the right hand side equals $W_\beta(1,0)$ given in 
(\ref{eq:Whittle-index-1n-discounted-IHR-DHR}), which is due to the fact 
that $\phi(0) = \infty$.

Let then $n \in \{0,1,\ldots\}$. 
Since 
\[
\bar p_1(i|n) \mu_1(n+i) = p_1(i|n) = \bar p_1(i|n) - \bar p_1(i+1|n), 
\]
it follows from 
(\ref{eq:howard-eqs-discounted-IHR-DHR-E1-case-9-solution}) that 
\begin{equation}
\begin{split}
& 
V_\beta^\pi(1,n;\nu) = 
(h + \nu) 
\Bigg( 
1 + \frac{\beta V_\beta^\pi(2,0;\nu)}{h + \nu} \; + \\
& \quad 
\sum_{i = 1}^\infty \beta^i \bar p_1(i|n) 
\bigg( 
1 - \frac{(1 - \beta) V_\beta^\pi(2,0;\nu)}{h + \nu} 
\bigg) 
\Bigg).
\end{split}
\label{eq:howard-eqs-discounted-IHR-DHR-E1-case-9-solution-cont}
\end{equation}
Now, since $\bar p_1(i|n)$ is an decreasing function of $n$ in the IHR-DHR 
case and we have above required that 
\[
V_\beta^\pi(2,0;\nu) \le \frac{h}{1 - \beta}, 
\]
we see from (\ref{eq:howard-eqs-discounted-IHR-DHR-E1-case-9-solution-cont}) 
that $V_\beta^\pi(1,n;\nu)$, as well, is a decreasing function of $n$. 
Thus, condition (\ref{eq:nu-req-discounted-IHR-DHR-E1-case-9b}) implies that 
the following condition for optimality of $\pi$ in state~$(1,n)$, 
\[
V_\beta^\pi(1,n;\nu) \le \frac{h}{1 - \beta}, 
\]
is satisfied for any $n \in \{0,1,\ldots\}$. 
This completes the proof of claim~$9^\circ$.

\vskip 3pt
$10^\circ$ 
Finally, we assume that $\nu \le 0$. 
However, the claim that the policy $\pi$ with activity set 
\[
{\mathcal B}^\pi = {\mathcal S} 
\]
is $(\nu,\beta)$-optimal for all 
\[
\nu \in (-\infty,0] 
\]
can be proved similarly as the corresponding claim~$5^\circ$ in the proof 
of Theorem~\ref{thm:Whittle-index-discounted-DHR-DHR} 
(the DHR-DHR-A subcase, see Appendix~\ref{app:DHR-DHR-A-proof}). 
Therefore we may omit the proof here.
\hfill $\Box$

\section{Proof of Theorem~\ref{thm:Whittle-index-discounted-IHR-DHR} 
in the IHR-DHR-E subcase with $n_1^* = \infty$}
\label{app:IHR-DHR-E2-proof}

\paragraph{Proof} 
Assume the IHR-DHR-E subcase defined in (\ref{eq:IHR-DHR-E}). 
As in Lemma~\ref{lem:Whittle-index-discounted-IHR-DHR-E-lemma-4}, 
let $n_1^*$ denote the greatest $\bar n_1$ satisfying (\ref{eq:IHR-DHR-E}). 
In this proof we assume that $n_1^* = \infty$. 
Under the assumption that $n_1^* < \infty$, the proof is slightly different 
and presented in Appendix~\ref{app:IHR-DHR-E1-proof}.

In this IHR-DHR-E subcase with $n_1^* = \infty$, we clearly have 
$\phi(n_1) = \infty$ for any $n_1 \in \{0,1,\ldots\}$. In addition, 
by Lemmas~\ref{lem:Whittle-index-discounted-IHR-DHR-lemma-1}, 
\ref{lem:Whittle-index-discounted-IHR-DHR-lemma-2}, and 
\ref{lem:Whittle-index-discounted-IHR-DHR-E-lemma-4}, 
we have the following ordering among the states: 
\begin{equation}
\begin{split}
& 
W_\beta(2,0) \ge W_\beta(2,1) \ge \ldots \ge W_\beta(2,\infty) \ge \\
& \quad \quad \quad 
W_\beta(1,\infty) \ge \ldots \ge W_\beta(1,1) \ge W_\beta(1,0) 
\ge 0, 
\end{split}
\label{eq:Wx-order-IHR-DHR-E2}
\end{equation}
where we have defined 
\begin{equation}
\begin{split}
& 
W_\beta(2,\infty) = w_2(\infty) = 
h \mu_2(\infty) \, 
\frac{\beta}{1 - \beta}, \\
& 
W_\beta(1,\infty) = w_1(\infty) = 
h \mu_1(\infty) \, 
\frac{\beta \sum_{i=0}^{\infty} \beta^i p_2(i)}
{1 + \beta \mu_1(\infty) \sum_{i=0}^{\infty} \beta^i \bar p_2(i)} \, 
\frac{\beta}{1 - \beta}.
\end{split}
\label{eq:Whittle-index-21-infty-discounted-IHR-DHR-E2}
\end{equation}

The main proof is now given in six parts ($1^\circ$--$6^\circ$). 
The idea is again to solve the relaxed optimization problem 
(\ref{eq:separable-discounted-costs}) for any $\nu$ by utilizing the 
optimality equations (\ref{eq:opt-eqs-discounted-general}). We partition 
the possible values of $\nu$, which is reflected by the six parts of 
the main proof. 
However, part $1^\circ$ is exactly the same as in 
the proof of Theorem~\ref{thm:Whittle-index-discounted-DHR-DHR} 
(the DHR-DHR-A subcase, see Appendix~\ref{app:DHR-DHR-A-proof}). 
Therefore, we omit it here and focus on the remaining parts 
$2^\circ$--$6^\circ$.

\vskip 3pt
$2^\circ$ 
Here we assume that $\nu \ge 0$, and the optimality equations 
(\ref{eq:opt-eqs-discounted-general}) read as given in 
(\ref{eq:opt-eqs-discounted-nu-pos}). 
Let $n_2 \in \{0,1,\ldots\}$. 
We prove that the policy $\pi$ with activity set 
\[
{\mathcal B}^\pi = \{(2,0),\ldots,(2,n_2)\} 
\]
is $(\nu,\beta)$-optimal for all 
\[
\nu \in [W_\beta(2,n_2+1),W_\beta(2,n_2)].
\]
It remains to prove that policy $\pi$ is optimal for these values of $\nu$ 
in any state $x \in {\mathcal S} \setminus \{*\}$. 
This is done below in two parts ($2.1^\circ$ and $2.2^\circ$).

\vskip 3pt
$2.1^\circ$ 
We start by first deriving the value function $V_\beta^\pi(x;\nu)$ for 
policy $\pi$ from the Howard equations: 
\begin{equation}
\begin{split}
& 
V_\beta^\pi(2,n;\nu) = 
h + \nu + \beta (1 - \mu_2(n)) V_\beta^\pi(2,n+1;\nu), 
\quad 
n \in \{0,1,\ldots,n_2\}, \\
& 
V_\beta^\pi(x;\nu) = 
h + \beta V_\beta^\pi(x;\nu), 
\quad x \in {\mathcal S} \setminus \{(2,0),\ldots,(2,n_2),*\}.
\end{split}
\label{eq:howard-eqs-discounted-IHR-DHR-E2-case-2}
\end{equation}
The unique solution of these linear equations is given by 
\begin{equation}
\begin{split}
& 
V_\beta^\pi(2,n;\nu) = 
(h + \nu) 
\left( 
\sum_{i = 0}^{n_2-n} \beta^i \bar p_2(i|n) 
\right) \; + \\
& \quad 
h \, \frac{\beta^{n_2-n+1} \bar p_2(n_2-n+1|n)}{1 - \beta}, 
\quad 
n \in \{0,1,\ldots,n_2\}, \\
& 
V_\beta^\pi(x;\nu) = 
\frac{h}{1 - \beta}, 
\quad x \in {\mathcal S} \setminus \{(2,0),\ldots,(2,n_2),*\}.
\end{split}
\label{eq:howard-eqs-discounted-IHR-DHR-E2-case-2-solution}
\end{equation}

By (\ref{eq:howard-eqs-discounted-IHR-DHR-E2-case-2-solution}) and some algebraic 
manipulations, the following condition for optimality of $\pi$ in 
state~$(2,n_2)$, 
\[
V_\beta^\pi(2,n_2;\nu) \le \frac{h}{1 - \beta}, 
\]
can be shown to be equivalent with 
\begin{equation}
\nu \le 
h \mu_2(n_2) \, \frac{\beta}{1 - \beta}, 
\label{eq:nu-req-discounted-IHR-DHR-E2-case-21}
\end{equation}
where the right hand side equals $W_\beta(2,n_2)$ given in 
(\ref{eq:Whittle-index-2n-discounted-IHR-DHR}).

Let then $n \in \{0,1,\ldots,n_2-1\}$. 
Again by (\ref{eq:howard-eqs-discounted-IHR-DHR-E2-case-2-solution}), 
the following condition for optimality of $\pi$ in state~$(2,n)$, 
\[
V_\beta^\pi(2,n;\nu) \le \frac{h}{1 - \beta}, 
\]
can be shown to be equivalent with 
\begin{equation}
\nu \le 
h \, 
\frac{\sum_{i = 0}^{n_2-n} \beta^i p_2(i|n)}
{\sum_{i = 0}^{n_2-n} \beta^i \bar p_2(i|n)} \, 
\frac{\beta}{1 - \beta}.
\label{eq:nu-req-discounted-IHR-DHR-E2-case-21-cont}
\end{equation}
Thus, condition (\ref{eq:nu-req-discounted-IHR-DHR-E2-case-21-cont}) 
follows from (\ref{eq:nu-req-discounted-IHR-DHR-E2-case-21}) 
by Lemma~\ref{lem:Whittle-index-discounted-IHR-DHR-lemma-1}(vi).

\vskip 3pt
$2.2^\circ$ 
By (\ref{eq:howard-eqs-discounted-IHR-DHR-E2-case-2-solution}), the following condition 
for optimality of $\pi$ in state~$(2,n_2+1)$, 
\[
\beta V_\beta^\pi(2,n_2+1;\nu) \le 
\nu + \beta (1 - \mu_2(n_2+1)) V_\beta^\pi(2,n_2+2;\nu), 
\]
is easily shown to be equivalent with 
\begin{equation}
\nu \ge 
h \mu_2(n_2+1) \, \frac{\beta}{1 - \beta}, 
\label{eq:nu-req-discounted-IHR-DHR-E2-case-22a}
\end{equation}
where the right hand side equals $W_\beta(2,n_2+1)$ given in 
(\ref{eq:Whittle-index-2n-discounted-IHR-DHR}).

Let then $n \in \{n_2+2,n_2+3,\ldots\}$. 
Again by (\ref{eq:howard-eqs-discounted-IHR-DHR-E2-case-2-solution}), 
the following condition for optimality of $\pi$ in state~$(2,n)$, 
\[
\beta V_\beta^\pi(2,n;\nu) \le 
\nu + \beta (1 - \mu_2(n)) V_\beta^\pi(2,n+1;\nu), 
\]
is easily shown to be equivalent with 
\[
\nu \ge 
h \mu_2(n) \, \frac{\beta}{1 - \beta}, 
\]
which follows from (\ref{eq:nu-req-discounted-IHR-DHR-E2-case-22a}) since 
$\mu_2(n)$ is decreasing.

Finally, let $n \in \{0,1,\ldots\}$. 
By (\ref{eq:howard-eqs-discounted-IHR-DHR-E2-case-2-solution}) 
and some algebraic manipulations, the following condition for optimality 
of $\pi$ in state~$(1,n)$, 
\[
\beta V_\beta^\pi(1,n;\nu) \le 
\nu + \beta \mu_1(n) V_\beta^\pi(2,0;\nu) + \beta (1 - \mu_1(n)) V_\beta^\pi(1,n+1;\nu), 
\]
can be shown to be equivalent with 
\begin{equation}
\nu \ge 
h \mu_1(n) \, 
\frac{\beta \sum_{i=0}^{n_2} \beta^i p_2(i)}
{1 + \beta \mu_1(n) \sum_{i=0}^{n_2} \beta^i \bar p_2(i)} \, 
\frac{\beta}{1 - \beta}.
\label{eq:nu-req-discounted-IHR-DHR-E2-case-22b}
\end{equation}
However, this inequality 
(\ref{eq:nu-req-discounted-IHR-DHR-E2-case-22b}) follows from 
requirement (\ref{eq:nu-req-discounted-IHR-DHR-E2-case-22a}), since 
\[
\psi(n,n_2) \ge 
h \mu_1(n) \, 
\frac{\beta \sum_{i=0}^{n_2} \beta^i p_2(i)}
{1 + \beta \mu_1(n) \sum_{i=0}^{n_2} \beta^i \bar p_2(i)} \, 
\frac{\beta}{1 - \beta} 
\]
by Lemma~\ref{lem:Whittle-index-discounted-IHR-DHR-lemma-1}(vii) and 
\[
\psi(n,n_2) \le w_2(n_2+1) = W_\beta(2,n_2+1), 
\]
which is due to the fact that $\phi(n) = \infty$. 
This completes the proof of claim~$2^\circ$.

\vskip 3pt
$3^\circ$ 
We still assume that $\nu \ge 0$ and utilize the optimality equations 
(\ref{eq:opt-eqs-discounted-nu-pos}). 
Now we prove that the policy $\pi$ with activity set 
\[
{\mathcal B}^\pi = 
\{(2,0),(2,1),\ldots\} 
\]
is $(\nu,\beta)$-optimal for all 
\[
\nu \in [W_\beta(1,\infty),W_\beta(2,\infty)], 
\]
where $W_\beta(1,\infty)$ and $W_\beta(2,\infty)$ are defined in 
(\ref{eq:Whittle-index-21-infty-discounted-IHR-DHR-E2}). 
It remains to prove that policy $\pi$ is optimal for these values of $\nu$ 
in any state $x \in {\mathcal S} \setminus \{*\}$. 
This is done below in two parts ($3.1^\circ$ and $3.2^\circ$).

\vskip 3pt
$3.1^\circ$ 
We start by first deriving the value function $V_\beta^\pi(x;\nu)$ for 
policy $\pi$ from the Howard equations: 
\begin{equation}
\begin{split}
& 
V_\beta^\pi(2,n;\nu) = 
h + \nu + \beta (1 - \mu_2(n)) V_\beta^\pi(2,n+1;\nu), 
\quad 
n \in \{0,1,\ldots\}, \\
& 
V_\beta^\pi(x;\nu) = 
h + \beta V_\beta^\pi(x;\nu), 
\quad 
x \in {\mathcal S} \setminus 
\{(2,0),(2,1),\ldots,*\}.
\end{split}
\label{eq:howard-eqs-discounted-IHR-DHR-E2-case-3}
\end{equation}
The unique solution of these linear equations is given by 
\begin{equation}
\begin{split}
& 
V_\beta^\pi(2,n;\nu) = 
(h + \nu) 
\left( 
\sum_{i = 0}^{\infty} \beta^i \bar p_2(i|n) 
\right), 
\quad 
n \in \{0,1,\ldots\}, \\
& 
V_\beta^\pi(x;\nu) = 
\frac{h}{1 - \beta}, 
\quad 
x \in {\mathcal S} \setminus 
\{(2,0),(2,1),\ldots,*\}.
\end{split}
\label{eq:howard-eqs-discounted-IHR-DHR-E2-case-3-solution}
\end{equation}

Let then $n \in \{0,1,\ldots\}$. 
Since $\bar p_2(i|n)$ is an increasing function of $n$ in the IHR-DHR 
case, we see from (\ref{eq:howard-eqs-discounted-IHR-DHR-E2-case-3-solution}) 
that $V_\beta^\pi(2,n;\nu)$, as well, is an increasing function of $n$ 
and approaches 
\[
\begin{split}
& 
\lim_{n \to \infty} V_\beta^\pi(2,n;\nu) = 
(h + \nu) 
\left( 
\sum_{i = 0}^\infty \beta^i (1 - \mu_2(\infty))^i 
\right) \\
& \quad = 
\frac{h + \nu}{1 - \beta (1 - \mu_2(\infty))}.
\end{split}
\]
Thus, the following condition for optimality of $\pi$ in state~$(2,n)$, 
\[
V_\beta^\pi(2,n;\nu) \le \frac{h}{1 - \beta}, 
\]
is satisfied for any $n$ if and only if 
\[
\lim_{n \to \infty} V_\beta^\pi(2,n;\nu) \le \frac{h}{1 - \beta}, 
\]
which is clearly equivalent with condition 
\begin{equation}
\nu \le 
h \mu_2(\infty) \, 
\frac{\beta}{1 - \beta}. 
\label{eq:nu-req-discounted-IHR-DHR-E2-case-31}
\end{equation}
Note that the right hand side equals $W_\beta(2,\infty)$ given in 
(\ref{eq:Whittle-index-21-infty-discounted-IHR-DHR-E2}).

$3.2^\circ$ 
Let $n \in \{0,1,\ldots\}$. 
By (\ref{eq:howard-eqs-discounted-IHR-DHR-E2-case-3-solution}) 
and some algebraic manipulations, the following condition for optimality 
of $\pi$ in state~$(1,n)$, 
\[
\beta V_\beta^\pi(1,n;\nu) \le 
\nu + \beta \mu_1(n) V_\beta^\pi(2,0;\nu) + \beta (1 - \mu_1(n)) V_\beta^\pi(1,n+1;\nu), 
\]
can be shown to be equivalent with 
\[
\nu \ge 
h \mu_1(n) \, 
\frac{\beta \sum_{i=0}^{\infty} \beta^i p_2(i)}
{1 + \beta \mu_1(n) \sum_{i=0}^{\infty} \beta^i \bar p_2(i)} \, 
\frac{\beta}{1 - \beta}.
\]
Since $\mu_1(n)$ is increasing, policy $\pi$ is, thus, optimal in any state $(1,n)$ 
if and only if 
\begin{equation}
\nu \ge 
h \mu_1(\infty) \, 
\frac{\beta \sum_{i=0}^{\infty} \beta^i p_2(i)}
{1 + \beta \mu_1(\infty) \sum_{i=0}^{\infty} \beta^i \bar p_2(i)} \, 
\frac{\beta}{1 - \beta}.
\label{eq:nu-req-discounted-IHR-DHR-E2-case-32}
\end{equation}
Note that the right hand side equals $W_\beta(1,\infty)$ given in 
(\ref{eq:Whittle-index-21-infty-discounted-IHR-DHR-E2}).
This completes the proof of claim~$3^\circ$.

\vskip 3pt
$4^\circ$ 
We still assume that $\nu \ge 0$ and utilize the optimality equations 
(\ref{eq:opt-eqs-discounted-nu-pos}). 
Let $n_1 \in \{1,2,\ldots\}$. 
We prove that the policy $\pi$ with activity set 
\[
{\mathcal B}^\pi = 
\{(2,0),(2,1),\ldots\} \cup \{(1,n_1),(1,n_1+1),\ldots\} 
\]
is $(\nu,\beta)$-optimal for all 
\[
\nu \in [W_\beta(1,n_1-1),W_\beta(1,n_1)].
\]
It remains to prove that policy $\pi$ is optimal for these values of $\nu$ 
in any state $x \in {\mathcal S} \setminus \{*\}$. 
This is done below in two parts ($4.1^\circ$ and $4.2^\circ$).

\vskip 3pt
$4.1^\circ$ 
We start by first deriving the value function $V_\beta^\pi(x;\nu)$ for 
policy $\pi$ from the Howard equations: 
\begin{equation}
\begin{split}
& 
V_\beta^\pi(1,n;\nu) = 
h + \nu + \beta \mu_1(n) V_\beta^\pi(2,0;\nu) \; + \\
& \quad 
\beta (1 - \mu_1(n)) V_\beta^\pi(1,n+1;\nu), 
\quad 
n \in \{n_1,n_1+1,\ldots\}, \\
& 
V_\beta^\pi(2,n;\nu) = 
h + \nu + \beta (1 - \mu_2(n)) V_\beta^\pi(2,n+1;\nu), \\
& \quad 
n \in \{0,1,\ldots\}, \\
& 
V_\beta^\pi(x;\nu) = 
h + \beta V_\beta^\pi(x;\nu), \\
& \quad 
x \in {\mathcal S} \setminus 
\{(2,0),(2,1),\ldots,(1,n_1),(1,n_1+1),\ldots,*\}.
\end{split}
\label{eq:howard-eqs-discounted-IHR-DHR-E2-case-4}
\end{equation}
The unique solution of these linear equations is given by 
\begin{equation}
\begin{split}
& 
V_\beta^\pi(2,n;\nu) = 
(h + \nu) 
\left( 
\sum_{i = 0}^{\infty} \beta^i \bar p_2(i|n) 
\right), 
\quad 
n \in \{0,1,\ldots\}, \\
& 
V_\beta^\pi(1,n;\nu) = 
(h + \nu) 
\left( 
\sum_{i = 0}^\infty \beta^i \bar p_1(i|n) 
\left( 
1 + \beta \mu_1(n+i) \frac{V_\beta^\pi(2,0;\nu)}{h + \nu} 
\right) 
\right), \\
& \quad 
n \in \{n_1,n_1+1,\ldots\},\\
& 
V_\beta^\pi(x;\nu) = 
\frac{h}{1 - \beta}, 
\quad 
x \in {\mathcal S} \setminus 
\{(2,0),(2,1),\ldots,(1,n_1),(1,n_1+1),\ldots,*\}.
\end{split}
\label{eq:howard-eqs-discounted-IHR-DHR-E2-case-4-solution}
\end{equation}

Let then $n \in \{0,1,\ldots\}$. 
Since $\bar p_2(i|n)$ is an increasing function of $n$ in the IHR-DHR 
case, we see from (\ref{eq:howard-eqs-discounted-IHR-DHR-E2-case-4-solution}) 
that $V_\beta^\pi(2,n;\nu)$, as well, is an increasing function of $n$ 
and approaches 
\[
\begin{split}
& 
\lim_{n \to \infty} V_\beta^\pi(2,n;\nu) = 
(h + \nu) 
\left( 
\sum_{i = 0}^\infty \beta^i (1 - \mu_2(\infty))^i 
\right) \\
& \quad = 
\frac{h + \nu}{1 - \beta (1 - \mu_2(\infty))}.
\end{split}
\]
Thus, the following condition for optimality of $\pi$ in state~$(2,n)$, 
\[
V_\beta^\pi(2,n;\nu) \le \frac{h}{1 - \beta}, 
\]
is satisfied for any $n$ if and only if 
\[
\lim_{n \to \infty} V_\beta^\pi(2,n;\nu) \le \frac{h}{1 - \beta}, 
\]
which is clearly equivalent with condition 
\begin{equation}
\nu \le 
h \mu_2(\infty) \, 
\frac{\beta}{1 - \beta}. 
\label{eq:nu-req-discounted-IHR-DHR-E2-case-41a}
\end{equation}
Note that the right hand side 
equals $ W_\beta(2,\infty)$ given in 
(\ref{eq:Whittle-index-21-infty-discounted-IHR-DHR-E2}). 
Note also that (\ref{eq:nu-req-discounted-IHR-DHR-E2-case-41a}) follows 
from the requirement that $\nu \le W_\beta(1,n_1)$ since 
$W_\beta(1,n_1) \le W_\beta(2,\infty)$ by (\ref{eq:Wx-order-IHR-DHR-E2}).

On the other hand, by (\ref{eq:howard-eqs-discounted-IHR-DHR-E2-case-4-solution}) 
and some algebraic manipulations, the following condition for optimality 
of $\pi$ in state~$(1,n_1)$, 
\[
V_\beta^\pi(1,n_1;\nu) \le \frac{h}{1 - \beta}, 
\]
can be shown to be equivalent with 
\begin{equation}
\nu \le 
h \, 
\frac{\frac{1}{1 - \beta} - \sum_{i=0}^{\infty} \beta^i \bar p_1(i|n_1) 
\left( 1 + \mu_1(n_1+i) \Big( 1 - \beta \sum_{j=0}^{\infty} \beta^{j} p_2(j) \Big) \, \frac{\beta}{1 - \beta} \right)}
{\sum_{i=0}^{\infty} \beta^i \bar p_1(i|n_1) \Big( 1 + \mu_1(n_1+i) \beta \sum_{j=0}^{\infty} \beta^{j} \bar p_2(j) \Big)}, 
\label{eq:nu-req-discounted-IHR-DHR-E2-case-41b}
\end{equation}
where the right hand side equals $W_\beta(1,n_1)$ given in 
(\ref{eq:Whittle-index-1n-discounted-IHR-DHR}), which is due to the fact 
that $\phi(n_1) = \infty$. 

Let then $n \in \{n_1,n_1+1,\ldots\}$. 
Since 
\[
\bar p_1(i|n) \mu_1(n+i) = p_1(i|n) = \bar p_1(i|n) - \bar p_1(i+1|n), 
\]
it follows from 
(\ref{eq:howard-eqs-discounted-IHR-DHR-E2-case-4-solution}) that 
\begin{equation}
\begin{split}
& 
V_\beta^\pi(1,n;\nu) = 
(h + \nu) 
\Bigg( 
1 + \frac{\beta V_\beta^\pi(2,0;\nu)}{h + \nu} \; + \\
& \quad 
\sum_{i = 1}^\infty \beta^i \bar p_1(i|n) 
\bigg( 
1 - \frac{(1 - \beta) V_\beta^\pi(2,0;\nu)}{h + \nu} 
\bigg) 
\Bigg).
\end{split}
\label{eq:howard-eqs-discounted-IHR-DHR-E2-case-4-solution-cont}
\end{equation}
Now, since $\bar p_1(i|n)$ is an decreasing function of $n$ in the IHR-DHR 
case and we have above required that 
\[
V_\beta^\pi(2,0;\nu) \le \frac{h}{1 - \beta}, 
\]
we see from (\ref{eq:howard-eqs-discounted-IHR-DHR-E2-case-4-solution-cont}) 
that $V_\beta^\pi(1,n;\nu)$, as well, is a decreasing function of $n$. 
Thus, condition (\ref{eq:nu-req-discounted-IHR-DHR-E2-case-41b}) implies that 
the following condition for optimality of $\pi$ in state~$(1,n)$, 
\[
V_\beta^\pi(1,n;\nu) \le \frac{h}{1 - \beta}, 
\]
is satisfied for any $n \in \{n_1,n_1+1,\ldots\}$.

\vskip 3pt
$4.2^\circ$ 
By (\ref{eq:howard-eqs-discounted-IHR-DHR-E2-case-4-solution}) 
and some algebraic manipulations, the following condition for optimality 
of $\pi$ in state~$(1,n_1-1)$, 
\[
\beta V_\beta^\pi(1,n_1-1;\nu) \le 
\nu + \beta \mu_1(n_1-1) V_\beta^\pi(2,0;\nu) + \beta (1 - \mu_1(n_1-1)) V_\beta^\pi(1,n_1;\nu), 
\]
can be shown to be equivalent with 
\begin{equation}
\nu \ge 
h \, 
\frac{\frac{1}{1 - \beta} - \sum_{i=0}^{\infty} \beta^i \bar p_1(i|n_1-1) 
\left( 1 + \mu_1(n_1-1+i) \Big( 1 - \beta \sum_{j=0}^{\infty} \beta^{j} p_2(j) \Big) \, \frac{\beta}{1 - \beta} \right)}
{\sum_{i=0}^{\infty} \beta^i \bar p_1(i|n_1-1) \Big( 1 + \mu_1(n_1-1+i) \beta \sum_{j=0}^{\infty} \beta^{j} \bar p_2(j) \Big)}, 
\label{eq:nu-req-discounted-IHR-DHR-E2-case-42}
\end{equation}
where the right hand side equals $W_\beta(1,n_1-1)$ given in 
(\ref{eq:Whittle-index-1n-discounted-IHR-DHR}), which is due to the fact 
that $\phi(n_1-1) = \infty$.

Let then $n \in \{0,1,\ldots,n_1-2\}$. 
By (\ref{eq:howard-eqs-discounted-IHR-DHR-E2-case-4-solution}) 
and some algebraic manipulations, the following condition for optimality 
of $\pi$ in state~$(1,n)$, 
\[
\beta V_\beta^\pi(1,n;\nu) \le 
\nu + \beta \mu_1(n) V_\beta^\pi(2,0;\nu) + \beta (1 - \mu_1(n)) V_\beta^\pi(1,n+1;\nu), 
\]
can be shown to be equivalent with 
\begin{equation}
\nu \ge 
h \mu_1(n) \, 
\frac{\beta \sum_{i=0}^{\infty} \beta^i p_2(i)}
{1 + \beta \mu_1(n) \sum_{i=0}^{\infty} \beta^i \bar p_2(i)} \, 
\frac{\beta}{1 - \beta}.
\label{eq:nu-req-discounted-IHR-DHR-E2-case-42-cont}
\end{equation}
However, this inequality 
(\ref{eq:nu-req-discounted-IHR-DHR-E2-case-42-cont}) follows from 
(\ref{eq:nu-req-discounted-IHR-DHR-E2-case-42}), since 
\[
W_\beta(1,n_1-1) \ge 
W_\beta(1,n) = 
\psi(n,\infty) \ge 
h \mu_1(n) \, 
\frac{\beta \sum_{i=0}^{\infty} \beta^i p_2(i)}
{1 + \beta \mu_1(n) \sum_{i=0}^{\infty} \beta^i \bar p_2(i)} \, 
\frac{\beta}{1 - \beta} 
\]
by (\ref{eq:Wx-order-IHR-DHR-E1}), the fact that $\phi(n) = \infty$, 
and Lemma~\ref{lem:Whittle-index-discounted-IHR-DHR-lemma-1}(vii). 
This completes the proof of claim~$4^\circ$.

\vskip 3pt
$5^\circ$ 
We still assume that $\nu \ge 0$. 
However, the claim that the policy $\pi$ with activity set 
\[
{\mathcal B}^\pi = {\mathcal S} \setminus \{*\} 
\]
is $(\nu,\beta)$-optimal for all 
\[
\nu \in [0,W_\beta(1,0)].
\]
can be proved exactly in the same way as the corresponding claim~$9^\circ$ 
in the proof of Theorem~\ref{thm:Whittle-index-discounted-IHR-DHR} in the 
IHR-DHR-E subcase with $n_1^* < \infty$, see 
Appendix~\ref{app:IHR-DHR-E1-proof}. 
Therefore we may omit the proof here.

\vskip 3pt
$6^\circ$ 
Finally, we assume that $\nu \le 0$. 
However, the claim that the policy $\pi$ with activity set 
\[
{\mathcal B}^\pi = {\mathcal S} 
\]
is $(\nu,\beta)$-optimal for all 
\[
\nu \in (-\infty,0] 
\]
can be proved similarly as the corresponding claim~$5^\circ$ in the proof 
of Theorem~\ref{thm:Whittle-index-discounted-DHR-DHR} 
(the DHR-DHR-A subcase, see Appendix~\ref{app:DHR-DHR-A-proof}). 
Therefore we may omit the proof here.
\hfill $\Box$

\section{Proof of Theorem~\ref{thm:Whittle-index-average}}
\label{app:average-proof}

\paragraph{Proof} 
{\em (i)} 
Assume the DHR-DHR case. 
By Theorem~\ref{thm:Whittle-index-discounted-DHR-DHR} together with 
Equation~(\ref{eq:w2-discounted-DHR-DHR}), we have 
\[
W(2,n) = 
\lim_{\beta \to 1} (1 - \beta) W_\beta(2,n) = 
\lim_{\beta \to 1} h \mu_2(n) \beta = 
h \mu_2(n). 
\]
In addition, by Theorem~\ref{thm:Whittle-index-discounted-DHR-DHR} 
together with Equations~(\ref{eq:w1-discounted-DHR-DHR}), 
(\ref{eq:phi-discounted-DHR-DHR}), and (\ref{eq:psi-discounted-DHR-DHR}), 
we have 
\[
\begin{split}
& 
W(1,n) = \lim_{\beta \to 1} (1 - \beta) W_\beta(1,n) \; = \\
& \quad 
\lim_{\beta \to 1} 
h \mu_1(n) \beta \, 
\frac{\beta \sum_{i=0}^{\phi(n)} \beta^i p_2(i)}
{1 + \beta \mu_1(n) \sum_{i=0}^{\phi(n)} \beta^i \bar p_2(i)} \; = \\
& \quad 
h \mu_1(n) \, 
\frac{\sum_{i=0}^{\bar \phi(n)} p_2(i)}
{1 + \mu_1(n) \sum_{i=0}^{\bar \phi(n)} \bar p_2(i)} = 
h \, 
\frac{P\{S_2 \le \bar \phi(n) + 1\}}
{\frac{1}{\mu_1(n)} + E[\min\{S_2, \bar \phi(n) + 1\}]}, 
\end{split}
\]
where 
\[
\bar \phi(n) = 
\min
\left \{ 
n_2 \in \{0,1,\ldots\} \cup \{\infty\} : 
\frac{P\{S_2 \le n_2 + 1\}}
{\frac{1}{\mu_1(n)} + E[\min\{S_2, n_2 + 1\}]} > \mu_2(n_2+1) 
\right \}.
\]
It remains to prove that 
\begin{equation}
\frac{P\{S_2 \le \bar \phi(n) + 1\}}
{\frac{1}{\mu_1(n)} + E[\min\{S_2, \bar \phi(n) + 1\}]} = 
\sup_{n_2 \ge 0} 
\left( 
\frac{P\{S_2 \le n_2 + 1\}}
{\frac{1}{\mu_1(n)} + E[\min\{S_2, n_2 + 1\}]} 
\right).
\label{eq:Whittle-index-average-result-DHR-DHR}
\end{equation}
However, since $p_2(i) = \bar p_2(i) \mu_2(i)$, it is easy to show that 
\[
\begin{split}
& 
\frac{P\{S_2 \le n_2 + 1\}}
{\frac{1}{\mu_1(n)} + E[\min\{S_2, n_2 + 1\}]} > \mu_2(n_2+1) 
\quad \Longleftrightarrow \\
& 
\sum_{i=0}^{n_2} \bar p_2(i) (\mu_2(i) - \mu_2(n_2+1)) > 
\frac{\mu_2(n_2+1)}{\mu_1(n)} 
\quad \Longleftrightarrow \\
& 
\frac{P\{S_2 \le n_2 + 1\}}
{\frac{1}{\mu_1(n)} + E[\min\{S_2, n_2 + 1\}]} > 
\frac{P\{S_2 \le n_2 + 2\}}
{\frac{1}{\mu_1(n)} + E[\min\{S_2, n_2 + 2\}]}, 
\end{split}
\]
which proves claim (\ref{eq:Whittle-index-average-result-DHR-DHR}).

\vskip 3pt
{\em (ii)} 
Assume the IHR-IHR case. 
By Theorem~\ref{thm:Whittle-index-discounted-IHR-IHR} together with 
Equation~(\ref{eq:w2-discounted-IHR-IHR}), we have 
\[
\begin{split}
& 
W(2,n) = 
\lim_{\beta \to 1} (1 - \beta) W_\beta(2,n) \; = \\
& \quad 
\lim_{\beta \to 1} h \, 
\frac{1 - (1 - \beta) \sum_{i=0}^{\infty} \beta^i \bar p_2(i|n)}
{\sum_{i=0}^{\infty} \beta^i \bar p_2(i|n)} = 
\frac{h}{\sum_{i=0}^{\infty} \bar p_2(i|n)} = 
\frac{h}{E[S_2 - n \mid S_2 \ge n + 1]}.
\end{split}
\]
In addition, by Theorem~\ref{thm:Whittle-index-discounted-IHR-IHR} 
together with Equation~(\ref{eq:w1-discounted-IHR-IHR}), 
we have 
\[
\begin{split}
& 
W(1,n) = \lim_{\beta \to 1} (1 - \beta) W_\beta(1,n) \; = \\
& \quad 
\lim_{\beta \to 1} 
h \, 
\frac{1 - (1 - \beta) \sum_{i=0}^{\infty} \beta^i \bar p_1(i|n) 
\Big( 1 + \beta \mu_1(n+i) \sum_{j=0}^{\infty} \beta^{j} \bar p_2(j) \Big)}
{\sum_{i=0}^{\infty} \beta^i \bar p_1(i|n) 
\Big( 1 + \beta \mu_1(n+i) \sum_{j=0}^{\infty} \beta^{j} \bar p_2(j) \Big)} \; = \\
& \quad 
\frac{h}{\sum_{i=0}^{\infty} \bar p_1(i|n) + \sum_{j=0}^{\infty} \bar p_2(j)} = 
\frac{h}{E[S_1 - n \mid S_1 \ge n + 1] + E[S_2]}.
\end{split}
\]

\vskip 3pt
{\em (iii)} 
Assume the DHR-IHR case. 
By Theorem~\ref{thm:Whittle-index-discounted-DHR-IHR} together with 
Equation~(\ref{eq:w2-discounted-DHR-IHR}), we have 
\[
\begin{split}
& 
W(2,n) = 
\lim_{\beta \to 1} (1 - \beta) W_\beta(2,n) \; = \\
& \quad 
\lim_{\beta \to 1} h \, 
\frac{1 - (1 - \beta) \sum_{i=0}^{\infty} \beta^i \bar p_2(i|n)}
{\sum_{i=0}^{\infty} \beta^i \bar p_2(i|n)} = 
\frac{h}{\sum_{i=0}^{\infty} \bar p_2(i|n)} = 
\frac{h}{E[S_2 - n \mid S_2 \ge n + 1]}.
\end{split}
\]
In addition, by Theorem~\ref{thm:Whittle-index-discounted-DHR-IHR} 
together with Equation~(\ref{eq:w1-discounted-DHR-IHR}), 
we have 
\[
\begin{split}
& 
W(1,n) = \lim_{\beta \to 1} (1 - \beta) W_\beta(1,n) \; = \\
& \quad 
\lim_{\beta \to 1} 
h \mu_1(n) \beta \, 
\frac{\beta \sum_{i=0}^{\infty} \beta^i p_2(i)}
{1 + \beta \mu_1(n) \sum_{i=0}^{\infty} \beta^i \bar p_2(i)} \; = \\
& \quad 
\frac{h \mu_1(n)}
{1 + \mu_1(n) \sum_{i=0}^{\infty} \bar p_2(i)} = 
\frac{h}{\frac{1}{\mu_1(n)} + E[S_2]}.
\end{split}
\]

\vskip 3pt
{\em (iv)} 
Assume the IHR-DHR case. 
By Theorem~\ref{thm:Whittle-index-discounted-IHR-DHR} together with 
Equation~(\ref{eq:w2-discounted-IHR-DHR}), we have 
\[
W(2,n) = 
\lim_{\beta \to 1} (1 - \beta) W_\beta(2,n) = 
\lim_{\beta \to 1} h \mu_2(n) \beta = 
h \mu_2(n). 
\]
In addition, by Theorem~\ref{thm:Whittle-index-discounted-IHR-DHR} 
together with Equations~(\ref{eq:w1-discounted-IHR-DHR}), 
(\ref{eq:phi-discounted-IHR-DHR}), and (\ref{eq:psi-discounted-IHR-DHR}), 
we have 
\[
\begin{split}
& 
W(1,n) = \lim_{\beta \to 1} (1 - \beta) W_\beta(1,n) \; = \\
& \quad 
\lim_{\beta \to 1} 
h \, 
\frac{1 - \sum_{i=0}^{\infty} \beta^i \bar p_1(i|n) 
\left( (1 - \beta)  + \mu_1(n+i) \beta 
\Big( 1 - \beta \sum_{j=0}^{\bar \phi(n)} \beta^{j} p_2(j) \Big) \right)}
{\sum_{i=0}^{\infty} \beta^i \bar p_1(i|n) \Big( 1 + \mu_1(n+i) \beta \sum_{j=0}^{\bar \phi(n)} \beta^{j} \bar p_2(j) \Big)} \; = \\
& \quad 
h \, 
\frac{\sum_{j=0}^{\bar \phi(n)} p_2(j)}
{\sum_{i=0}^{\infty} \bar p_1(i|n) + \sum_{j=0}^{\bar \phi(n)} \bar p_2(j)} = 
h \, 
\frac{P\{S_2 \le \bar \phi(n) + 1\}}
{E[S_1 - n \mid S_1 \ge n + 1] + E[\min\{S_2, \bar \phi(n) + 1\}]}, 
\end{split}
\]
where 
\[
\begin{split}
& 
\bar \phi(n) \; = \\
& \quad 
\min
\left \{ 
n_2 \in \{0,1,\ldots\} \cup \{\infty\} : 
\frac{P\{S_2 \le n_2 + 1\}}
{E[S_1 - n \mid S_1 \ge n + 1] + E[\min\{S_2, n_2 + 1\}]} > \mu_2(n_2+1) 
\right \}.
\end{split}
\]
It remains to prove that 
\begin{equation}
\begin{split}
& 
\frac{P\{S_2 \le \bar \phi(n) + 1\}}
{E[S_1 - n \mid S_1 \ge n + 1] + E[\min\{S_2, \bar \phi(n) + 1\}]} \; = \\
& \quad 
\sup_{n_2 \ge 0} 
\left( 
\frac{P\{S_2 \le n_2 + 1\}}
{E[S_1 - n \mid S_1 \ge n + 1] + E[\min\{S_2, n_2 + 1\}]} 
\right).
\end{split}
\label{eq:Whittle-index-average-result-IHR-DHR}
\end{equation}
However, since $p_2(i) = \bar p_2(i) \mu_2(i)$, it is easy to show that 
\[
\begin{split}
& 
\frac{P\{S_2 \le n_2 + 1\}}
{E[S_1 - n \mid S_1 \ge n + 1] + E[\min\{S_2, n_2 + 1\}]} > \mu_2(n_2+1) 
\quad \Longleftrightarrow \\
& 
\sum_{i=0}^{n_2} \bar p_2(i) (\mu_2(i) - \mu_2(n_2+1)) > 
E[S_1 - n \mid S_1 \ge n + 1] \, \mu_2(n_2+1) 
\quad \Longleftrightarrow \\
& 
\frac{P\{S_2 \le n_2 + 1\}}
{E[S_1 - n \mid S_1 \ge n + 1] + E[\min\{S_2, n_2 + 1\}]} \; > \\
& \quad 
\frac{P\{S_2 \le n_2 + 2\}}
{E[S_1 - n \mid S_1 \ge n + 1] + E[\min\{S_2, n_2 + 2\}]}, 
\end{split}
\]
which proves claim (\ref{eq:Whittle-index-average-result-IHR-DHR}).
\hfill $\Box$

\section{Proof of Theorem~\ref{thm:Gittins-index-average}}
\label{app:average-continuous-proof}

\paragraph{Proof} 
{\em (i)} 
Assume the DHR-DHR case. 
By Corollary~\ref{cor:Gittins-index-average}(i) and 
Equation~(\ref{eq:Gja-DHR}), we have 
\[
G(2,a) = h \mu_2(a) = G_2(a).
\]
and 
\[
G(1,a) = 
h \, 
\sup_{\Delta \ge 0} 
\left( 
\frac{P\{S_2 \le \Delta\}}
{\frac{1}{\mu_1(a)} + E[\min\{S_2,\Delta\}]} 
\right) = 
h \, 
\sup_{\Delta \ge 0} 
\left( 
\frac{P\{S_2 \le \Delta\}}
{\frac{h}{G_1(a)} + E[\min\{S_2,\Delta\}]} 
\right).
\]

\vskip 3pt
{\em (ii)} 
Assume the IHR-IHR case. 
By Corollary~\ref{cor:Gittins-index-average}(ii) and 
Equation~(\ref{eq:Gja-IHR}), we have 
\[
G(2,a) = \frac{h}{E[S_2 - a \mid S_2 \ge a]} = G_2(a).
\]
and 
\[
G(1,a) = 
\frac{h}
{E[S_1 - a \mid S_1 \ge a] + E[S_2]} = 
h \, 
\lim_{\Delta \to \infty} 
\left( 
\frac{P\{S_2 \le \Delta\}}
{\frac{h}{G_1(a)} + E[\min\{S_2,\Delta\}]} 
\right)
\]
It remains to prove that 
\begin{equation}
\lim_{\Delta \to \infty} 
\left( 
\frac{P\{S_2 \le \Delta\}}
{\frac{h}{G_1(a)} + E[\min\{S_2,\Delta\}]} 
\right) = 
\sup_{\Delta \ge 0} 
\left( 
\frac{P\{S_2 \le \Delta\}}
{\frac{h}{G_1(a)} + E[\min\{S_2,\Delta\}]} 
\right).
\label{eq:Whittle-index-average-result-IHR-IHR}
\end{equation}
Let $\Delta \ge 0$. Now, 
\[
\begin{split}
& 
\lim_{\Delta' \to \infty} 
\left( 
\frac{P\{S_2 \le \Delta'\}}
{\frac{h}{G_1(a)} + E[\min\{S_2,\Delta'\}]} 
\right) \ge 
\frac{P\{S_2 \le \Delta\}}
{\frac{h}{G_1(a)} + E[\min\{S_2,\Delta\}]} 
\quad \Longleftrightarrow \\
& 
\frac{1}
{E[S_1 - a \mid S_1 \ge a] + E[S_2]} \ge 
\frac{P\{S_2 \le \Delta\}}
{E[S_1 - a \mid S_1 \ge a] + E[\min\{S_2,\Delta\}]} 
\quad \Longleftrightarrow \\
& 
\frac{E[\min\{S_2,\Delta\}]}{P\{S_2 \le \Delta\}} - E[S_2] \ge 
E[S_1 - a \mid S_1 \ge a] 
\left( 1 - \frac{1}{P\{S_2 \le \Delta\}} \right) 
\quad \Longleftrightarrow \\
& 
\frac{1}{J_2(0,\Delta)} - \frac{1}{J_2(0,\infty)} \ge 
E[S_1 - a \mid S_1 \ge a] 
\left( 1 - \frac{1}{P\{S_2 \le \Delta\}} \right), 
\end{split}
\]
which is true, since the right hand side is clearly non-positive and 
the left hand side is non-negative by \cite[Corollary~7]{Aal09QS}. 
This completes the proof of claim 
(\ref{eq:Whittle-index-average-result-IHR-IHR}).

\vskip 3pt
{\em (iii)}
Assume the DHR-IHR case. 
By Corollary~\ref{cor:Gittins-index-average}(iii) together with 
Equations~(\ref{eq:Gja-DHR}) and (\ref{eq:Gja-IHR}), we have 
\[
G(2,a) = \frac{h}{E[S_2 - a \mid S_2 \ge a]} = G_2(a).
\]
and 
\[
G(1,a) = 
\frac{h}
{\frac{1}{\mu_1(a)} + E[S_2]} = 
h \, 
\lim_{\Delta \to \infty} 
\left( 
\frac{P\{S_2 \le \Delta\}}
{\frac{h}{G_1(a)} + E[\min\{S_2,\Delta\}]} 
\right)
\]
It remains to prove that 
\begin{equation}
\lim_{\Delta \to \infty} 
\left( 
\frac{P\{S_2 \le \Delta\}}
{\frac{h}{G_1(a)} + E[\min\{S_2,\Delta\}]} 
\right) = 
\sup_{\Delta \ge 0} 
\left( 
\frac{P\{S_2 \le \Delta\}}
{\frac{h}{G_1(a)} + E[\min\{S_2,\Delta\}]} 
\right).
\label{eq:Whittle-index-average-result-DHR-IHR}
\end{equation}
Let $\Delta \ge 0$. Now, 
\[
\begin{split}
& 
\lim_{\Delta' \to \infty} 
\left( 
\frac{P\{S_2 \le \Delta'\}}
{\frac{h}{G_1(a)} + E[\min\{S_2,\Delta'\}]} 
\right) \ge 
\frac{P\{S_2 \le \Delta\}}
{\frac{h}{G_1(a)} + E[\min\{S_2,\Delta\}]} 
\quad \Longleftrightarrow \\
& 
\frac{1}
{\frac{1}{\mu_1(a)} + E[S_2]} \ge 
\frac{P\{S_2 \le \Delta\}}
{\frac{1}{\mu_1(a)} + E[\min\{S_2,\Delta\}]} 
\quad \Longleftrightarrow \\
& 
\frac{E[\min\{S_2,\Delta\}]}{P\{S_2 \le \Delta\}} - E[S_2] \ge 
\frac{1}{\mu_1(a)} 
\left( 1 - \frac{1}{P\{S_2 \le \Delta\}} \right) 
\quad \Longleftrightarrow \\
& 
\frac{1}{J_2(0,\Delta)} - \frac{1}{J_2(0,\infty)} \ge 
\frac{1}{\mu_1(a)} 
\left( 1 - \frac{1}{P\{S_2 \le \Delta\}} \right), 
\end{split}
\]
which is true, since the right hand side is clearly non-positive 
the left hand side is non-negative by \cite[Corollary~7]{Aal09QS}. 
This completes the proof of claim 
(\ref{eq:Whittle-index-average-result-DHR-IHR}).

\vskip 3pt
{\em (iv)} 
Assume the IHR-DHR case. 
By Corollary~\ref{cor:Gittins-index-average}(iv) together with 
Equations~(\ref{eq:Gja-DHR}) and (\ref{eq:Gja-IHR}), we have 
\[
G(2,a) = h \mu_2(a) = G_2(a).
\]
and 
\[
\begin{split}
& 
G(1,a) = 
h \, 
\sup_{\Delta \ge 0} 
\left( 
\frac{P\{S_2 \le \Delta\}}
{E[S_1 - a \mid S_1 \ge a] + E[\min\{S_2,\Delta\}]} 
\right) \; = \\
& \quad 
h \, 
\sup_{\Delta \ge 0} 
\left( 
\frac{P\{S_2 \le \Delta\}}
{\frac{h}{G_1(a)} + E[\min\{S_2,\Delta\}]} 
\right), 
\end{split}
\]
which completes the proof.
\hfill $\Box$

\section{Examples on Whittle index values for discounted costs}
\label{app:whittle-index-examples-discounted-costs}

In this appendix, we give numerical examples on the discrete-time Whittle 
index related to the minimization of expected discounted holding costs in 
various cases of a sequential two-stage job with monotonous hazard rates 
in both stages. In these examples, the following parameters are kept fixed: 
$h = 1$ and $\beta = 0.9$.

\subsection{Case DHR-DHR}
\label{subapp:whittle-index-examples-DHR-DHR}

\paragraph{Subcase DHR-DHR-A} 
The first example belongs to the DHR-DHR-A subcase defined in 
Section~\ref{sec:DHR-DHR} by Equations~(\ref{eq:DHR-DHR-A1}) and 
(\ref{eq:DHR-DHR-A2}). The monotonous hazard rate functions for the 
two stages are given by 
\[
\mu_1(n) = \mu_2(n) = \alpha^{n+1}, 
\quad 
n \in \{0,1,\ldots\}.
\]
where $\alpha = 0.5$. 
See Table~\ref{tab:DHR-DHR-A} for the numerical values of the hazard 
rates $\mu_j(n)$ and Whittle indexes $W_\beta(j,n)$ when $j \in \{1,2\}$ 
and $n \in \{0,1,\ldots,6\}$. For $W_\beta(2,n_2)$, we have used 
Equations~(\ref{eq:Whittle-index-2n-discounted-DHR-DHR}) and 
(\ref{eq:w2-discounted-DHR-DHR}), and for $W_\beta(1,n_1)$ 
Equations~(\ref{eq:Whittle-index-1n-discounted-DHR-DHR}) and 
(\ref{eq:w1-discounted-DHR-DHR}). In addition, we give the related 
values of functions $\psi(n_1,n_2)$ and $\phi(n_1)$ utilizing 
Equations (\ref{eq:psi-discounted-DHR-DHR}) and 
(\ref{eq:phi-discounted-DHR-DHR}), respectively.

\begin{table}[h]
\caption{Subcase DHR-DHR-A: 
$\mu_2(n_2)$, $W_\beta(2,n_2)$, $\psi(n_1,n_2)$, 
$\mu_1(n_1)$, $\phi(n_1)$, and $W_\beta(1,n_1)$ 
for $n_1, n_2 \in \{0,1,\ldots,6\}$}
\begin{center}
\begin{tabular}{||c||c|c|c|c|c|c|c||}
\hline
$n_2$ & 
$0$ & $1$ & $2$ & $3$ & $4$ & $5$ & $6$ \\ 
\hline
$\mu_2(n_2)$ & 
$0.5000$ & $0.2500$ & $0.1250$ & $0.0625$ & $0.0313$ & $0.0156$ & $0.0078$ \\ 
\hline
$W_\beta(2,n_2)$ & 
$\mathit{4.0500}$ & $\mathit{2.2500}$ & $\mathit{1.1250}$ & $\mathit{0.5625}$ & $\mathit{0.2813}$ & $\mathit{0.1406}$ & $\mathit{0.0703}$ \\ 
\hline
$\psi(0,n_2)$ & 
$1.3966$ & $\mathit{1.5011}$ & $1.4724$ & $1.4208$ & $1.3687$ & $1.3217$ & $1.2806$ \\ 
\hline
$\psi(1,n_2)$ & 
$0.8265$ & $0.9352$ & $\mathit{0.9445}$ & $0.9303$ & $0.9106$ & $0.8907$ & $0.8724$ \\ 
\hline
$\psi(2,n_2)$ & 
$0.4551$ & $0.5332$ & $0.5501$ & $\mathit{0.5503}$ & $0.5454$ & $0.5391$ & $0.5327$ \\ 
\hline
$\psi(3,n_2)$ & 
$0.2396$ & $0.2867$ & $0.2997$ & $\mathit{0.3029}$ & $0.3027$ & $0.3013$ & $0.2995$ \\ 
\hline
$\psi(4,n_2)$ & 
$0.1231$ & $0.1490$ & $0.1569$ & $0.1595$ & $\mathit{0.1601}$ & $0.1601$ & $0.1597$ \\ 
\hline
$\psi(5,n_2)$ & 
$0.0624$ & $0.0760$ & $0.0803$ & $0.0819$ & $0.0825$ & $\mathit{0.0826}$ & $0.0826$ \\ 
\hline
$\psi(6,n_2)$ & 
$0.0314$ & $0.0384$ & $0.0407$ & $0.0415$ & $0.0419$ & $0.0420$ & $\mathit{0.0420}$ \\ 
\hline
\multicolumn{8}{c}{ } \\
\hline
$n_1$ & 
$0$ & $1$ & $2$ & $3$ & $4$ & $5$ & $6$ \\ 
\hline
$\mu_1(n_1)$ & 
$0.5000$ & $0.2500$ & $0.1250$ & $0.0625$ & $0.0313$ & $0.0156$ & $0.0078$ \\ 
\hline
$\phi(n_1)$ & 
$1$ & $2$ & $3$ & $3$ & $4$ & $5$ & $6$ \\ 
\hline
$W_\beta(1,n_1)$ & 
$\mathit{1.5011}$ & $\mathit{0.9445}$ & $\mathit{0.5503}$ & $\mathit{0.3029}$ & $\mathit{0.1601}$ & $\mathit{0.0826}$ & $\mathit{0.0420}$ \\ 
\hline
\end{tabular}
\end{center}
\label{tab:DHR-DHR-A}
\end{table}

\paragraph{Subcase DHR-DHR-B} 
Next example belongs to the DHR-DHR-B subcase defined in 
Section~\ref{sec:DHR-DHR} by Equation~(\ref{eq:DHR-DHR-B}). 
The monotonous hazard rate functions for the two stages are given by 
\[
\begin{split}
& 
\mu_1(n) = 
\frac{p_1 (1 - \mu_{11})^n \mu_{11} + (1 - p_1) (1 - \mu_{12})^n \mu_{12}}
{p_1 (1 - \mu_{11})^n + (1 - p_1) (1 - \mu_{12})^n}, 
\quad 
n \in \{0,1,\ldots\}. \\
& 
\mu_2(n) = 
\frac{p_2 (1 - \mu_{21})^n \mu_{21} + (1 - p_2) (1 - \mu_{22})^n \mu_{22}}
{p_2 (1 - \mu_{21})^n + (1 - p_2) (1 - \mu_{22})^n}, 
\quad 
n \in \{0,1,\ldots\}.
\end{split}
\]
where $p_1 = 0.5$, $\mu_{11} = 0.8$, $\mu_{12} = 0.3$, 
$p_2 = 0.5$, $\mu_{21} = 0.8$, and $\mu_{22} = 0.1$. 
See Table~\ref{tab:DHR-DHR-B} for the numerical values of the hazard 
rates $\mu_j(n)$ and Whittle indexes $W_\beta(j,n)$ when $j \in \{1,2\}$ 
and $n \in \{0,1,\ldots,6\}$. For $W_\beta(2,n_2)$, we have used 
Equations~(\ref{eq:Whittle-index-2n-discounted-DHR-DHR}) and 
(\ref{eq:w2-discounted-DHR-DHR}), and for $W_\beta(1,n_1)$ 
Equations~(\ref{eq:Whittle-index-1n-discounted-DHR-DHR}) and 
(\ref{eq:w1-discounted-DHR-DHR}). In addition, we give the related 
values of functions $\psi(n_1,n_2)$ and $\phi(n_1)$ utilizing 
Equations (\ref{eq:psi-discounted-DHR-DHR}) and 
(\ref{eq:phi-discounted-DHR-DHR}), respectively. 
In this example, $n_2^* = 2$, where $n_2^*$ denotes the smallest 
$\bar n_2$ satisfying (\ref{eq:DHR-DHR-B}).

\begin{table}[h]
\caption{Subcase DHR-DHR-B: 
$\mu_2(n_2)$, $W_\beta(2,n_2)$, $\psi(n_1,n_2)$, 
$\mu_1(n_1)$, $\phi(n_1)$, and $W_\beta(1,n_1)$ 
for $n_1, n_2 \in \{0,1,\ldots,6\}$}
\begin{center}
\begin{tabular}{||c||c|c|c|c|c|c|c||}
\hline
$n_2$ & 
$0$ & $1$ & $2$ & $3$ & $4$ & $5$ & $6$ \\ 
\hline
$\mu_2(n_2)$ & 
$0.4500$ & $0.2273$ & $0.1329$ & $0.1076$ & $0.1017$ & $0.1004$ & $0.1001$ \\ 
\hline
$W_\beta(2,n_2)$ & 
$\mathit{4.0500}$ & $\mathit{2.0455}$ & $\mathit{1.1965}$ & $\mathit{0.9684}$ & $\mathit{0.9153}$ & $\mathit{0.9034}$ & $\mathit{0.9008}$ \\ 
\hline
$\psi(0,n_2)$ & 
$1.3410$ & $\mathit{1.4402}$ & $1.4184$ & $1.3892$ & $1.3656$ & $1.3478$ & $1.3342$ \\ 
\hline
$\psi(1,n_2)$ & 
$1.0938$ & $\mathit{1.2060}$ & $1.2053$ & $1.1921$ & $1.1801$ & $1.1709$ & $1.1638$ \\ 
\hline
$\psi(2,n_2)$ & 
$0.9441$ & $1.0580$ & $\mathit{1.0673}$ & $1.0624$ & $1.0567$ & $1.0521$ & $1.0485$ \\ 
\hline
$\psi(3,n_2)$ & 
$0.8866$ & $0.9999$ & $\mathit{1.0124}$ & $1.0103$ & $1.0068$ & $1.0039$ & $1.0015$ \\ 
\hline
$\psi(4,n_2)$ & 
$0.8685$ & $0.9814$ & $\mathit{0.9949}$ & $0.9937$ & $0.9908$ & $0.9884$ & $0.9864$ \\ 
\hline
$\psi(5,n_2)$ & 
$0.8632$ & $0.9760$ & $\mathit{0.9897}$ & $0.9887$ & $0.9861$ & $0.9838$ & $0.9819$ \\ 
\hline
$\psi(6,n_2)$ & 
$0.8616$ & $0.9744$ & $\mathit{0.9882}$ & $0.9873$ & $0.9847$ & $0.9824$ & $0.9806$ \\ 
\hline
\multicolumn{8}{c}{ } \\
\hline
$n_1$ & 
$0$ & $1$ & $2$ & $3$ & $4$ & $5$ & $6$ \\ 
\hline
$\mu_1(n_1)$ & 
$0.5500$ & $0.4111$ & $0.3377$ & $0.3114$ & $0.3033$ & $0.3010$ & $0.3003$ \\ 
\hline
$\phi(n_1)$ & 
$1$ & $1$ & $2$ & $2$ & $2$ & $2$ & $2$ \\ 
\hline
$W_\beta(1,n_1)$ & 
$\mathit{1.4402}$ & $\mathit{1.2060}$ & $\mathit{1.0673}$ & $\mathit{1.0124}$ & $\mathit{0.9949}$ & $\mathit{0.9897}$ & $\mathit{0.9882}$ \\ 
\hline
\end{tabular}
\end{center}
\label{tab:DHR-DHR-B}
\end{table}

\paragraph{Subcase DHR-DHR-C} 
The last example in this section belongs to the DHR-DHR-C subcase defined in 
Section~\ref{sec:DHR-DHR} by Equation~(\ref{eq:DHR-DHR-C}). 
The monotonous hazard rate functions for the two stages are given by 
\[
\begin{split}
& 
\mu_1(n) = 
\frac{p_1 (1 - \mu_{11})^n \mu_{11} + (1 - p_1) (1 - \mu_{12})^n \mu_{12}}
{p_1 (1 - \mu_{11})^n + (1 - p_1) (1 - \mu_{12})^n}, 
\quad 
n \in \{0,1,\ldots\}. \\
& 
\mu_2(n) = 
\frac{p_2 (1 - \mu_{21})^n \mu_{21} + (1 - p_2) (1 - \mu_{22})^n \mu_{22}}
{p_2 (1 - \mu_{21})^n + (1 - p_2) (1 - \mu_{22})^n}, 
\quad 
n \in \{0,1,\ldots\}.
\end{split}
\]
where $p_1 = 0.5$, $\mu_{11} = 0.7$, $\mu_{12} = 0.2$, 
$p_2 = 0.5$, $\mu_{21} = 0.8$, and $\mu_{22} = 0.1$. 
See Table~\ref{tab:DHR-DHR-C} for the numerical values of the hazard 
rates $\mu_j(n)$ and Whittle indexes $W_\beta(j,n)$ when $j \in \{1,2\}$ 
and $n \in \{0,1,\ldots,6\}$. For $W_\beta(2,n_2)$, we have used 
Equations~(\ref{eq:Whittle-index-2n-discounted-DHR-DHR}) and 
(\ref{eq:w2-discounted-DHR-DHR}), and for $W_\beta(1,n_1)$ 
Equations~(\ref{eq:Whittle-index-1n-discounted-DHR-DHR}) and 
(\ref{eq:w1-discounted-DHR-DHR}). In addition, we give the related 
values of functions $\psi(n_1,n_2)$ and $\phi(n_1)$ utilizing 
Equations (\ref{eq:psi-discounted-DHR-DHR}) and 
(\ref{eq:phi-discounted-DHR-DHR}), respectively. 
In this example, $n_1^* = 2$, where $n_1^*$ denotes the smallest 
$\bar n_1$ satisfying (\ref{eq:DHR-DHR-C}).

\begin{table}[h]
\caption{Subcase DHR-DHR-C: 
$\mu_2(n_2)$, $W_\beta(2,n_2)$, $\psi(n_1,n_2)$, 
$\mu_1(n_1)$, $\phi(n_1)$, and $W_\beta(1,n_1)$ 
for $n_1, n_2 \in \{0,1,\ldots,6\}$}
\begin{center}
\begin{tabular}{||c||c|c|c|c|c|c|c||}
\hline
$n_2$ & 
$0$ & $1$ & $2$ & $3$ & $4$ & $5$ & $6$ \\ 
\hline
$\mu_2(n_2)$ & 
$0.4500$ & $0.2273$ & $0.1329$ & $0.1076$ & $0.1017$ & $0.1004$ & $0.1001$ \\ 
\hline
$W_\beta(2,n_2)$ & 
$\mathit{4.0500}$ & $\mathit{2.0455}$ & $\mathit{1.1965}$ & $\mathit{0.9684}$ & $\mathit{0.9153}$ & $\mathit{0.9034}$ & $\mathit{0.9008}$ \\ 
\hline
$\psi(0,n_2)$ & 
$1.1674$ & $\mathit{1.2771}$ & $1.2706$ & $1.2529$ & $1.2377$ & $1.2260$ & $1.2170$ \\ 
\hline
$\psi(1,n_2)$ & 
$0.9411$ & $1.0551$ & $\mathit{1.0645}$ & $1.0598$ & $1.0542$ & $1.0497$ & $1.0461$ \\ 
\hline
$\psi(2,n_2)$ & 
$0.7719$ & $0.8817$ & $0.8995$ & $0.9024$ & $0.9029$ & $\mathit{0.9029}$ & $0.9028$ \\ 
\hline
$\psi(3,n_2)$ & 
$0.6821$ & $0.7870$ & $0.8078$ & $0.8140$ & $0.8170$ & $0.8190$ & $0.8205$ \\ 
\hline
$\psi(4,n_2)$ & 
$0.6430$ & $0.7452$ & $0.7669$ & $0.7743$ & $0.7782$ & $0.7810$ & $0.7832$ \\ 
\hline
$\psi(5,n_2)$ & 
$0.6274$ & $0.7284$ & $0.7505$ & $0.7582$ & $0.7626$ & $0.7657$ & $0.7680$ \\ 
\hline
$\psi(6,n_2)$ & 
$0.6214$ & $0.7219$ & $0.7442$ & $0.7521$ & $0.7565$ & $0.7597$ & $0.7622$ \\ 
\hline
\multicolumn{8}{c}{ } \\
\hline
$n_1$ & 
$0$ & $1$ & $2$ & $3$ & $4$ & $5$ & $6$ \\ 
\hline
$\mu_1(n_1)$ & 
$0.4500$ & $0.3363$ & $0.2616$ & $0.2250$ & $0.2097$ & $0.2037$ & $0.2014$ \\ 
\hline
$\phi(n_1)$ & 
$1$ & $2$ & $5$ & $\infty$ & $\infty$ & $\infty$ & $\infty$ \\ 
\hline
$W_\beta(1,n_1)$ & 
$\mathit{1.2771}$ & $\mathit{1.0645}$ & $\mathit{0.9029}$ & $\mathit{0.8264}$ & $\mathit{0.7914}$ & $\mathit{0.7772}$ & $\mathit{0.7717}$ \\ 
\hline
\end{tabular}
\end{center}
\label{tab:DHR-DHR-C}
\end{table}

\subsection{Case IHR-IHR}
\label{subapp:whittle-index-examples-IHR-IHR}

This example belongs to the IHR-IHR class. The monotonous hazard rate functions 
for the two stages are given by 
\[
\mu_1(n) = \mu_2(n) = 1 - \alpha^{n+1}, 
\quad 
n \in \{0,1,\ldots\}.
\]
where $\alpha = 0.5$. 
See Table~\ref{tab:IHR-IHR} for the numerical values of the hazard 
rates $\mu_j(n)$ and Whittle indexes $W_\beta(j,n)$ when $j \in \{1,2\}$ 
and $n \in \{0,1,\ldots,6\}$. For $W_\beta(2,n_2)$, we have used 
Equations~(\ref{eq:Whittle-index-2n-discounted-IHR-IHR}) and 
(\ref{eq:w2-discounted-IHR-IHR}), and for $W_\beta(1,n_1)$ 
Equations~(\ref{eq:Whittle-index-1n-discounted-IHR-IHR}) and 
(\ref{eq:w1-discounted-IHR-IHR}).

\begin{table}[h]
\caption{Case IHR-IHR: 
$\mu_2(n_2)$, $W_\beta(2,n_2)$, $\mu_1(n_1)$, and $W_\beta(1,n_1)$ 
for $n_1, n_2 \in \{0,1,\ldots,6\}$}
\begin{center}
\begin{tabular}{||c||c|c|c|c|c|c|c||}
\hline
$n_2$ & 
$0$ & $1$ & $2$ & $3$ & $4$ & $5$ & $6$ \\ 
\hline
$\mu_2(n_2)$ & 
$0.5000$ & $0.7500$ & $0.8750$ & $0.9375$ & $0.9688$ & $0.9844$ & $0.9922$ \\ 
\hline
$W_\beta(2,n_2)$ & 
$\mathit{5.3967}$ & $\mathit{6.9886}$ & $\mathit{7.9364}$ & $\mathit{8.4531}$ & $\mathit{8.7227}$ & $\mathit{8.8605}$ & $\mathit{8.9299}$ \\ 
\hline
\multicolumn{8}{c}{ } \\
\hline
$n_1$ & 
$0$ & $1$ & $2$ & $3$ & $4$ & $5$ & $6$ \\ 
\hline
$\mu_1(n_1)$ & 
$0.5000$ & $0.7500$ & $0.8750$ & $0.9375$ & $0.9688$ & $0.9844$ & $0.9922$ \\ 
\hline
$W_\beta(1,n_1)$ & 
$\mathit{2.4696}$ & $\mathit{2.8177}$ & $\mathit{2.9882}$ & $\mathit{3.0720}$ & $\mathit{3.1135}$ & $\mathit{3.1341}$ & $\mathit{3.1444}$ \\ 
\hline
\end{tabular}
\end{center}
\label{tab:IHR-IHR}
\end{table}

\subsection{Case DHR-IHR}
\label{subapp:whittle-index-examples-DHR-IHR}

This example belongs to the DHR-IHR class. The monotonous hazard rate functions 
for the two stages are given by 
\[
\begin{split}
& 
\mu_1(n) = 
\frac{p_1 (1 - \mu_{11})^n \mu_{11} + (1 - p_1) (1 - \mu_{12})^n \mu_{12}}
{p_1 (1 - \mu_{11})^n + (1 - p_1) (1 - \mu_{12})^n}, 
\quad 
n \in \{0,1,\ldots\}. \\
& 
\mu_2(n) = 1 - \alpha_2^{n+1}, 
\quad 
n \in \{0,1,\ldots\}.
\end{split}
\]
where $p_1 = 0.5$, $\mu_{11} = 0.8$, $\mu_{12} = 0.1$, and $\alpha_2 = 0.5$. 
See Table~\ref{tab:DHR-IHR} for the numerical values of the hazard 
rates $\mu_j(n)$ and Whittle indexes $W_\beta(j,n)$ when $j \in \{1,2\}$ 
and $n \in \{0,1,\ldots,6\}$. For $W_\beta(2,n_2)$, we have used 
Equations~(\ref{eq:Whittle-index-2n-discounted-DHR-IHR}) and 
(\ref{eq:w2-discounted-DHR-IHR}), and for $W_\beta(1,n_1)$ 
Equations~(\ref{eq:Whittle-index-1n-discounted-DHR-IHR}) and 
(\ref{eq:w1-discounted-DHR-IHR}).

\begin{table}[h]
\caption{Case DHR-IHR: 
$\mu_2(n_2)$, $W_\beta(2,n_2)$, $\mu_1(n_1)$, and $W_\beta(1,n_1)$ 
for $n_1, n_2 \in \{0,1,\ldots,6\}$}
\begin{center}
\begin{tabular}{||c||c|c|c|c|c|c|c||}
\hline
$n_2$ & 
$0$ & $1$ & $2$ & $3$ & $4$ & $5$ & $6$ \\ 
\hline
$\mu_2(n_2)$ & 
$0.5000$ & $0.7500$ & $0.8750$ & $0.9375$ & $0.9688$ & $0.9844$ & $0.9922$ \\ 
\hline
$W_\beta(2,n_2)$ & 
$\mathit{5.3967}$ & $\mathit{6.9886}$ & $\mathit{7.9364}$ & $\mathit{8.4531}$ & $\mathit{8.7227}$ & $\mathit{8.8605}$ & $\mathit{8.9299}$ \\ 
\hline
\multicolumn{8}{c}{ } \\
\hline
$n_1$ & 
$0$ & $1$ & $2$ & $3$ & $4$ & $5$ & $6$ \\ 
\hline
$\mu_1(n_1)$ & 
$0.4500$ & $0.2273$ & $0.1329$ & $0.1076$ & $0.1017$ & $0.1004$ & $0.1001$ \\ 
\hline
$W_\beta(1,n_1)$ & 
$\mathit{2.0922}$ & $\mathit{1.3076}$ & $\mathit{0.8504}$ & $\mathit{0.7096}$ & $\mathit{0.6756}$ & $\mathit{0.6679}$ & $\mathit{0.6661}$ \\ 
\hline
\end{tabular}
\end{center}
\label{tab:DHR-IHR}
\end{table}

\subsection{Case IHR-DHR}
\label{subapp:whittle-index-examples-IHR-DHR}

\paragraph{Subcase IHR-DHR-D} 
The first example in this section belongs to the IHR-DHR-D subcase defined 
in Section~\ref{sec:IHR-DHR} by Equation~(\ref{eq:IHR-DHR-D}). The monotonous 
hazard rate functions for the two stages are given by 
\[
\begin{split}
& 
\mu_1(n) = 1 - \alpha_1^{n+1}, 
\quad 
n \in \{0,1,\ldots\}. \\
& 
\mu_2(n) = 
\frac{p_2 (1 - \mu_{21})^n \mu_{21} + (1 - p_2) (1 - \mu_{22})^n \mu_{22}}
{p_2 (1 - \mu_{21})^n + (1 - p_2) (1 - \mu_{22})^n}, 
\quad 
n \in \{0,1,\ldots\}.
\end{split}
\]
where $\alpha_1 = 0.5$, $p_2 = 0.5$, $\mu_{21} = 0.5$, and $\mu_{22} = 0.1$. 
See Table~\ref{tab:IHR-DHR-D} for the numerical values of the hazard 
rates $\mu_j(n)$ and Whittle indexes $W_\beta(j,n)$ when $j \in \{1,2\}$ 
and $n \in \{0,1,\ldots,6\}$. For $W_\beta(2,n_2)$, we have used 
Equations~(\ref{eq:Whittle-index-2n-discounted-IHR-DHR}) and 
(\ref{eq:w2-discounted-IHR-DHR}), and for $W_\beta(1,n_1)$ 
Equations~(\ref{eq:Whittle-index-1n-discounted-IHR-DHR}) and 
(\ref{eq:w1-discounted-IHR-DHR}). In addition, we give the related 
values of functions $\psi(n_1,n_2)$ and $\phi(n_1)$ utilizing 
Equations (\ref{eq:psi-discounted-IHR-DHR}) and 
(\ref{eq:phi-discounted-IHR-DHR}), respectively. 
In this example, $n_2^* = 3$, where $n_2^*$ denotes the smallest 
$\bar n_2$ satisfying (\ref{eq:IHR-DHR-D}).

\begin{table}[h]
\caption{Subcase IHR-DHR-D: 
$\mu_2(n_2)$, $W_\beta(2,n_2)$, $\psi(n_1,n_2)$, 
$\mu_1(n_1)$, $\phi(n_1)$, and $W_\beta(1,n_1)$ 
for $n_1, n_2 \in \{0,1,\ldots,6\}$}
\begin{center}
\begin{tabular}{||c||c|c|c|c|c|c|c||}
\hline
$n_2$ & 
$0$ & $1$ & $2$ & $3$ & $4$ & $5$ & $6$ \\ 
\hline
$\mu_2(n_2)$ & 
$0.3000$ & $0.2429$ & $0.1943$ & $0.1585$ & $0.1348$ & $0.1201$ & $0.1114$ \\ 
\hline
$W_\beta(2,n_2)$ & 
$\mathit{2.7000}$ & $\mathit{2.1857}$ & $\mathit{1.7491}$ & $\mathit{1.4269}$ & $\mathit{1.2131}$ & $\mathit{1.0809}$ & $\mathit{1.0028}$ \\ 
\hline
$\psi(6,n_2)$ & 
$1.2737$ & $1.4826$ & $\mathit{1.5186}$ & $1.5104$ & $1.4916$ & $1.4723$ & $1.4555$ \\ 
\hline
$\psi(5,n_2)$ & 
$1.2684$ & $1.4779$ & $\mathit{1.5144}$ & $1.5066$ & $1.4881$ & $1.4690$ & $1.4524$ \\ 
\hline
$\psi(4,n_2)$ & 
$1.2579$ & $1.4684$ & $\mathit{1.5060}$ & $1.4990$ & $1.4811$ & $1.4624$ & $1.4460$ \\ 
\hline
$\psi(3,n_2)$ & 
$1.2368$ & $1.4493$ & $\mathit{1.4890}$ & $1.4836$ & $1.4668$ & $1.4489$ & $1.4332$ \\ 
\hline
$\psi(2,n_2)$ & 
$1.1947$ & $1.4107$ & $\mathit{1.4545}$ & $1.4521$ & $1.4376$ & $1.4214$ & $1.4069$ \\ 
\hline
$\psi(1,n_2)$ & 
$1.1107$ & $1.3320$ & $1.3833$ & $\mathit{1.3868}$ & $1.3767$ & $1.3638$ & $1.3518$ \\ 
\hline
$\psi(0,n_2)$ & 
$0.9464$ & $1.1706$ & $1.2340$ & $\mathit{1.2483}$ & $1.2464$ & $1.2399$ & $1.2327$ \\ 
\hline
\multicolumn{8}{c}{ } \\
\hline
$n_1$ & 
$0$ & $1$ & $2$ & $3$ & $4$ & $5$ & $6$ \\ 
\hline
$\mu_1(n_1)$ & 
$0.5000$ & $0.7500$ & $0.8750$ & $0.9375$ & $0.9688$ & $0.9844$ & $0.9922$ \\ 
\hline
$\phi(n_1)$ & 
$3$ & $3$ & $2$ & $2$ & $2$ & $2$ & $2$ \\ 
\hline
$W_\beta(1,n_1)$ & 
$\mathit{1.2483}$ & $\mathit{1.3868}$ & $\mathit{1.4545}$ & $\mathit{1.4890}$ & $\mathit{1.5060}$ & $\mathit{1.5144}$ & $\mathit{1.5186}$ \\ 
\hline
\end{tabular}
\end{center}
\label{tab:IHR-DHR-D}
\end{table}

\paragraph{Subcase IHR-DHR-E} 
Next example belongs to the IHR-DHR-E subcase defined 
in Section~\ref{sec:IHR-DHR} by Equation~(\ref{eq:IHR-DHR-E}). The monotonous 
hazard rate functions for the two stages are given by 
\[
\begin{split}
& 
\mu_1(n) = 1 - \alpha_1^{n+1}, 
\quad 
n \in \{0,1,\ldots\}. \\
& 
\mu_2(n) = 
\frac{p_2 (1 - \mu_{21})^n \mu_{21} + (1 - p_2) (1 - \mu_{22})^n \mu_{22}}
{p_2 (1 - \mu_{21})^n + (1 - p_2) (1 - \mu_{22})^n}, 
\quad 
n \in \{0,1,\ldots\}.
\end{split}
\]
where $\alpha_1 = 0.8$, $p_2 = 0.5$, $\mu_{21} = 0.5$, and $\mu_{22} = 0.15$. 
See Table~\ref{tab:IHR-DHR-E1} for the numerical values of the hazard 
rates $\mu_j(n)$ and Whittle indexes $W_\beta(j,n)$ when $j \in \{1,2\}$ 
and $n \in \{0,1,\ldots,6\}$. For $W_\beta(2,n_2)$, we have used 
Equations~(\ref{eq:Whittle-index-2n-discounted-IHR-DHR}) and 
(\ref{eq:w2-discounted-IHR-DHR}), and for $W_\beta(1,n_1)$ 
Equations~(\ref{eq:Whittle-index-1n-discounted-IHR-DHR}) and 
(\ref{eq:w1-discounted-IHR-DHR}). In addition, we give the related 
values of functions $\psi(n_1,n_2)$ and $\phi(n_1)$ utilizing 
Equations (\ref{eq:psi-discounted-IHR-DHR}) and 
(\ref{eq:phi-discounted-IHR-DHR}), respectively. 
In this example, $n_1^* = 2$, where $n_1^*$ denotes the greatest 
$\bar n_1$ satisfying (\ref{eq:IHR-DHR-E}).

\begin{table}[h]
\caption{Subcase IHR-DHR-E with $n_1^* < \infty$: 
$\mu_2(n_2)$, $W_\beta(2,n_2)$, $\psi(n_1,n_2)$, 
$\mu_1(n_1)$, $\phi(n_1)$, and $W_\beta(1,n_1)$ 
for $n_1, n_2 \in \{0,1,\ldots,6\}$}
\begin{center}
\begin{tabular}{||c||c|c|c|c|c|c|c||}\hline
$n_2$ & 
$0$ & $1$ & $2$ & $3$ & $4$ & $5$ & $6$ \\ 
\hline
$\mu_2(n_2)$ & 
$0.3250$ & $0.2796$ & $0.2400$ & $0.2092$ & $0.1874$ & $0.1730$ & $0.1639$ \\ 
\hline
$W_\beta(2,n_2)$ & 
$\mathit{2.9250}$ & $\mathit{2.5167}$ & $\mathit{2.1598}$ & $\mathit{1.8827}$ & $\mathit{1.6868}$ & $\mathit{1.5573}$ & $\mathit{1.4753}$ \\ 
\hline
$\psi(6,n_2)$ & 
$1.2234$ & $1.4854$ & $1.5637$ & $1.5872$ & $\mathit{1.5922}$ & $1.5909$ & $1.5880$ \\ 
\hline
$\psi(5,n_2)$ & 
$1.1801$ & $1.4432$ & $1.5243$ & $1.5501$ & $1.5567$ & $\mathit{1.5568}$ & $1.5547$ \\ 
\hline
$\psi(4,n_2)$ & 
$1.1247$ & $1.3883$ & $1.4726$ & $1.5012$ & $1.5099$ & $\mathit{1.5115}$ & $1.5106$ \\ 
\hline
$\psi(3,n_2)$ & 
$1.0534$ & $1.3161$ & $1.4040$ & $1.4359$ & $1.4472$ & $1.4508$ & $\mathit{1.4513}$ \\ 
\hline
$\psi(2,n_2)$ & 
$0.9618$ & $1.2207$ & $1.3122$ & $1.3478$ & $1.3622$ & $1.3681$ & $1.3705$ \\ 
\hline
$\psi(1,n_2)$ & 
$0.8446$ & $1.0941$ & $1.1881$ & $1.2277$ & $1.2455$ & $1.2542$ & $1.2586$ \\ 
\hline
$\psi(0,n_2)$ & 
$0.6968$ & $0.9269$ & $1.0203$ & $1.0627$ & $1.0839$ & $1.0953$ & $1.1020$ \\ 
\hline
\multicolumn{8}{c}{ } \\
\hline
$n_1$ & 
$0$ & $1$ & $2$ & $3$ & $4$ & $5$ & $6$ \\ 
\hline
$\mu_1(n_1)$ & 
$0.2000$ & $0.3600$ & $0.4880$ & $0.5904$ & $0.6723$ & $0.7379$ & $0.7903$ \\ 
\hline
$\phi(n_1)$ & 
$\infty$ & $\infty$ & $9$ & $6$ & $5$ & $5$ & $4$ \\ 
\hline
$W_\beta(1,n_1)$ & 
$\mathit{1.1169}$ & $\mathit{1.2660}$ & $\mathit{1.3717}$ & $\mathit{1.4513}$ & $\mathit{1.5115}$ & $\mathit{1.5568}$ & $\mathit{1.5922}$ \\ 
\hline
\end{tabular}
\end{center}
\label{tab:IHR-DHR-E1}
\end{table}

The last example belongs also to the IHR-DHR-E subcase but now $n_1^* = \infty$. 
The monotonous hazard rate functions for the two stages are given by 
\[
\begin{split}
& 
\mu_1(n) = 1 - \alpha_1^{n+1}, 
\quad 
n \in \{0,1,\ldots\}. \\
& 
\mu_2(n) = 
\frac{p_2 (1 - \mu_{21})^n \mu_{21} + (1 - p_2) (1 - \mu_{22})^n \mu_{22}}
{p_2 (1 - \mu_{21})^n + (1 - p_2) (1 - \mu_{22})^n}, 
\quad 
n \in \{0,1,\ldots\}.
\end{split}
\]
where $\alpha_1 = 0.5$, $p_2 = 0.5$, $\mu_{21} = 0.5$, and $\mu_{22} = 0.3$. 
See Table~\ref{tab:IHR-DHR-E2} for the numerical values of the hazard 
rates $\mu_j(n)$ and Whittle indexes $W_\beta(j,n)$ when $j \in \{1,2\}$ 
and $n \in \{0,1,\ldots,6\}$. For $W_\beta(2,n_2)$, we have used 
Equations~(\ref{eq:Whittle-index-2n-discounted-IHR-DHR}) and 
(\ref{eq:w2-discounted-IHR-DHR}), and for $W_\beta(1,n_1)$ 
Equations~(\ref{eq:Whittle-index-1n-discounted-IHR-DHR}) and 
(\ref{eq:w1-discounted-IHR-DHR}). In addition, we give the related 
values of functions $\psi(n_1,n_2)$ and $\phi(n_1)$ utilizing 
Equations (\ref{eq:psi-discounted-IHR-DHR}) and 
(\ref{eq:phi-discounted-IHR-DHR}), respectively.

\begin{table}[h]
\caption{Subcase IHR-DHR-E with $n_1^* = \infty$: 
$\mu_2(n_2)$, $W_\beta(2,n_2)$, $\psi(n_1,n_2)$, 
$\mu_1(n_1)$, $\phi(n_1)$, and $W_\beta(1,n_1)$ 
for $n_1, n_2 \in \{0,1,\ldots,6\}$}
\begin{center}
\begin{tabular}{||c||c|c|c|c|c|c|c||}
\hline
$n_2$ & 
$0$ & $1$ & $2$ & $3$ & $4$ & $5$ & $6$ \\ 
\hline
$\mu_2(n_2)$ & 
$0.4000$ & $0.3833$ & $0.3676$ & $0.3534$ & $0.3413$ & $0.3314$ & $0.3234$ \\ 
\hline
$W_\beta(2,n_2)$ & 
$\mathit{3.6000}$ & $\mathit{3.4500}$ & $\mathit{3.3081}$ & $\mathit{3.1808}$ & $\mathit{3.0718}$ & $\mathit{2.9822}$ & $\mathit{2.9110}$ \\ 
\hline
$\psi(6,n_2)$ & 
$1.6983$ & $2.0539$ & $2.1809$ & $2.2354$ & $2.2611$ & $2.2740$ & $2.2808$ \\ 
\hline
$\psi(5,n_2)$ & 
$1.6912$ & $2.0471$ & $2.1744$ & $2.2291$ & $2.2550$ & $2.2679$ & $2.2748$ \\ 
\hline
$\psi(4,n_2)$ & 
$1.6772$ & $2.0336$ & $2.1615$ & $2.2166$ & $2.2426$ & $2.2558$ & $2.2627$ \\ 
\hline
$\psi(3,n_2)$ & 
$1.6491$ & $2.0062$ & $2.1353$ & $2.1912$ & $2.2177$ & $2.2312$ & $2.2383$ \\ 
\hline
$\psi(2,n_2)$ & 
$1.5929$ & $1.9511$ & $2.0823$ & $2.1396$ & $2.1670$ & $2.1810$ & $2.1885$ \\ 
\hline
$\psi(1,n_2)$ & 
$1.4809$ & $1.8388$ & $1.9735$ & $2.0333$ & $2.0624$ & $2.0775$ & $2.0856$ \\ 
\hline
$\psi(0,n_2)$ & 
$1.2618$ & $1.6101$ & $1.7479$ & $1.8112$ & $1.8428$ & $1.8594$ & $1.8686$ \\ 
\hline
\multicolumn{8}{c}{ } \\
\hline
$n_1$ & 
$0$ & $1$ & $2$ & $3$ & $4$ & $5$ & $6$ \\ 
\hline
$\mu_1(n_1)$ & 
$0.5000$ & $0.7500$ & $0.8750$ & $0.9375$ & $0.9688$ & $0.9844$ & $0.9922$ \\ 
\hline
$\phi(n_1)$ & 
$\infty$ & $\infty$ & $\infty$ & $\infty$ & $\infty$ & $\infty$ & $\infty$ \\ 
\hline
$W_\beta(1,n_1)$ & 
$\mathit{1.8815}$ & $\mathit{2.0967}$ & $\mathit{2.1985}$ & $\mathit{2.2476}$ & $\mathit{2.2717}$ & $\mathit{2.2837}$ & $\mathit{2.2896}$ \\ 
\hline
\end{tabular}
\end{center}
\label{tab:IHR-DHR-E2}
\end{table}

\section{Examples on Whittle index values for sequential multistage jobs}
\label{app:whittle-index-examples-average-multistage}

In this appendix, we give numerical examples that support our 
Conjectures~\ref{con:Gittins-index-average-two-stage} and 
\ref{con:Gittins-index-average-multistage} presented at the end of 
Section~\ref{sec:average-cost}. The idea is to confirm that the Gittins 
index values that are computed according to these conjectures are equal to 
the Gittins index values that are computed using the known method 
presented by Scully et al.\ in \cite{Scu18arXiv}.

\begin{example}
\label{ex:Gittins-index-average-two-stage-homogen}
{\rm 
We start with an example related to 
Conjecture~\ref{con:Gittins-index-average-two-stage}. 
Consider a sequential two-stage job with the following discrete-time 
nonmonotonous hazard rate $\mu_j(n)$ in both stages $j \in \{1,2\}$: 
\[
\mu_j(0) = 1/2, \quad 
\mu_j(1) = 0, \quad  
\mu_j(2) = 0, \quad  
\mu_j(3) = 1.
\]
It follows that, for both stages $j \in \{1,2\}$, 
\[
P\{S_j = 1\} = P\{S_j = 4\} = 1/2, \quad 
E[S_j] = 5/2, 
\]
and the total service time $S = S_1 + S_2$ satisfies 
\[
P\{S = 2\} = 1/4, \quad 
P\{S = 5\} = 1/2, \quad 
P\{S = 8\} = 1/4, \quad 
E[S] = 5.
\]

\vskip 6pt 

\noindent 
{\em \ref{ex:Gittins-index-average-two-stage-homogen}a) 
Gittins index values computed according to 
Conjecture~\ref{con:Gittins-index-average-two-stage}}
\ 

\vskip 6pt 

\noindent 
Gittins index $G_j(n_j)$ for individual stages $j \in \{1,2\}$: 
\[
\begin{split}
& 
G_j(0) = 
\max_{n \in \{1,2,3,4\}} 
\frac{P\{S_j \le n\}}{E[\min \{S_j, n\}]} = 
\max \left\{ \frac{1}{2}, \frac{1}{3}, \frac{1}{4}, \frac{2}{5} \right\} = 
\frac{1}{2}, \\
& 
G_j(1) = 
\max_{n \in \{1,2,3\}} 
\frac{P\{S_j - 1\le n \mid S_j > 1\}}{E[\min \{S_j - 1, n\} \mid S_j > 1]} = 
\max \left\{ 0, 0, \frac{1}{3} \right\} = 
\frac{1}{3}, \\
& 
G_j(2) = 
\max_{n \in \{1,2\}} 
\frac{P\{S_j - 2 \le n \mid S_j > 2\}}{E[\min \{S_j - 2, n\} \mid S_j > 2]} = 
\max \left\{ 0, \frac{1}{2} \right\} = 
\frac{1}{2}, \\
& 
G_j(3) = 
\frac{P\{S_j - 3 \le 1 \mid S_j > 3\}}{E[\min \{S_j - 3, 1\} \mid S_j > 3]} = 
1.
\end{split}
\]
Gittins index $G(2,n_2)$ for the second stage of the sequential two-stage job: 
\[
\begin{split}
& 
G(2,0) = G_2(0) = \frac{1}{2}, 
\quad 
G(2,1) = G_2(1) = \frac{1}{3}, \\
& 
G(2,2) = G_2(2) = \frac{1}{2}, 
\quad 
G(2,3) = G_2(3) = 1.
\end{split}
\]
Gittins index $G(1,n_1)$ for the first stage of the sequential two-stage job: 
\[
\begin{split}
& 
G(1,0) = 
\max_{n \in \{1,2,3,4\}} 
\frac{P\{S_2 \le n\}}{\frac{1}{G_1(0)} + E[\min \{S_2, n\}]} = 
\max \left\{ \frac{1}{6}, \frac{1}{7}, \frac{1}{8}, \frac{2}{9} \right\} = 
\frac{2}{9}, \\
& 
G(1,1) = 
\max_{n \in \{1,2,3,4\}} 
\frac{P\{S_2 \le n\}}{\frac{1}{G_1(1)} + E[\min \{S_2, n\}]} = 
\max \left\{ \frac{1}{8}, \frac{1}{9}, \frac{1}{10}, \frac{2}{11} \right\} = 
\frac{2}{11}, \\
& 
G(1,2) = 
\max_{n \in \{1,2,3,4\}} 
\frac{P\{S_2 \le n\}}{\frac{1}{G_1(2)} + E[\min \{S_2, n\}]} = 
\max \left\{ \frac{1}{6}, \frac{1}{7}, \frac{1}{8}, \frac{2}{9} \right\} = 
\frac{2}{9}, \\
& 
G(1,3) = 
\max_{n \in \{1,2,3,4\}} 
\frac{P\{S_2 \le n\}}{\frac{1}{G_1(3)} + E[\min \{S_2, n\}]} = 
\max \left\{ \frac{1}{4}, \frac{1}{5}, \frac{1}{6}, \frac{2}{7} \right\} = 
\frac{2}{7}.
\end{split}
\]

\vskip 6pt 

\noindent 
{\em \ref{ex:Gittins-index-average-two-stage-homogen}b) 
Gittins index values computed using the method presented 
in \cite{Scu18arXiv}}
\ 

\vskip 6pt 

\noindent 
SJP function $V_j^{\mathrm{SJP}}(r;n_j)$ for individual stages $j \in \{1,2\}$: 
\[
\begin{split}
& 
V_j^{\mathrm{SJP}}(r;0) = 
\max_{n \in \{1,2,3,4\}} 
\Big( 
r P\{S_j \le n\} - E[\min \{S_j, n\}] 
\Big) \\
& \quad = \; 
\max \left\{ \frac{r-2}{2}, \frac{r-3}{2}, \frac{r-4}{2}, \frac{2r-5}{2} \right\} = 
\max \left\{ \frac{r}{2} - 1, r - \frac{5}{2} \right\}, \\
& 
V_j^{\mathrm{SJP}}(r;1) = 
\max_{n \in \{1,2,3\}} 
\Big( 
r P\{S_j - 1 \le n \mid S_j > 1\} - E[\min \{S_j - 1, n\} \mid S_j > 1] 
\Big) \\
& \quad = \; 
\max \left\{ -1, -2, r-3 \right\} = 
\max \left\{ -1, r-3 \right\}, \\
& 
V_j^{\mathrm{SJP}}(r;2) = 
\max_{n \in \{1,2\}} 
\Big( 
r P\{S_j - 2 \le n \mid S_j > 2\} - E[\min \{S_j - 2, n\} \mid S_j > 2] 
\Big) \\
& \quad = \; 
\max \left\{ -1, r-2 \right\}, \\
& 
V_j^{\mathrm{SJP}}(r;3) = 
\Big( 
r P\{S_j - 3 \le 1 \mid S_j > 3\} - E[\min \{S_j - 3, 1\} \mid S_j > 3] 
\Big) = 
r-1.
\end{split}
\]
SJP function $V^{\mathrm{SJP}}(r;2,n_2)$ for the second stage of the sequential 
two-stage job: 
\[
\begin{split}
& 
V^{\mathrm{SJP}}(r;2,0) = V_2^{\mathrm{SJP}}(r;0) = 
\max \left\{ \frac{r}{2} - 1, r - \frac{5}{2} \right\}, \\
& 
V^{\mathrm{SJP}}(r;2,1) = V_2^{\mathrm{SJP}}(r;1) = 
\max \left\{ -1, r-3 \right\}, \\
& 
V^{\mathrm{SJP}}(r;2,2) = V_2^{\mathrm{SJP}}(r;2) = 
\max \left\{ -1, r-2 \right\}, \\
& 
V^{\mathrm{SJP}}(r;2,3) = V_2^{\mathrm{SJP}}(r;3) = 
r-1.
\end{split}
\]
SJP function $V^{\mathrm{SJP}}(r;1,n_1)$ for the first stage of the sequential 
two-stage job: 
\[
\begin{split}
& 
V^{\mathrm{SJP}}(r;1,0) = 
V_1^{\mathrm{SJP}}(V_2^{\mathrm{SJP}}(r;0);0) = 
\max \left\{ \frac{r}{4}-\frac{3}{2}, \frac{r}{2}-\frac{9}{4}, r-5 \right\}, \\
& 
V^{\mathrm{SJP}}(r;1,1) = 
V_1^{\mathrm{SJP}}(V_2^{\mathrm{SJP}}(r;0);1) = 
\max \left\{ -1, \frac{r}{2}-4, r-\frac{11}{2} \right\}, \\
& 
V^{\mathrm{SJP}}(r;1,2) = 
V_1^{\mathrm{SJP}}(V_2^{\mathrm{SJP}}(r;0);2) = 
\max \left\{ -1, \frac{r}{2}-3, r-\frac{9}{2} \right\}, \\
& 
V^{\mathrm{SJP}}(r;1,3) = 
V_1^{\mathrm{SJP}}(V_2^{\mathrm{SJP}}(r;0);3) = 
\max \left\{ \frac{r}{2}-2, r-\frac{7}{2} \right\}.
\end{split}
\]
Gittins index $G_j(n_j)$ for individual stages $j \in \{1,2\}$: 
\[
\begin{split}
& 
G_j(0) = 
\frac{1}{\inf \{r \ge 0 : V_j^{\mathrm{SJP}}(r;0) > 0\}} = \frac{1}{2}, \\
& 
G_j(1) = 
\frac{1}{\inf \{r \ge 0 : V_j^{\mathrm{SJP}}(r;1) > 0\}} = \frac{1}{3}, \\
& 
G_j(2) = 
\frac{1}{\inf \{r \ge 0 : V_j^{\mathrm{SJP}}(r;2) > 0\}} = \frac{1}{2}, \\
& 
G_j(3) = 
\frac{1}{\inf \{r \ge 0 : V_j^{\mathrm{SJP}}(r;3) > 0\}} = 1.
\end{split}
\]
Gittins index $G(2,n_2)$ for the second stage of the sequential two-stage job: 
\[
\begin{split}
& 
G(2,0) = 
\frac{1}{\inf \{r \ge 0 : V^{\mathrm{SJP}}(r;2,0) > 0\}} = \frac{1}{2}, \\
&  
G(2,1) = 
\frac{1}{\inf \{r \ge 0 : V^{\mathrm{SJP}}(r;2,1) > 0\}} = \frac{1}{3}, \\
& 
G(2,2) = 
\frac{1}{\inf \{r \ge 0 : V^{\mathrm{SJP}}(r;2,2) > 0\}} = \frac{1}{2}, \\
& 
G(2,3) = 
\frac{1}{\inf \{r \ge 0 : V^{\mathrm{SJP}}(r;2,3) > 0\}} = 1.
\end{split}
\]
Gittins index $G(1,n_1)$ for the first stage of the sequential two-stage job: 
\[
\begin{split}
& 
G(1,0) = 
\frac{1}{\inf \{r \ge 0 : V^{\mathrm{SJP}}(r;1,0) > 0\}} = \frac{2}{9}, \\
&  
G(1,1) = 
\frac{1}{\inf \{r \ge 0 : V^{\mathrm{SJP}}(r;1,1) > 0\}} = \frac{2}{11}, \\
& 
G(1,2) = 
\frac{1}{\inf \{r \ge 0 : V^{\mathrm{SJP}}(r;1,2) > 0\}} = \frac{2}{9}, \\
& 
G(1,3) = 
\frac{1}{\inf \{r \ge 0 : V^{\mathrm{SJP}}(r;1,3) > 0\}} = \frac{2}{7}.
\end{split}
\]
As seen from above, all the Gittins index values computed in the two different methods are equal.
}
\end{example}

\begin{example}
\label{ex:Gittins-index-average-two-stage-heterogen}
{\rm 
We continue with an example which is also related to 
Conjecture~\ref{con:Gittins-index-average-two-stage}. 
Consider a sequential two-stage job with the following discrete-time 
nonmonotonous hazard rate $\mu_2(n)$ in the second stage: 
\[
\mu_2(0) = 1/4, \quad 
\mu_2(1) = 3/4, \quad  
\mu_2(2) = 0,   \quad  
\mu_2(3) = 1.
\]
In the first stage, we use the same discrete-time nonmonotonous hazard rate 
$\mu_1(n)$ as in Example~\ref{ex:Gittins-index-average-two-stage-homogen}: 
\[
\mu_1(0) = 1/2, \quad 
\mu_1(1) = 0, \quad  
\mu_1(2) = 0, \quad  
\mu_1(3) = 1.
\]
Below we focus on the computation of the Gittins index for the first stage, 
which is the most interesting one from the conjecture point of view.

\vskip 6pt 

\noindent 
{\em \ref{ex:Gittins-index-average-two-stage-heterogen}a) 
Gittins index values computed according to 
Conjecture~\ref{con:Gittins-index-average-two-stage}}
\ 

\vskip 6pt 

\noindent 
Gittins index $G_1(n_1)$ for individual stage $1$ according to 
Example~\ref{ex:Gittins-index-average-two-stage-homogen}: 
\[
G_1(0) = \frac{1}{2}, \quad 
G_1(1) = \frac{1}{3}, \quad 
G_1(2) = \frac{1}{2}, \quad 
G_1(3) = 1.
\]
Gittins index $G(1,n_1)$ for the first stage of the sequential two-stage job: 
\[
\begin{split}
& 
G(1,0) = 
\max_{n \in \{1,2,3,4\}} 
\frac{P\{S_2 \le n\}}{\frac{1}{G_1(0)} + E[\min \{S_2, n\}]} = 
\max \left\{ \frac{1}{12}, \frac{13}{60}, \frac{13}{63}, \frac{8}{33} \right\} = 
\frac{8}{33}, \\
& 
G(1,1) = 
\max_{n \in \{1,2,3,4\}} 
\frac{P\{S_2 \le n\}}{\frac{1}{G_1(1)} + E[\min \{S_2, n\}]} = 
\max \left\{ \frac{1}{16}, \frac{13}{76}, \frac{13}{79}, \frac{8}{41} \right\} = 
\frac{8}{41}, \\
& 
G(1,2) = 
\max_{n \in \{1,2,3,4\}} 
\frac{P\{S_2 \le n\}}{\frac{1}{G_1(2)} + E[\min \{S_2, n\}]} = 
\max \left\{ \frac{1}{12}, \frac{13}{60}, \frac{13}{63}, \frac{8}{33} \right\} = 
\frac{8}{33}, \\
& 
G(1,3) = 
\max_{n \in \{1,2,3,4\}} 
\frac{P\{S_2 \le n\}}{\frac{1}{G_1(3)} + E[\min \{S_2, n\}]} = 
\max \left\{ \frac{1}{8}, \frac{13}{44}, \frac{13}{47}, \frac{8}{25} \right\} = 
\frac{8}{25}.
\end{split}
\]

\vskip 6pt 

\noindent 
{\em \ref{ex:Gittins-index-average-two-stage-heterogen}b) 
Gittins index values computed using the method presented 
in \cite{Scu18arXiv}}
\ 

\vskip 6pt 

\noindent 
SJP function $V_1^{\mathrm{SJP}}(r;n_1)$ for individual stage $1$ according to 
Example~\ref{ex:Gittins-index-average-two-stage-homogen}: 
\[
\begin{split}
& 
V_j^{\mathrm{SJP}}(r;0) = 
\max \left\{ \frac{r}{2} - 1, r - \frac{5}{2} \right\}, \quad 
V_j^{\mathrm{SJP}}(r;1) = 
\max \left\{ -1, r-3 \right\}, \\
& 
V_j^{\mathrm{SJP}}(r;2) = 
\max \left\{ -1, r-2 \right\}, \quad 
V_j^{\mathrm{SJP}}(r;3) = 
r-1.
\end{split}
\]
SJP function $V_2^{\mathrm{SJP}}(r;n_2)$ for individual stage $2$: 
\[
\begin{split}
& 
V_2^{\mathrm{SJP}}(r;0) = 
\max_{n \in \{1,2,3,4\}} 
\Big( 
r P\{S_2 \le n\} - E[\min \{S_2, n\}] 
\Big) \\
& \quad = \; 
\max \left\{ \frac{r-4}{4}, \frac{13r-28}{16}, \frac{13r-31}{16}, \frac{8r-17}{8} \right\} = 
\max \left\{ \frac{r}{4} - 1, \frac{13r}{16}-\frac{7}{4}, r - \frac{17}{8} \right\}, \\
& 
V_2^{\mathrm{SJP}}(r;1) = 
\max_{n \in \{1,2,3\}} 
\Big( 
r P\{S_2 - 1 \le n \mid S_2 > 1\} - E[\min \{S_2 - 1, n\} \mid S_2 > 1] 
\Big) \\
& \quad = \; 
\max \left\{ \frac{3r-4}{4}, \frac{3r-5}{4}, \frac{2r-3}{2} \right\} = 
\max \left\{ \frac{3r}{4} - 1, r - \frac{3}{2} \right\}, \\
& 
V_2^{\mathrm{SJP}}(r;2) = 
\max_{n \in \{1,2\}} 
\Big( 
r P\{S_2 - 2 \le n \mid S_2 > 2\} - E[\min \{S_2 - 2, n\} \mid S_2 > 2] 
\Big) \\
& \quad = \; 
\max \left\{ -1, r-2 \right\}, \\
& 
V_2^{\mathrm{SJP}}(r;3) = 
\Big( 
r P\{S_2 - 3 \le 1 \mid S_2 > 3\} - E[\min \{S_2 - 3, 1\} \mid S_2 > 3] 
\Big) = 
r-1.
\end{split}
\]
SJP function $V^{\mathrm{SJP}}(r;1,n_1)$ for the first stage of the sequential 
two-stage job: 
\[
\begin{split}
& 
V^{\mathrm{SJP}}(r;1,0) = 
V_1^{\mathrm{SJP}}(V_2^{\mathrm{SJP}}(r;0);0) \\
& \quad = \; 
\max \left\{ \frac{r}{8}-\frac{3}{2}, \frac{13r}{32}-\frac{15}{8}, \frac{r}{2}-\frac{33}{16}, 
\frac{r}{4}-\frac{7}{2}, \frac{13r}{16}-\frac{17}{4}, r-\frac{37}{8} \right\}, \\
& 
V^{\mathrm{SJP}}(r;1,1) = 
V_1^{\mathrm{SJP}}(V_2^{\mathrm{SJP}}(r;0);1) = 
\max \left\{ -1, \frac{r}{4}-4, \frac{13r}{16}-\frac{19}{4}, r-\frac{41}{8} \right\}, \\
& 
V^{\mathrm{SJP}}(r;1,2) = 
V_1^{\mathrm{SJP}}(V_2^{\mathrm{SJP}}(r;0);2) = 
\max \left\{ -1, \frac{r}{4}-3, \frac{13r}{16}-\frac{15}{4}, r-\frac{33}{8} \right\}, \\
& 
V^{\mathrm{SJP}}(r;1,3) = 
V_1^{\mathrm{SJP}}(V_2^{\mathrm{SJP}}(r;0);3) = 
\max \left\{ \frac{r}{4}-2, \frac{13r}{16}-\frac{11}{4}, r-\frac{25}{8} \right\}.
\end{split}
\]
Gittins index $G(1,n_1)$ for the first stage of the sequential two-stage job: 
\[
\begin{split}
& 
G(1,0) = 
\frac{1}{\inf \{r \ge 0 : V^{\mathrm{SJP}}(r;1,0) > 0\}} = \frac{8}{33}, \\
&  
G(1,1) = 
\frac{1}{\inf \{r \ge 0 : V^{\mathrm{SJP}}(r;1,1) > 0\}} = \frac{8}{41}, \\
& 
G(1,2) = 
\frac{1}{\inf \{r \ge 0 : V^{\mathrm{SJP}}(r;1,2) > 0\}} = \frac{8}{33}, \\
& 
G(1,3) = 
\frac{1}{\inf \{r \ge 0 : V^{\mathrm{SJP}}(r;1,3) > 0\}} = \frac{8}{25}.
\end{split}
\]
As seen from above, all the Gittins index values computed in the two different methods are equal.
}
\end{example}

\begin{example}
\label{ex:Gittins-index-average-multistage-homogen}
{\rm 
Next we give an example related to Conjecture~\ref{con:Gittins-index-average-multistage}. 
Consider a sequential three-stage job with the following discrete-time 
nonmonotonous hazard rate $\mu_j(n)$ in all stages $j \in \{1,2,3\}$: 
\[
\mu_j(0) = 1/2, \quad 
\mu_j(1) = 0, \quad  
\mu_j(2) = 0, \quad  
\mu_j(3) = 1.
\]
It follows that, for all stages $j \in \{1,2,3\}$, 
\[
P\{S_j = 1\} = P\{S_j = 4\} = 1/2, \quad 
E[S_j] = 5/2, 
\]
and the total service time $S = S_1 + S_2 + S_3$ satisfies 
\[
P\{S = 3\} = 1/8, \quad 
P\{S = 6\} = P\{S = 9\} = 3/8, \quad 
P\{S = 12\} = 1/8, \quad 
E[S] = 15/2.
\]
Note that the service time distribution in each individual stage is the same 
as in Example~\ref{ex:Gittins-index-average-two-stage-homogen}. The difference comes 
from the different number of stages. 
Below we focus on the computation of the Gittins index for the first stage, 
which is the most interesting one from the conjecture point of view.

\vskip 6pt 

\noindent 
{\em \ref{ex:Gittins-index-average-multistage-homogen}a) 
Gittins index values computed according to 
Conjecture~\ref{con:Gittins-index-average-multistage}}
\ 

\vskip 6pt 

\noindent 
Gittins index $G(1,n_1;1,2)$ for the first stage of the sequential two-stage job 
that consists of stages $1$ and $2$ of the sequential three-stage job (according to 
Example~\ref{ex:Gittins-index-average-two-stage-homogen}): 
\[
G(1,0;1,2) = \frac{2}{9}, \quad 
G(1,1;1,2) = \frac{2}{11}, \quad 
G(1,2;1,2) = \frac{2}{9}, \quad 
G(1,3;1,2) = \frac{2}{7}.
\]
Gittins index $G(1,n_1)$ for the first stage of the sequential three-stage job: 
\[
\begin{split}
& 
G(1,0) = 
\max_{n \in \{1,2,3,4\}} 
\frac{P\{S_3 \le n\}}{\frac{1}{G(1,0;1,2)} + E[\min \{S_3, n\}]} = 
\max \left\{ \frac{1}{11}, \frac{1}{12}, \frac{1}{13}, \frac{1}{7} \right\} = 
\frac{1}{7}, \\
& 
G(1,1) = 
\max_{n \in \{1,2,3,4\}} 
\frac{P\{S_3 \le n\}}{\frac{1}{G(1,1;1,2)} + E[\min \{S_3, n\}]} = 
\max \left\{ \frac{1}{13}, \frac{1}{14}, \frac{1}{15}, \frac{1}{8} \right\} = 
\frac{1}{8}, \\
& 
G(1,2) = 
\max_{n \in \{1,2,3,4\}} 
\frac{P\{S_3 \le n\}}{\frac{1}{G(1,2;1,2)} + E[\min \{S_3, n\}]} = 
\max \left\{ \frac{1}{11}, \frac{1}{12}, \frac{1}{13}, \frac{1}{7} \right\} = 
\frac{1}{7}, \\
& 
G(1,3) = 
\max_{n \in \{1,2,3,4\}} 
\frac{P\{S_3 \le n\}}{\frac{1}{G(1,3;1,2)} + E[\min \{S_3, n\}]} = 
\max \left\{ \frac{1}{9}, \frac{1}{10}, \frac{1}{11}, \frac{1}{6} \right\} = 
\frac{1}{6}.
\end{split}
\]

\vskip 6pt 

\noindent 
{\em \ref{ex:Gittins-index-average-multistage-homogen}b) 
Gittins index values computed using the method presented 
in \cite{Scu18arXiv}}
\ 

\vskip 6pt 

\noindent 
SJP function $V_3^{\mathrm{SJP}}(r;0)$ for individual stage $3$ and 
attained service $n_3 = 0$ (according to 
Example~\ref{ex:Gittins-index-average-two-stage-homogen}): 
\[
V_3^{\mathrm{SJP}}(r;0) = 
\max \left\{ \frac{r}{2} - 1, r - \frac{5}{2} \right\}.
\]
SJP function $V_{1,2}^{\mathrm{SJP}}(r;1,n_1)$ for the first stage of the 
sequential two-stage job that consists of stages $1$ and $2$ of the sequential 
three-stage job (according to 
Example~\ref{ex:Gittins-index-average-two-stage-homogen}): 
\[
\begin{split}
& 
V_{1,2}^{\mathrm{SJP}}(r;1,0) = 
\max \left\{ \frac{r}{4}-\frac{3}{2}, \frac{r}{2}-\frac{9}{4}, r-5 \right\}, \\
& 
V_{1,2}^{\mathrm{SJP}}(r;1,1) = 
\max \left\{ -1, \frac{r}{2}-4, r-\frac{11}{2} \right\}, \\
& 
V_{1,2}^{\mathrm{SJP}}(r;1,2) = 
\max \left\{ -1, \frac{r}{2}-3, r-\frac{9}{2} \right\}, \\
& 
V_{1,2}^{\mathrm{SJP}}(r;1,3) = 
\max \left\{ \frac{r}{2}-2, r-\frac{7}{2} \right\}.
\end{split}
\]
SJP function $V^{\mathrm{SJP}}(r;1,n_1)$ for the first stage of the sequential 
three-stage job: 
\[
\begin{split}
& 
V^{\mathrm{SJP}}(r;1,0) = 
V_{1,2}^{\mathrm{SJP}}(V_3^{\mathrm{SJP}}(r;0);1,0) = 
\max \left\{ \frac{r}{8}-\frac{7}{4}, \frac{r}{4}-\frac{17}{8}, \frac{r}{2}-\frac{7}{2}, r-\frac{15}{2} \right\}, \\
& 
V^{\mathrm{SJP}}(r;1,1) = 
V_{1,2}^{\mathrm{SJP}}(V_3^{\mathrm{SJP}}(r;0);1,1) = 
\max \left\{ -1, \frac{r}{4}-\frac{9}{2}, \frac{r}{2}-\frac{21}{4}, r-8 \right\}, \\
& 
V^{\mathrm{SJP}}(r;1,2) = 
V_{1,2}^{\mathrm{SJP}}(V_3^{\mathrm{SJP}}(r;0);1,2) = 
\max \left\{ -1, \frac{r}{4}-\frac{7}{2}, \frac{r}{2}-\frac{17}{4}, r-7 \right\}, \\
& 
V^{\mathrm{SJP}}(r;1,3) = 
V_{1,2}^{\mathrm{SJP}}(V_3^{\mathrm{SJP}}(r;0);1,3) = 
\max \left\{ \frac{r}{4}-\frac{5}{2}, \frac{r}{2}-\frac{13}{4}, r-6 \right\}.
\end{split}
\]
Gittins index $G(1,n_1)$ for the first stage of the sequential two-stage job: 
\[
\begin{split}
& 
G(1,0) = 
\frac{1}{\inf \{r \ge 0 : V^{\mathrm{SJP}}(r;1,0) > 0\}} = \frac{1}{7}, \\
&  
G(1,1) = 
\frac{1}{\inf \{r \ge 0 : V^{\mathrm{SJP}}(r;1,1) > 0\}} = \frac{1}{8}, \\
& 
G(1,2) = 
\frac{1}{\inf \{r \ge 0 : V^{\mathrm{SJP}}(r;1,2) > 0\}} = \frac{1}{7}, \\
& 
G(1,3) = 
\frac{1}{\inf \{r \ge 0 : V^{\mathrm{SJP}}(r;1,3) > 0\}} = \frac{1}{6}.
\end{split}
\]
As seen from above, all the Gittins index values computed in the two different methods are equal.
}
\end{example}

\begin{example}
\label{ex:Gittins-index-average-multistage-heterogen}
{\rm 
Our next example is also related to Conjecture~\ref{con:Gittins-index-average-multistage}. 
Consider a sequential three-stage job with the following discrete-time 
nonmonotonous hazard rate $\mu_3(n)$ in the third stage: 
\[
\mu_3(0) = 3/4, \quad 
\mu_3(1) = 1/4, \quad  
\mu_3(2) = 1/2,   \quad  
\mu_3(3) = 1.
\]
In the first two stages, we use the same discrete-time nonmonotonous hazard rates 
$\mu_1(n)$ and $\mu_2(n)$ as in Example~\ref{ex:Gittins-index-average-two-stage-heterogen}: 
\[
\begin{split}
& 
\mu_1(0) = 1/2, \quad 
\mu_1(1) = 0, \quad  
\mu_1(2) = 0, \quad  
\mu_1(3) = 1 \\
& 
\mu_2(0) = 1/4, \quad 
\mu_2(1) = 3/4, \quad  
\mu_2(2) = 0,   \quad  
\mu_2(3) = 1.
\end{split}
\]
Below we focus on the computation of the Gittins index for the first stage, 
which is the most interesting one from the conjecture point of view.

\vskip 6pt 

\noindent 
{\em \ref{ex:Gittins-index-average-multistage-heterogen}a) 
Gittins index values computed according to 
Conjecture~\ref{con:Gittins-index-average-multistage}}
\ 

\vskip 6pt 

\noindent 
Gittins index $G(1,n_1;1,2)$ for the first stage of the sequential two-stage job 
that consists of stages $1$ and $2$ of the sequential three-stage job (according to 
Example~\ref{ex:Gittins-index-average-two-stage-heterogen}): 
\[
G(1,0;1,2) = \frac{8}{33}, \quad 
G(1,1;1,2) = \frac{8}{41}, \quad 
G(1,2;1,2) = \frac{8}{33}, \quad 
G(1,3;1,2) = \frac{8}{25}.
\]
Gittins index $G(1,n_1)$ for the first stage of the sequential three-stage job: 
\[
\begin{split}
& 
G(1,0) = 
\max_{n \in \{1,2,3,4\}} 
\frac{P\{S_3 \le n\}}{\frac{1}{G(1,0;1,2)} + E[\min \{S_3, n\}]} = 
\max \left\{ \frac{6}{41}, \frac{13}{86}, \frac{29}{178}, \frac{32}{181} \right\} = 
\frac{32}{181}, \\
& 
G(1,1) = 
\max_{n \in \{1,2,3,4\}} 
\frac{P\{S_3 \le n\}}{\frac{1}{G(1,1;1,2)} + E[\min \{S_3, n\}]} = 
\max \left\{ \frac{6}{49}, \frac{13}{102}, \frac{29}{210}, \frac{32}{213} \right\} = 
\frac{32}{213}, \\
& 
G(1,2) = 
\max_{n \in \{1,2,3,4\}} 
\frac{P\{S_3 \le n\}}{\frac{1}{G(1,2;1,2)} + E[\min \{S_3, n\}]} = 
\max \left\{ \frac{6}{41}, \frac{13}{86}, \frac{29}{178}, \frac{32}{181} \right\} = 
\frac{32}{181}, \\
& 
G(1,3) = 
\max_{n \in \{1,2,3,4\}} 
\frac{P\{S_3 \le n\}}{\frac{1}{G(1,3;1,2)} + E[\min \{S_3, n\}]} = 
\max \left\{ \frac{2}{11}, \frac{13}{70}, \frac{29}{146}, \frac{32}{149} \right\} = 
\frac{32}{149}.
\end{split}
\]

\vskip 6pt 

\noindent 
{\em \ref{ex:Gittins-index-average-multistage-heterogen}b) 
Gittins index values computed using the method presented 
in \cite{Scu18arXiv}}
\ 

\vskip 6pt 

\noindent 
SJP function $V_3^{\mathrm{SJP}}(r;0)$ for individual stage $3$ with attained 
service $n_3 = 0$: 
\[
\begin{split}
& 
V_3^{\mathrm{SJP}}(r;0) = 
\max_{n \in \{1,2,3,4\}} 
\Big( 
r P\{S_3 \le n\} - E[\min \{S_3, n\}] 
\Big) \\
& \quad = \; 
\max \left\{ \frac{3r}{4}-1, \frac{13r}{16}-\frac{5}{4}, \frac{29r}{32}-\frac{23}{16}, r-\frac{49}{32} \right\}.
\end{split}
\]
SJP function $V_{1,2}^{\mathrm{SJP}}(r;1,n_1)$ for the first stage of the 
sequential two-stage job that consists of stages $1$ and $2$ of the sequential 
three-stage job (according to 
Example~\ref{ex:Gittins-index-average-two-stage-heterogen}): 
\[
\begin{split}
& 
V_{1,2}^{\mathrm{SJP}}(r;1,0) = 
\max \left\{ \frac{r}{8}-\frac{3}{2}, \frac{13r}{32}-\frac{15}{8}, \frac{r}{2}-\frac{33}{16}, 
\frac{r}{4}-\frac{7}{2}, \frac{13r}{16}-\frac{17}{4}, r-\frac{37}{8} \right\}, \\
& 
V_{1,2}^{\mathrm{SJP}}(r;1,1) = 
\max \left\{ -1, \frac{r}{4}-4, \frac{13r}{16}-\frac{19}{4}, r-\frac{41}{8} \right\}, \\
& 
V_{1,2}^{\mathrm{SJP}}(r;1,2) = 
\max \left\{ -1, \frac{r}{4}-3, \frac{13r}{16}-\frac{15}{4}, r-\frac{33}{8} \right\}, \\
& 
V_{1,2}^{\mathrm{SJP}}(r;1,3) = 
\max \left\{ \frac{r}{4}-2, \frac{13r}{16}-\frac{11}{4}, r-\frac{25}{8} \right\}.
\end{split}
\]
SJP function $V^{\mathrm{SJP}}(r;1,n_1)$ for the first stage of the sequential 
three-stage job: 
\[
\begin{split}
& 
V^{\mathrm{SJP}}(r;1,0) = 
V_{1,2}^{\mathrm{SJP}}(V_3^{\mathrm{SJP}}(r;0);1,0) \; = \\
& \quad 
\max \left\{ 
\frac{3r}{32}-\frac{13}{8}, \frac{13r}{128}-\frac{53}{32}, 
\frac{29r}{256}-\frac{215}{128}, \frac{r}{8}-\frac{433}{256}, 
\frac{3r}{16}-\frac{15}{4}, \frac{13r}{64}-\frac{61}{16}, 
\right. \\
& \quad \quad 
\left. 
\frac{29r}{128}-\frac{247}{64}, \frac{r}{4}-\frac{497}{128}, 
\frac{39r}{128}-\frac{73}{32}, \frac{169r}{512}-\frac{305}{128}, 
\frac{377r}{1024}-\frac{1259}{512}, \frac{3r}{8}-\frac{41}{16}, 
\right. \\
& \quad \quad 
\left. 
\frac{13r}{32}-\frac{2557}{1024}, \frac{29r}{64}-\frac{89}{32}, 
\frac{r}{2}-\frac{181}{64}, \frac{39r}{64}-\frac{81}{16}, 
\frac{169r}{256}-\frac{337}{64}, \frac{377r}{512}-\frac{1387}{256}, 
\right. \\
& \quad \quad 
\left. 
\frac{3r}{4}-\frac{45}{8}, \frac{13r}{16}-\frac{2813}{512}, 
\frac{29r}{32}-\frac{97}{16}, r-\frac{197}{32} 
\right\}, \\
& 
V^{\mathrm{SJP}}(r;1,1) = 
V_{1,2}^{\mathrm{SJP}}(V_3^{\mathrm{SJP}}(r;0);1,1) \; = \\
& \quad 
\max \left\{ 
-1, 
\frac{3r}{16}-\frac{17}{4}, \frac{13r}{64}-\frac{69}{16}, 
\frac{29r}{128}-\frac{279}{64}, \frac{r}{4}-\frac{561}{128}, 
\frac{39r}{64}-\frac{89}{16}, \frac{169r}{256}-\frac{369}{64}, 
\right. \\
& \quad \quad 
\left. 
\frac{377r}{512}-\frac{1515}{256}, \frac{3r}{4}-\frac{49}{8}, 
\frac{13r}{16}-\frac{3069}{512}, \frac{29r}{32}-\frac{105}{16}, 
r-\frac{213}{32} 
\right\}, 
\end{split}
\]
\[
\begin{split}
& 
V^{\mathrm{SJP}}(r;1,2) = 
V_{1,2}^{\mathrm{SJP}}(V_3^{\mathrm{SJP}}(r;0);1,2) \; = \\
& \quad 
\max \left\{ 
-1, 
\frac{3r}{16}-\frac{13}{4}, \frac{13r}{64}-\frac{53}{16}, 
\frac{29r}{128}-\frac{215}{64}, \frac{r}{4}-\frac{433}{128}, 
\frac{39r}{64}-\frac{73}{16}, \frac{169r}{256}-\frac{305}{64}, 
\right. \\
& \quad \quad 
\left. 
\frac{377r}{512}-\frac{1259}{256}, \frac{3r}{4}-\frac{41}{8}, 
\frac{13r}{16}-\frac{2557}{512}, \frac{29r}{32}-\frac{89}{16}, 
r-\frac{181}{32} 
\right\}, \\
& 
V^{\mathrm{SJP}}(r;1,3) = 
V_{1,2}^{\mathrm{SJP}}(V_3^{\mathrm{SJP}}(r;0);1,3) \; = \\
& \quad 
\max \left\{ 
\frac{3r}{16}-\frac{9}{4}, \frac{13r}{64}-\frac{37}{16}, 
\frac{29r}{128}-\frac{151}{64}, \frac{r}{4}-\frac{305}{128}, 
\frac{39r}{64}-\frac{57}{16}, \frac{169r}{256}-\frac{241}{64}, 
\right. \\
& \quad \quad 
\left. 
\frac{377r}{512}-\frac{1003}{256}, \frac{3r}{4}-\frac{33}{8}, 
\frac{13r}{16}-\frac{2045}{512}, \frac{29r}{32}-\frac{73}{16}, 
r-\frac{149}{32} 
\right\}.
\end{split}
\]
Gittins index $G(1,n_1)$ for the first stage of the sequential three-stage job: 
\[
\begin{split}
& 
G(1,0) = 
\frac{1}{\inf \{r \ge 0 : V^{\mathrm{SJP}}(r;1,0) > 0\}} = \frac{32}{181}, \\
&  
G(1,1) = 
\frac{1}{\inf \{r \ge 0 : V^{\mathrm{SJP}}(r;1,1) > 0\}} = \frac{32}{213}, \\
& 
G(1,2) = 
\frac{1}{\inf \{r \ge 0 : V^{\mathrm{SJP}}(r;1,2) > 0\}} = \frac{32}{181}, \\
& 
G(1,3) = 
\frac{1}{\inf \{r \ge 0 : V^{\mathrm{SJP}}(r;1,3) > 0\}} = \frac{32}{149}.
\end{split}
\]
As seen from above, all the Gittins index values computed in the two different methods are equal.
}
\end{example}

\begin{example}
\label{ex:Gittins-index-average-two-stage-long}
{\rm 
In our last example, which is related to  
Conjecture~\ref{con:Gittins-index-average-two-stage}, 
we consider a sequential two-stage job with the following discrete-time 
nonmonotonous hazard rate $\mu_j(n)$ in both stages $j \in \{1,2\}$: 
\[
\mu_j(0) = 0, \quad 
\mu_j(1) = 9/10, \quad 
\mu_j(2) = \ldots = \mu_j(8) = 0, \quad 
\mu_j(9) = 1.
\]
It follows that, for both stages $j \in \{1,2\}$, 
\[
P\{S_j = 2\} = 9/10, \quad P\{S_j = 10\} = 1/10, \quad 
E[S_j] = 14/5, 
\]
and the total service time $S = S_1 + S_2$ satisfies 
\[
P\{S = 4\} = 81/100, \quad 
P\{S = 12\} = 9/50, \quad 
P\{S = 20\} = 1/100, \quad 
E[S] = 28/5.
\]

\vskip 6pt 

\noindent 
{\em \ref{ex:Gittins-index-average-two-stage-long}a) 
Gittins index values computed according to 
Conjecture~\ref{con:Gittins-index-average-two-stage}}
\ 

\vskip 6pt 

\noindent 
Gittins index $G_j(n_j)$ for individual stages $j \in \{1,2\}$: 
\[
\begin{split}
& 
G_j(0) = 
\max_{n \in \{1,\ldots,10\}} 
\frac{P\{S_j \le n\}}{E[\min \{S_j, n\}]} \; = \\
& \quad 
\max \{ 0, \frac{9}{20}, \frac{9}{21}, \frac{9}{22}, \frac{9}{23}, 
\frac{9}{24}, \frac{9}{25}, \frac{9}{26}, \frac{9}{27}, \frac{5}{14} \} = 
\frac{9}{20}, \\
& 
G_j(1) = 
\max_{n \in \{1,\ldots,9\}} 
\frac{P\{S_j - 1\le n \mid S_j > 1\}}{E[\min \{S_j - 1, n\} \mid S_j > 1]} \; = \\
& \quad 
\max \{ \frac{9}{10}, \frac{9}{11}, \frac{9}{12}, \frac{9}{13}, 
\frac{9}{14}, \frac{9}{15}, \frac{9}{16}, \frac{9}{17}, \frac{5}{9} \} = 
\frac{9}{10}, \\
& 
G_j(2) = 
\max_{n \in \{1,\ldots,8\}} 
\frac{P\{S_j - 2 \le n \mid S_j > 2\}}{E[\min \{S_j - 2, n\} \mid S_j > 2]} = 
\max \{ 0, \frac{1}{8} \} = 
\frac{1}{8}, \\
& 
G_j(3) = 
\max_{n \in \{1,\ldots,7\}} 
\frac{P\{S_j - 3 \le n \mid S_j > 3\}}{E[\min \{S_j - 3, n\} \mid S_j > 3]} = 
\max \{ 0, \frac{1}{7} \} = 
\frac{1}{7}, \\
& 
G_j(4) = 
\max_{n \in \{1,\ldots,6\}} 
\frac{P\{S_j - 4 \le n \mid S_j > 4\}}{E[\min \{S_j - 4, n\} \mid S_j > 4]} = 
\max \{ 0, \frac{1}{6} \} = 
\frac{1}{6}, \\
& 
G_j(5) = 
\max_{n \in \{1,\ldots,5\}} 
\frac{P\{S_j - 5 \le n \mid S_j > 5\}}{E[\min \{S_j - 5, n\} \mid S_j > 5]} = 
\max \{ 0, \frac{1}{5} \} = 
\frac{1}{5}, \\
& 
G_j(6) = 
\max_{n \in \{1,\ldots,4\}} 
\frac{P\{S_j - 6 \le n \mid S_j > 6\}}{E[\min \{S_j - 6, n\} \mid S_j > 6]} = 
\max \{ 0, \frac{1}{4} \} = 
\frac{1}{4}, \\
& 
G_j(7) = 
\max_{n \in \{1,2,3\}} 
\frac{P\{S_j - 7 \le n \mid S_j > 7\}}{E[\min \{S_j - 7, n\} \mid S_j > 7]} = 
\max \{ 0, \frac{1}{3} \} = 
\frac{1}{3}, \\
& 
G_j(8) = 
\max_{n \in \{1,2\}} 
\frac{P\{S_j - 8 \le n \mid S_j > 8\}}{E[\min \{S_j - 8, n\} \mid S_j > 8]} = 
\max \{ 0, \frac{1}{2} \} = 
\frac{1}{2}, \\
& 
G_j(9) = 
\frac{P\{S_j - 9 \le n \mid S_j > 9\}}{E[\min \{S_j - 9, n\} \mid S_j > 9]} = 
1.
\end{split}
\]

Gittins index $G(1,n_1)$ for the first stage of the sequential two-stage job: 
\[
\begin{split}
& 
G(1,0) = 
\max_{n \in \{1,\ldots,10\}} 
\frac{P\{S_2 \le n\}}{\frac{1}{G_1(0)} + E[\min \{S_2, n\}]} \; = \\
& \quad 
\max \{ 0, \frac{81}{380}, \frac{81}{389}, \frac{81}{398}, \frac{81}{407}, 
\frac{81}{416}, \frac{81}{425}, \frac{81}{434}, \frac{81}{443}, \frac{45}{226} \} = 
\frac{81}{380}, \\
& 
G(1,1) = 
\max_{n \in \{1,\ldots,10\}} 
\frac{P\{S_2 \le n\}}{\frac{1}{G_1(1)} + E[\min \{S_2, n\}]} \; = \\
& \quad 
\max \{ 0, \frac{81}{280}, \frac{81}{289}, \frac{81}{298}, \frac{81}{307}, 
\frac{81}{316}, \frac{81}{325}, \frac{81}{334}, \frac{81}{343}, \frac{45}{176} \} = 
\frac{81}{280}, \\
& 
G(1,2) = 
\max_{n \in \{1,\ldots,10\}} 
\frac{P\{S_2 \le n\}}{\frac{1}{G_1(2)} + E[\min \{S_2, n\}]} \; = \\
& \quad 
\max \{ 0, \frac{9}{100}, \frac{9}{101}, \frac{9}{102}, \frac{9}{103}, 
\frac{9}{104}, \frac{9}{105}, \frac{9}{106}, \frac{9}{107}, \frac{5}{54} \} = 
\frac{5}{54}, \\
& 
G(1,3) = 
\max_{n \in \{1,\ldots,10\}} 
\frac{P\{S_2 \le n\}}{\frac{1}{G_1(3)} + E[\min \{S_2, n\}]} \; = \\
& \quad 
\max \{ 0, \frac{9}{90}, \frac{9}{91}, \frac{9}{92}, \frac{9}{93}, 
\frac{9}{94}, \frac{9}{95}, \frac{9}{96}, \frac{9}{97}, \frac{5}{49} \} = 
\frac{5}{49}, \\
& 
G(1,4) = 
\max_{n \in \{1,\ldots,10\}} 
\frac{P\{S_2 \le n\}}{\frac{1}{G_1(4)} + E[\min \{S_2, n\}]} \; = \\
& \quad 
\max \{ 0, \frac{9}{80}, \frac{9}{81}, \frac{9}{82}, \frac{9}{83}, 
\frac{9}{84}, \frac{9}{85}, \frac{9}{86}, \frac{9}{87}, \frac{5}{44} \} = 
\frac{5}{44}, \\
& 
G(1,5) = 
\max_{n \in \{1,\ldots,10\}} 
\frac{P\{S_2 \le n\}}{\frac{1}{G_1(5)} + E[\min \{S_2, n\}]} \; = \\
& \quad 
\max \{ 0, \frac{9}{70}, \frac{9}{71}, \frac{9}{72}, \frac{9}{73}, 
\frac{9}{74}, \frac{9}{75}, \frac{9}{76}, \frac{9}{77}, \frac{5}{39} \} = 
\frac{9}{70}, \\
& 
G(1,6) = 
\max_{n \in \{1,\ldots,10\}} 
\frac{P\{S_2 \le n\}}{\frac{1}{G_1(6)} + E[\min \{S_2, n\}]} \; = \\
& \quad 
\max \{ 0, \frac{9}{60}, \frac{9}{61}, \frac{9}{62}, \frac{9}{63}, 
\frac{9}{64}, \frac{9}{65}, \frac{9}{66}, \frac{9}{67}, \frac{5}{34} \} = 
\frac{9}{60} = \frac{3}{20}, \\
& 
G(1,7) = 
\max_{n \in \{1,\ldots,10\}} 
\frac{P\{S_2 \le n\}}{\frac{1}{G_1(7)} + E[\min \{S_2, n\}]} \; = \\
& \quad 
\max \{ 0, \frac{9}{50}, \frac{9}{51}, \frac{9}{52}, \frac{9}{53}, 
\frac{9}{54}, \frac{9}{55}, \frac{9}{56}, \frac{9}{57}, \frac{5}{29} \} = 
\frac{9}{50}, \\
& 
G(1,8) = 
\max_{n \in \{1,\ldots,10\}} 
\frac{P\{S_2 \le n\}}{\frac{1}{G_1(8)} + E[\min \{S_2, n\}]} \; = \\
& \quad 
\max \{ 0, \frac{9}{40}, \frac{9}{41}, \frac{9}{42}, \frac{9}{43}, 
\frac{9}{44}, \frac{9}{45}, \frac{9}{46}, \frac{9}{47}, \frac{5}{24} \} = 
\frac{9}{40}, \\
& 
G(1,9) = 
\max_{n \in \{1,\ldots,10\}} 
\frac{P\{S_2 \le n\}}{\frac{1}{G_1(9)} + E[\min \{S_2, n\}]} \; = \\
& \quad 
\max \{ 0, \frac{9}{30}, \frac{9}{31}, \frac{9}{32}, \frac{9}{33}, 
\frac{9}{34}, \frac{9}{35}, \frac{9}{36}, \frac{9}{37}, \frac{5}{19} \} = 
\frac{9}{30} = \frac{3}{10}.
\end{split}
\]

\vskip 6pt 

\noindent 
{\em \ref{ex:Gittins-index-average-two-stage-long}b) 
Gittins index values computed using the method presented 
in \cite{Scu18arXiv}}
\ 

\vskip 6pt 

\noindent 
SJP function $V_j^{\mathrm{SJP}}(r;n_j)$ for individual stages $j \in \{1,2\}$: 
\[
\begin{split}
& 
V_j^{\mathrm{SJP}}(r;0) = 
\max_{n \in \{1,\ldots,10\}} 
\Big( 
r P\{S_j \le n\} - E[\min \{S_j, n\}] 
\Big) \\
& \quad = \; 
\max \left\{ -1, \frac{9r}{10}-2, r-\frac{14}{5} \right\}, \\
& 
V_j^{\mathrm{SJP}}(r;1) = 
\max_{n \in \{1,\ldots,9\}} 
\Big( 
r P\{S_j - 1 \le n \mid S_j > 1\} - E[\min \{S_j - 1, n\} \mid S_j > 1] 
\Big) \\
& \quad = \; 
\max \left\{ \frac{9r}{10}-1, r-\frac{9}{5} \right\}, \\
& 
V_j^{\mathrm{SJP}}(r;2) = 
\max_{n \in \{1,\ldots,8\}} 
\Big( 
r P\{S_j - 2 \le n \mid S_j > 2\} - E[\min \{S_j - 2, n\} \mid S_j > 2] 
\Big) \\
& \quad = \; 
\max \left\{ -1, r-8 \right\}, \\
& 
V_j^{\mathrm{SJP}}(r;3) = 
\max_{n \in \{1,\ldots,7\}} 
\Big( 
r P\{S_j - 3 \le n \mid S_j > 3\} - E[\min \{S_j - 3, n\} \mid S_j > 3] 
\Big) \\
& \quad = \; 
\max \left\{ -1, r-7 \right\}, \\
& 
V_j^{\mathrm{SJP}}(r;4) = 
\max_{n \in \{1,\ldots,6\}} 
\Big( 
r P\{S_j - 4 \le n \mid S_j > 4\} - E[\min \{S_j - 4, n\} \mid S_j > 4] 
\Big) \\
& \quad = \; 
\max \left\{ -1, r-6 \right\}, \\
& 
V_j^{\mathrm{SJP}}(r;5) = 
\max_{n \in \{1,\ldots,5\}} 
\Big( 
r P\{S_j - 5 \le n \mid S_j > 5\} - E[\min \{S_j - 5, n\} \mid S_j > 5] 
\Big) \\
& \quad = \; 
\max \left\{ -1, r-5 \right\}, \\
& 
V_j^{\mathrm{SJP}}(r;6) = 
\max_{n \in \{1,\ldots,4\}} 
\Big( 
r P\{S_j - 6 \le n \mid S_j > 6\} - E[\min \{S_j - 6, n\} \mid S_j > 6] 
\Big) \\
& \quad = \; 
\max \left\{ -1, r-4 \right\}, \\
& 
V_j^{\mathrm{SJP}}(r;7) = 
\max_{n \in \{1,2,3\}} 
\Big( 
r P\{S_j - 7 \le n \mid S_j > 7\} - E[\min \{S_j - 7, n\} \mid S_j > 7] 
\Big) \\
& \quad = \; 
\max \left\{ -1, r-3 \right\}, \\
& 
V_j^{\mathrm{SJP}}(r;8) = 
\max_{n \in \{1,2\}} 
\Big( 
r P\{S_j - 8 \le n \mid S_j > 8\} - E[\min \{S_j - 8, n\} \mid S_j > 8] 
\Big) \\
& \quad = \; 
\max \left\{ -1, r-2 \right\}, \\
& 
V_j^{\mathrm{SJP}}(r;9) = 
\Big( 
r P\{S_j - 9 \le 1 \mid S_j > 9\} - E[\min \{S_j - 9, n\} \mid S_j > 9] 
\Big) = 
r-1.
\end{split}
\]
SJP function $V^{\mathrm{SJP}}(r;1,n_1)$ for the first stage of the sequential 
two-stage job: 
\[
\begin{split}
& 
V^{\mathrm{SJP}}(r;1,0) = 
V_1^{\mathrm{SJP}}(V_2^{\mathrm{SJP}}(r;0);0) = 
\max \left\{ -1, 
\frac{81r}{100}-\frac{19}{5}, \frac{9r}{10}-\frac{113}{25}, 
r-\frac{28}{5} \right\}, \\
& 
V^{\mathrm{SJP}}(r;1,1) = 
V_1^{\mathrm{SJP}}(V_2^{\mathrm{SJP}}(r;0);1) = 
\max \left\{ -\frac{19}{10}, 
\frac{81r}{100}-\frac{14}{5}, \frac{9r}{10}-\frac{88}{25}, 
r-\frac{23}{5} \right\}, \\
& 
V^{\mathrm{SJP}}(r;1,2) = 
V_1^{\mathrm{SJP}}(V_2^{\mathrm{SJP}}(r;0);2) = 
\max \left\{ -1, 
\frac{9r}{10}-10, r-\frac{54}{5} \right\}, \\
& 
V^{\mathrm{SJP}}(r;1,3) = 
V_1^{\mathrm{SJP}}(V_2^{\mathrm{SJP}}(r;0);3) = 
\max \left\{ -1, 
\frac{9r}{10}-9, r-\frac{49}{5} \right\}, \\
& 
V^{\mathrm{SJP}}(r;1,4) = 
V_1^{\mathrm{SJP}}(V_2^{\mathrm{SJP}}(r;0);4) = 
\max \left\{ -1, 
\frac{9r}{10}-8, r-\frac{44}{5} \right\}, \\
& 
V^{\mathrm{SJP}}(r;1,5) = 
V_1^{\mathrm{SJP}}(V_2^{\mathrm{SJP}}(r;0);5) = 
\max \left\{ -1, 
\frac{9r}{10}-7, r-\frac{39}{5} \right\}, \\
& 
V^{\mathrm{SJP}}(r;1,6) = 
V_1^{\mathrm{SJP}}(V_2^{\mathrm{SJP}}(r;0);6) = 
\max \left\{ -1, 
\frac{9r}{10}-6, r-\frac{34}{5} \right\}, \\
& 
V^{\mathrm{SJP}}(r;1,7) = 
V_1^{\mathrm{SJP}}(V_2^{\mathrm{SJP}}(r;0);7) = 
\max \left\{ -1, 
\frac{9r}{10}-5, r-\frac{29}{5} \right\}, \\
& 
V^{\mathrm{SJP}}(r;1,8) = 
V_1^{\mathrm{SJP}}(V_2^{\mathrm{SJP}}(r;0);8) = 
\max \left\{ -1, 
\frac{9r}{10}-4, r-\frac{24}{5} \right\}, \\
& 
V^{\mathrm{SJP}}(r;1,9) = 
V_1^{\mathrm{SJP}}(V_2^{\mathrm{SJP}}(r;0);9) = 
\max \left\{ -2, 
\frac{9r}{10}-3, r-\frac{19}{5} \right\}.
\end{split}
\]
Gittins index $G_j(n_j)$ for individual stages $j \in \{1,2\}$: 
\[
\begin{split}
& 
G_j(0) = 
\frac{1}{\inf \{r \ge 0 : V_j^{\mathrm{SJP}}(r;0) > 0\}} = \frac{9}{20}, \\
& 
G_j(1) = 
\frac{1}{\inf \{r \ge 0 : V_j^{\mathrm{SJP}}(r;1) > 0\}} = \frac{9}{10}, \\
& 
G_j(2) = 
\frac{1}{\inf \{r \ge 0 : V_j^{\mathrm{SJP}}(r;2) > 0\}} = \frac{1}{8}, \\
& 
G_j(3) = 
\frac{1}{\inf \{r \ge 0 : V_j^{\mathrm{SJP}}(r;3) > 0\}} = \frac{1}{7}, \\
& 
G_j(4) = 
\frac{1}{\inf \{r \ge 0 : V_j^{\mathrm{SJP}}(r;4) > 0\}} = \frac{1}{6}, \\
& 
G_j(5) = 
\frac{1}{\inf \{r \ge 0 : V_j^{\mathrm{SJP}}(r;5) > 0\}} = \frac{1}{5}, 
\end{split}
\]
\[
\begin{split}
& 
G_j(6) = 
\frac{1}{\inf \{r \ge 0 : V_j^{\mathrm{SJP}}(r;6) > 0\}} = \frac{1}{4}, \\
& 
G_j(7) = 
\frac{1}{\inf \{r \ge 0 : V_j^{\mathrm{SJP}}(r;7) > 0\}} = \frac{1}{3}, \\
& 
G_j(8) = 
\frac{1}{\inf \{r \ge 0 : V_j^{\mathrm{SJP}}(r;8) > 0\}} = \frac{1}{2}, \\
& 
G_j(9) = 
\frac{1}{\inf \{r \ge 0 : V_j^{\mathrm{SJP}}(r;9) > 0\}} = 1.
\end{split}
\]
Gittins index $G(1,n_1)$ for the first stage of the sequential two-stage job: 
\[
\begin{split}
& 
G(1,0) = 
\frac{1}{\inf \{r \ge 0 : V^{\mathrm{SJP}}(r;1,0) > 0\}} = \frac{81}{380}, \\
&  
G(1,1) = 
\frac{1}{\inf \{r \ge 0 : V^{\mathrm{SJP}}(r;1,1) > 0\}} = \frac{81}{280}, \\
& 
G(1,2) = 
\frac{1}{\inf \{r \ge 0 : V^{\mathrm{SJP}}(r;1,2) > 0\}} = \frac{5}{54}, \\
& 
G(1,3) = 
\frac{1}{\inf \{r \ge 0 : V^{\mathrm{SJP}}(r;1,3) > 0\}} = \frac{5}{49}, \\
& 
G(1,4) = 
\frac{1}{\inf \{r \ge 0 : V^{\mathrm{SJP}}(r;1,4) > 0\}} = \frac{5}{44}, \\
& 
G(1,5) = 
\frac{1}{\inf \{r \ge 0 : V^{\mathrm{SJP}}(r;1,5) > 0\}} = \frac{9}{70}, \\
& 
G(1,6) = 
\frac{1}{\inf \{r \ge 0 : V^{\mathrm{SJP}}(r;1,6) > 0\}} = \frac{3}{20}, \\
& 
G(1,7) = 
\frac{1}{\inf \{r \ge 0 : V^{\mathrm{SJP}}(r;1,7) > 0\}} = \frac{9}{50}, \\
& 
G(1,8) = 
\frac{1}{\inf \{r \ge 0 : V^{\mathrm{SJP}}(r;1,8) > 0\}} = \frac{9}{40}, \\
& 
G(1,9) = 
\frac{1}{\inf \{r \ge 0 : V^{\mathrm{SJP}}(r;1,9) > 0\}} = \frac{3}{10}.
\end{split}
\]
As seen from above, all the Gittins index values computed in the two different methods are equal.
}
\end{example}



\begin{thebibliography}{99}
\bibitem{Aal09QS}
S. Aalto, U. Ayesta, and R. Righter, 
On the {Gittins} index in the {M/G/1} queue, 
Queueing Systems 63, 437--458, 2009
\bibitem{Aal11PEIS}
S. Aalto, U. Ayesta, and R. Righter, 
Properties of the {Gittins} index with application to optimal scheduling, 
Probability in the Engineering and Informational Sciences 25, 269--288, 2011
\bibitem{Aal19aPEVA} 
S. Aalto and P. Lassila, 
Near-optimal dispatching policy for energy-aware server clusters, 
Performance Evaluation 135, article 102034, 2019
\bibitem{Aal16QS}
S. Aalto, P. Lassila, and P. Osti, 
Whittle index approach to size-aware scheduling for time-varying channels with multiple states, 
Queueing Systems 83, 195--225, 2016
\bibitem{Aal17PEVA} 
S. Aalto, P. Lassila, and P. Osti, 
Opportunistic scheduling with flow size information for {Markovian} time-varying channels, 
Performance Evaluation 112, 27--52, 2017
\bibitem{Aal19bPEVA} 
S. Aalto, P. Lassila, and I. Taboada, 
Whittle index approach to opportunistic scheduling with partial channel information, 
Performance Evaluation 136, article 102052, 2019
\bibitem{Ana18POMACS} 
A. Anand and G. {de Veciana}, 
A {Whittle’s} index based approach for {QoE} optimization in wireless networks, 
Proceedings of the ACM on Measurement and Analysis of Computing Systems 
2, 1, article 15, 2018
\bibitem{Arg09PEIS} 
N.T. Argon, L. Ding, K.D. Glazebrook, and S. Ziya, 
Dynamic routing of customers with general delay costs in a multiserver queuing system, 
Probability in the Engineering and Informational Sciences 23, 175--203, 2009
\bibitem{Aye10PEVA} 
U. Ayesta, M. Erasquin, and P. Jacko, 
A modeling framework for optimizing the flow-level scheduling with time-varying channels, 
Performance Evaluation 67, 1014--1029, 2010
\bibitem{Cec16PEVA} 
F. Cecchi and P. Jacko, 
Nearly-optimal scheduling of users with {Markovian} time-varying transmission rates, 
Performance Evaluation 99-100, 16--36, 2016
\bibitem{Cox61}
D.R. Cox and W.L. Smith, 
Queues, Methuen, 1961
\bibitem{Git11}
J. Gittins, K. Glazebrook, and R. Weber, 
Multi-armed Bandit Allocation Indices, Second edition, Wiley, 2011
\bibitem{Git89}
J.C. Gittins, 
Multi-armed Bandit Allocation Indices, Wiley, 1989
\bibitem{Jac11PEVA} 
P. Jacko, 
Value of information in optimal flow-level scheduling of users with {Markovian} time-varying channels, 
Performance Evaluation 68, 1022--1036, 2011
\bibitem{Lar16TNET} 
M. {Larra\~{n}aga}, U. Ayesta, and I.M. Verloop, 
Dynamic control of birth-and-death restless bandits: Application to resource-allocation problems, 
{IEEE/ACM} Transactions on Networking 24, 3812--3825, 2016
\bibitem{Nin02MP}
J. {Ni\~{n}o-Mora}, 
Dynamic allocation indices for restless projects and queueing admission control: A polyhedral approach, 
Mathematical Programming 93, 361--413, 2002
\bibitem{Rig89PEIS} 
R. Righter and J.G. Shanthikumar, 
Scheduling multiclass single server queueing systems to stochastically maximize the number of successful departures, 
Probability in the Engineering and Informational Sciences 3, 323--334, 1989
\bibitem{Rig90JAP} 
R. Righter, J.G. Shanthikumar, and G. Yamazaki, 
On extremal service disciplines in single-stage queueing systems, 
Journal of Applied Probability 27, 409--416, 1990
\bibitem{Ros70} 
S.M. Ross, 
Applied Probability Models with Optimization Applications, Holden-Day, 1970
\bibitem{Sch68} 
L.E. Schrage, 
A proof of the optimality of the shortest remaining processing time discipline, 
Operations Research 16, 687--690, 1968
\bibitem{Scu17PER} 
Z. Scully, G. Blelloch, M. Harchol-Balter, and A. Scheller-Wolf, 
Optimally scheduling jobs with multiple tasks, 
{ACM} {SIGMETRICS} Performance Evaluation Review 45, 2, 36--38, 2017
\bibitem{Scu18arXiv} 
Z. Scully, M. Harchol-Balter, and A. Scheller-Wolf, 
Optimal Scheduling and Exact Response Time Analysis for Multistage Jobs, 
{arXiv:1805.06865v2}, 2018
\bibitem{Sha95COR} 
M. Shaked, J.G. Shanthikumar, and J.B. Valdez-Torres, 
Discrete hazard rate functions, 
Computers {\&} Operations Research 22, 391--402, 1995
\bibitem{Smi78} 
D.R. Smith, 
A new proof of the optimality of the shortest remaining processing time discipline, 
Operations Research 26, 197--199, 1978
\bibitem{Tab14PEVA} 
I. Taboada, F. Liberal, and P. Jacko, 
An opportunistic and non-anticipating size-aware scheduling proposal for mean holding cost minimization in time-varying channels, 
Performance Evaluation 79, 90--103, 2014
\bibitem{Whi88JAP} 
P. Whittle, 
Restless bandits: Activity allocation in a changing world, 
Journal of Applied Probability 25A, 287--298, 1988
\bibitem{Whi02OpnsRes} 
P. Whittle, 
Applied probability in {Great} {Britain}, 
Operations Research 50, 227--239, 2002
\bibitem{Yas87QS}
S.F. Yashkov, 
Processor sharing queues: Some progress in analysis, 
Queueing Systems 2, 1--17, 1987
\end{thebibliography}
\end{document}